# Le Monde arabe
# face au défi de l’eau

*Enjeux et Conflits*

Georges MUTIN

Professeur honoraire de géographie

Institut d’Etudes Politiques de Lyon

2009

Sommaire

# Introduction

L’eau est devenue ces dernières années un sujet de préoccupation à l’échelle planétaire. Cette ressource indispensable et irremplaçable est particulièrement mal répartie. Sur la carte des disponibilités mondiales l’Afrique du Nord et le Moyen-Orient apparaissent comme la zone la plus menacée. Un constat s’impose d’emblée : 4,3% de la population mondiale ne dispose que de 0,67% des ressources en eau douce renouvelable.

Après ces dernières décennies de volontarisme, d’engouement développementaliste et technicien vient le temps des inquiétudes. Après les années 1950, dans le Monde Arabe tout l’effort a été tendu vers la mobilisation de volumes croissants, on a loué les avantages attendus de l’agriculture irriguée, de la production d’électricité, de l’extension des réseaux de distribution dans les quartiers des villes et dans les campagnes. C’est l’époque des grandes réalisations qui contribuent à la légitimation des équipes en place : le haut barrage d’Assouan en Égypte, qui, en son temps a été le plus vaste chantier du Monde, le barrage syrien de Tabqa, le slogan marocain du million d’hectares irrigués. Sous le triple choc de la sécheresse, des pollutions, de la croissance spectaculaire des besoins consécutifs à l’augmentation de la population et à la croissance urbaine, la ressource naturelle que l’on croyait disponible à jamais devient un bien économique rare. Le discours change radicalement : rareté, pénurie, pollution, affrontements sont les mots clefs d’une nouvelle problématique. Envolées les certitudes d’hier, le temps des bilans et des interrogations s’impose à tous. Peut-être faut-il se garder en ce domaine d’adopter des attitudes trop tranchées, de critiquer systématiquement ce qui, il y a peu, passait pour la voie du progrès.

L’analyse de la question hydraulique ne peut pas se résumer à des données purement techniques et économiques, à de simples analyses de volumes et de flux. L’eau raconte la société. Les facteurs sociaux et politiques sont aussi déterminants. L’utilisation de la ressource, sa destination compte autant que le simple décompte des quantités consommées. Le partage d’une ressource médiocre et irrégulièrement répartie pose de multiples problèmes de tous ordres. Concurrences et conflits, déjà anciens, ne font que s’aviver, s’exacerber à l’intérieur des espaces nationaux entre la ville, l’usine et les champs mais aussi entre les États. Les arbitrages sont de plus en plus difficiles à rendre. L’eau, son usage, son appropriation sont plus que jamais un enjeu dans le Monde Arabe.

Nous avons tenté de dresser un tableau, d'élaborer une synthèse aussi claire et précise que possible d'une situation complexe, aux multiples facettes. Les sources sont constituées par les nombreuses publications spécialisées de chercheurs français, maghrébins mais aussi anglo-saxons. Les informations sur le sujet sont fort dispersées, parfois contradictoires ou peu fiables car elles peuvent refléter les intérêts des parties en cause. Aussi avons-nous privilégié les données émanant d'organismes internationaux : ONU, Banque Mondiale, World Ressources.

L'analyse de la ressource a tout particulièrement retenu notre attention et le recours à des résultats concrets d'enquêtes de terrain privilégié dans toute la mesure du possible. Les problèmes sont toujours exposés dans un cadre physique bien précis : le bassin fluvial est le cadre naturel étudié.

Les divers aspects du problème de l'eau sont par nature étroitement imbriqués. Nous avons pourtant tenté de les traiter séparément. C'est ainsi que le premier chapitre est consacré à une présentation fouillée des conditions naturelles, essentiellement climatiques, et à une approche globale qui conduit à la détermination de la pénurie qui caractérise le Monde Arabe.

Deux chapitres mettent l'accent sur les résultats d'une mise en valeur fondée sur l'eau et les conflits interétatiques qui menacent en raison de la dépendance dans laquelle se trouvent placés les pays arabes : c'est le cas du bassin du Nil et de celui du Tigre et de l'Euphrate.

Un quatrième chapitre est consacré aux redoutables problèmes posés par un partage inégal de la ressource au Proche-Orient (Israël, Territoires autonomes palestiniens, Jordanie, Syrie).

Enfin le cadre maghrébin nous a permis d'analyser toute l'acuité du partage de la ressource  entre les différents utilisateurs : la ville, l'usine et les champs.

# I. Une ressource rare...trop souvent gaspillée

Les populations d'Afrique du Nord et du Moyen-Orient, en forte croissance démographique, doivent partager des ressources en eau qui sont médiocres et très irrégulièrement réparties. Les données climatiques constituent une contrainte de première grandeur pour la mise en valeur auxquelles s'ajoutent trop souvent les contraintes hypsométriques ou pédologiques. L'analyse de la ressource fait apparaître de sérieuses difficultés pour sa mobilisation. Avec l'accroissement démographique que connaît le Monde Arabe, la rareté est désormais bien installée.

# 1. De sévères contraintes naturelles

## 1.1 Les traits d'ensemble de la circulation atmosphérique

La circulation atmosphérique qui caractérise le Maghreb et le Machreq est conditionnée par deux éléments :

- la position en latitude : la région est comprise entre le 36e parallèle au nord à la frontière syro-turque, et le 12e au sud sur le littoral méridional de la péninsule Arabique.
- la présence de la Méditerranée, vaste espace marin qui pénètre très profondément à l'est dans la masse continentale eurafricaine et laisse pénétrer les dépressions d'Ouest.

Cet ensemble de 14 millions de km2 n'est pas soumis à un seul régime climatique c'est un **espace de transition** entre deux zones :

- la zone tropicale et subtropicale qui se caractérise par la présence constante ou quasi constante de hautes pressions dynamiques très stables.
- la zone méditerranéenne qui se rattache au domaine tempéré et se caractérise par une circulation ouest est de dépressions cycloniques.
- Le front polaire limite les deux domaines tropical et tempéré et se déplace au cours de l'année en phase avec les oscillations, en très haute altitude, du «jet stream». Il remonte en latitude en été, il descend en hiver jusqu'au nord de l'Afrique permettant ainsi le passage des dépressions cycloniques jusqu'en Méditerranée orientale.

Une région échappe à ce schéma général : le Sud de la péninsule Arabique et notamment le Yémen qui reçoit en été des pluies de mousson.

## 1.2 Faiblesse des précipitations

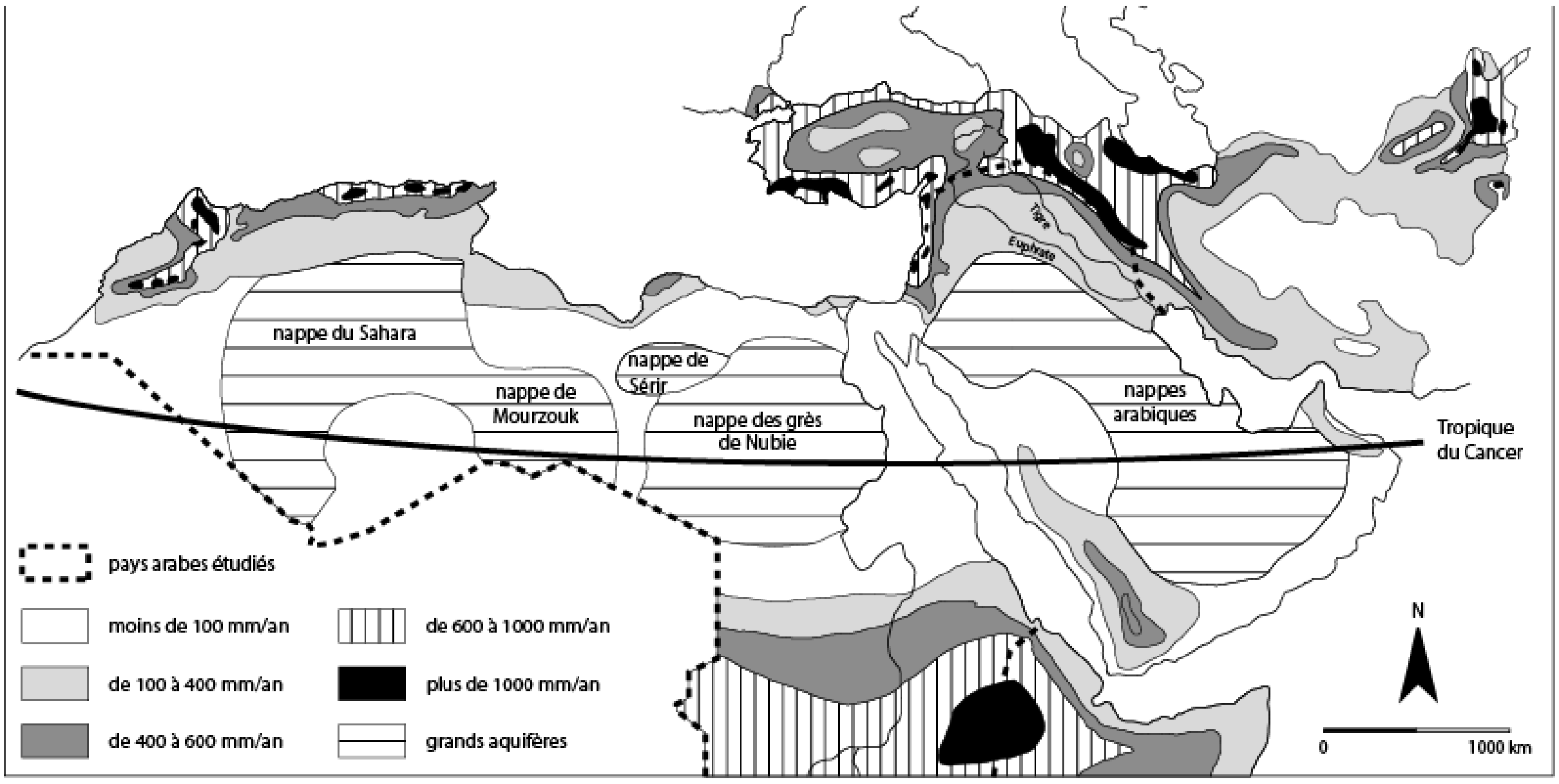

*Figure 1 - Précipitations annuelles et grands aquifères dans le Monde Arabe*

Les précipitations sont de **type méditerranéen**. Elles le sont dans leur rythme : elles se placent essentiellement en saison froide ou fraîche : en automne et en hiver. Leur répartition dans l'espace est très variable. Elle est commandée par trois facteurs : la position en latitude mais aussi la continentalité et le relief. Les pluies sont les plus abondantes sur le littoral, elles déclinent très rapidement dès que l'on s'enfonce dans les terres soit vers le sud en Afrique du Nord, soit vers l'est au Moyen-Orient. Dans cette évolution zonale, l'altitude introduit des différences sensibles renforçant ou contrariant les effets de la continentalité selon l'orientation des reliefs.

Quel bilan peut-on dresser?

**Les pluies sont insuffisantes** *(carte 1)*.

Elles ne dépassent que très localement un total annuel de 600 mm, lui-même bien moyen. Ces zones privilégiées se rencontrent essentiellement dans les régions littorales où les effets de l'altitude sont particulièrement bien observés et peuvent porter le total pluviométrique annuel à plus de 1 000 mm. Au Maghreb, les pluies dépassent 600 mm dans les montagnes du Rif (plus de 1 000 mm dans les parties les plus hautes), le Moyen Atlas et le Haut Atlas, dans l'Atlas tellien à l'est d'Alger où le total peut atteindre 1352 mm à Yakouren en Kabylie et 1773 dans l'Edough. On retrouve ces régions privilégiées au Liban et en Syrie (Mont Liban et Anti Liban, Jebel Ansarié). Ces zones humides ne représentent pourtant qu'une part infime de l'espace : moins de 7%. En Algérie, seul le quart des terres cultivables reçoit plus de 600 mm par an soit moins de 1% du territoire national! La zone comprise entre 400 et 600 mm n'est guère plus étendue que la précédente sauf peut être au Maghreb et dans la partie méridionale du Soudan où les précipitations relèvent du régime tropical.

Les territoires où les moyennes annuelles sont comprises entre 400 et 100 mm se rencontrent à l'intérieur du Croissant fertile, sur le littoral libyen et dans les hautes plaines maghrébines. Enfin, il existe d'immenses zones désertiques où il tombe moins de 100 mm de pluies : le Sahara, en quasi totalité compris dans les pays arabes, le désert arabique. Le désert est toujours très proche du littoral à 400 km des côtes algériennes et parfois comme en Libye ou en Égypte il atteint le bord de mer. A l'exception du Liban, tous les États comptent à l'intérieur de leurs frontières un pourcentage notable de terres désertiques. Les espaces désertiques s'étendent sur 7 000 000 de km2 soit la moitié de la superficie totale des pays arabes. Koweït City ne reçoit que 111 mm de pluies, Riyad 82, Jedda 25, Aden 39, Le Caire 22, Touggourt en Algérie 60!

## 1.3 Des précipitations irrégulières

L'irrégularité des précipitations présente un double aspect.

Elle est **intraannuelle** (à l'intérieur de l'année agricole). Il y a d'une façon générale opposition entre une saison sèche et chaude et une saison humide plus fraîche. Le maximum pluviométrique intervient en hiver, en décembre et janvier, au Maghreb et sur le littoral du Levant. Au Moyen-Orient, à l'intérieur des terres, s'affirment progressivement deux maxima : à Mossoul comme à Bagdad, le maximum en mars avril est aussi important que celui de décembre janvier dont le sépare un creux net en février. Ainsi les pluies tombent-elles quand la végétation est ralentie autrement dit quand on en a le moins besoin. Il y a donc nécessité de **stocker l'eau d'hiver pour l'utiliser l'été**. L'été, toujours sec, dure de 3 à 5 mois mais dépasse 6 mois dès que l'on s'éloigne du littoral : 5 mois à Beyrouth, mais 6 à Mossoul, 8 à Damas, 10 à Bagdad.

**L'irrégularité est aussi interannuelle**. D'une année sur l'autre le total pluviométrique peut varier dans de très fortes proportions du simple au double et parfois de 1 à 9. L'irrégularité est d'autant plus forte que le total annuel moyen est faible. Il y a donc nécessité de **stocker l'eau d'une année sur l'autre**.

| Station | Pays | Moyenne | Maximum | Minimum | variation |
|---|---|---|---|---|---|
| Kénitra | Maroc | 595 | 822 | 330 | 2,5 |
| Jérusalem | Israël | 529 | 1134 | 273 | 3,5 |
| Tunis | Tunisie | 415 | 820 | 221 | 3,7 |
| Amman | Jordanie | 273 | 476 | 128 | 3,7 |
| Alexandrie | Égypte | 169 | 313 | 33 | 9,4 |
| Bagdad | Irak | 151 | 336 | 72 | 4,6 |
| Touggourt | Algérie | 60 | 126 | 14 | 9 |
| Le Caire | Égypte | 22 | 63 | 1,5 | 42 |

*Tableau 1 : L'irrégularité interannuelle des précipitations en mm*

Cette irrégularité interannuelle des pluies est tout à fait spécifique à la région. Le volume d'eau fourni par les précipitations sur lequel on peut compter d'une année sur l'autre n'est que de 10% de la moyenne. A titre comparatif, en Amérique du Nord, il est de 60 à 80% à l'est et de 30% à l'ouest!

Cette double irrégularité de la pluviométrie présente de très graves conséquences. L'idée d'un climat capricieux s'impose de même que celle de l'inconsistance du calendrier agricole. Cette variabilité extrême s'observe plus particulièrement dans les espaces compris entre les isohyètes de 100 et 400 mm. Les isohyètes de 300 ou 350 mm au Maghreb, 200 à 250 mm en Orient limitent deux zones. Au dessus de ce total pluviométrique il est possible d'entreprendre avec quelque chance de succès la culture des céréales en sec, au dessous, le recours à l'irrigation est indispensable.

On observe des **cycles d'années sèches** dont on cherche d'ailleurs à établir la périodicité. Deux années sèches successives voire 3 sont situations fréquentes. Au cours des deux dernières décennies, la Tunisie a dû affronter une longue sécheresse de 1987 à 1989; le Maroc a connu cinq années de sécheresse consécutives de 1979 à 1984 et à nouveau une longue période de déficit pluviométrique de 1992 à 1994. En 1980 et 1981 le déficit pluviométrique marocain a été de 40%. Le déficit s'est maintenu en 1982 et en 1983. C'est la plus grande sécheresse enregistrée par le pays depuis le début du siècle. Les conséquences ont été catastrophiques :

- La récolte des céréales tombe à 20 millions de qx contre 35 en année normale, le rendement moyen à 5qx/ha. Le déficit céréalier par rapport à la consommation est de 30 millions de qx, ce qui entraîne un doublement des importations par rapport à une année normale. Il en coûte chaque année 2 milliards de dirhams soit le 1/4 des importations pétrolières. Les exportations d'agrumes sont menacées ce qui contribue encore à aggraver le déficit de la balance commerciale. L'endettement du pays pour financer toutes ces importations a beaucoup augmenté.

- Le cheptel a diminué de 40%. Les petits éleveurs incapables d'acheter des aliments pour bétail dont le prix a doublé sont contraints de se défaire d'une partie de leur troupeau. Les petits fellahs, ruinés, vendent parfois leurs terres : la concentration foncière s'accentue et l'exode rural connaît une forte poussée.
- Partout le niveau des lacs diminue. Les retenues des barrages si nombreux au Maroc, connaissent des étiages-record. Le coefficient de remplissage de la plupart d'entre eux se situe autour de 20%. Seule la retenue du Loukkos enregistre encore en 1982 un niveau acceptable de 65%. La production électrique est compromise et l'approvisionnement des grandes villes comme Casablanca pose problème. Les usines de Tanger ont du arrêter toute activité durant l'été 1983 par manque d'eau. En ville, les prix des fruits et légumes ont considérablement augmenté.
- Dix ans plus tard, de 1992 à 1994, la nouvelle période de sécheresse a des effets identiques.

Grâce aux méthodes de la dendrochronologie, l'étude des périodes de sécheresse au cours du dernier millénaire a pu être conduite par Charles W. Stockton. Les travaux consistent à analyser les anneaux des arbres pour reconstituer les variations du climat marocain. Les résultats mettent en évidence la fréquence des sécheresses au cours de la période 1000/1984.

| durée des périodes de sécheresse | Nombre d'occurrences | Fréquence moyenne |
|---|---|---|
| 1 an | 89 | 11 |
| 2 ans | 35 | 28,5 |
| 3 ans | 9 | 113,7 |
| 4 ans | 6 | 182 |
| 5 ans | 4 | 303 |
| 6 ans | 3 | 455 |

Ainsi la longue sécheresse récente du Maroc (1979/84) se retrouve dans le passé en 1069/74 et 1626/32. Les phénomènes de sécheresse sont un fait structurel. Leur fréquence est à prendre en compte pour la mobilisation des eaux qu'elle rend beaucoup plus difficile. C'est un handicap de taille. Les données statistiques dont on dispose, toujours fondées sur des moyennes, sont toujours à manier avec prudence en raison de la très grande variabilité des précipitations.

# 1.4 Les pluies présentent un caractère excessif

Elles sont très violentes et tombent dans un petit nombre de jours dans l'année. Dans les régions les plus favorisées, on peut compter de 80 à 100 jours de pluies/an mais dans les régions steppiques qui occupent d'immenses superficies le nombre de jours de pluie se situe entre 20 et 50!

**Un climat agressif**

Plusieurs dizaines de millimètres peuvent s'abattre en trombes d'eau en quelques minutes :

- à Oujda (Maroc oriental), 50 mm en 24 heures alors que la moyenne annuelle est de 350 mm.
- dans le Rif ou en Kabylie, il peut tomber de 1 000 à 2 000 mm en quelques semaines, en en une de périodes de 3 à 4 jours.
- à Rabat : il tombe 150 mm en 9 heures en 1959.
- à l'Arba Nath Irathen (Kabylie), il tombe 83 mm en une demi-heure le 13 septembre 1931.
- à Blida (dans la région algéroise) 156 mm en un jour et 228 mm le 17 décembre 1957.
- à Aïn Oussera (Hautes Plaines algériennes) : 60 mm en un jour alors que la moyenne annue a est de 250 mm.

L'abondance des précipitations peut revêtir un caractère catastrophique. En Tunisie steppique, l'automne 1 jalonné par trois crises orageuses réparties sur un mois, de fin septembre à fin fin fin octobre. Les abats ont été de l'ordre de 800 à 900 mm : 300 mm sont tombés en seul seul jour! Les dégâts sont considérables et routes emportés, maisons détruites, champs
et pâturages ravagés. Ils sont à peine réparés qu'en mars 1973 de nouveaux orages tout aussi violents dévastent la région.

L'idée d'un climat «agressif» s'impose. De telles trombes d'eau n'apportent que malheurs : elles ruissellent, ne pénètrent pas dans le sol et accélèrent l'érosion sans alimenter les nappes phréatiques. Leur action est d'autant plus dévastatrice que le couvert végétal est réduit, la lithologie fragile.

Au total, sur les terres arabes situées aux marges de l'aridité ou dans le domaine franchement aride, la recherche et la maîtrise de l'eau constituent une préoccupation majeure. L'irrigation doit pallier l'insuffisance des précipitations. Il faut aussi remédier à un régime des pluies peu favorable à l'activité agricole même lorsque le total annuel est satisfaisant. Elles tombent en saison froide quand la végétation en a le moins besoin : il y a distorsion entre saison humide et saison végétative. Enfin l'apport d'eau est indispensable pour atténuer les très fortes irrégularités interannuelles de la pluviométrie.

## 1.5 Les températures et l'aridité

Les températures sont souvent très élevées avec une très forte amplitude qui augmente au fur et à mesure que l'on s'éloigne du littoral. Dans les régions en bordure de mer la température moyenne de janvier est de 10°, celle de l'été autour de 25° (amplitude de 15), mais cette faible amplitude est essentiellement due à la relative douceur hivernale plus qu'au caractère modéré des températures d'été. Dès que l'on pénètre vers l'intérieur l'amplitude augmente liée à la chaleur des étés torrides et surtout à l'apparition du gel hivernal. Arbre typiquement méditerranéen, l'olivier est arrêté vers l'intérieur par l'aridité et surtout le froid. On le trouve jusqu'à 1 200 m dans les montagnes du Liban Sud, il joue encore un rôle important dans la Ghouta de Damas et l'oasis de Palmyre; il disparaît plus à l'est. Le palmier dattier redoute le froid hivernal : il est absent de la Ghouta de Damas et on ne le trouve au Moyen-Orient que dans les points bas ou le sud de la Mésopotamie. Ainsi sur les hautes plaines algériennes l'amplitude est de 25° et la neige tombe toutes les années. Dans les déserts la température moyenne annuelle dépasse 20°, elle atteint 25° à Assouan et l'amplitude est du même ordre. Autre trait caractéristique des déserts  : une importante amplitude diurne : le sol surchauffé le jour rayonne intensément la nuit.

En dehors de ces valeurs moyennes, les maxima de température sont partout impressionnants. A l'intérieur des terres, ils peuvent s'élever, selon les stations, entre 40 et 50°  : 44° à Damas, 49°8 à Mossoul, 50°2 à Bagdad! De même les minima peuvent être rigoureux  : -7°8 à Bagdad, -4°4 à Bassora, -8° dans le Sahara algérien. L'insolation annuelle est par ailleurs très forte : elle est de 2952 heures à Beyrouth au bord de la mer, 3244 heures à Bagdad, elle atteint 3500 heures dans les zones désertiques et même 3863 heures à Assouan!

Dans ces conditions l'évapotranspiration est très forte. Les plus fortes valeurs d'évapotranspiration potentielle (plus de 1140 mm/an selon le calcul de Thornthwaite) s'enregistrent dans les zones désertiques. Des valeurs moyennes (entre 570 et 1140 mm) sont relevées uniquement dans les régions proches du littoral méditerranéen. Partout ailleurs l'évaporation potentielle annuelle dépasse le total pluviométrique et on enregistre un déficit hydrique.

Ainsi, en raison de la forte chaleur qui entraîne une forte évaporation et de la faiblesse et de l'irrégularité des précipitations, l'aridité est une menace constante. Permanente ou saisonnière, elle conditionne la mise en valeur et les grands types d'aménagement de l'espace. Rechercher et maîtriser les eaux sont un impératif pour les populations, qui doivent s'adapter aux vastes étendues désertiques ou insuffisamment pourvues en eau.

Le cycle de l'eau au Maroc nous permet d'illustrer ce point de vue. Les précipitations atmosphériques annuelles sont évaluées à 150 milliards de m3, l'évapotranspiration à 120. Il reste donc 30 milliards de m3 de pluies utiles (le 1/5 des précipitations) qui, pour les 2/3 ruissellent et pour 1/3 s'infiltrent. En fin de compte le potentiel hydraulique mobilisable à partir des cours d'eau et des nappes n'est que de 21 milliards de m3! En Algérie sur un total de précipitations de 65 milliards de m3, 85% s'évapore, il ne reste que 12 milliards de m3 de pluie utile. En Tunisie sur 33 milliards de m3 de précipitations, il reste en fin de course 4,35 milliards de pluie utile.

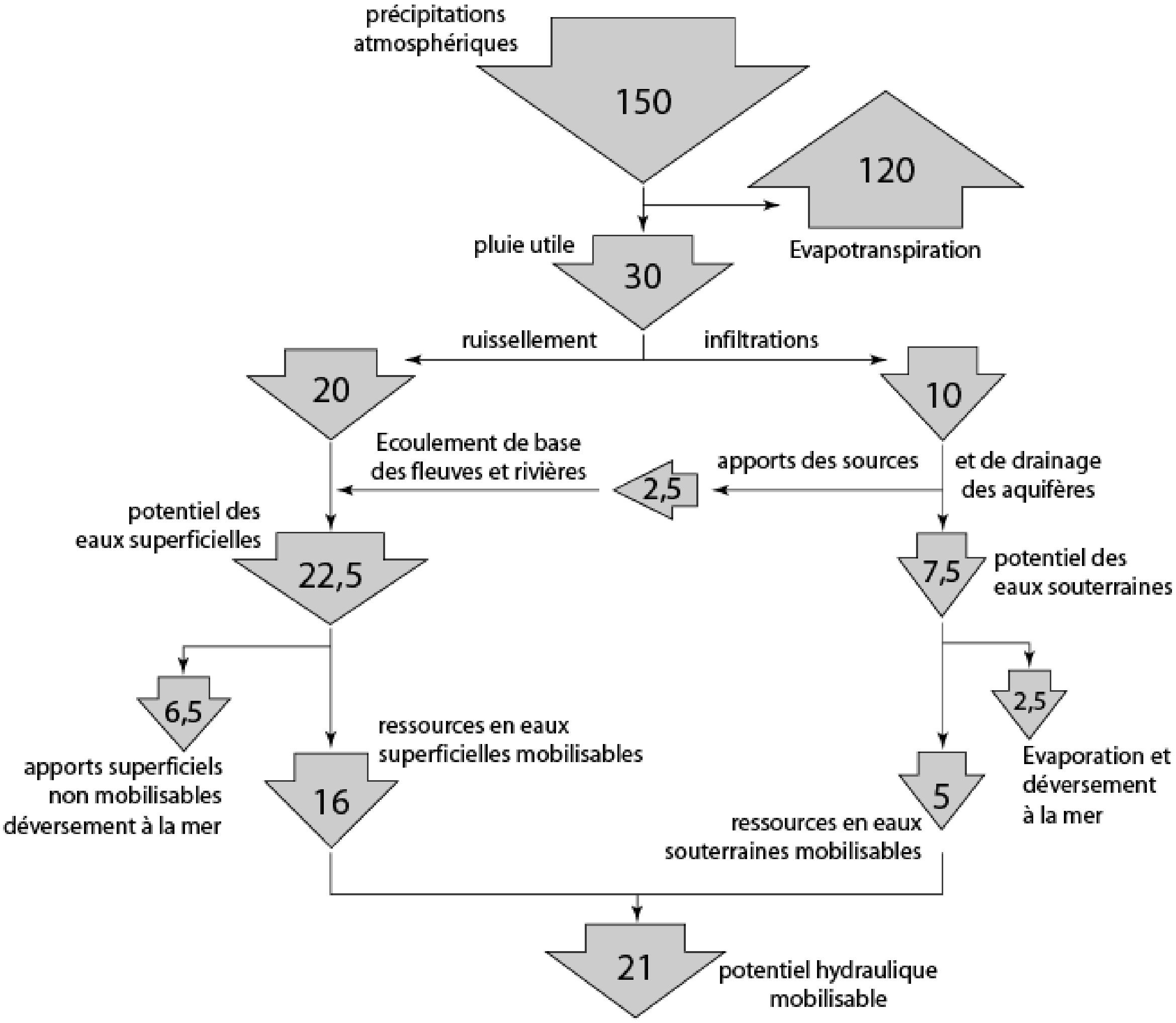

*Figure 2 – Le cycle de l'eau au Maroc (d'après Perennes, 1993)*

## 1.6 Des écoulements difficilement maîtrisables

Les conditions climatiques influent directement sur les écoulements. Il existe dans le Monde Arabe d'immenses territoires où aucun écoulement n'est organisé. C'est le cas des déserts. Dans les régions steppiques, l'endoréisme domine : les oueds intermittents se jettent dans des dépressions fermées. La faiblesse des précipitations se conjugue avec l'importance de l'évaporation pour rendre compte de cette situation. Ce n'est que dans la faible frange humide que se forment des réseaux hydrographiques mais le relief, morcelé, fractionné, montagnard n'autorise pas la constitution de réseaux hydrographiques de grande ampleur et bien structurés. Les montagnes mieux arrosées assurent des écoulements permanents mais les régimes des cours d'eau sont très irréguliers : de très fortes crues, extrêmement brutales, soudaines, s'opposent à des étiages très creusés en été. Les oueds écoulent environ les 3/4 de leur débit annuel au cours de 2 à 3 mois d'hiver. En outre, les rivières transportent une masse considérable de débris solides. Ces eaux permanentes sont, on peut le deviner, très difficiles à mobiliser en raison même des modalités de leur écoulement. Une partie importante du potentiel, dans ces conditions, va directement à la mer.

**L'irrégularité des écoulements au Maghreb**

L'oued Sebou qui écoule à peu près 40% des ressources en eau courante du Maroc a un débit annuel moyen de 137 m3/s, son étiage moyen est de 17 m3/s mais ses crues peuvent fréquemment atteindre 2 à 4 000 m3/s et on a enregistré des débits extrêmes de 10 000 m3/s.! Un de ses affluents, l'Ouergha, nous fournit un excellent exemple de la soudaineté des crues et en même temps de la rapidité de la décrue. A la fin du mois de décembre 1950, le débit est passé par les valeurs extrêmes suivantes :

- 28/12 12 h 65 m3/s
- 28/12 24 h 970 m3/s
- 29/12 6 h 1350 m3/s
- 29/12 18 h 4850 m3/s
- 29/12 20 h 6000 m3/s
- 30/12 6 h 3350 m3/s
- 30/12 24 h 1000 m3/s
- 31/12 18 h 500 m3/s

Toujours au Maroc, la Moulouya qui draine tout l'Oriental aux conditions climatiques plus sévères présente encore plus d'irrégularité avec les données suivantes :

- débit annuel moyen 34 m3/s
- étiages moyens 5 m3/s
- étiages extrêmes 1 m3/s
- crues moyennes 5 000 m3/s
- crue exceptionnelle 8000 m3/s

A l'image du régime des précipitations, les écoulements présentent de notables différences d'une année sur l'autre. Ainsi le Chélif, le plus important des cours d'eau algérien, a roulé 60 millions de m3 en 1926 et seulement 1,3 en 1927!

Par contre, de **substantielles ressources en eau courante proviennent de fleuves allogènes** qui prennent naissance en dehors de la zone arabe. Les abondantes pluies méditerranéennes de l'Anatolie orientale alimentent le Tigre et l'Euphrate et permettent la mise en valeur du désert syrien et surtout du désert mésopotamien. De même ce sont les pluies équatoriales d'Afrique orientale et les abondantes précipitations de l'Éthiopie tropicale qui rendent compte du régime du Nil et autorisent la mise en valeur du désert égyptien.

# 2. Les techniques de mobilisation des eaux

La mobilisation de l'eau est désormais un problème d'une extraordinaire acuité : les disponibilités ne sont pas très abondantes, les concurrences pour son utilisation sont très vives. L'eau dans cette vaste zone du monde devient enjeu du développement pour lequel sont engagées des sommes importantes mises au service de techniques nouvelles. Les systèmes d'exploitation diffèrent profondément selon qu'il s'agît des eaux de surface ou des nappes phréatiques.

## 2.1 L'utilisation des eaux souterraines

L'utilisation des eaux souterraines est ancienne : nappes phréatiques et grands aquifères viennent compléter les ressources fournies par les eaux de surface. Elles sont d'importance fort diverse, situées à des profondeurs très variables.

Les nappes d'**inféro-flux** sont exploitées dans les déserts ou pendant la saison sèche dans les régions plus arrosées. Les **galeries drainantes** qui permettent d'amener par gravité l'eau des nappes souterraines à la surface sont la création d'une collectivité. Cette technique de mobilisation de l'eau est fort ancienne; elle exige des travaux considérables effectués dans les zones de piémont. Les nappes sont captées par des puits en série réunis par une galerie souterraine construite en pente plus faible que la surface topographique qui débouche à quelques kilomètres de la montagne, là où les alluvions plus fines sont plus aisées à travailler. On les rencontre encore en Syrie (*kanawat*) au Sahara algérien notamment dans le Gourara et le Touat (ce sont les *foggara*) ou bien encore dans le Haouz de Marrakech et le Sahara marocain (*khettara*). En surface la présence de ces galeries drainantes se repère par l'alignement des puits d'évacuation des déblais lors du creusement. Toutefois, ces techniques traditionnelles ne permettent pas de mobiliser des volumes importants et de très nombreuses galeries drainantes sont désormais abandonnées.

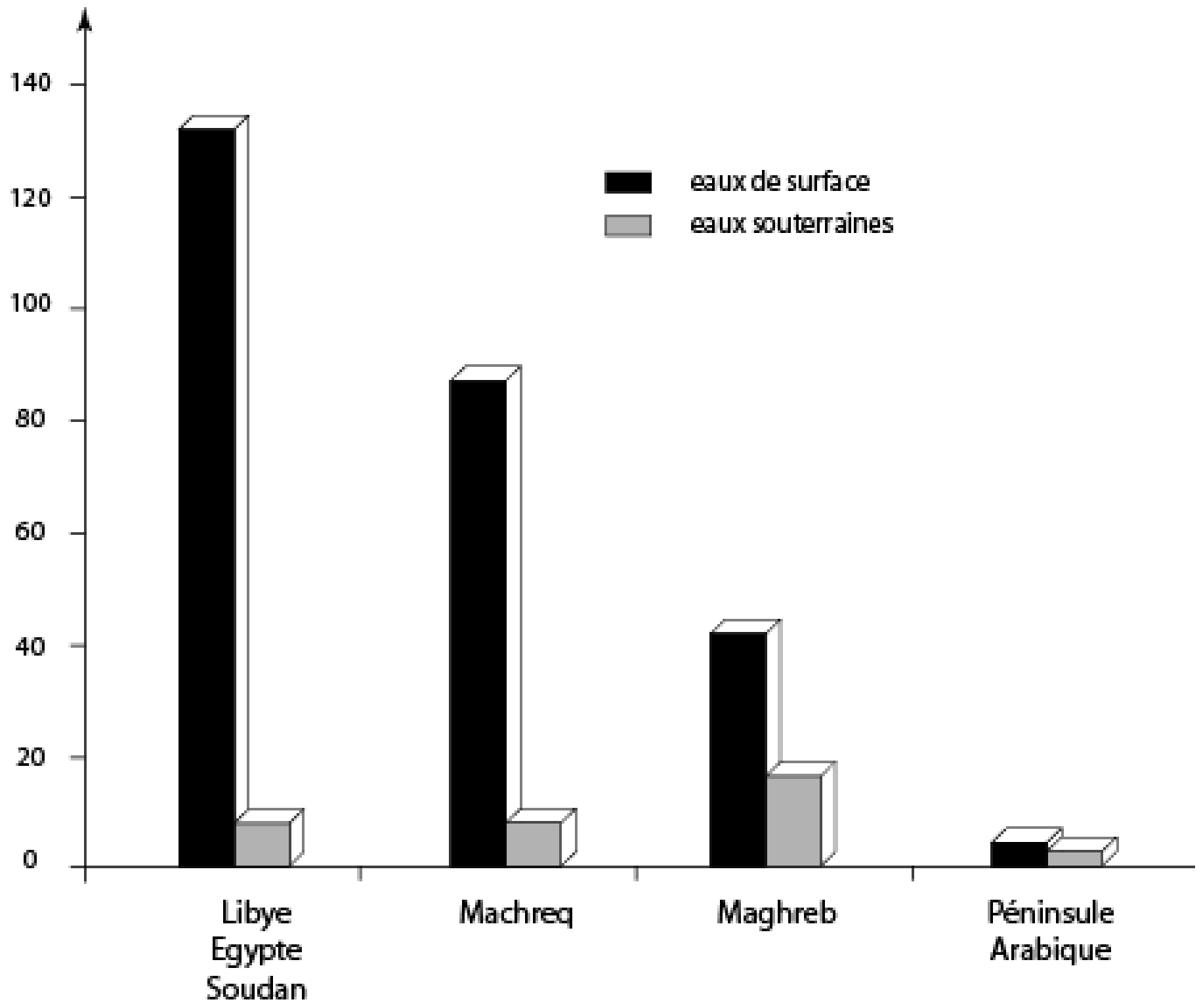

*Figure 3 : Ressources en eau douce : nappes et eaux de surface (en milliards de m3)*

Les **nappes phréatiques** se localisent à des profondeurs variables, de quelques mètres à plusieurs dizaines (une cinquantaine), dans les vallées, les plaines et les dépressions. La structure géologique compartimentée du bâti du Maghreb septentrional et des régions littorales du Levant a favorisé l'existence d'un très grand nombre de ces nappes dans ces zones relativement humides : chaque bassin renferme un aquifère de petite ou moyenne taille. Ces nappes phréatiques sont alimentées par les pluies hivernales et éventuellement par des infiltrations à partir des cours d'eau. Le plus souvent, on accède à ces ressources par des moyens individuels. La variété des appareils élévatoires est surprenante : puits à balancier, puits à poulie, puits à godets fixés à une chaîne sans fin animée par un tambour mû par un animal. Tous ces engins élévatoires traditionnels sont aujourd'hui supplantés par les motopompes.

Les **nappes profondes** sont situées à plus de 50 mètres et parfois à 300 ou 400 mètres de profondeur; elles sont exploitées par des forages. On les rencontre très souvent, notamment au Maghreb, dans les couches calcaires ou gréseuses du tertiaire. Dans le Bas Sahara (Tunisie et Algérie), plusieurs nappes communiquant entre elles forment le Complexe terminal (Continental terminal) qui s'étend sur plus de 350 000 km2 à des profondeurs variant entre 100 et 400 mètres. L'eau est froide, assez faiblement minéralisée. La nappe est alimentée par des infiltrations d'eau de ruissellement mais les flux sortants sont plus importants (évaporation dans les chotts, forages) si bien qu'en 1970, on estimait qu'il y avait un déstockage du Continental terminal de l'ordre de 100 millions de m3/an.

Au cours des dernières années, en liaison avec les prospections pétrolières, on a découvert dans le sous-sol désertique d'**immenses aquifères** dans le matériel sédimentaire qui recouvre le vieux socle sur d'énormes épaisseurs. Les techniques récentes de forages profonds, parfois à plusieurs centaines de mètres, ont modifié les conditions de mobilisation de l'eau. L'avenir repose en partie sur l'existence de trois immenses aquifères : la **nappe albienne dans le Sahara** algérien qui se prolonge à l'ouest de la Libye, **l'aquifère des grès de Nubie** en Libye orientale et en Égypte, les **aquifères de la péninsule Arabique**.

Dans le Bas Sahara, on exploite depuis 1948 la nappe albienne (dite du Continental intercalaire) profonde, selon les secteurs de 800 à 1 500 mètres *(carte 4)*. C'est en fait un système aquifère contenu dans les grès et sables du secondaire qui affleurent dans la région du Tademaït et s'enfonce plus au nord, constituant une sorte d'entonnoir dont le point le plus profond se localise dans la région de Biskra. Les eaux que l'on utilise aujourd'hui correspondraient à des pluies tombées au cours des périodes pluviales du Quaternaire (Holocène). L'incertitude demeure quant à son renouvellement actuel mais la taille du réservoir est gigantesque (60 000 milliards de m3) et autorise une exploitation sur un long terme sans risque de rabattement notable. Le réservoir s'étire sur 600 000 km2 entre la bordure sud de l'Atlas au nord et les escarpements qui, au Sahara central, soulignent la limite des grès du Continental intercalaire : Tinrhert et Tademaït. Cet aquifère est bien connu : plus de 600 forages ont été pratiqués. D'après une estimation, publiée par l'UNESCO en 1972, il pourrait fournir un débit de 1 000 m3/s pendant deux mille ans! Mais ces chiffres sont illusoires dans la mesure où l'on ne vide pas une nappe aquifère comme un lac et, dès que l'eau n'est plus sous pression, le coût du pompage devient prohibitif. En général, on peut espérer exploiter tout au plus 1/10 000 de ces réserves théoriques. Les eaux de la nappe albienne sont chaudes (55 à 66°C), fortement minéralisées (jusqu'à 7 g/litre). L'aquifère est très faiblement alimenté par des infiltrations d'eau de ruissellement en périphérie (270 millions de m3/an). En 1970, les flux sortants étaient estimées à 350 millions de m3/an (ils ont beaucoup augmenté depuis). Il y a donc un déstockage de quelque 80 millions de m3/an. Plus à l'est, en Libye, le bassin de Mourzouk s'étend sur une superficie équivalente avec un réservoir saturé de plus de 1 000 mètres de grès dans sa partie centrale.

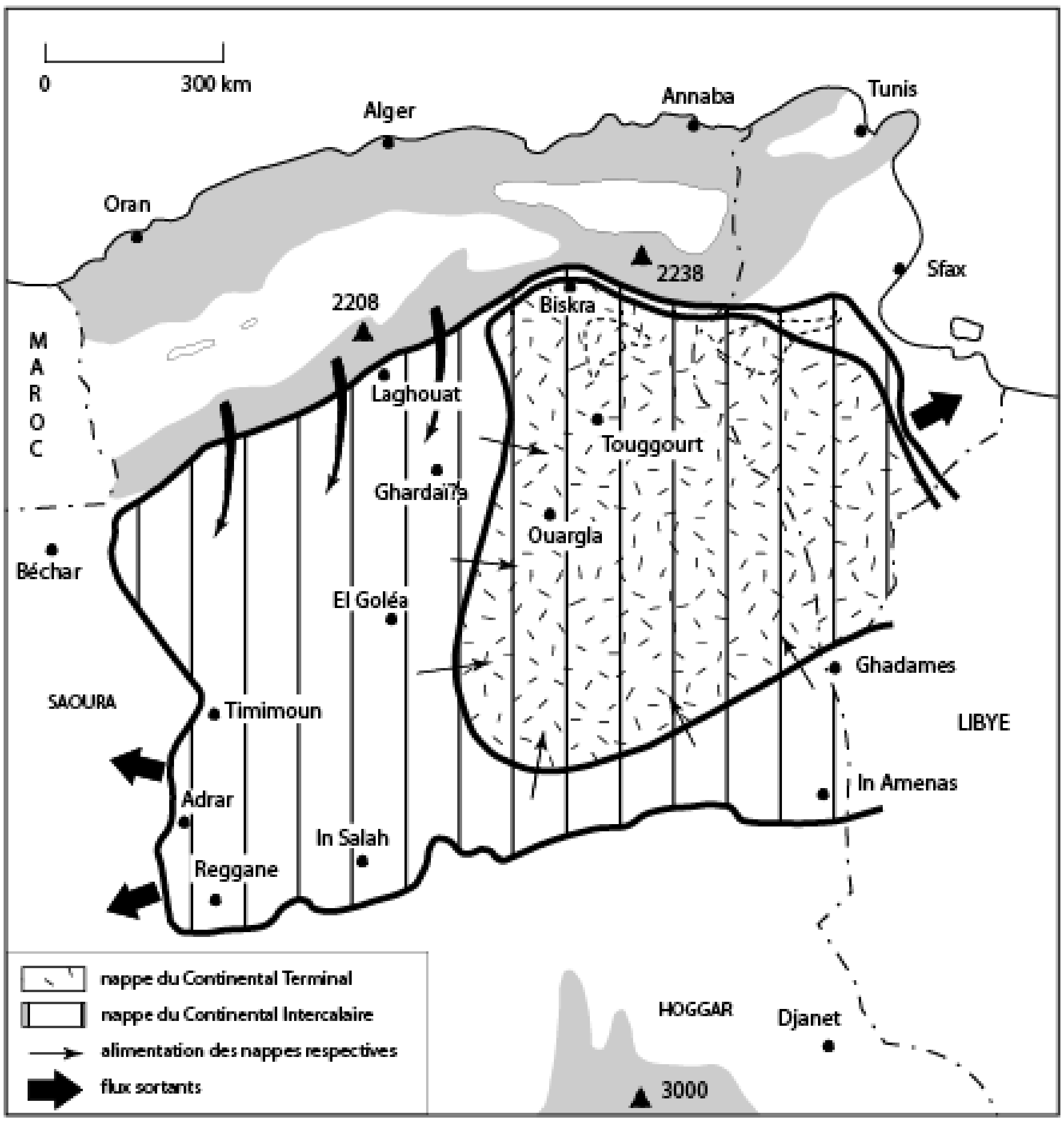

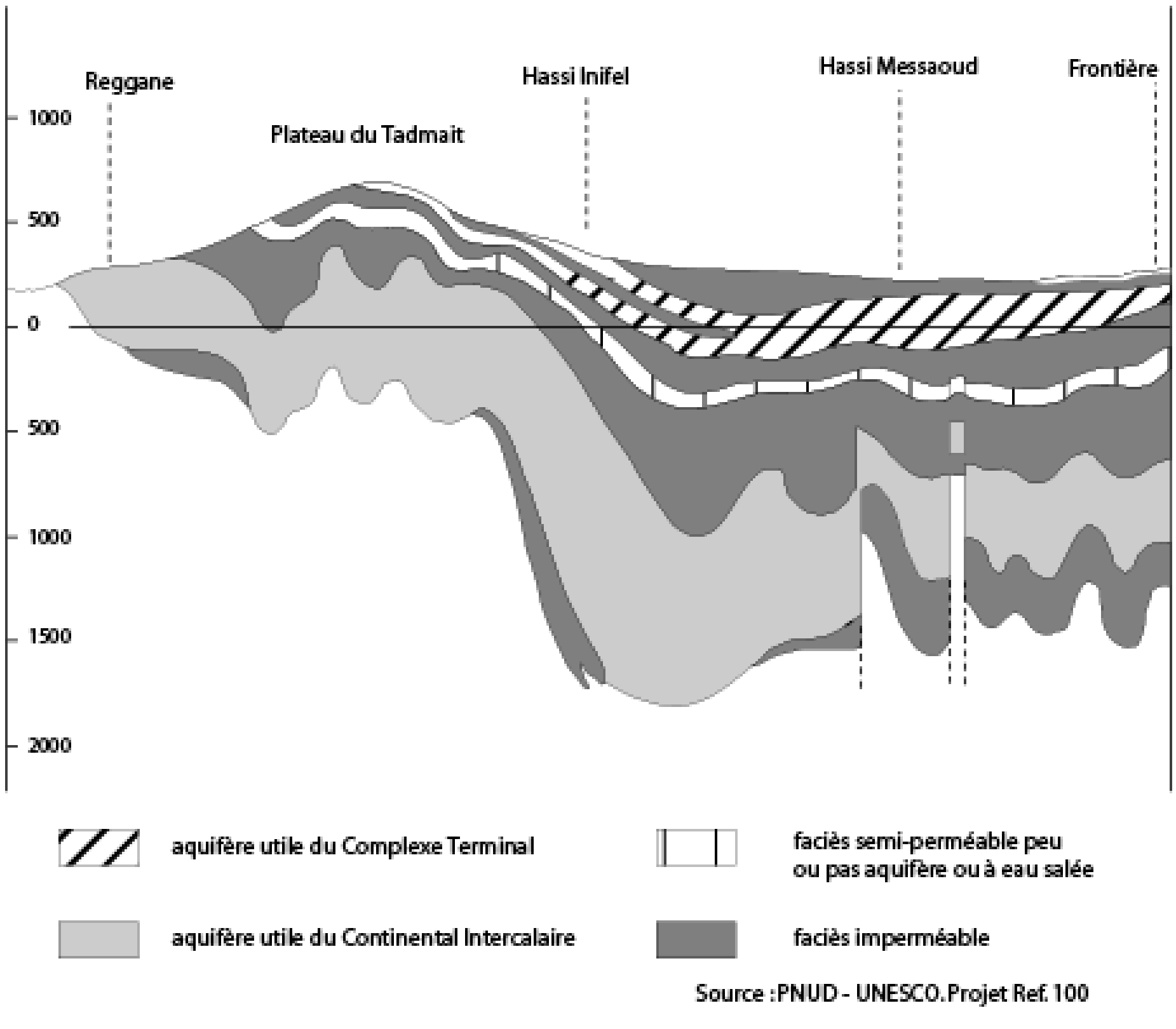

Source : PNUD - UNESCO. Projet Ref. 100

*Figure 4 : Les nappes fossiles au Sahara*

L'aquifère nubien est un des plus grands bassins artésiens du monde : épais de 3 500 mètres, il couvre environ 2 500 000 km2 (Égypte, Libye orientale, Nord du Soudan et du Tchad). De structure complexe, il fonctionne en fait comme un ensemble unique s'écoulant vers le nord-est. Est-il alimenté par les précipitations tropicales du Soudan et du nord du Tchad?. Est-il aussi rechargé par les eaux du Nil en amont d'Assouan. Rien n'est moins sûr. Il fournit actuellement le chapelet d'oasis à l'ouest de la vallée du Nil : Swa, Farafra, Bahariya, Dakhla, Kharga et la dépression de Qattara. De grands projets sont fondés sur l'exploitation de ce réservoir de 50 000 milliards de m3 dont la roche magasin est constituée par des calcaires sablonneux. La profondeur de la nappe est de 800 mètres au sud et 2 000 mètres au nord. L'eau, peu salée, est à une température comprise entre 25 et 45°C.

En péninsule Arabique, l'aquifère est encore plus complexe *(carte 5)*. Les eaux souterraines sont stockées dans les épaisses séries sédimentaires qui occupent les 2/3 de la Péninsule. Depuis le socle surélevé à l'est, elles plongent en pente douce vers le Golfe, la Jordanie et le Yémen. L'ensemble est composé de 12 nappes superposées depuis les formations du Trias et du Jurassique jusqu'au Néogène. Tous les États de la péninsule recèlent dans leur sous-sol une partie de ces nappes. Elles sont exploitées jusqu'à 300 mètres de profondeur et leur stock ainsi défini est estimé à 2 175 milliards de m3. L'ensemble très faiblement alimenté (recharge de 2 700 millions de m3/an) depuis les hauteurs du Hejaz, de l'Asir et du Yémen s'écoule vers l'est et le nord-est de la Péninsule. L'Arabie saoudite contrôle la plus grande partie de ces nappes (1 919 milliards de m3) qui peuvent toutefois s'étendre jusqu'en Jordanie, Syrie et même en Irak. La qualité des eaux est très variable. Certaines sont saumâtres. La salinité varie entre 300 et 15 000 ppm[1]. La température est fonction de la profondeur : elle est comprise entre 40 et 65°. Dans certains cas l'eau peut être utilisée directement pour l'irrigation mais le plus souvent un traitement est nécessaire.

---

[1] ppm : particules par millions; cette unité est équivalente à mg/litre. Ces unités font référence à la quantité de particules solides, en particulier les chlorures, présente dans un volume d'eau. L'OMS fixe un taux de 250 mg/l ou 250 ppm de chlorures pour considérer l'eau comme potable. L'eau saumâtre contient de 5000 à 10000 ppm de chlorures.

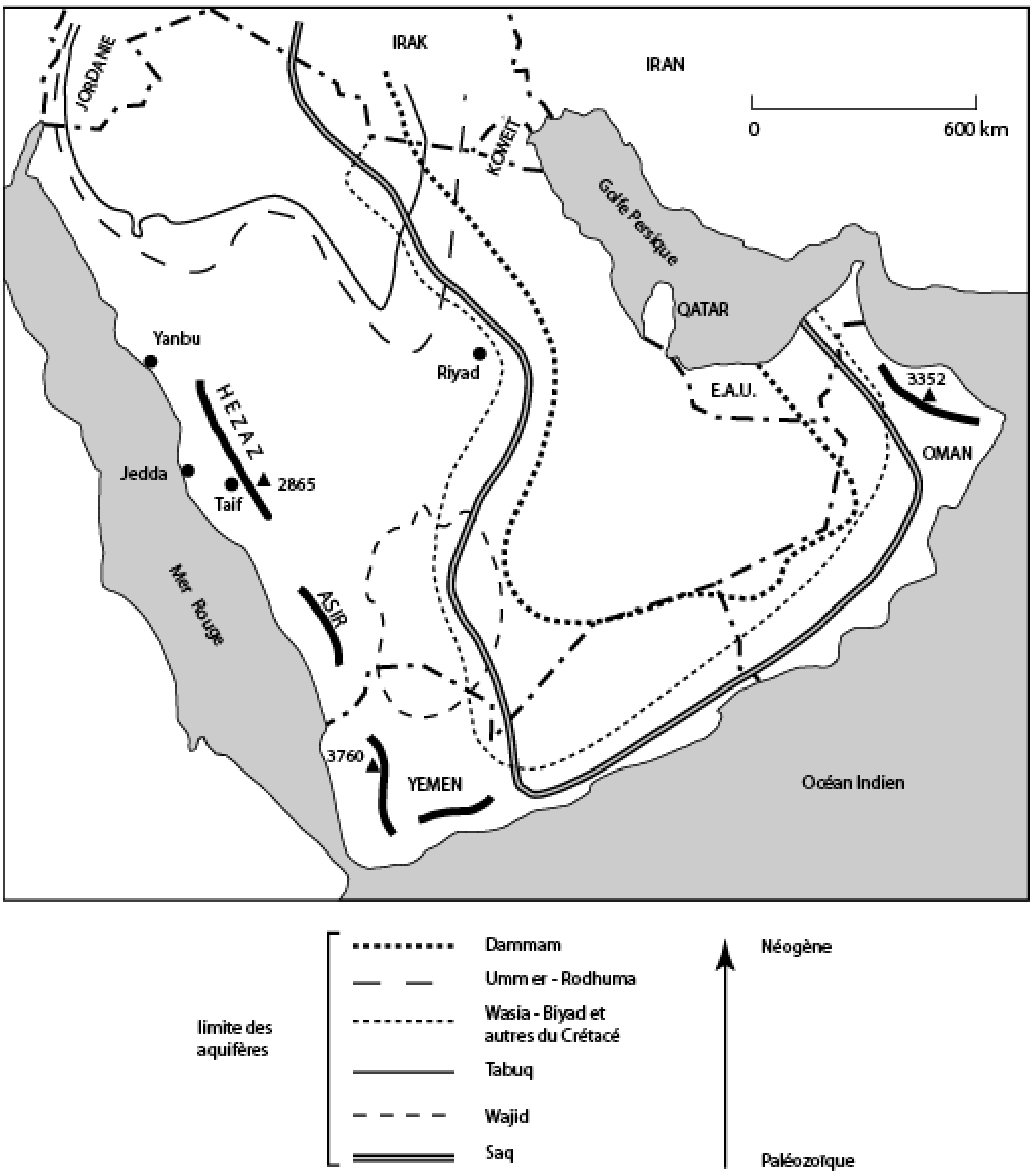

*Figure 5 : Les nappes fossiles dans la péninsule Arabique (d'après Al Alawi & Abdulrazzak, 1994*

# 2.2 La mobilisation des cours d'eau

En dehors des immenses espaces désertiques, les eaux de surface constituent l'essentiel du potentiel et des prélèvements réalisés en dépit de la faiblesse et de l'irrégularité des écoulements. Les techniques traditionnelles, encore largement pratiquées, visent à l'utilisation optimale de l'eau, notamment au moment des crues. A partir d'aménagements légers, de **petits barrages de dérivation**, on cherche à répartir la crue sur la plus grande superficie possible. L'importance de l'épandage dépend de la crue elle-même, l'irrégularité de l'écoulement rend les résultats aléatoires. Cette technique est mise en oeuvre par de petits collectifs d'exploitants alors que le pompage au fil de l'eau relève d'initiatives individuelles. Toute irrigation efficace nécessite cependant le stockage de l'eau aussi bien au cours de l'année agricole que d'une année sur l'autre. Les **barrages collinaires** peuvent fournir une solution mais leur retenue est de faible ampleur.

L'image la plus générale nous est donnée par le **barrage-réservoir** et, éventuellement, le périmètre d'irrigation qui lui est associé construit le long des fleuves dont l'alimentation est suffisante. L'importance des capitaux engagés a toujours fait de la construction du barrage une entreprise étatique. Le nombre de ces barrages s'est multiplié depuis un demi-siècle en même temps que leur capacité s'est considérablement accrue. Dans certains cas les volumes retenus peuvent dépasser le milliard de $m^3$ ce qui permet l'aménagement de plusieurs dizaines de milliers d'hectares. Barrages et périmètres irrigués constituent la pièce essentielle des politiques hydrauliques de beaucoup de pays arabes. Au Maghreb, leur construction a été très largement initiée sous la colonisation et accélérée ces dernières années. L'équipement de nouveaux sites, la construction de nouveaux barrages ne paraissent pas ouvrir des perspectives de grande envergure. Beaucoup a été fait. Par ailleurs, les critiques ne manquent pas à l'égard de ce qu'il est convenu d'appeler la grande hydraulique. L'investissement paraît très lourd en comparaison des résultats obtenus. Le procès fait aux grands barrages est aussi d'ordre technique. Sous le climat aride du Monde Arabe les retenues en amont du barrage perdent une partie de leur eau par évaporation. Par ailleurs l'envasement est souvent très important. Au Maroc 54 millions de m3 sont perdus chaque année par envasement. En 20 ou 30 ans des barrages peuvent perdre jusqu'au tiers de leur capacité : 35% pour le barrage Mohamed V au Maroc depuis 1967 et les exemples peuvent se multiplier à l'infini. On estime que le taux d'envasement toujours au Maroc est de 2%/an et que l'investissement pour maintenir la capacité initiale du barrage représente jusqu'à 20% du coût total de l'ouvrage! Dans le domaine économique et social le procès fait à la grande hydraulique est tout aussi virulent. Trop souvent l'aménagement foncier indispensable en aval pour donner toute son efficacité à l'irrigation est différé et, trop souvent, l'eau ne bénéficie qu'aux agriculteurs les plus riches seuls capables d'apporter les fonds nécessaires à une bonne organisation de l'irrigation.

Le Moyen-Orient est parcouru par de **grands fleuves allogènes** qui apportent la vie dans ces régions sèches, en prenant leur source dans des régions plus humides : les montagnes de Turquie orientale alimentées par d'importantes précipitations de type méditerranéen constituent un château d'eau pour le Tigre et l'Euphrate. Ce sont les pluies équatoriales et surtout tropicales d'Éthiopie qui alimentent le Nil. Ces fleuves sont assez puissants pour traverser d'immenses étendues désertiques sans s'y épuiser. De tout temps, le Nil, le Tigre et l'Euphrate ont été utilisés. Pour répondre à une demande en terres irriguées qui ne fait que croître avec l'évolution démographique, ils ont fait et continuent de faire l'objet de gigantesques travaux d'aménagement. Ces grandes entreprises, dans lesquels s'identifient les régimes politiques, revêtent une dimension internationale : tous les pays traversés par ces grands organismes fluviaux sont concernés. Source de conflits et de tensions d'autant plus qu'en droit international, les dispositions, en ce domaine sont pratiquement inexistantes.

## 2.3 Les ressources non conventionnelles

La rareté de la ressource renouvelable rend inévitable le recours aux eaux non conventionnelles. Sous ce terme on entend la régénération des eaux usées, le dessalement de l'eau de mer, les transferts d'eau. Les pays arabes ont recours à ces différentes techniques de façon très différenciée.

**2.3.1** Les **techniques de recyclage des eaux** sont encore très peu utilisées. Des préjugés ont longtemps dissuadé les autorités de ces pays de réutiliser «l'eau impure». Il est vrai aussi que les réseaux d'assainissement urbain sont en très mauvais état. Les attitudes évoluent toutefois. Ainsi le nouveau réseau d'assainissement du Caire vient seulement achevé, on pense, dès lors, pouvoir retraiter chaque année 2 milliards de m3. La Tunisie recycle également une partie de ses eaux usées (100 millions de m3 en 1994, 200 en 2 002). La Syrie, le Maroc et la Jordanie ont aussi recours à ces technologies : une partie des eaux d'Amman est régénérée. Les États de la péninsule Arabique réutilisent actuellement 479 millions de m3/an. Toutefois, le retard pris est important, il est d'autant plus regrettable que le coût de l'eau retraitée est trois à quatre fois inférieur à celui de l'eau dessalée. Pour l'irrigation qui absorbe 88% de la consommation d'eau dans le Monde Arabe, on pourrait aussi utiliser de façon beaucoup plus performante les eaux de drainage après les avoir traitées.

**2.3.2** Par contre, le recours aux techniques de **dessalement de l'eau de mer** est beaucoup plus largement pratiqué. Il l'est notamment dans les pays de la péninsule Arabique. Les capacités de traitement sont importantes : 7,3 milliards de m3/an en 2008. La Péninsule est de très loin le leader mondial en ce domaine (40% des capacités mondiales en 2008). Les unités de dessalement fournissent 70 à 80 % des besoins en eau potable des villes du Golfe.

| 2008 | Millions m3/an | % |
|---|---|---|
| Arabie saoudite | 2 700 | 37,7 |
| E.A.U | 2 300 | 31 |
| Koweït | 760 | 10 |
| Libye | 220 | 3 |
| Qatar | 328 | 4,5 |
| Tunisie | 100 | 1,3 |
| Irak | 80 | 1 |
| Bahreïn | 182 | 2,5 |
| Oman | 182 | 2,5 |
| Algérie | 480 | 6,5 |
| **Total** | 7332 | 100 |

*Tableau 2 : Dessalement de l'eau de mer : volumes traités par les pays arabe*

L'Arabie saoudite a mis en œuvre depuis les années 60 le programme de dessalement le plus ambitieux. Elle compte 23 complexes de dessalement : 17 sur la Mer rouge et 6 sur le Golfe avec 15 % des capacités mondiales. Elle a investi de 1980 à 1985 plus de 10 milliards de $ dans la construction d'unités de dessalement. Elle est le premier producteur mondial. La plupart des villes saoudiennes sont alimentées en eau dessalée grâce à un réseau de canalisations (plus de 3 000 km) qui acheminent l'eau depuis les usines de la mer Rouge ou du Golfe. 365 millions de m3 sont acheminés chaque année à Ryad à partir de l'usine de Jubail par une conduite de 300 km. La Mecque et Médine sont approvisionnées depuis les usines de Jedda et Yanbu sur la mer Rouge grâce à une canalisation de 240 km qui s'élève jusqu'à 2 400 mètres pour traverser les montagnes de l'Asir. Au total, la Mecque est approvisionnée à 40% en eau dessalée, Jedda à 90%, Damman à 95% et Ryad à 60%.

**Koweït**, qui a construit la première usine au cours des années 50, est presqu’entièrement dépendant de quatre usines de dessalement pour sa consommation urbaine et industrielle (760 millions m3/an). Les ressources locales sont très faibles : ce sont des eaux souterraines. Koweït pouvait assurer en 1990 une distribution d’eau importante à ses citadins : 560 litres/jour. Le **Qatar** dépend essentiellement de l'eau dessalée pour la satisfaction de ses besoins (328 millions de m3/an), il en est de même pour Bahreïn. Les **Emirats Arabes Unis** disposent désormais d'une importante capacité de traitement. Le sultanat **d'Oman** est le dernier à avoir recours à ces techniques. Une dizaine d'unités sont achevées et déjà l'eau dessalée assure 80% de la consommation de la capitale, Mascate (182 millions de m3/an).

**Le dessalement par distillation**
C'est le procédé le plus répandu mais le plus gourmand en énergie. Il consiste à faire passer de l'eau de mer dans une série de chambres, soumises à des températures et à des pressions différentes. La circulation d'une chambre à l'autre permet la vaporisation de l'eau qui est recueillie en fin de course débarrassée de ses sels. Cette technologie couvre en 2008 45 % du marché mondial. Le prix du m3 produit est compris entre 0,65 e 1,8 €

**Le dessalement par osmose inverse**
L'osmose inverse vise à faire traverser les molécules d'eau au travers d'une membrane qui retient les sels dissous grâce à l'augmentation de la pression. Cette technologie mise au point dans les années 70 a un grand avenir. Elle est deux fois plus gourmandes en énergie que le procédé par distillation. Au début, elle n'était utilisée que pour le traitement des eaux saumâtres; elle l'est maintenant pour l'eau de mer. Elle couvre 55 % du marché mondial en 2008. Le coût du m3 en sortie d'usine est compris entre 0,4 et 0,8 €

**Le dessalement par électrodialyse**
Un champ électrique permet de séparer les sels et de les filtrer à travers des membranes sélectives que les molécules d'eau ne peuvent pas traverser. Cette technologie est surtout utilisée pour le traitement des eaux saumâtres et dans de petites unités (ex : hôtels).

Cette eau dessalée à grands frais n'est pas destinée à la seule consommation domestique. Dans ces villes de la Péninsule, le nouvel urbanisme a multiplié en milieu désertique espaces verts et jardins d'agrément qui exigent pour leur entretien d'énormes quantités d'eau. Koweït consacre 45% de l'eau disponible à cet usage, Jedda : 20% etc...

Ce choix ne sera pas remis en cause car il représente la solution la mieux adaptée aux besoins de la Péninsule, riche en pétrole et qui dispose de l'énorme quantité d'énergie nécessaire à bon marché. Le coût est très élevé : il ne peut être supporté que par les riches monarchies du Golfe. Par contre, les usines de dessalement présentent une grande vulnérabilité : en cas de conflit elles peuvent être la cible de bombardements ou peuvent pâtir d'une marée noire provoquée.

Ailleurs la production d'eau dessalée est restée longtemps marginale : elle est liée à l'activité pétrochimique comme en Algérie (à Arzew ou Skikda) et en Libye (130 millions de m3/an). L'activité touristique a aussi recours à cette ressource : c'est le cas des hôtels qui se multiplient sur le littoral égyptien du golfe d'Akaba ou de certaines installations tunisiennes (81 millions de m3/an). Dans l'ex-Sahara espagnol, la capitale prestige que le Maroc édifie, Laayoune, est aussi alimentée par de l'eau dessalée. Il n'en est plus de même aujourd'hui et les Etats maghrébins notamment **l'Algérie** ont lancé de gigantesques programmes pour tenter de mettre fin aux multiples pénuries de la distribution d'eau potable dans leurs grandes villes. Ainsi depuis février 2009, l’ouverture d’une  usine de dessalement permet la distribution désormais régulière de l’eau dans l’agglomération algéroise. Par ailleurs, un programme de construction de 13 stations de dessalement est en chantier. Si ces projets se concrétisent c’est 10 % de l’alimentation en eau potable du pays qui relèvera du dessalement de l’eau de mer.

**2.3.3** Les **transferts d'eau** peuvent s'organiser à l'intérieur des espaces nationaux parfois sur une grande échelle. En Tunisie, par exemple, 80% des eaux de surface sont concentrées dans les espaces montagnards du Nord-Ouest alors que 91% de la consommation s'effectue le long de l'axe littoral oriental. Cette dysharmonie a entraîné de massifs transferts d'eau. Les réseaux comprennent à la fois un système de canalisations et une interconnexion des grands barrages. L'eau des montagnes du Nord-Ouest ravitaille Tunis et sa région mais tout récemment la desserte vers le Centre et la ville de Sfax a été réalisée. Au Maroc, le 1/6 des prélèvements effectués actuellement fait l'objet de transferts interrégionaux et ils vont se multiplier à l'avenir.

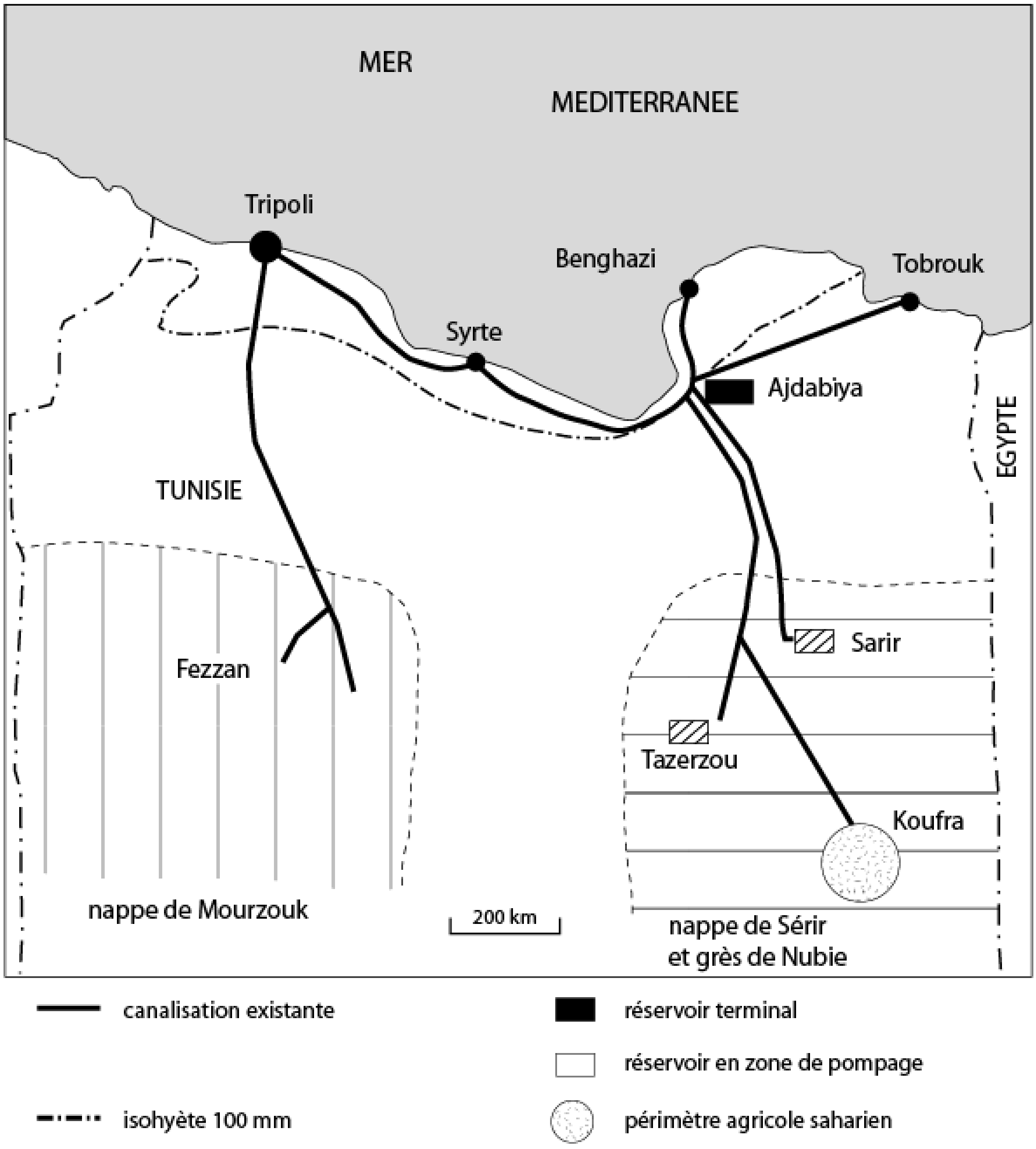

*Figure 6  La Grande Rivière artificielle libyenne*

L'entreprise libyenne est encore plus spectaculaire. La Libye a essayé sans grand succès de créer un secteur agricole alimenté par les eaux souterraines dans le désert. Faute de pouvoir utiliser l'eau sur place, elle a décidé de la transférer vers le littoral pour faire renaître une agriculture sur les anciennes fermes des colons italiens et alimenter les villes en pleine croissance *(carte 6)*. L'entreprise est gigantesque, l'investissement énorme : 30 milliards de dollars. La Grande Rivière, mise en chantier en 1983, est achevée et permet le transfert, par 4 000 km de canalisations enterrées, de 2 milliards de m3 d'eau par an, pompés à 600 m. de profondeur !. Déjà, sur 2 000 km, des conduites de béton armé de 4 m de diamètre acheminent un milliard de m3/an depuis les nappes de Tazerbo, Sarir et Koufra vers le réservoir géant d'Ajdabiya. Les villes de Benghazi, Brega et Syrte sont désormais approvisionnées avec l'eau du désert, Tripoli l'est depuis septembre 1996 mais les périmètres agricoles prévus ne fonctionnent pas encore. L'apport d'eau aura pour avantage de recharger la nappe littorale déjà infiltrée par l'eau de mer. Une seconde liaison est terminée à l'ouest entre est le Fezzan et Tripoli (1 000 km). En phase finale, 6,5 millions de m3 sont pompés chaque jour, chiffre considérable qui correspond au débit d'une rivière moyenne dans les régions tempérées ! A son terme l'opération devrait permettre, outre le ravitaillement des villes, l'irrigation de 250 000 ha dans les régions de Benghazi, Syrte et Jeffara : 180 000 hectares seront aménagés en nouveaux périmètres tandis qu'en Jeffara il s'agit surtout de revitaliser des zones autrefois irriguées. La production agricole aussi bien végétale qu'animale devrait augmenter considérablement... à condition que les Libyens soient intéressés ce qui est incertain. Il est alors probable que des fellahs égyptiens constitueront une main d'œuvre de substitution. Aujourd'hui 70% des eaux utilisées en Libye proviennent de ressources non renouvelables, même si leur acheminement sur le littoral est fort coûteux : 1 à 1,2 $/m3. Cet aménagement est souvent contesté en raison de son gigantisme et de l'ampleur des prélèvements effectués dans des nappes fossiles. La durée de vie du grand fleuve serait comprise entre un demi-siècle et un siècle. Sa réalisation inquiète fortement les pays voisins. Les prélèvements sont effectués sur des aquifères transfrontaliers. L'Égypte craint que des pompages excessifs dans les grès de Nubie ne créent une dépression provoquant un abaissement du niveau du Nil et de son aquifère. La position de l'Algérie est identique pour les futurs pompages du Fezzan. En réalité, la situation actuelle est bien éloignée du projet initial. L'eau du Sahara est exclusivement réservée au ravitaillement des villes littorales. Pour assurer sa sécurité alimentaire, la Libye a le projet d'investir massivement au Tchad avec la mise en valeur de 50 000 hectares irrigués en bordure du fleuve Chari !

En 1987 la Turquie a lancé un projet encore plus ambitieux de transfert international : l'«aqueduc de la paix» qui pourrait fournir 2 milliards de m3 d'eau douce par an à la Syrie, la Jordanie, l'Arabie saoudite et aux autres États du Golfe. L'aqueduc serait constitué de deux conduites qui achemineraient l'eau de deux rivières turques, issues du Taurus, le Ceyhan et le Seyhan dont les eaux abondantes se perdent en Méditerranée. Le projet fut accueilli avec beaucoup de réserves par les États arabes pour des raisons géopolitiques. Mais surtout la position turque a beaucoup évolué et le pays ne veut plus se présenter aux yeux de ses voisins arabes comme le château d'eau de la région. On voit bien avec ce projet, en quelque sorte mort-né, la difficulté à établir une coopération régionale en ce domaine. La perspective des transferts d’eau n‘est toutefois pas totalement abandonnée. Des discussions ont été engagées entre les Turcs et Israël et également avec la Jordanie.

| conduite Ouest | m3/jour | conduite du Golfe | $m^3$/jour |
|---|---|---|---|
| Turquie | 300 000 | Koweït | 600 000 |
| Syrie | 1 100 000 | Arabie Saoudite | 800 000 |
| Jordanie | 600 000 | Bahreïn | 200 000 |
| Arabie saoudite | 1 500 000 | Qatar | 100 000 |
| | | E.A.U. | 600 000 |
| | | Oman | 200 000 |
| TOTAL | 3 5000 000 | | 2 500 000 |

*Tableau 3 : le projet d’ aqueduc de la paix*

# 3. Une situation critique : vers la pénurie

C'est une véritable course contre la montre qui est engagée et la menace d'une pénurie à long terme n'est pas à écarter.

## 3.1 Le point en 2009

La Banque Mondiale classe le Maghreb et le Moyen-Orient comme la région du monde la plus pauvre en ressources hydrauliques naturelles renouvelables. Les pays arabes, tous inclus dans cette grande aire régionale, sont encore plus mal pourvus**. Ils rassemblent en 2007 4,8 % de la population mondiale mais ne disposent que de 0,67% des ressources hydrauliques naturelles.**

**3.1.1** Les **ressources en eau renouvelable** qui étaient de l’ordre de 3000 m3/an/habitant en 1960 ne s’élèvent plus qu’ à **884 m3 /an/habitant en 2009** . A titre de comparaison, le niveau de la ressource est de 36 619 m3/an/habitant en Océanie, 23 103 en Amérique latine, 18 742 en Amérique du Nord, 14 659 en Europe orientale et en Asie centrale, 7 485 en Afrique au Sud du Sahara, 5 183 en Europe occidentale, 3 283 en Asie. On mesure bien avec cette première approximation l'ampleur du handicap du Monde Arabe. L'eau est sans aucun doute une des ressources naturelles les plus mal réparties à la surface de la planète.

Des normes internationales ont été établies pour juger plus exactement les différentes situations :

- Le seuil de pénurie est fixé à 1 000 m3/an/habitant.
- En dessous de 1 000 m3/habitant on estime qu'un pays peut être confronté à des pénuries régionales.
- A 500 m3/an/habitant la situation est considérée comme critique.
- En dessous de 100, le recours massif à de coûteuses ressources non conventionnelles est inévitable.

| | Ressources totales Km3 | Population 2009 Millions | Ressources totales m3/habitant | Population 2O25 millions | Ressources totales m3/hab | Population 2050 Millions | Ressources 2050 m3/hab |
|---|---|---|---|---|---|---|---|
| Koweït | 0,02 | 3 | 7 | 3,9 | 5 | 5 | 4 |
| Qatar | 0,05 | 1,4 | 36 | 1,1 | 45 | 2,3 | 22 |
| Arabie | 2,4 | 28,7 | 84 | 35,7 | 67 | 50 | 48 |
| Libye | 0,6 | 6,3 | 95 | 8,1 | 74 | 10 | 60 |
| EAU | 0,5 | 5,1 | 98 | 6,2 | 81 | 9 | 56 |
| Bahreïn | 0,12 | 1,2 | 100 | 1 | 120 | 2 | 60 |
| Jordanie | 0,9 | 5,9 | 153 | 7,8 | 115 | 10 | 90 |
| Yémen | 4,1 | 22,9 | 179 | 36,6 | 112 | 52 | 79 |
| Territ occupés | 0,8 | 4 | 200 | 6,2 | 129 | 9 | 89 |
| Algérie | 12 | 35,4 | 339 | 43,2 | 278 | 51 | 235 |
| Oman | 1 | 3,1 | 323 | 3,1 | 323 | 5 | 200 |
| Tunisie | 4,6 | 10,4 | 442 | 12,1 | 380 | 14 | 329 |
| Egypte | 58 | 78,6 | 738 | 95,9 | 605 | 122 | 475 |
| Maroc* | 29 | 32 | 906 | 38,9 | 746 | 43 | 674 |
| Liban | 4,8 | 3,9 | 1231 | 4,6 | 1043 | 5 | 960 |
| Syrie | 26,2 | 21,9 | 1196 | 27,5 | 953 | 37 | 708 |
| Soudan | 77 | 42,3 | 1820 | 54,3 | 1418 | 76 | 1013 |
| Irak | 75 | 30 | 2500 | 43,3 | 1732 | 62 | 1210 |
| Total/Moyenne | **297** | **336** | **884** | **430** | **692** | **564** | **526** |

* y compris l'ex Sahara espagnol

*Tableau 4 : Disponibilités en eau douce par habitant (m3/an) en 2007 et prévisions pour 2025 et 2050*

La situation des pays arabes que l'on peut observer au tableau 4 apparaît fort préoccupante. La moyenne arabe se situe en dessous du seuil de pénurie en 2009 avec 884 m3/an/habitant. Cette situation ira en s'aggravant dans les années à venir : les ressources évidemment resteront les mêmes alors que la population va inéluctablement augmenter. En retenant les projections démographiques habituelles qui tiennent compte de la récente inflexion de la fécondité, on peut estimer que la dotation moyenne d'ensemble s'abaissera à 692 m3/an/habitant vers les années 2025/2030 et 526 en 2050 ! En se fiant à ces normes internationales, le Monde arabe va, globalement, entrer dans une situation de pénurie!

Ces moyennes d'ensemble cachent en fait des situations très contrastées entre pays arabes comme nous l'indique l'observation du tableau 5 :

- seuls 4 pays rassemblant 24% de la population se situent au dessus du seuil de 1 000 m3/an/habitant en 2009 (Liban, Syrie, Soudan, Irak).
- 2 pays (Egypte, Maroc) se trouvent déjà dans une situation de pénuries régionales.
- enfin 12 pays (37 % de la population) se trouvent placés au dessous du seuil critique de 500 m3/an/habitant et 5 d'entre eux disposent de moins de 100 m3/an/habitant!

| 2009 | Ressources | | Population | Ressources | | Prélèvements | |
|---|---|---|---|---|---|---|---|
| | totales Km3 | internes km3 | 2009 Millions | totales m3/habitant | internes m3/habitant | km3 | m3/hab |
| Algérie | 12 | 11,4 | 35,4 | 339 | 322 | 6 | 169 |
| Arabie | 2,4 | 2,4 | 28,7 | 84 | 84 | 16,3 | 568 |
| Bahreïn | 0,12 | 0,12 | 1,2 | 100 | 100 | 0,21 | 175 |
| EAU | 0,5 | 0,5 | 5,1 | 98 | 98 | 1,4 | 275 |
| Egypte | 58 | 1,8 | 78,6 | 738 | 23 | 59,4 | 756 |
| Irak | 75 | 34 | 30 | 2500 | 1133 | 42,8 | 1427 |
| Jordanie | 0,9 | 0,7 | 5,9 | 153 | 119 | 0,88 | 149 |
| Koweït | 0,02 | 0,02 | 3 | 7 | 7 | 0,4 | 133 |
| Liban | 4,8 | 4,8 | 3,9 | 1231 | 1231 | 0,86 | 221 |
| Libye | 0,6 | 0,6 | 6,3 | 95 | 95 | 2,83 | 449 |
| Maroc | 29 | 29 | 32 | 906 | 906 | 12,5 | 391 |
| Oman | 1 | 1 | 3,1 | 323 | 323 | 1,4 | 452 |
| Qatar | 0,05 | 0,05 | 1,4 | 36 | 36 | 0,19 | 136 |
| Soudan | 77 | 30 | 42,3 | 1820 | 709 | 18,6 | 440 |
| Syrie | 26,2 | 7 | 21,9 | 1196 | 320 | 8,8 | 402 |
| Territ occup | 0,8 | 0,8 | 4 | 200 | 200 | 0,73 | 183 |
| Tunisie | 4,6 | 4,2 | 10,4 | 442 | 404 | 3 | 288 |
| Yémen | 4,1 | 4,1 | 22,9 | 179 | 179 | 3,2 | 140 |
| **Total/Moyenne** | **297** | **132** | **336** | **884** | **394** | **180** | **534** |

*Tableau 5 :. Les ressources en eau douce renouvelable d'origine interne et externe par pays en 2008*

*Source : Ined et FAO /aquastat*

**NB**. Les prélèvements prennent en compte à la fois les eaux renouvelables et les eaux fossiles.

Pour l'Irak et la Syrie, il est tenu compte de l'accord de 1987 concernant le partage des eaux de l'Euphrate. Il en est de même pour l'Egypte et le Soudan et les eaux du Nil.

La Mauritanie, Djibouti, la Somalie sont des Etats arabes au sens politique du terme (ils sont membres de la Ligue arabe). Ils n'ont pas été retenus pour l'étude car ils sont rattachés par leurs conditions naturelles au Sahel africain plutôt qu'au Nord de l'Afrique

Le tableau fait donc apparaître la très grande inégalité de la ressource à l'intérieur de l'aire arabe. Trois pays : l'Irak, l'Égypte et le Soudan concentrent 70% des eaux (210 milliards de m3 sur 297) alors qu'ils ne rassemblent que 44% de la population. Il y a bien, toutes proportions gardées, des pays peu nombreux, relativement mieux dotés, et beaucoup d'autres qui sont affrontés à des situations de sérieuse pénurie.

**3.1.2** Apparaît également avec force la **dépendance du Monde Arabe**. Les ressources internes en eau douce s'élèvent à 132 milliards de m3 seulement (pas plus de 394 m3/an/habitant)! Les ressources totales *(colonne 1 du tableau 5)* qui tiennent compte des apports extérieurs s'élèvent à 297 milliards de m3 (soit 884 m3/an/habitant). Ainsi 44 % de la ressource est d'origine externe ! Ce taux de dépendance est très élevé. La dépendance hydraulique du Monde arabe se pose à une double échelle.

- Elle pèse de façon globale entre les pays arabes et l'Afrique orientale d'une part (le Nil) et la Turquie d'autre part (Tigre et Euphrate)
- Elle pèse entre pays arabes : ainsi entre la Syrie et l'Irak (Euphrate) entre la Syrie et le Liban (Oronte) ou entre les pays riverains du bassin du Jourdain

**3.1.3** Les **prélèvements annuels** (y compris les eaux fossiles) mesurent d'une certaine manière la **consommation** d'eau. Ils ne s'élèvent qu'à 180 milliards de m3/an; ils sont bien inférieurs à la ressource (61%) : l'eau, en effet, n'est pas totalement mobilisable pour des raisons bien souvent naturelles mais aussi techniques ou politiques. Le tableau de ce point fait apparaître des situations très contrastées entre des pays qui mobilisent pratiquement la totalité de la ressource (Égypte, Jordanie) et des pays qui sont loin d'avoir mobilisé toutes leurs potentialités (Soudan et Oman 25%).

Dans un grand nombre de pays par contre, les prélèvements sont supérieurs et, parfois, dans de très fortes proportions aux ressources totales ou aux ressources internes. Cette catégorie recouvre deux types de situations :

• soit les pays qui recourent à des sources non conventionnelles comme le dessalement de l'eau de mer. Entrent dans cette catégorie tous les pays du Golfe mal pourvus.

• soit les pays qui surexploitent leurs nappes souterraines (Yémen, les Territoires occupés) ou qui ont recours massivement à l'eau fossile des grands aquifères (Libye, Arabie saoudite, Algérie dans une moindre mesure).

Enfin le **niveau des prélèvements par habitant varie dans des proportions considérables** dans le rapport de 1 à 10. La valeur moyenne de 534 m3/an/habitant n'a pas grande signification. Le volume des prélèvements rend compte de plusieurs facteurs : le niveau de la ressource ne paraît pas le plus important il est en effet très souvent corrigé par le recours à des apports non conventionnels. Par contre l'importance des surfaces irriguées est déterminant ce qui nous confirme que pour l'instant l'agriculture reste bien le secteur utilisateur principal.

On peut le constater la situation est déjà très tendue or, fait plus inquiétant, dans les années à venir les consommations d'eau vont croître inéluctablement dans de très fortes proportions. Enfin la très grande diversité des situations que l'on peut constater incitent à examiner avec une certaine prudence les données d'ensemble et à étudier la situation en nous appuyant sur des cas précis. Si tous les pays sont menacés de pénurie, la situation égyptienne, par exemple, a peu à voir avec les tensions qui règnent entre les pays riverains du Jourdain ou avec les concurrences entre utilisateurs qui s'aiguisent dans les pays du Maghreb!

### **3.1.4.** Le recours massif aux **ressources non conventionnelles dans la péninsule Arabique**

• Dans cette partie du Monde la consommation d'eau a connu une impressionnante augmentation. De 6 milliards de m3 en 1980, elle est passée à 23 en 1990; elle est actuellement 31 milliards de m3, on prévoit 35 en 2010. Comment les 7 États de la Péninsule qui, à l'exception du Yémen, sont riches en capital et en énergie répondent-ils à cette situation?

| millions de m3 en 1990 | eau dessalée | Eau surface. | Aquifères al | Eau recyclée | aquifères fossiles | Total |
|---|---|---|---|---|---|---|
| Arabie Saoudite | 795 | 900 | 950 | 217 | 13480 | 16342 |
| Koweït | 240 | | | 83 | 80 | 403 |
| Bahreïn | 56 | | | 107 | 53 | 216 |
| Qatar | 83 | | 50 | | 61 | 194 |
| E.A.U. | 342 | 75 | 400 | 62 | 500 | 1379 |
| Oman | 32 | 650 | 645 | 10 | | 1337 |
| Yémen | 9 | 1450 | 1700 | | | 3159 |
| Total | 1557 | 3075 | 3745 | 479 | 14174 | 23030 |
| % | 6,8% | 13,4% | 16,2% | 2% | 61,6% | 100 |

- en partie fossiles

- *Tableau 8 : L'origine des eaux consommées dans les États de la péninsule Arabique en 1990*

• En 1990, l'importance relative des grands secteurs de consommation n'est pas très éloignée de celle de l'ensemble du Monde Arabe : 12% pour l'eau municipale, 1,5% pour l'industrie et 86,5% pour l'irrigation. On note l'importance relative de l'eau urbaine et la faiblesse de la consommation industrielle. C'est surtout l'origine de l'eau consommée qui fait l'originalité de la Péninsule : près de 70% sont assurés par des eaux fossiles. La très grande faiblesse de la part prise par les eaux de surface s'expliquent par les conditions climatiques. Cette eau renouvelable se localise dans des régions bien déterminées : les montagnes d'Oman, du Yémen et de l'ouest de l'Arabie saoudite ou dans certaines nappes souterraines (aquifères alluviaux).

• Pour satisfaire une demande qui, d'ici 2010 augmentera de 50% par rapport à 1990, les États de la Péninsule vont recourir de façon encore plus importante aux eaux fossiles qui, avec 28,5 milliards de m3, couvriront 80% de la consommation. Un effort sera bien tenté pour accroître les capacités de dessalement ou du traitement des eaux usées mais leur rôle restera limité (respectivement 8,3 et 4,4 % de l'eau consommée).

• La poursuite de cette politique n'est pas sans risque pour l'environnement. Le surpompage entraîne le rabattement et une salinisation des eaux. La pollution ne fait que s'accroître : elle est liée à la forte consommation d'engrais et de pesticides dans l'agriculture irriguée et à l'abondance des eaux usées entraînée par la surconsommation urbaine. Pour l'instant il n'existe pas de solutions alternatives... sinon l'importation d'eaux douces depuis des pays voisins (Turquie ou Iran) mais la situation géopolitique actuelle n'autorise guère d'espoir en ce domaine.

• Une telle gestion purement minière des eaux, sans aucun souci de l'avenir et des générations futures peut-elle continuer à de tels rythmes pendant des années? En tout cas, la situation de la Péninsule est tout a fait particulière dans le Monde Arabe : ce n'est pas un "modèle" qui peut se généraliser.

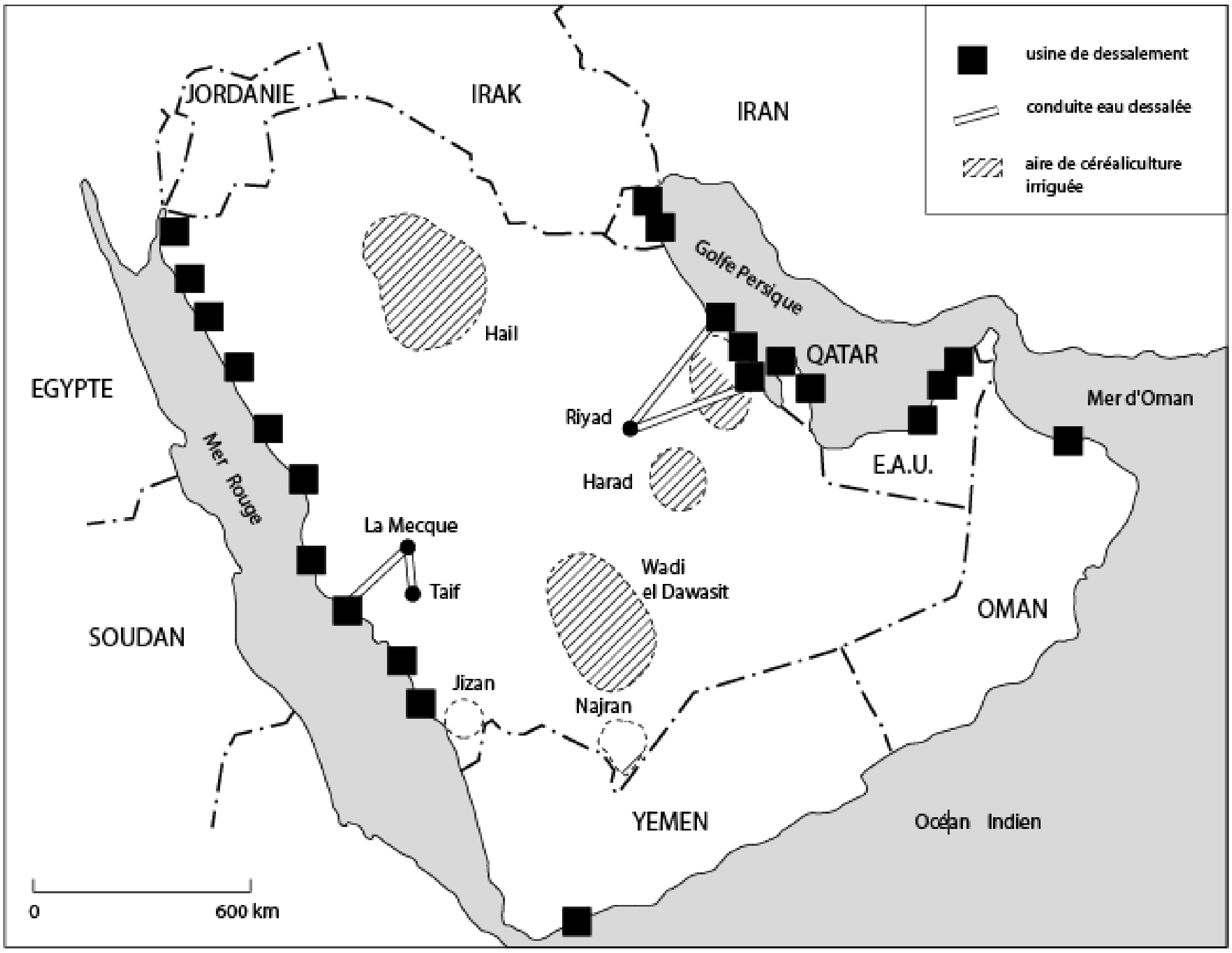

*Figure 7 – Dessalement de l'eau de mer en péninsule Arabique*

## 3.2 L'inéluctable augmentation de la demande d'eau

Elle tient à **l'accroissement démographique** et aux conséquences qui en découlent. La zone arabe est encore en pleine transition démographique. L'accroissement maximum s'est placé au cours de la décennie 80 (3 %/an) et, actuellement, il est en recul inégal selon les pays. Il est en moyenne de l'ordre de 2 à 2,5% l'an. Toutes les projections concordent pour estimer que la population de 297 millions en 2007 se situera à quelque 429 millions d'habitants en 2 025. (tableau 4) Si les tendances actuelles de la chute de la fécondité se confirment cette évaluation semble être réellement un maximum et sans doute ne sera-t-elle pas atteinte.

| | Prélèvements | Eau urbaine | | Industrie | | Irrigation | |
|---|---|---|---|---|---|---|---|
| | annuels km3 | % | km3 | % | km3 | % | km3 |
| Algérie | 6 | 22 | 1,32 | 14 | 0,84 | 64 | 3,84 |
| Arabie | 16,3 | 9,3 | 1,52 | 1,2 | 0,20 | 89,5 | 14,59 |
| Bahreïn | 0,21 | 39,8 | 0,08 | 7,9 | 0,02 | 52,3 | 0,11 |
| Égypte | 59,4 | 7 | 4,16 | 5 | 2,97 | 88 | 52,27 |
| Émirats | 1,4 | 34,4 | 0,48 | 1,8 | 0,03 | 63,8 | 0,89 |
| Iraq | 42,8 | 3 | 1,28 | 5 | 2,14 | 92 | 39,38 |
| Jordanie | 0,88 | 29 | 0,26 | 6 | 0,05 | 65 | 0,57 |
| Koweït | 0,4 | 77 | 0,31 | 2 | 0,01 | 21 | 0,08 |
| Liban | 0,86 | 11 | 0,09 | 4 | 0,03 | 85 | 0,73 |
| Libye | 2,83 | 15 | 0,42 | 10 | 0,28 | 75 | 2,12 |
| Maroc | 12,5 | 6 | 0,75 | 3 | 0,38 | 91 | 11,38 |
| Oman | 1,4 | 6,6 | 0,09 | 0,4 | 0,01 | 93 | 1,30 |
| Qatar | 0,19 | 36 | 0,07 | 26 | 0,05 | 38 | 0,07 |
| Soudan | 18,6 | 1 | 0,19 | 0 | 0,00 | 99 | 18,41 |
| Syrie | 8,8 | 7 | 0,62 | 10 | 0,88 | 83 | 7,30 |
| Terr Occ | 0,73 | 23 | 0,17 | 1 | 0,01 | 76 | 0,55 |
| Tunisie | 3 | 13 | 0,39 | 7 | 0,21 | 80 | 2,40 |
| Yémen | 3,2 | 5,4 | 0,17 | 2,6 | 0,08 | 92 | 2,94 |
| **TOTAL** | **180** | 7 | **12,60** | 4 | **7,4** | 89 | **160** |

*Tableau 6 : Consommation d’eau par les grands secteurs utilisateurs*

Mais il faut s'attendre à une demande d'eau beaucoup plus importante que le simple accroissement démographique en raison des **mutations économiques et sociales qu’enregistre la société arabe : l'augmentation des superficies irriguées et la croissance urbaine.**

Tous les États sont confrontés aux problèmes de la dépendance alimentaire et contraints d'importer de grandes quantités de biens alimentaires. Au début de la décennie 1990, ils devaient importer la moitié des céréales consommées, 70% du sucre et des oléagineux, 25% de la viande. Ils ont tous pour objectif sinon d'atteindre l'autosuffisance du moins une certaine sécurité alimentaire. L’amélioration des performances dans le domaine agricole est une priorité afin d’alléger la charge financière que représentent des importations massives. L’espace agricole est très limité, très peu de terres nouvelles ont été mises en valeur au cours des dernières décennies, et bien souvent il s’agissait de terres marginales. Ainsi de 1961 à 1996 alors que la population passait de 92 millions d’habitants à 255 (presqu’un triplement!), les terres cultivées n’ont progressé que de 23% en passant de 47 à 58 millions d’hectares. Ainsi en 35 ans, la superficie arable par habitant a-t-elle diminué de plus de la moitié passant de 0,51 hectare à 0,22! Faute de pouvoir conquérir de nouvelles terres, l'objectif prioritaire en matière agricole est donc l'intensification pour améliorer la production et, de ce point de vue, la mobilisation des eaux et l'extension de l'irrigation apparaissent essentielles. La progression des cultures irriguées constatée ces 35 dernières années a été de 90% en passant de 7 millions d'hectares irrigués à 14,7. Pour atteindre cet objectif les pays arabes font appel à 89 % de leurs ressources en eau (160 milliards de m3).

La mobilisation des eaux se pose désormais en termes nouveaux. La multiplication des hommes, la croissance des villes, le développement des activités industrielles et touristiques ont fait apparaître de nouvelles demandes. La demande urbaine est exponentielle en raison de la croissance urbaine proprement dite qui ces dernières années a été très forte : plus de 50% des Arabes sont des citadins. La croissance urbaine s'effectue à des rythmes de 4 à 5% l'an. Les 150 millions de citadins de 2007 deviendront 270 voire 300 millions. La progression de la demande urbaine sera impressionnante car actuellement le niveau de consommation de l'eau dans les villes reste relativement bas (120 à 150 litres/jour/citadin), il parviendra vraisemblablement à un niveau de 200 ou 250 litres. Par ailleurs croissance de la ville signifie aussi multiplication des raccordements au réseau public d'eau. Bref tous ces facteurs sont cumulatifs. Une rétrospective de la consommation d'eau d'Alger et de sa périphérie proche au cours des dernières années nous donne une idée de l'ampleur de la demande. En 1972 les besoins étaient estimés à 72 millions de m3/an, ils sont passés à 200 en 1990 et vraisemblablement plus de 300 actuellement. Déjà, dans de nombreux pays, villes et campagnes se disputent une eau de plus en plus rare. Comment gérer cette pénurie? On mesure bien à ces quelques notations l'ampleur de l'enjeu pour les années à venir.

Des projections ont été tentées (Rogers 1994) pour l'horizon 2025. Bien que ce genre de calcul reste aléatoire, il peut fixer des ordres de grandeur pour déterminer l'ampleur de la demande à venir. Trois variantes ont été envisagées :

• la première suppose que la dotation d'eau agricole en 2025 reste la même qu'en 1990 et que la dotation d'eau urbaine augmente sous le seul effet de la croissance des effectifs citadins.

• la seconde suppose une dotation d'eau agricole toujours inchangée mais la dotation d'eau urbaine augmente sous l'effet de la croissance des effectifs mais aussi de l'amélioration des revenus qui induisent une amélioration de la dotation par habitant.

• la troisième variante porte sur les bases de la seconde variante mais tient compte d'une amélioration de la dotation d'eau agricole de 3%/an.

• enfin dans les trois cas de figure envisagés la consommation d'eau industrielle reste identique : elle évolue en fonction de la croissance envisagée pour le PNB.

Les résultats pour les pays étudiés figurent dans le tableau suivant :

| milliards de m3 | Eau urbaine | Eau industrielle | Eau agricole | Total |
|---|---|---|---|---|
| situation 1990 | 11,8 | 7,8 | 157 | 177 |
| situation 2025 variable 1 | 30,2 | 19,4 | 143,5 | 193,1 |
| situation 2025 variable 2 | 45,3 | 19,4 | 143,5 | 208,2 |
| situation 2025 variable 3 | 45,3 | 19,4 | 416,4 | 481,1 |

*Tableau 7 : Prévisions de la demande d'eau à l'horizon 2025*

Alors que les ressources totales d'eau douce (qui on l'a vu ne sont pas toutes mobilisables!) sont de 297 milliards de m3, la demande dans les années à venir va les dépasser. Ces projections font apparaître trois données essentielles :

• le quadruplement de la demande urbaine paraît inéluctable mais cela ne signifie pas comme on l'écrit trop souvent que la ville deviendra le plus important consommateur.

• l'importance de la demande d'eau agricole : le devenir de l'agriculture irriguée est en question. Il paraît exclu que le Monde Arabe puisse consacrer de nouveaux volumes d'eau à ses campagnes au rythme des années qui viennent de s'écouler. L'autosuffisance alimentaire est véritablement utopique. La dépendance alimentaire ne pourra que s'accroître.

• enfin les ressources internes au Monde Arabe ne sont que de 132 milliards de m3 : il restera très largement dépendant de ses voisins mieux pourvus pour son alimentation en eau. Ce n'est pas assurément une position très confortable.

## 3.3 Une meilleure gestion du potentiel existant

Il y a sans doute peu à espérer sur la mobilisation de nouveaux volumes capables de faire face à une demande croissante. Si les prélèvements effectués actuellement apparaissent inférieurs aux ressources totales : 61% (*tableau 5*) cela ne signifie pas que l'on puisse pour autant les augmenter de façon significative en raison du régime des précipitations et des contraintes liées à la mobilisation des eaux. D'un point de vue technique des possibilités sont offertes mais les coûts sont prohibitifs pour l'instant tel est le cas du dessalement de l'eau de mer envisageable uniquement pour la fourniture d'eau urbaine. L'exploitation des aquifères fossiles incite à la prudence. Il n'est que le retraitement des eaux usées qui peut offrir quelques perspectives dans les années à venir. Dans la plupart des cas, le coût du retraitement revient moins cher que la plupart des infrastructures pour l'amenée des eaux vierges.

Dans ces conditions, la **mise en place de politiques d'économie des eaux est impérative.** De l'avis de tous les experts, on pourrait obtenir des gains tout à fait significatifs en luttant contre les multiples aspects de **gaspillage des eaux** que l'on peut constater. Le **domaine** **agricole** est de ce point de vue particulièrement concerné. Les économies pourraient être substantielles. En estimation globale, les États arabes consacrent à l'irrigation environ 11 000 m3 à l'hectare ce qui apparaît très élevé (160 milliards de m3 d'eau sont consacrés à l'irrigation de 14,7 millions d'hectares). On estime qu'il y ainsi une importante surconsommation. Les charges d'irrigation sont bien supérieures à ce qu'elles devraient être. La technique d'irrigation par gravité, la plus utilisée est très forte consommatrice. On pourrait avoir recours, beaucoup plus largement qu'on ne le fait à l'aspersion ou au goutte à goutte. L'irrigation par aspersion n'excède pas 21% des terres irriguées en Égypte, 16% au Maroc et 11% en Tunisie. En Syrie le recours aux techniques modernes d'irrigation (aspersion ou goutte à goutte ne concerne que 17 % des terres irriguées. La généralisation de l'aspersion pourrait réduire la consommation d'eau de 20% environ. Le goutte à goutte pourrait améliorer la productivité et réduire encore plus fortement la consommation d'eau jusqu'à 45%. Il faut aussi mettre en cause **l'inefficacité des équipements** : beaucoup de pertes au cours du transport dans les canalisations ou bien par évaporation. L'arrosage est trop souvent inefficace : on évalue à 70% la quantité d'eau qui est consommée ou s'évapore en pure perte. La Banque mondiale estime que dans cette région du Monde on dépense beaucoup d'eau pour une faible production. L'Afrique du Nord et le Moyen Orient consacrent 88% de leurs ressources en eau à l'irrigation alors que la moyenne mondiale se situe à 67%. Pour les experts de la Banque Mondiale, on pourrait multiplier les rendements avec deux fois moins d'eau. L'eau ainsi économisée permettrait de faire face en partie aux nouveaux besoins urbains et industriels. Une réduction de 15% du volume d'eau consacrée à l'irrigation permettrait de doubler la quantité disponible pour les citadins et l'industrie. En fin de compte toutes ces améliorations sont théoriquement possibles, l'obstacle principal est bien évidemment de changer les habitudes, les pratiques des paysans..... ce qui est une autre affaire, jamais réellement prise en compte par les experts. Le gaspillage de l'eau ne concerne pas que le seul secteur agricole. Dans les villes, les canalisations sont souvent très vétustes et mal entretenues. Les pertes d'eau en réseau sont impressionnantes : 30, 40 voire 50%.

Au total pour la plupart des experts des marges de progrès sont possibles. Avec une réduction de moitié des pertes d'eau en réseau et une meilleure efficience de l'irrigation, ils estiment que le Maroc pourrait réduire de 35 % sa consommation actuelle, l'Algérie de 32 %, la Tunisie de 27%, l'Egypte de 36%, le Liban de 27% !

La question **du prix de l'eau, de sa tarification est désormais posé en termes pressants par** de nombreux experts notamment les spécialistes de la Banque Mondiale dans la logique d'économie libérale du F.M.I.. Elle se heurte en Terre d'islam à une autre conception. L'eau

“don de Dieu” est considérée comme un bien naturel, gratuit, inépuisable. Pendant trop longtemps on est resté attaché à une conception minière de la ressource. Il est vrai que l'eau a été et reste souvent facturée à des prix dérisoires ou même distribuée gratuitement. Dans les stratégies de mobilisation de l'eau l'approche techniciste est partout privilégiée. Il s'agit de mobiliser des volumes sans cesse croissants pour tenter de satisfaire la demande. **L’eau est administrée en terme de distribution et non de conservation et d’économie**.

La rareté de la ressource, les perspectives d'une réelle pénurie ont conduit les organismes internationaux à agir auprès des autorités pour tenter de modifier cet état d'esprit. On pousse à une meilleure gestion de la ressource, à une vision plus économiste à une modification de la politique de tarification. Les mentalités évoluent peu à peu chez les responsables. La production et la distribution de l'eau sont désormais considérées comme une activité économique, mais de très gros efforts restent à faire pour faire accepter cette nouvelle conception auprès des utilisateurs. Des **ajustements tarifaires ont été tentés surtout pour la consommation urbaine**, ils sont beaucoup plus difficiles à imposer pour l'eau d'irrigation qui reste souvent gratuite ou bien facturée à un coût bien inférieur à son prix de revient. Ainsi au Maroc, l'eau d'irrigation est cédée 10 centimes le m3 alors que l'eau potable est vendue entre 2,40F et 6,5F. De même en Jordanie, si l'eau d'irrigation est passée de 5 à 25 centimes le m3, son prix reste encore inférieur de moitié au coût d'entretien du réseau alors que l'eau potable est livrée à 1,90F soit 1/3 de son prix de revient. Les Tunisiens ne payent leur eau d'irrigation qu'au 1/7e de son coût; dans les villes l'eau est cédée à 0,17 dinars alors que son prix de revient s'élève à 0,64 dinars. En Algérie on note aussi des décalages de même ampleur : le prix de revient de l'eau urbaine est estimé à un équivalent de 3 F le m3, elle est vendue à 0,70 centimes; pour l'eau d'irrigation l'écart est encore plus considérable : 1,80F et 0,10 centimes!

Dans les sociétés, notamment rurales, d’Afrique du Nord et du Moyen-Orient arabe où- à l’exception de la petite minorité profitant de la rente pétrolière- les revenus sont très bas, cette idée de créer en fait un marché de l’eau est assez irréaliste. Elle ne peut avoir qu’un effet limité : on ne peut pas résoudre les problèmes de l’eau par les lois du marché. L’eau a surtout une valeur d’usage qui dépend de son mode d’utilisation. Le marché ne peut vraiment jouer que pour l’eau potable qui ne représente qu’une proportion réduite de l’eau consommée. L’irrigation suppose de l’eau à faible coût et en grande quantité.

## La pollution : une menace qui ne cesse de s'aggraver

Au cours des dernières années les phénomènes de pollution des eaux ne cessent de s'aggraver et dans maintes régions la situation peut prendre l'allure d'une crise. Trois grands facteurs de pollution des eaux renouvelables sont à l'œuvre.

Les **effluents urbains et industriels** sont partout à l'origine d'une importante dégradation. C'est le résultat d'une croissance urbaine mal maîtrisée (de nombreuses constructions sont illicites). Une fraction importante de la population citadine n'est pas reliée à un réseau d'assainissement: entre le 1/3 et la moitié! Quand il y a un réseau, leseaux sont rarement traitées et les effluents urbains et industriels se déversent directement dans les rivières ou envahissent les nappes qui servent à l'irrigation ou à l'alimentation domestique. Les exemples peuvent être multipliés à l'infini. En Jordanie, la rivière Zarka reçoit les effluents d'Amman et de ses industries avant de s'accumuler dans le réservoir du roi Tahal et de se déverser dans le canal du Ghor qui permet l'irrigation de la vallée du Jourdain. La pollution y a atteint un degré gravissime à tel point qu'au cours des années 1988/90 l'utilisation des eaux du barrage a été interdite pour l'irrigation. Le cas de l'oasis de Damas (la Ghouta) est aussi très révélateur (Bianquis 1977). La Ghouta, alimentée par les eaux du Barada descendues de l'Anti-Liban était une oasis fonctionnant en parfaite harmonie avec la ville de Damas. En outre, une partie des eaux tombant dans l'Anti Atlas ne sont pas drainées par le Barada mais s'infiltrent et assurent la recharge d'une importante nappe phréatique de l'oasis. A partir des années 50, avec la croissance urbaine et industrielle, l'agglomération damascène (plus de 2 millions d'habitants actuellement) assure son alimentation en eau à partir du Barada et en puisant abondamment dans la nappe phréatique privant l'oasis d'une bonne partie de ses eaux. Surtout la ville et ses usines déversent directement dans le Barada leurs effluents tandis que le pompage excessif entraîne une salinité des eaux. La dégradation agricole de l'oasis est rapide et la Ghouta se meurt. Ce n'est que récemment en 1998 qu'une usine de traitement des eaux a été construite

Les **nappes phréatiques**, d'un accès facile et peu coûteux sont trop souvent **surexploitées** et elles peuvent connaître des rabattements importants. Cela entraîne une salinisation des eaux. Quand elles sont en position littorale il y appel au vide et intrusion de l'eau de mer : c'est le phénomène du biseau salé des hydrogéologues. Les exploitants pompent alors dans les puits les moins profonds de l'eau salée ou saumâtre. Le phénomène est connu à Gaza, en Palestine, où l'eau de la nappe devient peu à peu impropre à la consommation humaine. Le phénomène est aussi apparu sur le littoral libyen et algérien où la nappe de la Mitidja est pénétrée par le biseau salé. La même nappe de la Mitidja subit la pollution provenant des effluents industriels et urbains de l'agglomération algéroise. C'est une évolution catastrophique dont la solution passe par un contrôle rigoureux des nappes et une limitation des pompages. Des essais de recharge des nappes ont aussi été tentés.

- La **pollution d'origine agricole** prend aussi des dimensions inquiétantes. Le drainagedes eaux qui ont servi à l'irrigation est très souvent mal assuré. Ces eaux de drainage fortement chargées de nitrates et de sels divers polluent les nappes phréatiques. La multiplication des élevages industriels de volailles est aussi un important facteur de pollution.

# II. L'Égypte et le bassin nilotique

Pendant des millénaires, grâce aux eaux abondantes et limoneuses du grand fleuve, les Égyptiens ont construit leur pays. “L'Égypte est un don du Nil”, la formule a été sans cesse répétée depuis Hérodote. Elle reste toujours vraie mais, de nos jours, les eaux nilotiques sont aussi revendiquées par d’autres pays riverains. L’axe nilotique, le cœur du Monde Arabe, est devenu un enjeu géopolitique.

# 1 Le fleuve

## 1.1 Son cours

Le plus long fleuve du monde draine sur ses 6 671 km un immense bassin (2 870 000 km2) partagé de façon fort inégale entre 11 pays (Egypte, Ethiopie, Soudan, Burundi, Rwanda, Erythrée, Kenya, Ouganda, République Centre-Afrique, République démocratique du Congo, Tanzanie). Du sud au nord, il traverse successivement trois grands domaines climatiques : la zone équatoriale en Afrique orientale, la zone tropicale avec sa double variante humide et sèche et le désert saharien avant de se jeter dans la Méditerranée.

Il prend naissance en Afrique orientale dans la région des Grands Lacs. Deux sources se conjuguent pour donner naissance au Nil. De la source Kasumo (2 050 m.) sur les pentes du Mont Moujoumbiro au Burundi est issue la Kagéra qui, après avoir reçu un affluent ruandais de rive gauche se jette dans le lac Victoria (1 300 m.). Au sortir du lac, le Nil Victoria se fraye un passage parsemé de nombreuses chutes et rapides, traverse le lac Kyoga et après avoir franchi les impressionnantes chutes Murchinson, parvient au lac Mobutu (ex-Albert). La seconde source du Nil est localisée à plus de 4 000 m. dans les monts Ruwenzori en Ouganda. Nés de la fonte des glaces et du ruissellement des eaux de pluie, des torrents abondants viennent alimenter la rivière Semliki qui, après avoir traversé les lacs Edouard et George, mêle ses eaux au Nil Victoria au sortir du lac Mobutu. Le Nil, désormais devenu Nil Albert, au cours impétueux, dévale ensuite vers la cuvette soudanaise. Dans tout ce parcours, dans les montagnes d’Afrique orientale partagées entre la Tanzanie, le Burundi, le Ruanda, le Zaïre, le Kenya et surtout l'Ouganda, le fleuve est alimenté par d'abondantes **pluies de régime équatorial** (800 à 1200 mm/an).

En pénétrant au Soudan à Nimulé, il devient le Bahr el-Gebel (le "Nil de la Montagne") dont le cours est toujours parsemé de rapides. Ce n'est qu'à partir de Mongalla à 4 740 km de son embouchure que la pente diminue. Le fleuve s'étale alors dans la cuvette soudanaise en un lacis inextricable de lacs aux contours mal définis, de marais, de bras divergents, d'îles flottantes et d'amas de jacinthes d'eau : c'est la région des **Sudd**, alimentés aussi par de nombreux organismes fluviaux issus, en rive gauche, de la dorsale Nil-Congo. Cette zone mal drainée est immense : 300 000 km2 dans son extension maximale. L'**évaporation** y est intense : 14 milliards de m3 sont perdus chaque année : plus de la moitié du débit du fleuve.

Au nord de cette cuvette, le fleuve reçoit, en rive gauche, le Bahr el-Ghazal, une rivière permanente et devient, après le confluent, **le Nil Blanc** qui pénètre dans une **zone tropicale semi-aride puis aride**. Il est renforcé par les apports, en rive droite, des rivières éthiopiennes, torrentielles et limoneuses. Ces affluents vont à la fois considérablement augmenter le module du Nil et surtout en modifier le régime. La Sobat qui draine les eaux du Sud du plateau éthiopien et des marais du Sud-Est soudanais ne modifie que légèrement le régime du fleuve. A Malakal, au confluent avec la Sobat, le Nil n'est plus qu'à 386 m d'altitude et il lui reste 3 825 km à parcourir. Pendant 800 km jusqu'à Khartoum, le fleuve ne reçoit aucun affluent. Venu du lac Tana à 1840 m d'altitude, au cœur du massif éthiopien, le **Nil Bleu**, parvient au Soudan à travers des gorges vertigineuses et inaccessibles et fournit au Nil Blanc 59% de son débit. Ses impressionnantes crues d'été et ses énormes apports de sédiments boueux donnent au fleuve ses caractéristiques essentielles. Au delà de Khartoum (à 2 700 km de son embouchure) le Nil trace un gigantesque S qui, à travers le désert le conduit jusqu'à la frontière égyptienne. Il roule ses eaux tantôt bouillonnantes tantôt apaisées dans une vallée souvent étroite et encaissée (les 6 cataractes) mais dont les berges peuvent aussi s'élargir. Au delà de la 6e cataracte, la plus en amont, le Nil reçoit son dernier affluent **l'Atbara** (13% du débit), au régime comparable, mais au débit moins puissant que le Nil Bleu.

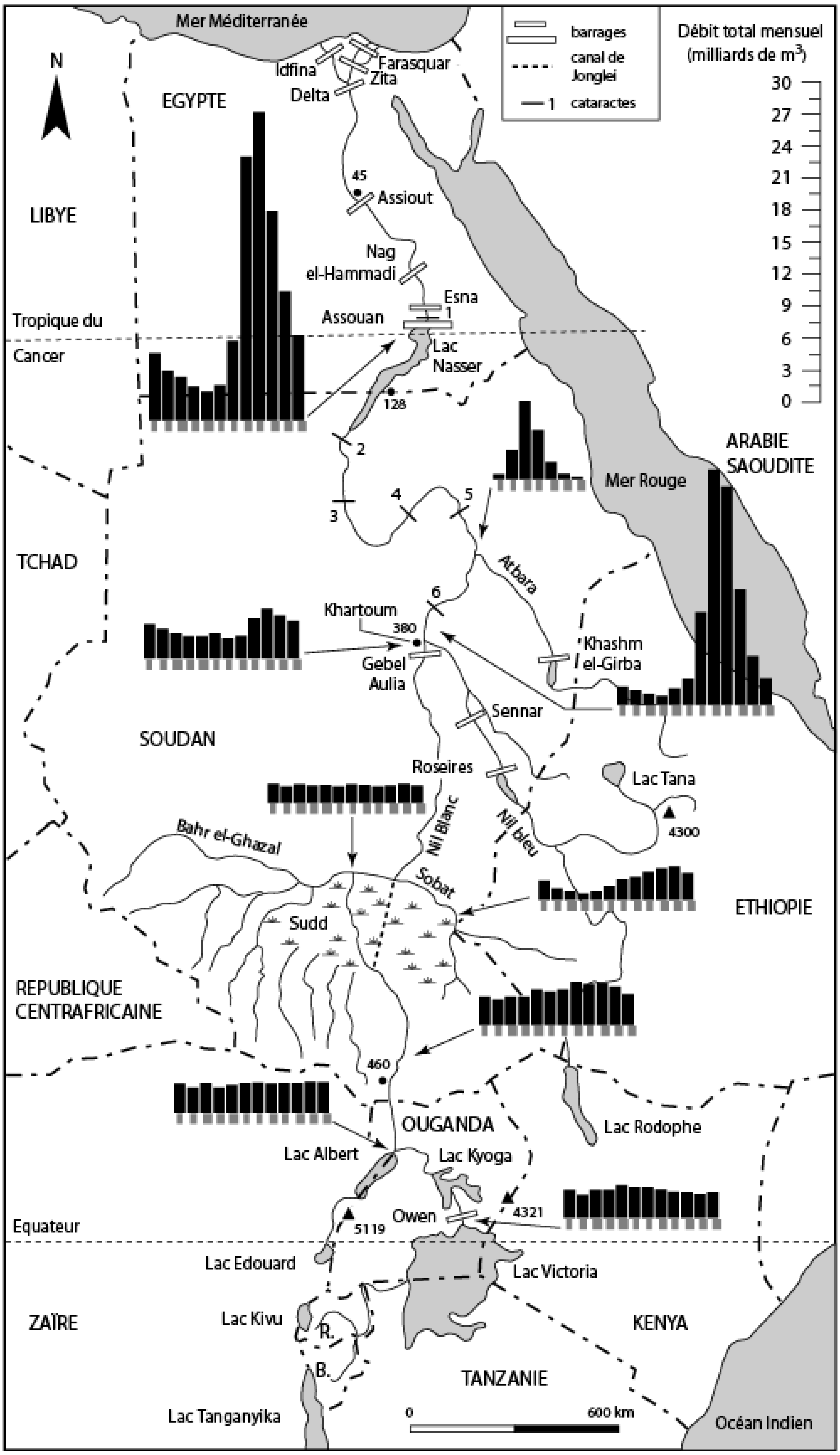

*Figure 8 – Le bassin du Nil : débit naturel et aménagement*

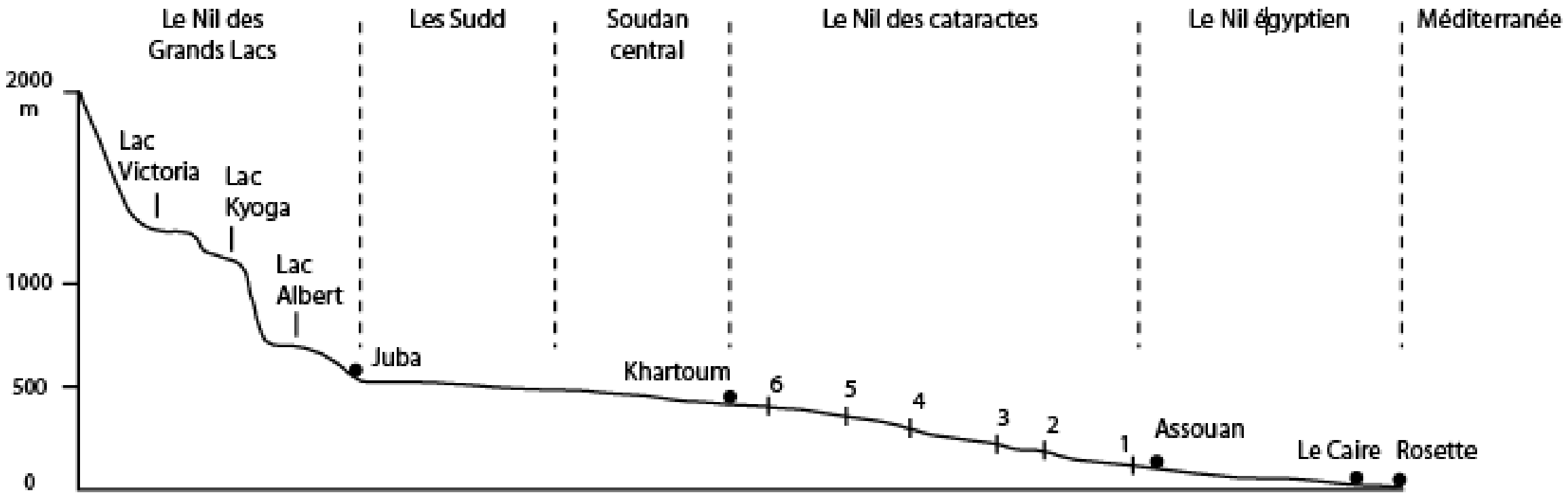

*Figure 9 – Profil longitudinal du Nil*

Le cours du Nil se prolonge par une ultime et **longue traversée du désert** : 2 700 km dont 1 250 km en territoire égyptien. La dernière cataracte est franchie peu après l'entrée en territoire égyptien à Assouan. Le fleuve est à 87 m d'altitude alors qu'il lui reste encore 1 180 km à parcourir. Entièrement domestiqué, puissant et majestueux, indifférent au désert, le Nil coule dans une vallée encaissée d'une centaine de mètres qui canalise la crue. Peu après le Caire, les eaux résiduelles -le 1/3 des eaux de la partie supérieure- se partagent entre 9 bras principaux enserrant un **delta de 24 000 km2** entre les bouches des branches de Rosette et Damiette.

# 1.2 Le régime (naturel) du Nil à Assouan

Le débit des eaux est impressionnant de régularité dans les rythmes des crues sinon dans leur volume. **Le module moyen annuel,** à l'enrée en Egypte**, est de 84 milliards de m3/an soit 2 700 m3/s** (il est faible si on le compare aux 6 000 milliards de m3 de l'Amazone aux 1 400 du Congo ou aux 600 du Mississippi!).Le Nil ne cesse de s'affaiblir pendant son parcours égyptien : il n'est plus que de 2000 m3/s au Caire.

Les **irrégularités de l'écoulement** d'une année à l'autre peuvent être importantes. Les extrêmes enregistrés ont été de 150 milliards de m3 en 1878 et 34 seulement en 1987 (soit un rapport de 1 à 4,4). Le fait essentiel est la **«crue» du Nil**, une véritable submersion de la vallée qui intervient à un bon moment : au cœur et à la fin de l'été. Le débit en mai est de 520 m3/s, il monte à 8 500 m3 en septembre (rapport 1 à 16). Avant les aménagements récents, l'élévation du plan d'eau était de 8 à 10 mètres à Assouan, 6 à 8 mètres au Caire. N'échappaient à l'inondation que les levées de terres et les collines artificiellement exhaussées sur lesquelles se blottissent encore la plupart des villages.

> "La vue de l'Égypte, dans le temps de l'inondation, est, sans contredit, un des spectacles les plus charmants au Monde. Du haut de ces montagnes (les rebords du plateau qui enserre la vallée) on découvre une vaste mer, d'où s'élèvent des villes et des bourgades sans nombre qui n'ont de communication entre elles que par des chaussées. La communication se fait par bateau et ce n'est pas un médiocre agrément de voir tout le pays couvert d'un nombre infini de ces maisons flottantes. La scène change à la fin de novembre. Alors les yeux se promènent sur une universelle qui au mois d'avril fait place aux moissons"
>
> De Maillet, consul de Louis XIV au Caire

La retombée des eaux est rapide (3 000 m3/s en novembre) puis elle se ralentit selon une courbe décroissante. Dans l'ensemble, le Nil se présente comme un fleuve aux pulsations simples, puissantes et raisonnablement régulières. La période où le débit est inférieur à la moyenne de l'année s'étend de la mi-novembre à la fin de la première décade de juillet, soit pendant près de 8 mois. La montée des hautes eaux s'effectue en une cinquantaine de jours tandis que la redescente en exige près de 80! Au cours des trois mois d'août, septembre, et octobre, la période des hautes eaux le fleuve charrie les 2/3 du total annuel. Enfin, pour l'essentiel, le Nil est alimenté par les **affluents venus, en rive droite, de l'Éthiopie : ils fournissent environ 86% du débit annuel et participent à 95% à la crue.**

| | | | |
|---|---|---|---|
| Pluies tropicales | Nil Bleu | 59 | 68 |
| Pluies tropicales | Sobat | 14 | 5 |
| Pluies tropicales | Atbara | 13 | 22 |
| Pluies équatoriales | Bahr el-Gebel | 14 | 9 |

## 1.3 Les aménagements successifs au cours de l'histoire

L'irrigation traditionnelle depuis l'Antiquité s'effectuait par les *hods* ou bassins d'irrigation qui fournissaient de l'eau durant 3 à 4 mois après l'inondation. Les *hods* sont des casiers à fond nivelé, alimentés par des canaux branchés sur le Nil. Leur superficie varie de 400 à 17 000 ha. Les *hods* sont normalement organisés en chaîne : ils dépendent les uns des autres et sont alimentés par gravité. Les bassins terminaux de la chaîne ont des superficies beaucoup plus vastes afin d'amortir les excédents éventuels de la crue. Tous ces travaux ne pouvaient être réalisés par de petites communautés villageoises : ils sont l'expression géographique d'un État centralisé capable de penser, de réaliser et de maintenir une œuvre gigantesque d'intérêt général. Aussi n'est-il pas étonnant que l'Antiquité ait connu une administration spécialisée. Il existait, une "administration des canaux des bas cantons", un "scribe chargé des canaux", un "chef du grand bras du Nil".

La submersion de chaque bassin dure de 40 à 70 jours, un temps nécessaire pour que l'eau imprègne le sol. C'est dans ces bassins que se dépose le limon avec la décrue.

Après le retrait des eaux on pratiquait :

- une culture d'hiver dite *chetoui* de céréales, fourrages ou légumineuses (les cultures directement liées à la crue et à la submersion de la vallée sont les cultures *nili*.).
- puis une jachère *charaqî* où la surface du *hod* mise à nu se craquelle en d'innombrables fissures de dessiccation constituant autant de voies d'accès pour l'eau limoneuse lors de la crue suivante.

Un grand nombre d'engins élévatoires permettaient d'atteindre la nappe phréatique le long des berges du fleuve et de prolonger l'irrigation longtemps après la décrue (*châdouf* ou puits à balancier, *sâqayé* ou chaîne à godets, vis d'Archimède etc..).

Au cours du XIXe siècle, de notables innovations techniques permettent d'étendre considérablement la période d'irrigation rendue nécessaire avec l'extension de la culture du coton. A partir de 1861 sont édifiés des barrages sans stockage afin de distribuer l'eau sur une aire plus vaste. Le delta est d'abord équipé : un double barrage à son apex commande les branches de Damiette et de Rosette. Plus en aval, sont édifiés les barrages de Zita, Idfina et Farasquour. Dans la vallée proprement dite sont achevés successivement les barrages de dérivation d'Assiout (1902), Esna (1909), Nag el Hammadi (1930). Ils permettent de relever le plan d'eau, d'améliorer ainsi l'alimentation des *hods*. et d'assurer un écoulement satisfaisant dans les canaux d'irrigation. Mais, les progrès sont insuffisants : une grande partie des eaux de crue s'écoule vers la mer pendant les hautes eaux et le déficit du printemps et de l'été n'est pas comblé. On vise alors à édifier des barrages-réservoirs.

Assis sur le granite de la première cataracte, le vieux barrage d'Asssouan est construit au début du siècle (1902). Les surélévations de 1912 et 1933 portent la retenue de 1 à 5 milliards de m3. Le barrage laisse filer le plus fort de la crue car sa capacité de stockage est trop faible, il ne retient que les eaux de fin de crue, faiblement chargées en limons ce qui permet une redistribution des eaux pendant la période de maigre. Cette capacité ne permet d'assurer ni une régularisation interannuelle ni l'extension souhaitée des superficies irriguées. C'est alors que sont envisagés des projets de correction conçus à l'échelle du bassin : des négociations difficiles sont conduites avec le Soudan. Le *Nile Water Agreement* de 1929 détermine les volumes d'eau attribués à chacun des deux pays (48 milliards de m3 pour l'Égypte, 4 pour le Soudan) et prohibe la construction sur le fleuve de tout ouvrage susceptible d'altérer ou de diminuer la valeur de l'écoulement naturel mesuré à l'entrée du territoire égyptien. Dans le cadre de cet accord sont construits deux barrages de régulation au Soudan. Le barrage de **Sennar**, sur le Nil Bleu, avec sa retenue de 800 millions de m3, permet de stocker, pendant les crues, l'eau dont a besoin l'agriculture soudanaise mais en contrepartie, le Soudan renonce à tout prélèvement pendant la période de maigre entre janvier et juillet. **Gebel Aulia** (3,5 milliards de m3), édifié sur le Nil Blanc (1937), compense les effets du barrage de Sennar : il retient la crue de la Sobat et permet de soutenir le débit du Nil en Égypte d'octobre à février.

Toutefois la mobilisation des eaux enregistre un relatif insuccès dans la mesure où l'aménagement n'est pas complètement maîtrisé. La crue n'est pas entièrement stockée, l'irrigation pérenne n'est pas généralisée.

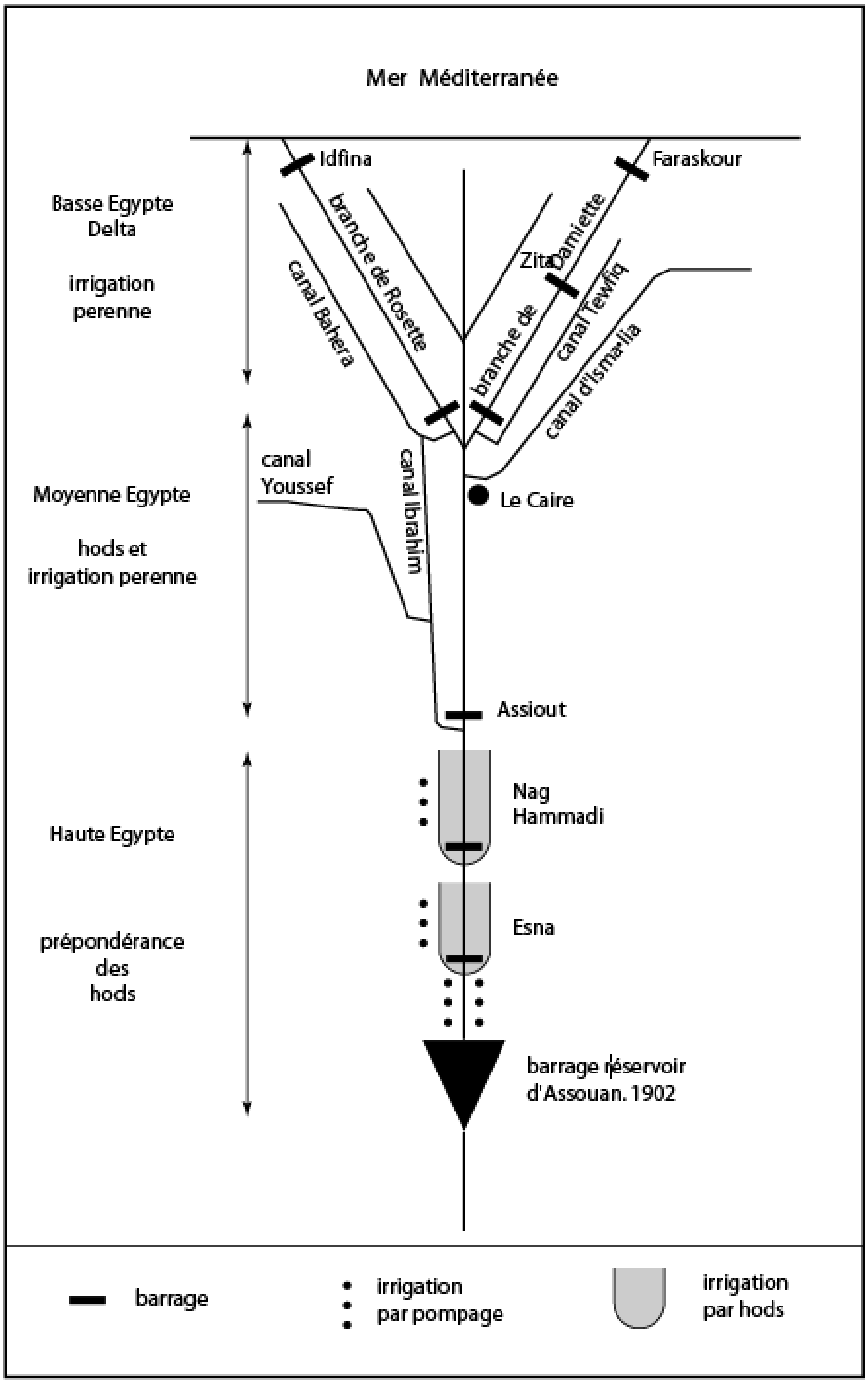

*Figure 10 – Le système d'irrigation en Egypte avant la construction du Haut Barrage d'Assouan (d'après Birot & Dresch, 1953*)

# 2. Le Haut Barrage d'Assouan et ses implications

De nombreux projets ont été élaborés tout au long du XXe siècle afin d’assurer une capacité de stockage autorisant une irrigation pérenne en Egypte. Le projet retenu, le haut barrage d'Assouan ou Sadd el ali a une portée qui dépasse de loin le plan économique. C'est une grandiose entreprise dans laquelle s'est identifié le nouveau régime mis en place après le renversement de la monarchie par les Officiers libres en 1952. En 1955, l'Egypte essuie un refus de financement de la BIRD car les Américains ne veulent pas apporter leur caution. En quête de moyens de financement, Nasser lie formellement la nationalisation du canal de Suez en 1956 à la réalisation du barrage (les droits de péage serviront à son financement). En 1958, l'URSS propose à l'Egypte son aide financière et technique. Dans cette période de décolonisation, le haut Barrage devient symbole de la lutte contre le néo-colonialisme et un chantier du socialisme.

## 2.1 Un barrage gigantesque

A l'époque où il a été construit il faisait partie des grands travaux mondiaux. Ancré sur une étroite gorge granitique, à 7 km en amont du premier barrage d’Assouan, Sadd el ali est un barrage poids de 3600 mètres de long, 980 mètres d'épaisseur à la base, 40 mètres au sommet, 111 mètres de haut. Son réservoir s'étend sur plus de 500 km en amont et constitue le lac Nasser de 12 km de large et d’une superficie de 6500 km2 (soit 11 fois celle du Léman).

**Les paramètres d’exploitation du barrage d’Assouan**

de la base à la cote 147 : stockage des sédiments sur 500 ans : 31 km3
de la cote 147 à 175 : capacité utile de 90 km3
de la cote175 à 182 : capacité de stockage des crues : 41 km3
196 : cote maximale

Le **volume de la retenue est impressionnant: 162 km3** : on voulait pouvoir stocker une quantité égale au double de la crue moyenne annuelle. Mais tout n'est pas utilisable. Il faut tenir compte notamment des nécessités de stockage des crues (41 km3), de l'envasement prévisible sur 500 ans (31 km3) si bien que le volume utile se réduit à 90 km3. Si, dans ce total, on tient compte de l'évaporation et des infiltrations, l'eau effectivement disponible n'est plus que de 74 km3/an partagés selon un accord international entre le Soudan (18,5 km3) et l'Egypte (55,5 km3). L’investissement est énorme : 450 milliards de livres égyptiennes.

La dérivation du Nil est réalisée en 1964, les travaux du barrage proprement dit s’engagent, il est inauguré en 1971 mais n’entre en pleine production qu’en 1975. Il permet grâce à sa capacité de stockage d’effectuer une péréquation entre années excédentaires et déficitaires. Le régime du Nil s’en trouve profondément modifié. En aval du barrage, un régime artificiel avec de très faibles variations interannuelles se substitue au régime naturel du fleuve marqué par l’importance de la crue d’août, septembre, octobre. Comme dans tous les grands ouvrages du genre, trois buts sont recherchés :

l’extension des superficies irriguées

l’amélioration de la navigation

la production d’électricité

## 2.2 Les effets d’impact

La réalisation du barrage a entraîné de très vives polémiques. On redoutait de nombreux effets négatifs. Après un quart de siècle d’exploitation, il est possible de faire le point avec sérénité.

Le Lac Nasser sous le climat désertique joue le rôle d’une gigantesque **machine évaporatoire**. On estime le volume évaporé à près de 12 km3-an. Cela entraîne une légère salinisation des eaux qui passe de 160 à 225 mg/litre. Pourtant ce n’est pas là la cause essentielle de la salinisation des sols que l’on a pu constater en Egypte. Avec le barrage, le niveau de l’eau est non seulement relevé mais reste stable toute l’année. La remontée des eaux entraîne des difficultés de drainage et conduit à une salinisation des terres notamment dans le delta. On estime que 2 500 000 feddans seraient atteints (1 feddan=0,42 ha) soit 35% de la superficie irriguée.

La polémique a aussi été vive à propos de la **perte des limons**, désormais retenus par le barrage et auxquels était liée la fertilité traditionnelle de la terre égyptienne. En fait, il s’agissait, en partie, d’un mythe. Sur les quelque 135 millions de tonnes de matières en suspension apportées annuellement par le Nil, 125 étaient étaient concentrées pendant les trois mois de la crue et gagnaient directement la mer. La masse déposée dans la vallée et le delta ne dépassait pas 16 millions de tonnes. Cette perte a été compensée par des apports d’engrais.

On constate une réelle **reprise d'érosion tout au long de la vallée**. Les eaux du fleuve, désormais beaucoup moins chargées, retrouvent un potentiel érosif d'autant plus que le débit reste relativement constant tout au long de l'année. Il est, pour l'instant, difficile de mesurer les effets de cette reprise d'érosion (1,7 cm/an pour certains), ils ne se feront sentir qu'à l'échelle séculaire. Par contre, les ruptures d'équilibre constatés sur le littoral sont bien effectives Il y a une réelle reprise de l'érosion sur les berges du delta où l'action de la mer n'est plus compensée par les apports alluviaux provenant du Nil. Le littoral recule alors que, jusque là, il avançait. Il existe également une menace d'invasion de l'eau de mer dans les nappes aquifères du delta. La situation paraît critique dans la région de Rosette mais on peut y remédier par des travaux qui, ni du point de vue technique ni du point de vue financier, sont hors de portée.

Autrefois la crue nettoyait les canaux d'irrigation. L'arrêt des mouvements naturels du Nil a favorisé l'accumulation des eaux stagnantes. Elle conduit à une **aggravation de la bilharziose** auprès des fellahs qui restent beaucoup plus longtemps, jambes nues, dans les canaux d'irrigation. Les bilharzies, vers parasites, ne peuvent eêtre détruites que par un assèchement prolongé, d'au moins trois semaines, des canaux d'irrigation.

Les conséquences de la construction du barrage sur la **pêche**, une activité en plein essor en Egypte, ont été sérieuses. Les prises des pêcheries traditionnelles ont notablement diminué dans le delta et en Méditerranée à partir de la fin des années 60. Mais parallèlement une compensation a été réalisée grâce à la très forte production de lac Naser qui compte pour 10% dans les prises nationales Il y a eu une nouvelle répartition des activités halieutiques sans que le bilan global soit affecté.

La **centrale électrique associée au barrage a considérablement amélioré le bilan énergétique du pays**. Avec un équipement de douze turbines, sa puissance installée de2,1 millions de kw autorise une production annuelle de 10 milliards de kw/h d'éléctricité (en 1960, l'Egypte n'en produisait que 7 !). Actuellement le barrage fournit 20% de l'électricité du pays. Il alimente directement trois de ses bases industrielles : l'usine d'engrais azotés d'Assouan, l'usine d'aluminium de Nag Hammadi et l'aciérie d'Hélouan.

Grâce au barrage, il est désormais possible de maintenir tout au long de l'année le débit du fleuve au seuil minimal pour permettre la navigation. **L'activité touristique**, fondée essentiellement sur les croisières sur le Nil a connu ces dernières années un essor considérable : plus de 300 bateaux-hôtels sillonnent le grand fleuve.

## 2.3 L'extension des terres cultivées : intensification et bonification

L'amélioration de la production agricole constitue l'objectif essentiel du barrage. L'évaluation des progrès enregistrés depuis 1975 peut s'effectuer selon deux critères. Il y a eu, en même temps, généralisation de l'irrigation pérenne et conquête de nouvelles terres agricoles (bonification).

**La généralisation de l'irrigation pérenne, "l'extension verticale" des terres selon l'expression des agronomes égyptiens est désormais acquise.** L'eau est disponible toute l'année et, sous le climat égyptien, il n'existe pas de contrainte thermique pour les cultures. Chaque parcelle peut porter deux parfois trois récoltes/an. Dans les statistiques égyptiennes, la superficie récoltée est le double de la superficie cultivée. Cette extension verticale des terres cultivées équivaut à la mise en culture de 1 000 000 de feddans nouveaux.

| Année | Superficie cultivée en feddans | | Superficie récoltée en feddans | |
|---|---|---|---|---|
| | totale en.000 | fed/hab. | totale en.000 | fed/hab. |
| 1907 | 5374 | 0,48 | 7595 | 0,69 |
| 1960 | 5900 | 0,22 | 10200 | 0,39 |
| 1980 | 5820 | 0,14 | 11135 | 0,26 |
| 1990 | 6223 | 0,11 | 12100 | 0,22 |
| 2000 | 7500 | 0,10 | 13900 | 0.18 |

*Tableau 10 : Évolution des superficies cultivées en Egypte de 1907 à 1998*

**"L'extension horizontale", c'est à dire les opérations de bonification, de conquête de nouvelles terres** est une entreprise beaucoup plus complexe. Elle est organisée par l'État qui traditionnellement détient la propriété des terres désertiques. En dépit d'une incontestable lourdeur bureaucratique des résultats positifs ont été acquis : 2 300 000 feddans ont été bonifiés jusqu'en 1993. Cette conquête des terres nouvelles a été irrégulière et a beaucoup varié selon les circonstances. Le mouvement se poursuit actuellement et on avance le chiffre de 2 800 000 feddans. Toutefois terre bonifiée ne signifie pas terre cultivée. La progression de la superficie cultivée est loin de suivre celle de la bonification. En 1998, on estime que la superficie cultivée n'est que de 7 000 000 feddans soit un gain de 1 100 000 feddans depuis 1960. Pourquoi un tel décalage?

Toutes les terres bonifiées n'ont pas été distribuées. Il y a des terres qui ne sont pas destinées à la culture mais à des infrastructures : routes, canaux. Surtout il s’écoule un grand laps de temps entre la conquête de la terre et la mise en culture. La terre conquise doit être amendée, un sol agricole doit être constitué. Ces opérations peuvent exiger plusieurs années. Il faut, en outre, défalquer du bilan les terres agricoles perdues en raison de la salinisation et de la progression de l'urbanisation. La croissance des villes, notamment celle du Caire, qui ne s'effectue pas sur les espaces désertiques mais dans la vallée, les extensions villageoises, l'argile utilisée en quantité par les briqueteries réduisent aussi le terroir agricole. D'une façon globale on estime que depuis 1960 c'est, sans doute, 850 000 feddans qui ont été "mangés" par une croissance urbaine et villageoise mal contrôlée.

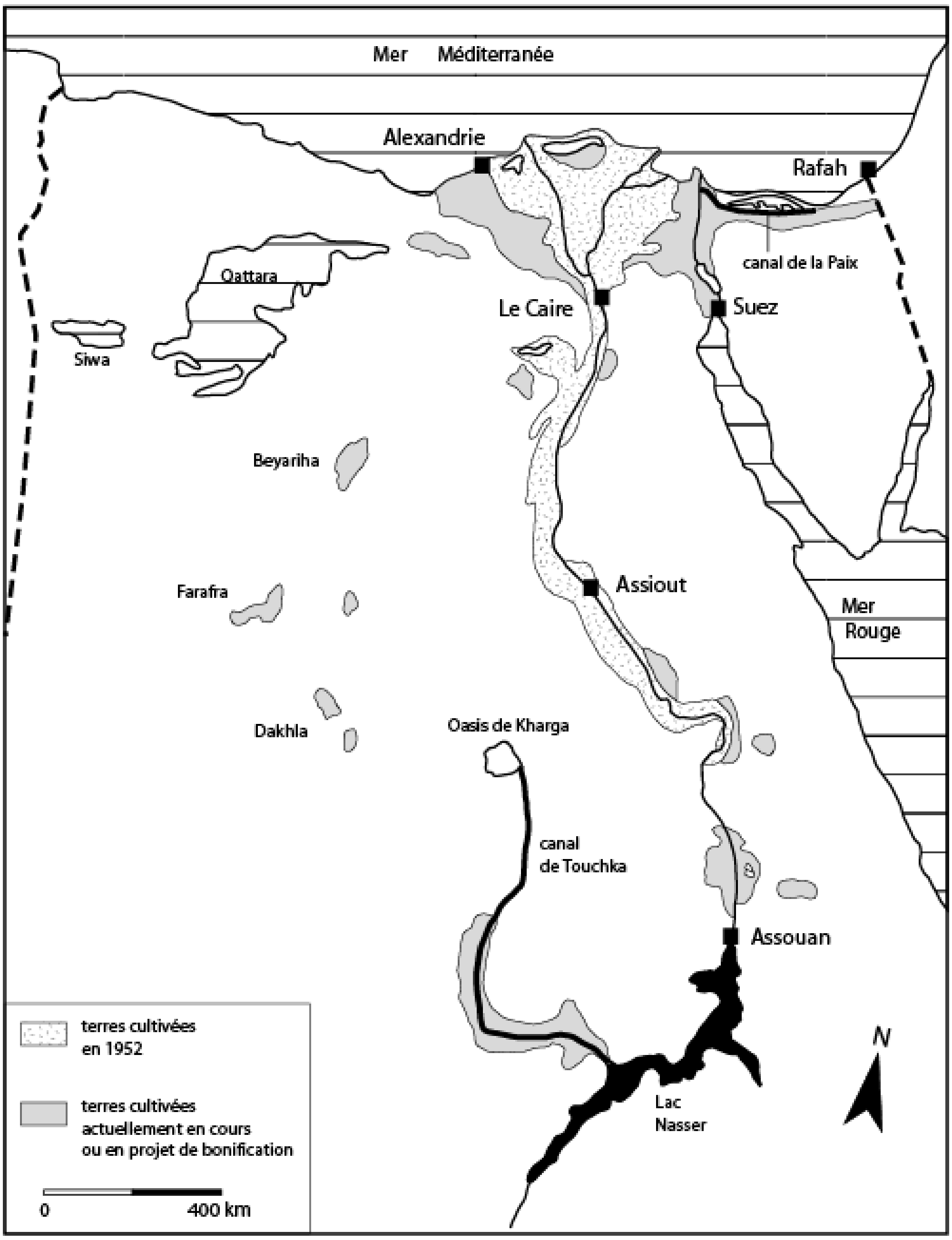

*Figure11 – Espace cultivé ou en cours de bonification en Egypte (d'après Ayeb, 1998)*

**L’utilisation des terres bonifiées entre 1952 et 1970**

Sur les 912 000 feddans bonifiés entre 1952 et 1970 :

- •139 000 sont utilisés pour les infrastructures (15%)
- •sur les 773 000 restants (85%)
  - •40% produisent suffisamment pour faire vivre les colons
  - •28% sont cultivés à perte
  - •12% sont abandonnés
  - •20% ne disposent pas des infrastructures nécessaires

La distribution des terres a permis la mise en place d'une **politique de colonisation pendant la période nassérienne** : 400 000 feddans ont été distribués à 110 000 familles installées dans des lots de 2 à 5 feddans. Les colons sont regroupés en coopératives de bonification. Depuis la décennie 1970, cette politique est abandonnée et les terres bonifiées sont mises en valeur dans le cadre de grandes exploitations modernes privées ou parapubliques

La répartition géographique fait apparaître la prépondérance du delta, notamment sur ses marges et au nord. En Haute Égypte, la majeure partie des terres bonifiées appartiennent à la Nouvelle Nubie où ont été recasés 55 000 Nubiens dont les 36 villages ont été submergés par le lac Nasser. Deux périmètres ont été aménagés (150 000 feddans) à Esna et surtout Kom Ombo. Le transfert de population a été effectué dans le cadre d'un projet global qui devait améliorer les conditions de vie matérielle et sociale des Nubiens. La bonification a, pour l'essentiel, permis d'étendre les superficies cultivées dans le delta : soit par aménagement après désalinisation des marécages soit par conquête de nouvelles superficies sur le désert. Les terres bonifiées en 1991 représentent 20% de la superficie du delta (environ 1 300 000 feddans récupérées pour les 2/3 sur le désert et 1/3 sur les marécages) mais 1 000 000 de feddans seulement sont réellement cultivés.

## 2.4 Sortir de la vallée du Nil : les nouveaux projets

**L'extension des nouvelles terres se poursuit actuellement à l'est du delta**. Un projet d’envergure déjà ancien est en chantier. Le canal as-Salam, canal de la paix, acheminera vers le **Sinaï** l'eau captée de la branche de Damiette. Un don saoudien, en 1990, a permis d'entreprendre les travaux de la première phase qui achemine l'eau sur 84 km jusqu'au canal de Suez et a autorisé la conquête de 200 000 feddans. La deuxième phase du projet est en exécution depuis 1994. Après un passage en siphon sous le canal de Suez, le canal de la paix devenu canal Cheikh Jaber (le Fonds koweïtien de développement finance les travaux!) conduira les eaux du Nil, sur 155 km, jusqu'à El Arish. Complétant le dispositif, en parallèle au canal de Suez, un second canal est en cours d'exécution. Il est financé par l'émirat d'Abou Dhabi et porte le nom de Cheikh Sayed. Ces deux ouvrages devraient permettre la bonification et l'irrigation de 500 000 feddans supplémentaires (dont 400 000 pour le Nord Sinaï). L'innovation tient moins à la conquête proprement dite de nouvelles terres qu'aux techniques qui seront mises en œuvre pour leur utilisation : goutte à goutte pour les cultures arbustives, serres pour les productions maraîchères, aspersion pour les céréales devraient permettre une répartition optimale des ressources en tranchant radicalement avec les techniques de submersion des parcelles qui prévaut sur les vieilles terres de la vallée du Nil. Plus de la moitié des terres seront attribuées à de grandes compagnies d'investissement sous la forme de lots de plus de 500 feddans. Le projet national pour le développement du Sinaï prévoit le peuplement de la Péninsule par 3 millions de personnes (elle ne compte actuellement que 350 000 habitants) et la création de 27 agglomérations nouvelles. L'Égypte dispose-t-elle des quantités d'eau nécessaires pour mener à bien l'opération? Elle nécessitera 4 milliards de m3 d'eau : 2 proviendraient du Nil, 2 de la récupération d'eaux de drainage que l'on mélangerait à l'eau douce. Le pays pourrait, semble-t-il répondre à cette nouvelle demande mais s'ajoute plus récemment la mise en chantier d'un nouveau projet, celui de la Nouvelle Vallée, qui posera, sans aucun doute, la question du volume d'eau disponible.

Le **projet de la Nouvelle Vallée** est avancé depuis plus de 25 ans. Il s'agit de mettre en culture toute la zone où s'égrène un chapelet d'oasis à l'ouest de la vallée du Nil (El Kharga, El Dakhla, El Farafra, El Bahira, Swa) qui jalonnent un ancien tracé du Nil. C'est peut être 2 à 3 millions de feddans qui pourraient être conquis à un horizon de 20 ou 30 ans. Actuellement ces oasis sont alimentées par une nappe souterraine, elle est insuffisante et il faudrait la recharger par un apport des eaux du Nil pour conduire l'entreprise à bien. On songe aussi à l'utilisation massive de l'eau de l'aquifère de Nubie. Le projet avance. Les travaux de creusement du canal de Touchka ont débuté en 1997. Ce canal relie le lac Nasser par une prise d'eau à 170 km au sud d'Assouan, à une dépression située à 25 km où sera aménagé un lac artificiel de 120 milliards de m3, point de départ du canal définitif vers les oasis de Kharja et Dakhla. Dans une première étape, on songe à mettre en valeur 500 000 feddans qui exigeraient un apport d'eau en provenance du Nil. En l'état actuel l'Égypte ne dispose pas de suffisamment d'eau pour mener à bien tous ces projets et doit impérativement songer à la mobilisation de nouvelles ressources. Le projet de la Nouvelle Vallée exigerait s'il était conduit à bien un prélèvement de 5 km3

Ces deux projets posent la question de leur crédibilité. Où trouver le financement de travaux gigantesques ? Où trouver les 9 ou 10 km3 d'eau nécessaires à leur utilisation. Les perspectives les plus réalistes laissent entendre que l'Egypte a peu de chances d e voir sa dotation en eau augmenter à supposer même que cette dotation ne vienne pas à diminuer !

**Consommation annuelle d'eau**

Eau disponible : 63,1 milliards de m3 dont
55,5 du Nil,
4,7 de la réutilisation des eaux de drainage
2,9 de pompages dans la nappe phréatique du delta et de la vallée

Consommation environ 62 milliards de m3
eau potable : 3,7 milliards de m3
eau industrielle : 2,8 milliards de m3
eau pour l'irrigation : 51,6 milliards de m3
maintien des eaux du fleuve fin janvier-début février (nettoyage des canaux) p production d'électricité et la navigation : 4 milliards de m3
Besoins en eau en 2025 : 118 km3 dont 95 pour le secteur agricole

# 3. Le barrage insuffisant

Ce projet gigantesque, construit à l'échelle du pays entier il y a un quart de siècle, se révèle aujourd'hui insuffisant face à la croissance démographique et aux risques de sécheresse.

## 3.1 Les gains du barrage annihilés par la croissance démographique

Lors du lancement de la construction du barrage d'Assouan, l'Égypte était un pays de 30 millions d'habitants, en 2009, on dénombre 78 millions d'Égyptiens, ils seront vraisemblablement 96 millions en 2025. Selon les projections démographiques de l'ONU, la population pourrait se stabiliser à 122 millions d'habitants autour de 2040/2050. Avec un tel rythme d'accroissement, en dépit des progrès réalisés, la superficie agricole par habitant se réduit d'année en année telle une peau de chagrin! Entre 1960 et 1998, elle a diminué de moitié

: de 0,22 feddan, elle est passée à 0,11! En dépit des incontestables progrès de l'agriculture, et d'une orientation vers les cultures vivrières aux dépens du coton le pays ne peut faire face à une demande alimentaire en très forte augmentation. Alors qu'il s'engageait dans ce gigantesque effort de bonification, de conquêtes de terres nouvelles, la dépendance alimentaire ne cessait d'augmenter. Actuellement, le secteur agricole ne peut satisfaire que la moitié de la demande. Le recours à de massives importations (3 à 4 milliards de $ par an) est un facteur de déstabilisation de la balance commerciale (plus du quart du montant total des importations). L'Égypte est un très gros importateur de céréales (3e au monde) : 8 millions de tonnes en moyenne annuelle au cours de la présente décennie. De 1985 à 1990, les importations moyennes de produits alimentaires ont été de 10 millions de tonnes/an soit 200 kg/hab/an! L'augmentation des volumes importés est spectaculaire : en 1990, l'Égypte achète 4 fois plus de céréales qu'en 1960 pour un coût dix fois supérieur!.

| en % | 1960 | 1986/87 | 1988/89 | 1989/90 | 1990/91 |
|---|---|---|---|---|---|
| Blé | 66 | 22 | 31 | 37 | 45 |
| maïs | nd | 66 | 75 | 77 | 89 |
| Riz | 124 | 100 | 101 | 105 | 107 |
| sucre | 118 | 52 | 61 | 65 | 68 |
| huile végétale | 105 | 31 | 20 | 23,5 | 44,5 |
| poulet | nd | 75 | 68 | 78 | nd |

*Tableau 11 : Évolution de quelques ratios d'autosuffisance alimentaire en Egypte*

Autant dire que l'effort doit être poursuivi mais, en même temps, de sérieuses alertes apparaissent pour les disponibilités en eau. Depuis la mise en eau du barrage d'Assouan, les disponibilités du pays n'ont pas augmenté. En raison du croît démographique, un Égyptien qui disposait de 1604 m3 d'eau/an, ne peut plus compter que sur 790 m3 en 2007 et seulement 605 en 2025 !.

## 3.2. L'alerte de la récente sécheresse des années 80

De 1979 à 1988, l'Afrique orientale a connu des précipitations inférieures à la moyenne. Le **débit du Nil** s’en et ressenti : il a été **inférieur de 33% à la moyenne** ! Les apports du Nil ont été insuffisants et il a été nécessaire de puiser dans les réserves du lac Nasser. En 1984, la crue, sans les réserves du lac Nasser, aurait permis l'irrigation de 5 millions de feddans sur les 11 récoltés. En début 1988 il ne restait que 7 milliards de m3 de réserves utiles et le niveau du réservoir s'était abaissé jusqu' à la cote 150! Par bonheur les fortes pluies soudanaises de l'été sont venues mettre un terme à cet enchaînement de la sécheresse. Il n'en demeure pas moins que le problème obère l'avenir : on note en effet une baisse tendancielle de l'alimentation du Nil. De toute évidence en raison de la pression démographique et des variations climatiques, le barrage actuel ne peut plus assurer la sécurité de l'approvisionnement en eau du pays.

# 4. L'impossible aménagement de l'axe nilotique

## 4.1 Le plan Hurst

Cette situation n'a pas surpris les techniciens. Avant même d'entreprendre la construction du barrage, on pensait que, très rapidement, la retenue deviendrait insuffisante. Le calcul théorique montre que la capacité du réservoir capable à la fois de fournir un volume d'eau inchangé pour une année très déficitaire et de retenir la crue centennale devrait être de 300 milliards de m3 en tenant compte des pertes normales dues à l'infiltration et à l'évaporation. Ainsi avec les 162 milliards de m3 du lac Nasser on se trouve loin du compte. Aucun site ne convient pour l'établissement d'un tel ouvrage dans la vallée du Nil. Il est donc nécessaire d'aménager l'ensemble du bassin. Les propositions des techniciens ont été formulées dans ce qu'il est convenu d'appeler le plan Hurst :

- sur le Nil Blanc : le réservoir pluriannuel du Lac Albert en Ouganda et, au Soudan : le réservoir à capacité saisonnière de Nimule, le creusement du canal de Jongleï, le barrage de Gambela
- sur le Nil Bleu **:** le réservoir de stockage pluriannuel du lac Tana en Éthiopie, l'élévation du barrage de Rosières au Soudan
- sur le Nil en aval de Khartoum : le barrage de Mérowé à l'emplacement de la 4e cataracte

Mais la situation politique et économique est loin d'être favorable à la réalisation d'un tel ensemble qui suppose l'accord de 9 pays riverains (300 millions d'habitants) et notamment celui des trois pays les plus concernés : Égypte, Éthiopie et Soudan (160 millions d'habitants). En outre, la discussion est rendue d'autant plus difficile que le droit international est très imprécis pour le partage des eaux des fleuves ou des nappes souterraines entre deux ou plusieurs États souverains.

**Droit international, grands fleuves et pays riverains**

En droit international, il n'existe pas de régime uniforme qui organise le partage de l'eau des fleuves ou des nappes souterraines entre deux ou plusieurs États riverains. Les fleuves interna sont définis uniquement comme des cours d'eau qui, dans leur cours navigable, traversent des territoires dépendant de plusieurs États. Seule la navigation international régie par la Convention de Barcelone (1921).
Ces règles sont aujourd'hui insuffisantes et inadaptées en raison de la multiplication des usages de l'eau, des nouvelles technologies, du développement économique, de la concentration urbaine.
En 1966, l'Association de Droit international réunie à Helsinki a défini la notion de "bassin de drainage international" comme une zone géographique s'étendant sur deux ou plusieurs États et déterminée par les limites de l'aire d'alimentation du système hydrographique, eaux de surface et eaux souterraines comprises, s'écoulant dans un collecteur commun.

En 1970, l'Assemblée générale de l'ONU a chargé la Commission du Droit international d'entrepren «l'étude du droit relatif aux utilisations des voies d'eaux internationales à des fins autres que la navigation, en vue du développement progressif et de la codification de ce droit». Les travaux sont difficiles et n'ont pas encore abouti. Un premier projet a été présenté en 1992 avec les principes suivants

- La définition du bassin de drainage : "un bassin de drainage international est une zone géographi s'étendant sur deux ou plusieurs États et déterminée par les limites de l'aire d'alimentation du systè hydrographique, eaux de surface et eaux souterraines comprises, s'écoulant dans un collecteur commun"
- "Tout État du bassin a droit sur son territoire à une part raisonnable et équitable à l'utilisation avantageuse des eaux du bassin de drainage international"
- "La détermination de ce qu'est une part raisonnable et équitable se fait à la lumière de tous les facteurs pertinents dans chaque cas particulier"

En l'absence de toute convention internationale ces principes sont l'objet d'interprétations qui varient en fonction des intérêts des États. C'est ainsi qu'Addis Abeba invoque l que le Nil n'est pas navigable sur toute sa longueur pour lui refuser le statut de fleuve international l'inverse, ne fait pas de doute pour l'Egypte.

## 4.2 La position éthiopienne : une sérieuse menace à moyen terme

L'État éthiopien considère comme nuls et non avenus les accords de partage des eaux du Nil entre le Soudan et l'Égypte de 1959. Il refuse d'envisager une quelconque politique hydraulique commune entre les autres États riverains. Il ne reconnaît pas au Nil le statut de fleuve international et tend à exploiter unilatéralement le Nil Bleu et les affluents qui prennent naissance sur son territoire. La position éthiopienne peut se résumer ainsi : ne rien accepter des États de l'aval, ne rien leur accorder tant que le pays n'a pas réalisé ses propres projets. **L'Éthiopie** fait valoir que, si plus de 80% des débits du Nil égyptien se forment sur son territoire, elle **n'utilise actuellement que 0,3% du débit** correspondant. Elle fait tout pour empêcher la réalisation de projets qui se transformeraient en droits acquis. Elle reste opposée à toute entreprise d'aménagement hydraulique commune du Nil tant qu'un nouveau partage des eaux n'est pas négocié entre les trois États principaux de la vallée, (Egypte, Soudan, Ethiopie).

Le pays qui compte 82 millions d'habitants en 2009 et frôlera en 2025 les 110 millions et 150 en 2050 (il sera alors plus peuplé que l'Égypte), est confronté à de très graves séquences de sécheresse. Pour faire face à ce défi menace, le développement des superficies irriguées apparaît inévitable. Le potentiel d'irrigation des rivières éthiopiennes est estimé à 3 500 000 hectares, seules 4,6 % de ces surfaces sont irriguées ! Les projets actuels retenus par l'Éthiopie prévoient la bonification d'environ 90 000 ha dans le bassin versant du Nil Bleu grâce à l'aménagement du réservoir de stockage pluriannuel (7,5 milliards de m3) du lac Tana. A plus long terme, les prévisions sont beaucoup plus impressionnantes : elles portent sur 1 500 000 hectares! Irriguer une telle superficie supposerait un diminution en aval d'environ 9 milliards de m3 annuels ce qui diminuerait de 16% les volumes reçus annuellement par l'Egypte. Mais rien n'est entrepris pour des raisons politiques et économiques. Si ces plans venaient à exécution ils exigeraient un prélèvement important sur les eaux du Nil ce qui serait une catastrophe pour le Soudan et l'Égypte.

Pour l'instant l'Ethiopie voit toutes ses demandes de financement international de barrages systématiquement refusées par la Banque Mondiale ou le FMI où l'influence égyptienne fonctionne à plein.

## 4.3 La position soudanienne

Le Soudan est lié à l'Égypte par l'accord de 1959 négocié après la décision de construire Assouan. Dans ce cadre, le Soudan dispose d'une dotation de 18,5 milliards de m3 ce qui lui a permis de construire, dans la décennie 60 deux importants barrages réservoirs. **Roseires**, achevé en 1966, sur le Nil Bleu a une capacité de retenue de 3 milliards de m3. Il permet une extension de l'irrigation dans la Gézira et la production d'électricité pour alimenter Khartoum. **Khashm el Girba**, sur l'Atbara a été achevé en 1964. Sa retenue de 1,2 milliards de m3 autorise l'irrigation du périmètre de la Nouvelle-Halfa où se réinstallent 50 000 Nubiens soudanais chassés de leurs terres par la montée des eaux du lac Nasser et des tribus nomades de la région poussées à se sédentariser.

Ces deux ouvrages ont entraîné, l'extension des superficies irriguées qui couvrent actuellement au Soudan 4 millions de feddans (soit 57% de la superficie irriguée en Égypte). Toutefois ces deux barrages édifiés sur des fleuves abyssins très limoneux s'envasent très vite. Entre 1966 et 1975 la profondeur du réservoir de Roseires est passée de 50 à 17 m.! Leur capacité d'emmagasinage et de production électrique s'en trouve fortement réduite. L'accord de 1959 a laissé au Soudan un surplus inutilisé de 1,5 milliards de m3 jusqu'en 1977. Depuis cette date, la création de nouveaux périmètres et la sécheresse récente ont réduit à néant les disponibilités soudanaises. Comme l'Égypte, le Soudan est à la recherche de nouvelles disponibilités.

L'accord de 1959 avait prévu la réalisation de travaux ultérieurs pour répondre à ce besoin. Les travaux doivent être financés conjointement par les deux États, avec partage des volumes d'eau éventuellement récupérés. Dans cette perspective, refait surface en 1972, le vieux projet de mise en valeur de la cuvette du Haut Nil (les Sudd). Le premier élément est le creusement du canal de Jongleï sur 360 km de Bor à Malakal. En coupant la grande boucle du Nil Blanc dans les marais de Bahr el Ghazal, le canal, favoriserait l'écoulement et limiterait les pertes par évaporation de moitié : plus de 4 milliards de m3 pourraient être récupérés dont 2,3 pour le Soudan et 1,9 pour l'Égypte. En outre, il faciliterait la navigation nilotique et la circulation sur l'axe nord-sud de Khartoum à Juba. Commencés depuis 1978, les travaux ont été arrêtés en 1983 aux 2/3 du parcours en raison de la guerre civile qui règne dans le sud du Soudan : il reste 93 kilomètres à creuser.

Pour les populations du Sud Soudan, en opposition avec le régime de Khartoum, l'aménagement du Haut Nil porte, en lui-même, un projet géopolitique redoutable. Un axe de communication facile (voie navigable, grande route, voie ferrée) entre le Nord et le Sud est d'abord perçu comme un axe de pénétration à usage militaire. Les populations s'y opposent farouchement soutenues par l'opposition sudiste et les milieux écologistes internationaux. Certains y voient le prélude à l'arrivée de milliers de colons venus du Nord; d'autres craignent que l'assèchement partiel des marais et l'obstacle formé par le canal aux migrations saisonnières des troupeaux ne viennent bouleverser les modes de vie traditionnels; d'autres, enfin, refusent de voir une nouvelle fois, les ressources du Sud exploitées au profit exclusif du Nord du Soudan et de l'Égypte. Derrière le projet du canal de Jongleï, se trouve l'Égypte. Derrière l'arrêt des travaux de ce même canal se trouve l'Éthiopie. Entre les deux le Soudan partagé : le Nord historiquement lié à l'Égypte, le Sud par nécessité allié à l'Éthiopie. Le Nil voué à assurer la liaison entre les États du bassin hydrographique est devenu entre eux un objet de division.

La mobilisation de nouvelles ressources paraît donc exclue dans l'immédiat. Si les différents projets prouvent que le potentiel du Nil au Soudan n'est pas encore entièrement exploité, ses limites sont nettement perceptibles. On peut, tout au plus, compter sur 20 milliards de m3 à partager avec l'Égypte. On mesure dès lors l'ampleur du problème égyptien. En raison de l'hostilité éthiopienne, de l'arrêt de toute nouvelle mise en chantier au Soudan, l'Égypte a peu de chances d'augmenter ses ressources en eau. Les **tensions** seront encore plus fortes avec les **autres pays amont le jour** où l'Ouganda, la Tanzanie et le Kenya qui enregistrent une forte croissance démographique se décideront à tirer un meilleur profit des eaux du lac Victoria où le Nil Blanc prend sa source. Des projets portent, en effet, sur l’irrigation de 130 000 hectares en Ouganda, 57 000 au Kenya et 200 000 en Tanzanie! L’Erythrée, de son côté, a des projets sur l’Atbara.

Il ne reste qu'une voie pour l'avenir égyptien dans le court terme : économiser l'eau dont le pays dispose.

# 5. Une gestion plus rigoureuse des eaux

## 5.1 De substantielles économies d'eau sont possibles

Tous les rapports techniques le montrent à l'évidence, l'Égypte peut faire de substantielles économies dans sa consommation. L'effort doit porter sur plusieurs registres.

"Un genou bien noyé, c'est un champ bien irrigué", ce vieux proverbe égyptien valable pour les cultures de crue est appliqué à la situation actuelle où l'eau est en permanence à la disposition du fellah. Les paysans utilisent l'eau sans aucune limite. En première approche, très grossière, 51 milliards de m3 sont actuellement consacrés à l'irrigation pour une superficie de 7 millions de feddans : soit, en principe, 17 000 m3/an/ha, une quantité énorme! Les études de terrain montrent que les charges d'irrigation réelles se situent entre 7 et 8 000 $m^3$/an alors que, le plus souvent, les besoins sont estimés entre 4 et 5 000 m3. Le nombre de rotations d'irrigation est par ailleurs très élevé : entre 10 et 16/an. Il a tendance à augmenter avec la généralisation de l'usage des motopompes (souvent achetées grâce à l'argent envoyé par les émigrés du Golfe) encouragé par le très faible coût du gazole. Il y a donc un évident gaspillage de l'eau. **Une** **étude de la Fao pour l'année 1996 a montré que les besoins théoriques de l'agriculture s'élevaient à 28 km3, or, cette même années, il a été prélevé 48 km3**. Plus grave encore : cet excès d'irrigation se traduit par une dilapidation du patrimoine : elle entraîne une remontée des nappes et une forte salinisation des sols.

**Excès d'irrigation : l'exemple de l'oasis du Fayoum**

superficie cultivable : 365 000 feddans
eau d'irrigation mesurée à l'entrée des parcelles : 2,4 milliards de m3
6 600 m3/feddan ou 15 800/ha = 1580 mm pluies/an!
15% des sols perdus par remontée de la nappe et salinisation

Il faut aussi prendre en compte les techniques utilisées. Sur les vieilles terres de la vallée on pratique uniquement l'irrigation par l'inondation des parcelles. Le recours à l'aspersion et surtout au goutte à goutte permettrait d'importantes économies en réduisant des 2/3 les consommations d'eau. Mais cela suppose l'adoption par les fellahs égyptiens de nouvelles technologies qui ne sont, pour l'instant, pratiquées que sur les terres récemment bonifiées. Actuellement l'aspersion n'est utilisée que sur le 1/5 des terres seulement.

On peut aussi, dans un autre ordre d'idées, mettre en cause le système de culture. Les paysans égyptiens répugnent à pratiquer des cultures dont le prix est fixé par l'État : le blé par exemple. Par contre, les superficies consacrées aux cultures dont le prix est libre ont tendance à augmenter plus qu'il ne faudrait! C'est le cas du riz et du bersim (trèfle d'Alexandrie) très exigeants en eau. Les autorités estimaient à 800 000 feddans les superficies rizicoles nécessaires, elles s'élèvent à 1 000 000 feddans! Quant au bersim, il occupe près de 70% de la superficie des cultures d'hiver (3 500 000 feddans). Tôt ou tard, le pays devra adopter des systèmes de culture plus compatibles avec les ressources dont il dispose.

En fin de compte, de très nombreux techniciens estiment qu'il est indispensable de passer d'un système où l'eau est gratuite et utilisée à profusion à une situation où l'eau est payante et son utilisation contrôlée. Mais quelle révolution dans l'univers et la mentalité du fellah égyptien! Est-ce même envisageable à court terme? Faire payer l'eau suppose une refonte totale du système agricole et de l'environnement dans lequel évolue la paysannerie des bords du Nil.

Mieux maîtriser la distribution de l'eau, éviter les pertes en circuit permettraient également de faire de notables économies. On songe à revenir à l'irrigation nocturne afin de diminuer les rotations. L'entretien des dizaines de milliers de kilomètres de canaux de toutes sortes (y compris les canaux secondaires) qui constituent la chaîne hydraulique de l'Égypte est de ce point de vue une priorité. Partout s'observent les signes de laisser-aller : canaux mal curés, ouvertures mal contrôlées, pertes en tout genre. Les jacinthes du Nil couvrent des canaux entiers : elles activent l'évaporation et constituent des foyers où pullulent toutes sortes de parasites. On estime à 10 milliards de m3 les pertes dans les canalisations autant que l'évaporation sur le lac Nasser!

On a pensé, enfin, à récupérer les eaux actuellement utilisées aux seules fins de la navigation et de la production d'électricité. Il est nécessaire une fois par an de mettre à sec les canaux pour différentes opérations d'entretien, essentiellement leur curage, et de réparation. Le mois de janvier était traditionnellement consacré à ces opérations. Depuis une dizaine d'années, cette période a été prolongée de deux semaines précisément par souci d'économie. Pendant ces travaux il est nécessaire de maintenir un niveau minimal pour le Nil afin de permettre la navigation sur le fleuve et la production d'électricité au barrage. Ainsi de 3 à 4 milliards de m3/an sont réservés à ces usages. L'idée est de récupérer en aval, dans des lacs au nord du delta, cette eau qui s'écoule vers la mer, de la stocker et de l'utiliser pour l'irrigation. Ce projet techniquement réalisable est une farouche pomme de discorde entre le ministère de l'Irrigation qui soutient l'idée et le ministère de l'Agriculture qui la combat essentiellement pour des raisons écologiques.

## 5.2 Le recours à des ressources non conventionnelles

L'Egypte réutilise déjà les eaux de drainage à hauteur de 4,7 milliards de m3. Il ne semble pas que ce volume puisse augmenter dans de fortes proportions si le pays adopte une politique d'économie d'eau. Le volume des eaux de drainage est, en effet, lié à celui des eaux utilisées pour l'irrigation et au niveau de la nappe pour les terres non équipées de système de drainage souterrain. Par ailleurs, l'utilisation des eaux de drainage ne peut pas se faire sans précaution sous peine de salinisation. Au sortir du barrage d'Assouan, la salinité des eaux est de 225 mg/litre, elle atteint déjà sur les bords du delta 2 000 mg. Elles doivent être mélangées à parts égales avec de l'eau douce. Il est donc évident que leur utilisation ne peut pas se pratiquer sur une très grande échelle.

La **réutilisation des eaux usées après** traitement offre des perspectives plus intéressantes. Jusque là, leur emploi a été ridiculement faible : pas plus de 200 millions de m3/an. La situation pourrait changer si les chantiers d'assainissement des grandes villes sont menés à bien.

L’achèvement des travaux d’assainissement du Caire permettent de retraiter 2 milliards de m3/an.

L'exploitation des nappes souterraines pose problème. En fait il faut distinguer entre deux nappes. La nappe phréatique à faible profondeur dans la vallée du Nil et surtout le Delta et le grand aquifère nubien. La nappe phréatique est déjà très fortement sollicitée : on y puise quelque 2,6 milliards de m3/an. Cette nappe est directement alimentée par les infiltration des eaux du Nil ou des eaux d'irrigation. Toute amélioration des conditions de l'irrigation et du drainage se répercutera sur la nappe et rendra son exploitation plus difficile. Par contre, de grands espoirs sont fondés sur **l'aquifère nubien** de 50 000 milliards de m3 dont 20 000 dans le sous-sol égyptien. Seuls 5 millions de m3/an sont exploités actuellement. La création de la Nouvelle Vallée ne peut pas se concevoir sans un recours massif à cette abondante ressource. Mais cette eau est fossile donc non renouvelable. Si un risque d'épuisement à court terme est à écarter en raison du volume considérable de l'aquifère, n'y-a-t-il pas un réel risque à moyen terme? Les analyses prospectives divergent. L'Égypte souhaite mettre en valeur dans cette Nouvelle Vallée 2 à 3 millions de feddans. Les estimations de la Banque Mondiale sont beaucoup moins optimistes : pas plus de 800 000 feddans en raison de la limitation des ressources en eau et de l'ampleur des investissements nécessaires.

Le Haut Barrage a procuré un sursis de trois décennies aux autorités du Caire. Les Égyptiens aujourd'hui se retrouvent devant un problème de disponibilité des ressources hydrauliques que la croissance démographique rend de jour en jour plus urgent. De nouvelles stratégies doivent être mises en œuvre qui supposent, à l'échelle du bassin nilotique, un accord géopolitique qui, pour l'instant, est bien difficile à négocier dans cette région du globe où l'emportent antagonismes, rivalités, conflits sur l'indispensable concertation. Récemment quelques progrès ont été faits. Une structure de concertation a été mise en place sous l’égide de l’Agence pour le Développement des Nations Unies : l’Initiative du Bassin du Nil (***Nile Basin Initiative***). Le Conseil des Ministres de l’organisation a entériné un accord soulignant la nécessité de renforcer la coopération mutuelle. L’idée est de « *promouvoir un développement socio-économique durable par une utilisation équitable des eaux et une juste répartition des avantages de cette ressource commune* ». Des études ont été lancées avec l’aide d’un financement international. Mais on est encore très loin d’une réelle mise en œuvre de projets.

En attendant l'Égypte ne peut compter que sur ses propres forces et gérer au plus près les ressources dont elle dispose.

# III. Le Tigre et l'Euphrate de la discorde

L'exploitation des eaux du Tigre et de l'Euphrate oppose de façon de plus en plus ouverte ces dernières années les trois pays riverains : Turquie, Syrie et Irak. Les aménagements que chacun d'entre eux a entrepris ou va entreprendre sont spectaculaires et les rivalités très vives pour le partage des eaux.

# 1. Les données hydrographiques

Le Tigre et l'Euphrate, les deux grands fleuves du Moyen-Orient et leurs affluents, prennent naissance dans les hautes terres enneigées de l'Anatolie orientale (Taurus oriental) et les montagnes du Zagros avant de se déverser dans les basses terres de Mésopotamie. Dans ce "château d'eau", les pluies sont abondantes (plus 1 000 mm) et de type méditerranéen avec des précipitations (pluie ou neige) hivernales et une sécheresse estivale.

## 1.1 Les tracés

**L'Euphrate**, long de 2 700 km, prend naissance au nord du lac de Van aux confins de l'Arménie : en fait il résulte de la confluence de deux rivières le Kara Sou (450 km) et le Murat Sou (650 km), qui prennent leur source à plus de 3 000 m. d'altitude. Le fleuve dessine ensuite une grande courbe en Turquie (420 km) et pénètre en Syrie à Karkemich où il s'encaisse légèrement dans un plateau désertique qu'il parcourt sur 680 km. Il n'y reçoit, en rive gauche, que deux affluents le Balikh et le Khabour. Puis il pénètre en territoire irakien et, rapidement, c'est l'entrée dans la plaine mésopotamienne parcourue sur 1 235 km en territoire irakien. Il n'est plus qu'une artère d'évacuation et ne reçoit aucun affluent.

Le **Tigre** (1899 km) qui prend naissance au sud du lac de Van coule en Turquie (455 km) mais ne pénètre pas en Syrie (il est fleuve frontalier sur 44 km), il s'écoule ensuite directement en Irak où il reçoit en rive gauche de très nombreux affluents bien alimentés issus des monts Zagros notamment le Grand et le Petit Zab (392 et 400 km), l'Adhaïm (230 km) la Diyala (386 km). Le Tigre arrose Bagdad qui n'est qu'à 32 mètres d'altitude alors qu'il lui reste 550 km à parcourir.

En Basse Mésopotamie, il se jette dans l'Euphrate à Garmat Ali. Les eaux mêlées des deux fleuves constituent sur 170 km environ le **Chott el Arab** qui débouche dans le golfe Arabo-Persique. Le Chott el Arab reçoit en rive gauche, les eaux abondantes, tumultueuses et limoneuses du Karun (16 milliards de m3), au parcours entièrement iranien.

## 1.2 Les régimes

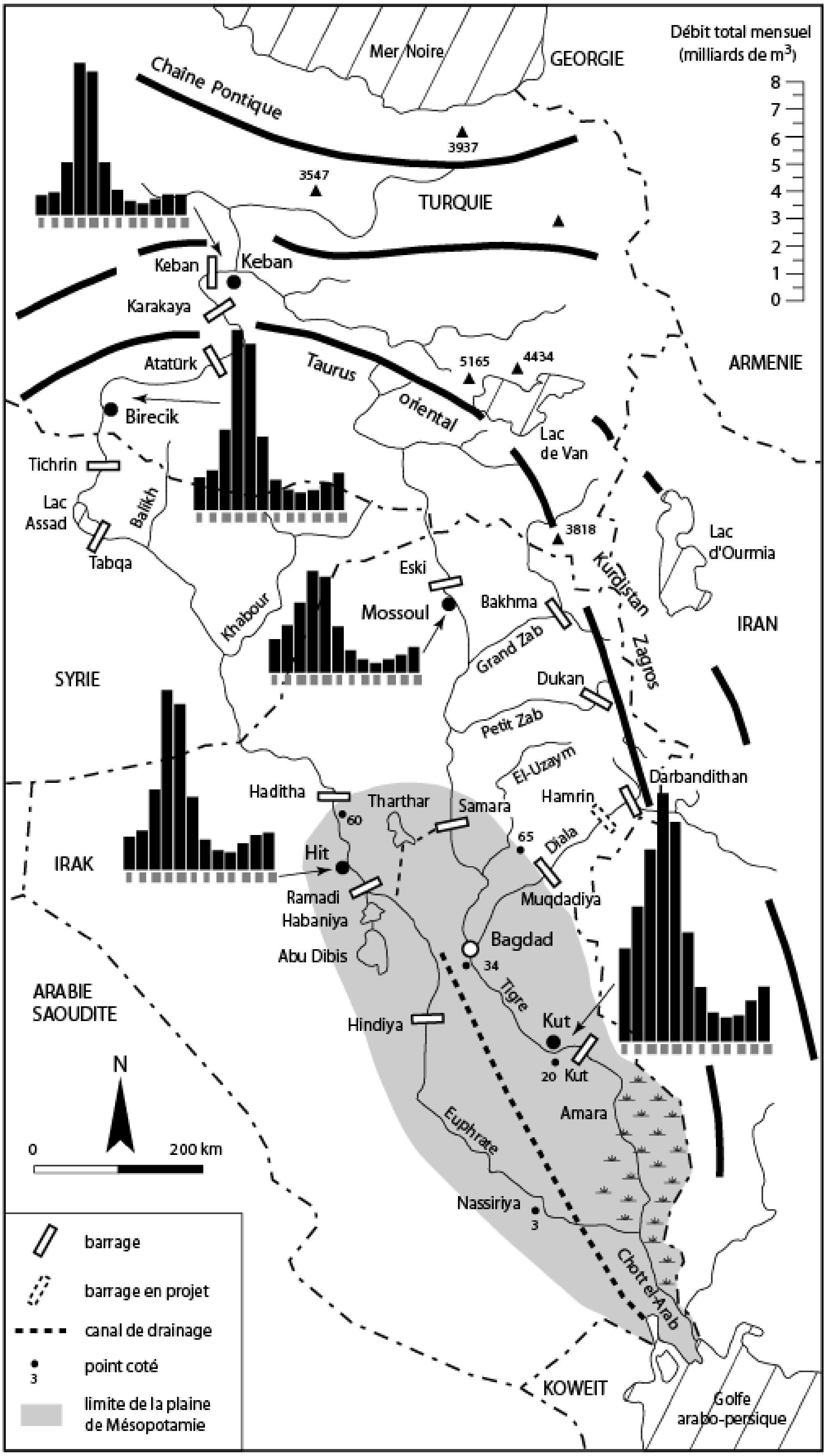

*Figure12 – Les bassins du Tigre et de l'Euphrate : débit naturel et aménagement*

Les régimes des deux fleuves sont très comparables : ils sont de **type pluvio-nival, marqués par les pluies méditerranéennes** de saison froide et la fonte des neiges des montagnes de Turquie orientale et du Zagros iranien. Le régime est donc très différent de celui du Nil : les hautes eaux sont moins abondantes et surtout ce sont des crues printanières, trop tardives pour les cultures d'hiver, trop précoces pour les cultures d'été.

D'une façon générale, il y a déphasage entre les périodes de hautes et basses eaux et les phases de cultures. Les hautes eaux du printemps gênent les moissons des céréales (blé et orge) et les ravagent parfois dans la plaine mésopotamienne. Elles entravent aussi les travaux agricoles des cultures d'été. Par contre la période des basses eaux de juillet à novembre correspond à celle où l'agriculture a le plus grand besoin d'eau.

Les écoulements du Tigre et de l'Euphrate présentent trois grandes caractéristiques :

• Leur **irrégularité** est très forte et présente un double aspect. Elle est **interannuelle**. Déjà, en amont, en Turquie, le module annuel peut varier dans le rapport de 1 à 4 aussi bien pour le Tigre que pour l'Euphrate. Plus en aval, les écarts sont à peine atténués. Le débit moyen annuel peut varier dans de fortes proportions. Pour l'Euphrate à Hit, les deux extrêmes sont 382m3/s en 1930 et 1140 en 1941 (1 à 3). Pour le Tigre à Bagdad : 499 en 1930 et 1670 en 1946 (rapport 1 à 3,3). En outre d'une année à l'autre les hautes eaux peuvent être avancées dès janvier. En fait, elles peuvent se placer durant une période de 5 mois. De même, les étiages peuvent s'étaler jusqu'en décembre. **L'irrégularité est aussi saisonnière**. 53 % des écoulements s'effectuent en trois mois (mars, avril, mai). Les étiages estivaux sont très prononcés : 300 m3/s (débit moyen pour l'Euphrate 830 à l'entrée en Irak) et 360 pour le Tigre (débit moyen de 1410 m3/s à Bagdad) et, surtout, ils interviennent à la fin de l'été (août et septembre) alors que les besoins en eau sont encore élevés.

• **Le débit décroît de façon notable d'amont en aval, notamment en Mésopotamie**. À l'entrée en Syrie, le débit moyen de **l'Euphrate** est de 830 m3/s (écoulement annuel de 28 milliards de m3). Le débit diminue légèrement pendant la traversée syrienne, les apports du Khabour (1,5 milliards de m3/an) et du Balikh (150 millions de m3/an) ne compensent pas l'évaporation durant la traversée, il n'est que 775 m3/s à la frontière irakienne. Le débit s'affaiblit considérablement en aval en raison de l'évaporation et de la difficulté de l'écoulement : il n'est plus que de 458 m3/s à Nasiriya. Le **Tigre** lors de son entrée en Irak a un débit annuel de 18 milliards de m3 mais, à l'inverse de l'Euphrate, il s'enrichit considérablement avec les apports des affluents venus du Zagros : Grand Zab (13 milliards de m3), petit Zab (7 milliards de m3), Adhaïm, Diyala (5 milliards de m3). Ces apports marquent très fortement le régime du Tigre : cours d'eau montagnards à forte pente, ils transportent une très importante charge alluviale et comptent des crues fréquentes, brutales et violentes. En aval de Bagdad le débit moyen est de 1410 m3/s (écoulement annuel de 46 milliards de m3) mais pour les mêmes raisons que l'Euphrate, il n'est plus que de 218 m3/s à Amara en Basse Mésopotamie et 78 m3/s à Qalat Saleh.

En fin de parcours, après la confluence des deux fleuves, le débit moyen annuel du chott el Arab, qui a reçu en rive gauche les apports de la Karun, s’élève à 43 milliards de m3.

• **L'ampleur et la brutalité des crues sont spectaculaires**. Alors que le débit moyen du Tigre est de 1410 m3/s à Bagdad, le fleuve a enregistré des crues de 13 000 m3/s. La crue maximale théorique est de 26 000 m3/s. Pour l'Euphrate à Hit, ces valeurs sont respectivement de 775 et 5 200. La crue maximale théorique est estimée à 8 000 m3/s. Ces crues sont très supérieures aux possibilités d'évacuation des lits qui ne dépassent pas 2 000 m3/s pour l'Euphrate et 8 000 pour le Tigre. La gravité de ces crues est renforcée par le fait qu'elles se produisent dans un véritable delta intérieur où les chenaux des fleuves sont sujets à des variations constantes et où il n'existe aucune vallée au sens topographique du terme. Nous sommes bien loin des conditions égyptiennes où une vallée très nettement encaissée guide, canalise l'écoulement de la crue. Aussi déviations et changements de cours apparaissent-ils comme la norme. L'insécurité est le lot des fellah mésopotamiens : les ravages des fleuves peuvent réduire à néant le travail humain, digues et canaux d'irrigation. On garde le souvenir de la crue de 1831 du Tigre qui en une nuit emporta Bagdad et anéantit 7 000 maisons.

## 1.3 Le système traditionnel de défense contre les eaux en Mésopotamie

Soumise à la menace permanente des inondations, en but à des défluviations de grande envergure, la Mésopotamie n'est habitable et cultivable que dans la mesure où les hommes ont réussi à se protéger des eaux. Un contrôle même imparfait des débits est indispensable. Le système traditionnel de défense contre les eaux repose sur deux techniques : la construction des digues et l'inondation dirigée.

La **Mésopotamie est le pays des digues**. Elles sont indispensables car les fleuves divaguent et coulent sur leurs propres alluvions. Ainsi à Bagdad les terres sont de 2 à 3 mètres en dessous du niveau des hautes eaux. La construction des digues est nécessaire très en amont, très loin de la mer. La Mésopotamie manque de matériaux durs, elle ne dispose que de terres alluviales et de briques crues. Les digues sont de simples levées de terre longeant les fleuves sans autre renforcement que quelques claies de branchages. Leurs crêtes ne dépassent pas les hautes eaux de plus de 1m ou 1,50 pour les plus récentes. Elles sont donc très fragiles, souvent rompues quand le courant devient fort. Au total au milieu du XXe siècle, la longueur des digues de l'Euphrate est de 1 500 km, sur le Tigre 1 000 km. Si les digues canalisent les fleuves tant que les débits demeurent dans des limites raisonnables, elles sont impuissantes à évacuer les crues dès que celles ci dépassent 2 000 m3/s sur l'Euphrate et 7 ou 8 000 m3/s sur le Tigre.

Dans ces conditions, la solution réside dans l'inondation naturelle ou dirigée. En recourant à l'inondation les hommes ne font qu'obéir à la nature, essayant seulement d'en modifier les processus en leur faveur. Ouvrir volontairement des percées dans les digues est le seul moyen en période de crue d'orienter tant bien que mal le flot vers certains secteurs plutôt que d'autres. L'eau des crues est dirigée vers des déversoirs topographiques, des cuvettes, de petites dépressions. Des canaux de dérivation permettent d'orienter les eaux de crue. C'est ainsi qu'avant les aménagements récents 85% des terres cultivées étaient susceptibles d'être recouvertes. Les conditions topographiques sont meilleures pour les eaux de l'Euphrate que pour celles du Tigre dont les crues ont toujours été redoutables. C'est ainsi que de 1944 à 1954, le Tigre a rompu ses digues 4 années sur 10 en amont de Bagdad, 8 fois à l'aval. L'efficacité de la dérivation a toujours été faible. Jusqu'au milieu du XXe siècle, la Mésopotamie malgré sa richesse en terres et en eau a été incapable de les mettre en valeur de façon autre que restreinte et précaire. L'impression dominante en Irak est celle d'une très grande fragilité des liens qui unissent l'homme et la terre qu'il cultive.

L'utilisation des eaux gardait un caractère empirique et décousu, sans aucun plan d'ensemble. Dans ce contexte, des facteurs historiques sont venus bien souvent renforcer le poids des contraintes naturelles et se sont conjugués avec elles pour expliquer la médiocrité de la mise en valeur. L'ancienneté des civilisations de la Mésopotamie ne doit pas faire illusion pas plus que les périodes brillantes qu'elle a pu connaître à certains moments de son histoire. On s'accorde sur la relative prospérité des époques babylonienne et séleucide et sous le califat abbasside. Le déclin se dessine dès la fin des Abassides; s'amorcent alors une régression des superficies cultivées et une dégradation des travaux d'irrigation qui culmineront lors des dévastations des invasions mongoles aux XII et XIIIe siècles. La Mésopotamie tombe sous l'influence des tribus bédouines qui annexent peu à peu la vallée à leurs pâturages et n'y pratiquent que des cultures épisodiques avec des techniques rudimentaires. En 1913, les superficies cultivées dans la plaine étaient tombées à 315 000 hectares soit moins de 5% de la superficie cultivable. La population de l'Irak, à la même époque, comptait un million et demi d'habitants, elle a, sans doute, été beaucoup plus importante lors des périodes de prospérité. Plus que tout autre foyer irrigué, l'Irak avait besoin de paix et d'une forte autorité apte à organiser l'entretien des canaux et des digues.

# 2. Maîtrise des eaux et équipement actuel en Irak

Ce n'est qu'au début du XXe siècle que l'on envisage sérieusement de discipliner définitivement le Tigre et l'Euphrate. La première tentative remonte à la période ottomane quand, en 1911, la Sublime Porte fait appel à un expert britannique William Willcocks qui avait acquis une solide expérience aux Indes et en Égypte. Sous le Mandat britannique, un département de l'irrigation est créé; les premiers travaux inspirés des plans de Willcocks sont entrepris. En 1950 le Bureau de l'équipement qui bénéficie des premiers financements d'origine pétrolière impulse une réelle dynamique à l'entreprise.

## 2.1 Les contraintes à surmonter

L'Irak, désormais en pleine croissance démographique (1947 : 4,5 millions d'habitants, 10 en 1972, 29 en 2007), doit mobiliser des quantités croissantes d'eau pour faire face à une forte demande. La quantité minimale sur laquelle on peut tabler 9 années sur 10 n'est que de 45 milliards de m3, et les débits les plus faibles jamais enregistrés ont été de 25 milliards de m3 (le débit moyen est de 74 milliards de m3/an). Or les quantités d'eau utilisées pour l'irrigation ne font que s'accroître : 19 milliards de m3 ont été prélevés, en moyenne annuelle, pour la période 1940-49, 28 entre 1950 et 1959, 49 maintenant.

Le sel pose de très graves problèmes. Le drainage est très difficile et à mesure que l'on se dirige vers l'aval la salinisation augmente : teneur normale de 250 mg/l à la frontière turque, plus de 600 dans la partie inférieure du cours irakien et 5 000 au débouché sur le Golfe.

L'importance de l'alluvionnement nécessite de très importants travaux de curage. La Mésopotamie est une immense plaine alluviale où les eaux coulent souvent au dessus du niveau de la plaine.

## 2.2 La construction de nombreux barrages

caractérise les travaux initiés tout au long du XXe siècle. Dans ces opérations on peut distinguer trois étapes :

Dans un premier temps, entre les deux guerres, des **barrages de dérivation** sont édifiés : ils orientent les eaux vers des canaux d'irrigation. Le barrage d'Hindiya sur l'Euphrate est construit de 1911 à 1913 et modernisé en 1927. Sur le Tigre on réalise le barrage de Kut de 1937 à 1939 et celui de Muqdadiya sur la Diyala. De ces barrages partent toute une série de canaux qui permettent l'extension de l'irrigation. Les progrès de l'occupation du sol sont rapides : on passe de 1 700 000 hectares irrigués à 3 000 000. Dans cette phase de l'expansion une place capitale est tenue par les procédés d'irrigation individuels : machines élévatoires (norias) et surtout les pompes à moteur qui en 1950 ont en grande partie supplanté les engins traditionnels.

Au lendemain de la seconde guerre mondiale, le dispositif se complète : on veut protéger la plaine des **inondations**. À partir du barrage de Ramadi (achevé en 1956), les crues de l'Euphrate sont détournées vers les dépressions naturelles d'Habaniya et d'Abu Dibis dont les capacités de stockage s'élèvent à 6,7 milliards de m3. Les eaux du Tigre sont orientées vers l'immense dépression endoréïque de l'oued Tharthar (85 milliards de m3) grâce au barrage de Samara (1956). Le contrôle des eaux du Tigre et de l'Euphrate est désormais assuré. La dernière crue destructrice date de 1954.

Dans une nouvelle phase, on cherche à lutter contre l'**irrégularité interannuelle** en construisant des barrages de retenue en dehors de la plaine mésopotamienne soit sur le plateau de la Jeziré irakienne soit dans les régions montagneuses parcourues par les affluents de rive gauche du Tigre. Un stockage de 40 milliards de m3 est prévu grâce à 6 barrages qui sont aussi producteurs d'électricité. Tel est le cas du barrage d'Haditha sur l'Euphrate, achevé en 1985 ainsi que le barrage d'Eski, sur le Tigre, en amont de Mossoul. Dans les montagnes du Zagros, le long des affluents du Tigre, 2 sites sont équipés : Dokan (1961) sur le petit Zab, Darbandithan (1962) sur la Diala. Depuis 4 autres sites ont été retenus. Dans la même perspective, le canal Tharthar-Euphrate permet depuis 1976 de réutiliser les eaux accumulées dans le lac Tharthar et de pallier dans une certaine mesure la faible alimentation de l'Euphrate après les travaux entrepris en amont en Syrie et en Turquie. L'aménagement des deux grands fleuves du Moyen-Orient, dans leur partie irakienne, est donc en passe de s'achever.

## 2.3 Un frein à la mise en valeur : la salinisation

Par contre, la salinisation des sols reste un problème entier, et un frein considérable à la mise en valeur. Elle a, en partie, des causes naturelles : les eaux du Tigre et de l'Euphrate contiennent une charge non négligeable de sels dissous et ne peuvent être utilisées sans précaution; l'écoulement des eaux, notamment celle des nappes phréatiques, s'effectue très difficilement et favorise ainsi la salinisation. La salinité touche de 70 à 80% des terres cultivées. 25 000 hectares sont perdus chaque année. Dès l'entrée dans la plaine en amont de Bagdad, le Tigre et l'Euphrate coulent entre des digues dans un lit exhaussé par rapport à la plaine qui les environne. L'irrigation par gravité ne pose aucun problème : tous les canaux s'étirent entre des digues au-dessus des champs. En revanche, l'écoulement des nappes est très lent; d'immenses marais jalonnent la plaine, surtout en Basse Mésopotamie.

L'action humaine contribue également à l'extension de la salinisation. L'eau est utilisée sans contrôle et, à certains égards, gaspillée. Il n'est pas certain que l'intensification des systèmes de cultures soit écologiquement acceptable. L'économie agricole traditionnelle avec jachère maintenait un certain équilibre. Une intensification qui tend à une occupation pérenne des terroirs accroît le rythme des arrosages et aboutit à une dégradation accrue de l'environnement dans l'état présent de la technique.

L'extension de l'irrigation ne peut s'envisager qu'avec la mise en place d'un système de drainage. L'entreprise est difficile. De Bagdad au Golfe a été mise en chantier la construction d'un grand canal de drainage pour évacuer vers la mer les eaux salées au lieu de les rejeter dans les fleuves. Long de 565 km, ce «troisième fleuve», qui passe en siphon sous l'Euphrate, vient d'être achevé en 1992. Il pourrait permettre de gagner par la désalinisation 1,5 million d'hectares, de limiter les inondations en période de crue afin d'assurer une production agricole moins aléatoire et aussi, pour certains, d'améliorer la navigation vers le Chott el Arab. Sans conteste le projet répond à une rationalité technologique et économique mais sa réalisation suscite polémiques et contestations. En arrière plan apparaît une dimension sociale et économique. Le troisième fleuve est perçu aussi comme une opération politique dirigée contre la communauté chiite locale et contre les Arabes des marais dont le cadre de vie et les conditions d'existence seraient totalement transformés. Il risque de conduire les populations à l'exode rural et permettrait au pouvoir central d'asseoir son autorité sur une région trop souvent rebelle.

Mais l'Irak n'est plus seul maître de ses choix, les deux autres pays riverains en amont ont entrepris de spectaculaires travaux d'aménagement qui, inévitablement, influeront sur l'alimentation de l'Euphrate.

# 3. Les aménagements syrien et turc

Ces deux dernières décennies la Syrie d'abord et la Turquie ensuite ont entrepris la construction d'importants barrages en amont sur l'Euphrate qui peuvent provoquer certaines incertitudes sur les disponibilités en eau dont pourra disposer l'Irak.

## 3.1 Le barrage de Tabqa sur l'Euphrate et l'équipement du Khabour

Opération symbole à laquelle s'identifie le régime alaouite, la construction du barrage de **Tabqa** en Syrie a été conduite de 1968 à 1976 avec l'assistance soviétique. Ce barrage-poids crée une retenue, le lac Assad, qui couvre 640 km2 et emmagasine 12 milliards de m3. La puissance installée permet de produire 5,6 milliards de kw/h, mais l'intérêt principal du barrage est d'augmenter les superficies irriguées en Jeziré. Le barrage régulateur al-Bath complète le dispositif tandis que, plus en amont, le barrage de Tichrin (1991) a une finalité purement énergétique.

Le projet, dont la mise en œuvre souffre de nombreux retards, prévoyait l'irrigation de 640 000 ha nouveaux répartis en six grandes zones, le long de l'Euphrate jusqu'à la frontière irakienne et le long des deux affluents de rive gauche, le Balikh et le Khabour. On vise à irriguer 450 000 hectares de terres sèches sur la steppe et à bonifier le long des rives de l'Euphrate 160 000 hectares de terres déjà irriguées. Ainsi, les superficies irriguées syriennes pourraient être doublées. Le système agricole de la vallée de l'Euphrate pourrait être intensifié. Les rendements des cultures traditionnelles (blé, orge et coton) devraient être améliorés, de nouvelles cultures introduites (plantes fourragères, légumes, riz et surtout betterave à sucre).

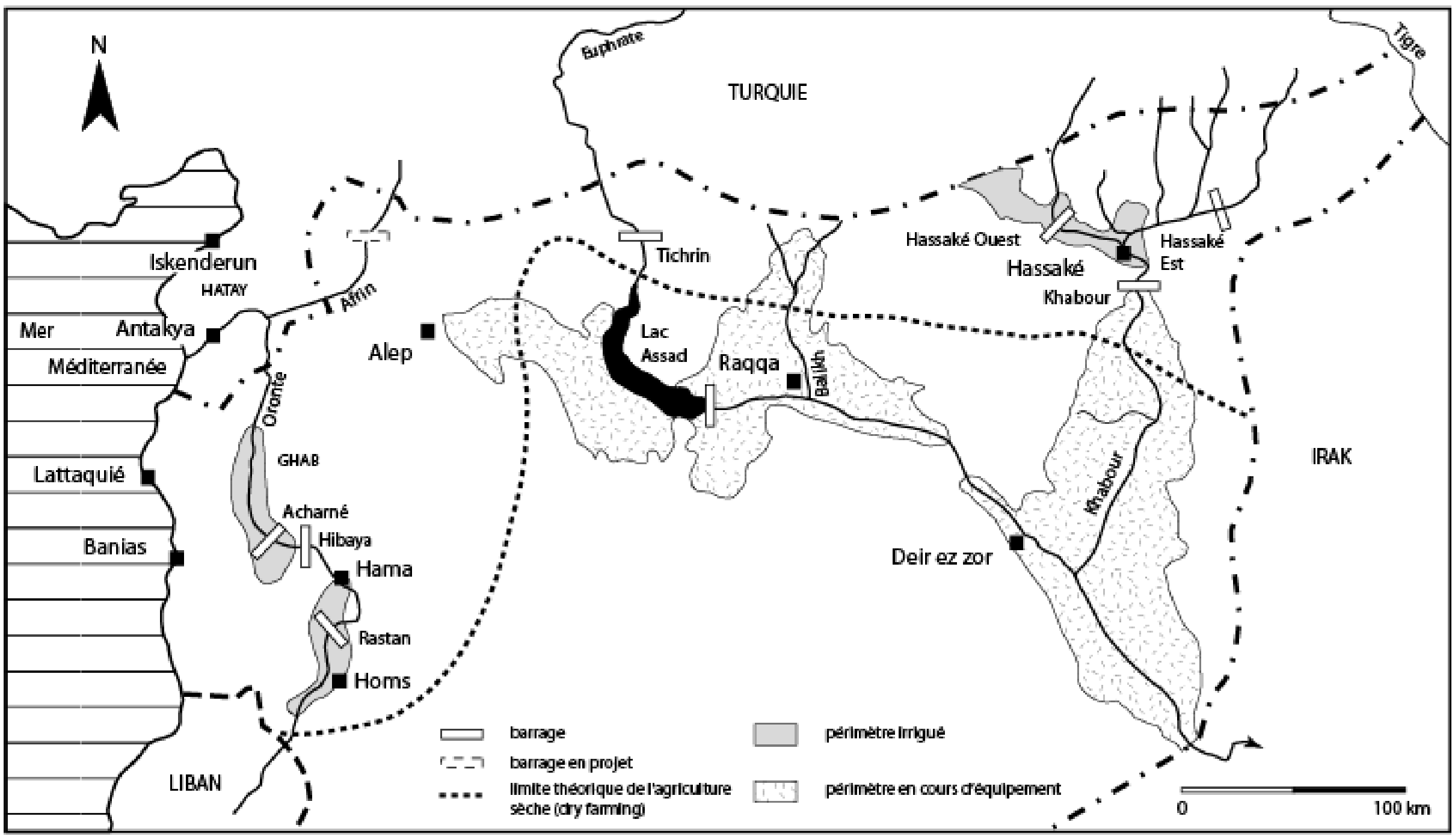

*Figure 13 – Périmètres irrigués en Syrie*

Après plus de 20 ans d'efforts, le **bilan des réalisations n'est pas à la hauteur des espérances placées dans le projet**. L'intensification des systèmes de culture est lente à venir. La mise sous irrigation se heurte à de très sérieux problèmes techniques : salinisation des terres due au surpompage, trop forte concentration de gypse dans le sol, affaissement des canaux d'irrigation, pertes d'eau d'irrigation en réseau de l'ordre de 50%! L'objectif fixé ne sera certainement pas réalisé entièrement. Les nouveaux colons, qui sont astreints à un système contraignant de coopératives, se recrutent avec difficulté : une nouvelle paysannerie a du mal à s'enraciner. Au total, le long de la vallée de l'Euphrate, l'irrigation, en 2008, porte sur 430 000 hectares (y compris les 160 000 hectares irrigués avant la construction du barrage.

L'aménagement de la haute **vallée du Khabour** doit compléter le dispositif mis en place dans la vallée de l'Euphrate. Le plan vise à l'irrigation à terme de 360 000 hectares. Il repose sur deux types d'intervention. D'une part une dizaine de petits barrages et de prises d'eau ont été réalisés le long des petits affluents de la section amont du Khabour. La retenue globale pour cet ensemble est de 100 millions de m3. Par ailleurs, l'aménagement de la haute et de la moyenne vallée du fleuve se poursuit actuellement à une autre échelle. Trois ouvrages de moyenne capacité sont achevés : le barrage d'Hassaké-ouest a une capacité de retenue de 91 millions de m3, celui d'Hassaké-est 232 millions de m3 et celui du Khabour en moyenne vallée a une retenue beaucoup plus importante : 665 millions de m3. Au total c'est plus du milliard de m3 qui sont ou vont être mobilisés dans cette vallée du Khabour. Actuellement les superficies irriguées s’élèvent à 260 000 hectares ;

Enfin le long du cours frontalier du Tigre, les Syriens envisagent l'aménagement de stations de pompage pour la fourniture d'eau potable des villes de la région.

Au total, les infrastructures réalisées au cours de ces deux dernières décennies par la Syrie le long de l'Euphrate et de ses affluents autorisent une mobilisation d'au moins 13 milliards de m3. Symétriquement les Turcs, plus en amont, procèdent à la mobilisation d'énormes volumes d'eau ce qui ne sera pas sans effet sur le débit de l'Euphrate à son entrée en Syrie et par voie de conséquence en Irak.

## 3.2 Un projet colossal : le GAP (Güneydogu Anadolu Projesi)

L’Euphrate représente, à lui seul, environ 45% des potentialités hydroélectriques de la Turquie. A partir d'un aménagement hydraulique du Tigre et de l'Euphrate, le Programme Régional de Développement de l’Anatolie du Sud-Est vise à un développement intégré d'une vaste zone de 75 000 km2 incluant 9 départements d'Anatolie orientale peuplés de 10 millions d'habitants. La phase de réalisation est déjà largement entamée.

Sur l'Euphrate, le barrage de Keban -le plus en amont- dont la retenue est de 30 milliards de m3 est terminé depuis 1974; il fournit exclusivement de l'électricité (1,2 milliard de kw/h). Le projet global, en aval de Keban, est beaucoup plus ambitieux. Une gigantesque opération hydraulique se décompose en treize sous-projets : sept sur l'Euphrate et ses affluents et six dans le bassin du Tigre. Une dizaine de centrales hydro-électriques produiront 26 milliards de kw/h, dont 8,1 pour Atatürk et 7,3 pour Karakaya. Le **barrage Atatürk**, la pièce essentielle, (48 milliards de m3, soit deux fois le débit moyen annuel du fleuve) est entré en service en 1992. Outre Ataturk, ont été achevés, sur l’Euphrate, les barrages de Karakaya (1987), Camgazi (1998), Karkamès (1999), Birecik (2000), Hançagiz (2002), Kayacik et sur le Tigre : Kralkizi (1997) Dicle (1998) Batman (1998) et plus récemment Cizre. Actuellement est mis en chantier sur le Tigre le barrage d’Ilusu, qui, après son achèvement en 2013 sera encore plus important qu’Atatürk. Toutefois cette nouvelle construction est très controversée. Les détracteurs dénoncent un « génocide culturel ». Il engloutirait la bourgade d’Hasankeyf (55 000 habitants) avec ses vestiges archéologiques : un pont médiéval, une mosquée ottomane de l'époque ayyoubide et des centaines de grottes troglodytiques habitées pendant des siècles. On dénonce également les impacts négatifs sur l'environnement. Le financement de ce projet coûteux (1,2 milliard d'€) est fragilisé par les réticences des financiers internationaux. Par ailleurs, un recours contre cette entreprise a été déposé devant la Cour européenne des droits de l'homme.

L'eau ainsi mobilisée doit allier la **production d'énergie et l'irrigation**. Sur une superficie cultivée de 3 000 000 hectares, 1 700 000 seront irrigués et consommeront 22 milliards de m3 d'eau/an. A partir de la retenue Atatürk, le tunnel hydraulique le plus long du Monde permettra l'écoulement de 328 m3/s (le tiers du débit de l'Euphrate) et l'irrigation de la plaine d'Urfa-Harran. Des canaux assureront, en outre, un transfert sur plusieurs dizaines de kilomètres de l'eau nécessaire à l'irrigation des régions limitrophes de la Syrie et notamment la plaine de Mardin-Ceylanpinar. Des pompages à partir de retenues le long du Tigre permettront la conquête de nouvelles superficies irriguées plus à l'est. Actuellement, la production électrique atteint 16 milliards de kw/h et 120 000 hectares sont effectivement irrigués et 200 000 prêts à l'être. Quand tous les projets (22 barrages capables de stocker 60 milliards de m3 : 14 sur l'Euphrate, 9 sur le Tigre et 19 centrales) qui intéressent aussi bien la vallée de l'Euphrate que celle du Tigre viendront à terme, on estime qu' entre 17 et 34% du débit sera absorbé. **Si tout se passe comme prévu le débit de l'Euphrate en Syrie devrait être réduit de 11 milliards de m3 et celui du Tigre de 6**. En outre, les **risques de pollution** sont prévisibles. Les eaux usées du GAP vont se déverser dans la zone où se forme la source du Khabour, l'affluent syrien de l'Euphrate. Les experts estiment qu'en fin d'aménagement, les eaux de l'Euphrate à l'entrée en Syrie seront polluées à 40%, celles du Tigre à l'entrée en Irak à 25% et les eaux de l'Euphrate à leur entrée en Irak à 50%. On peut deviner la vigueur des réactions syrienne et irakienne !

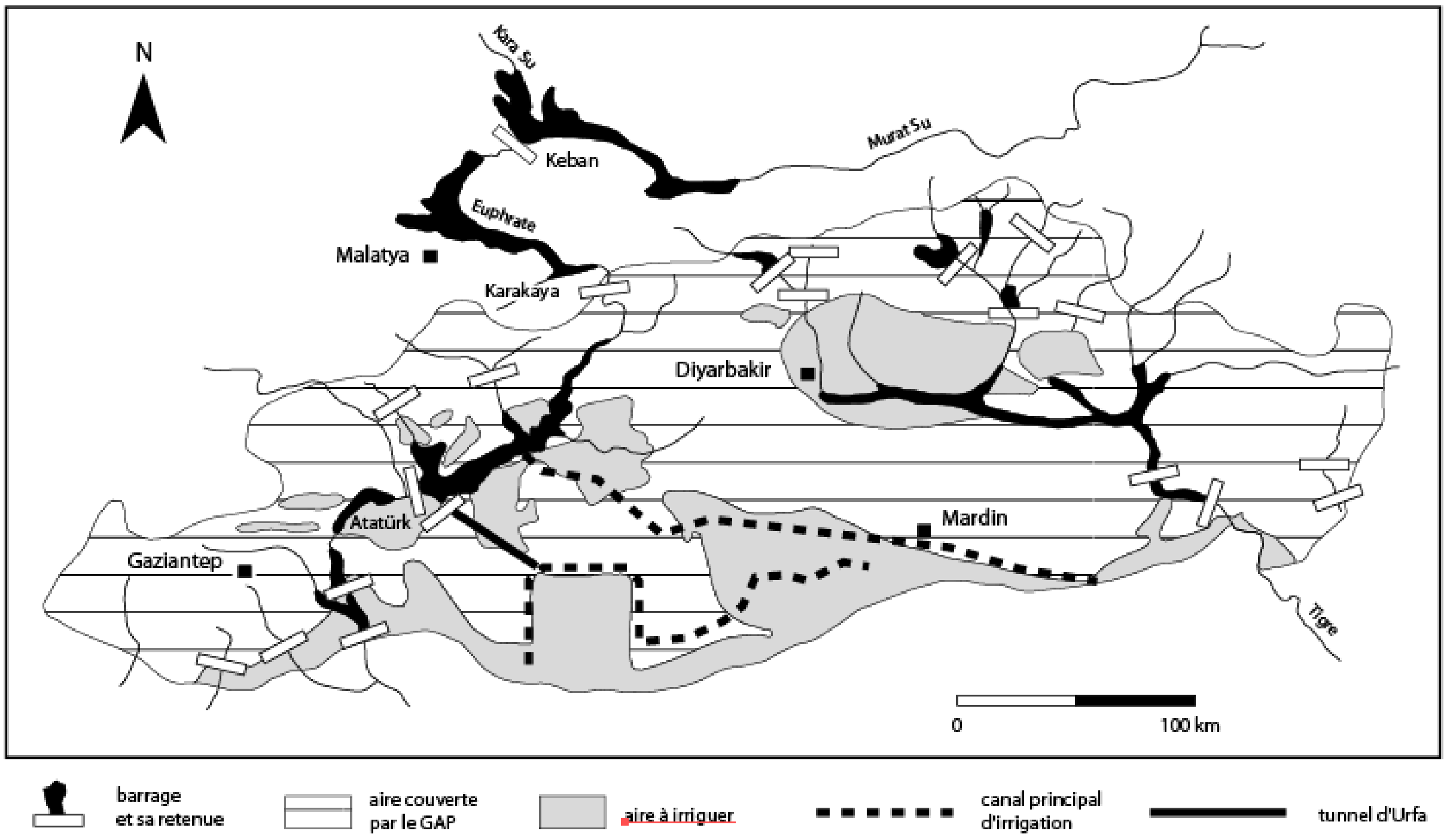

*Figure14 – Le G.A.P. turc*

La politique gouvernementale en faveur de l'Est s'est concentrée sur ce projet gigantesque, érigé en véritable mythe du développement national. Le GAP est pour les autorités turques conçu comme une solution au sous développement de la partie kurde du pays et une réponse économique aux demandes d'autodétermination de ses habitants. Les effets d'impact sont assez spectaculaires. Le projet, qui inclut le transfert de la population de plusieurs centaines de villages et de la petite ville de Samsat, l'antique Samosate, et plusieurs dizaines de chantiers de fouilles archéologiques de sauvetage, est considérable. Les travaux d'infrastructure représentent l'équivalent du budget annuel de la Turquie. Le coût total est estimé à 32 milliards de $, soit le 1/5 du PNB annuel du pays. On souhaite donc rentabiliser au mieux ces investissements, en substituant à la céréaliculture extensive une agriculture irriguée intensive tournée vers les cultures industrielles, en premier lieu le coton. L'irrigation permettra aussi l'augmentation du rendement des céréales et des vergers et l'introduction de nouvelles cultures : soja, maïs, arachide, riz. L'électricité des barrages doit alimenter de nouvelles usines sur place au lieu d'être expédiée vers l'Ouest industrialisé. L'amélioration de l'habitat rural et le développement d'activités touristiques sont également programmés. Le but de ce plan ambitieux est d'arrêter le flux d'émigration en fixant la population avec des activités économiquement efficaces.

# 4. Conflits et tensions entre les pays riverains

Avec la poursuite des aménagements hydrauliques dans les cours syrien et turc du Tigre et surtout de l'Euphrate, les relations interétatiques, déjà fort délicates dans cette partie du Moyen Orient, se compliquent dangereusement. La question du partage de l'eau se greffe sur les autres questions en suspens (question kurde, non reconnaissance de certains tracés frontaliers) et contribue sérieusement à aggraver le contexte géopolitique. Les deux pays arabes d'aval : la Syrie et l'Irak se trouvent placés dans une inconfortable position de dépendance à l'égard de la Turquie *(tableau 12)*. L'Euphrate, le Tigre et ses affluents coulent bien en Irak mais ils sont alimentés par des précipitations extérieures : 70% de l'alimentation est turque, 7% iranienne et 23% seulement irakienne. Cette situation ne posait pas de problème jusqu'alors dans la mesure où l'Irak était, de fait, le seul utilisateur. Il n'en est pas de même aujourd'hui avec les projets syrien et turc.

| % | Euphrate | | Tigre | |
|---|---|---|---|---|
| | Bassin | Débit | Bassin | Débit |
| Turquie | 28 | 88 | 12 | 40 |
| Syrie | 17 | 12 | 2 | 0 |
| Irak | 40 | 0 | 54 | 51 |
| Iran | | | 34 | 9 |

*Tableau 12 : Répartition de la superficie des bassins et du volume des débits entre les pays riverains du Tigre et de l'Euphrate*

## 4.1 Hydropolitique et crises interétatiques

Elles ont été fort nombreuses depuis une trentaine d'années. Elles opposent évidemment la Turquie aux deux autres pays arabes. Mais les frères arabes ennemis (Syrie et Irak) s'opposent aussi violemment entre eux. Les premières discussions entre États riverains remontent à la décennie 1960 mais la réunion tripartite de 1965 aboutit à un échec.

La construction du barrage syrien de Tabqa a provoqué une vive réaction de la part de l'Irak d’autant plus, qu’au même moment, la Turquie mettait en eau le barrage hydroélectrique de Keban. L'Euphrate fournit en effet 37% des eaux d'irrigation de l'Irak. Le remplissage du lac Assad priva temporairement l'Irak d'une partie des eaux de l'Euphrate mais les évaluations des deux pays différent. L'Irak prétendait n'avoir disposé en 1975 que de 9,4 milliards de m3 (moins du 1/3 du débit habituel) alors que la Syrie avançait le chiffre de 12,8 l'équivalent de la consommation annuelle de l'Irak à l'époque. Devant la détérioration des relations entre les deux pays une médiation saoudite fut tentée mais le projet saoudien de répartition proportionnelle des eaux n'eut jamais de suite. Il fallut l'intervention soviétique pour que la Syrie accepte de laisser s'écouler une quantité d'eau supplémentaire. Pendant la période de sécheresse des années 1980, l'Irak accusa plusieurs fois la Syrie de retenir les eaux de l'Euphrate.

Les tensions entre la Turquie et ses voisins arabes sont récurrentes. Avec la Syrie, elles sont les plus fortes. La Turquie établit un lien avec le problème de l'Oronte. Entre la Turquie et la Syrie il existe, en effet, un contentieux de fond lié à l'annexion du Sandjak d'Alexandrette devenu le Hatay turc. En 1939, la France, puissance mandataire en Syrie, céda le Hatay à la Turquie pour s'assurer sa neutralité dans le conflit à venir avec l'Allemagne. La Syrie n'a jamais reconnu cette annexion du Sandjak d'Alexandrette parcouru par la partie aval de l'Oronte. L'eau de l'Oronte est actuellement, dans la partie amont du fleuve, mobilisée par la Syrie à plus de 90%. Depuis 1964, la Turquie propose à la Syrie un accord sur tous les cours d'eau communs aux deux États, en particulier sur l'Oronte, ce qui reviendrait à une reconnaissance syrienne indirecte de la souveraineté turque sur Alexandrette. Damas qui persiste dans sa revendication du Sandjak d'Alexandrette n'obtient pas de règlement satisfaisant à propos de l'Euphrate.

**L'Oronte, un autre différend entre la Turquie et la Syrie**

L'Oronte (570 km) traverse trois pays *(carte 14)* dans les régions les plus arrosées du Proche-Orient. Le "fleuve rebelle", bien alimenté, prend sa source au Liban dans la plaine de la Bekaa qu'il parcourt sur 35 km seulement. A son entrée en Syrie, son module annuel est de 370 millions de m3. Son tracé est alors nettement guidé par les donnée orographiques. Il traverse les plateaux syriens de Homs, du Hama où autrefois son courant animait les célèbres norias; puis il s'écoule dans le fossé tectonique du Ghab entre le Jbel Alaouite à l'ouest et le Jbel Zawiyé à l'est. La plaine du Ghab a longtemps été marécageuse : un verrou basaltique empêchait l'écoulement des eaux fort abondantes car des affluents et surtout les sources karstiques renforçaient considérablement le débit du fleuve. Les apports supplémentaires reçus au cours des 325 km du parcours syrien sont évalués à quelque 430 millions de m3 ce qui porte le débit naturel - théorique - à plus de 800 millions de m3. L'Oronte pénètre ensuite en Turquie dans la province du Hatay. Avant de se jeter en Méditerranée, l'Oronte reçoit l'Afrin un affluent de rive droite qui a pris naissance en Turquie mais traverse une partie de la Syrie du Nord-Ouest avant d'atteindre à nouveau le territoire t il mêle ses 230 millions de m3 au débit de l'Oronte.
La Syrie a conduit d'importants projets d'irrigation à partir des eaux du fleuve. Les superficies irriguées se sont multipliées autour de Homs et de Hama. Surtout, pièce essentielle, le fossé du Ghab a fait l'objet au cours de la décennie 60 d'un aménagement spectaculai Après avoir assuré le drainage de la plaine en faisant sauter en aval le verrou du Khabour, trois ouvrages permettent une mobilisation importante des eaux de l'Oronte pour l'irrigation : le barrage de Rastan (250 millions de m3) entre Homs et Hama, le barrage d'Hifaya-Mehardé (65 millions de m3) dans les gorges de Cheizar tandis que le distributeur d'Acharn répartit les eaux sur la surface à irriguer entre trois canaux principaux.
L'opération d'aménagement est présentée comme une réussite : près de 100 000 hectares sont effectivement irrigués, l'ancienne zone marécageuse est repeuplée par plus de 200 000 habitants. Le Ghab est désormais une des régions pionnières de la Syrie. Il fournit 24% des produits agricoles du pays. Cette extension de l'irrigation prélève beaucoup d'eau (630 millions de m3) et à son entrée en Turquie le débit de l'Oronte n 'est plus que de 170 millions de m3! La Syrie construit un barrage à des fins d'irrigation sur le parcours s de la rivière Afrin qui prélèverait 130 millions de m3 sur son débit naturel.
Au total, y compris les apports de l'Afrin, la Turquie ne disposerait plus que 270 millions de m3 qu'elle estime insuffisants pour desservir les villes d'Iskenderun (ex-Alexandrette), d'Antakya et leur région. La Syrie prélèverait 760 millions de m3 sur la partie amont du bassin de l'Oronte. On a là une position symétriquement inverse à celle de l'Euphrate pour les deux pays : la Syrie est placée en ce cas dans la position avantageuse d'amont! Cet inégal partage des eaux de l'Oronte ajoute encore au contentieux entre les deux pays.

Plus récemment la décision unilatérale de la Turquie d'entreprendre le GAP a été perçue par ses voisins d'aval comme agressive et indélicate. La construction du barrage de Keban suscite, en 1972, des protestations officielles de la Syrie non pas à cause d'une baisse effective du débit (le barrage produit de l'électricité et doit régulariser le fleuve) mais parce que la Turquie démontrait qu'elle était capable de contrôler l'Euphrate en amont. L'affrontement le plus sérieux qui opposa la Turquie et ses deux voisins eut lieu lors du remplissage du lac de retenue du barrage Atatürk au début de 1990. La Turquie est accusée non sans raison de ne pas avoir honoré les engagements antérieurs (celui de 1987). Il y a eu effectivement rupture de l'alimentation en eau de l'Euphrate durant le mois de janvier 1990. En Irak, l'interruption de l'écoulement a conduit à une perte de 15% des récoltes. Depuis la mise en eau des nouveaux barrages est toujours une source de tension. Au total, le remplissage des différents barrages turcs sur l'Euphrate a nécessité 100 km3 auxquels il faut ajouter une perte évaporatoire de 3 km3/an depuis 1990, alors que le module du fleuve n'est que de 29 km3

## 4.2 Le difficile accord de 1987 :

Il n'existe aucun traité tripartite sur l'exploitation et la répartition des eaux entre les États riverains du bassin du Tigre et de l'Euphrate. Le traité de Lausanne de 1923 contenait une clause stipulant que la Turquie devait consulter l'Irak avant d'entreprendre des travaux hydrauliques. En 1962, la Syrie et l'Irak créèrent une commission mixte mais son rôle resta limité du fait de l'absence de travaux hydrauliques importants. Vers 1972/73 les deux mêmes pays firent des tentatives infructueuses pour négocier un accord sur l'Euphrate. L'imprécision du droit international en ce domaine ne facilite pas les choses.

Le seul arrangement consenti par la Turquie, en **1987**, est un **accord bilatéral avec la Syrie** portant sur les quotas, la Syrie reçoit 500 m3/s (soit 15,75 milliards de m3) alors que le débit naturel de l'Euphrate à l'entrée en Turquie est de 28 milliards de m3. Un autre **accord bilatéral syro irakien (avril 1990**) prévoit une répartition proportionnelle des eaux de l'Euphrate entre les deux pays (42% pour la Syrie, 58% pour l'Irak) quel que soit le débit du fleuve soit en année «normale» 6,6 milliards de m3 pour la Syrie et 9 pour l'Irak

Toutefois les crises ont été nombreuses entre les trois pays concernés que ce soit avant ou après la signature de ces accords.

## 4.3 Quelles perspectives pour la prochaine décennie?

Un règlement satisfaisant pour les trois parties en présence paraît très difficile sinon impossible tant les positions de principe sont éloignées.

**L'Irak estime que les deux fleuves sont internationaux et demande le respect des droits acquis**. Cette position sous-entend le respect de la consommation antérieure de chacun des États riverains et le partage équitable des ressources supplémentaires obtenus par des aménagements ultérieurs. **L'Irak demande aussi que soit reconnue l'indépendance des** **bassins versants et s'oppose à la position turque mais aussi syrienne qui considère que le Tigre et l'Euphrate constituent deux branches d'un même bassin hydrographique**. En optant pour l'unicité du bassin, la Turquie et la Syrie proposent que l'Irak prenne sa part de ressources sur le Tigre difficilement aménageable dans sa partie amont laissant ainsi à la Turquie et à la Syrie le bénéfice exclusif des eaux de l'Euphrate. Pour l'Irak au contraire les deux fleuves doivent être considérés séparément et un partage équitable des eaux de chacun d'eux doit être envisagé entre les trois États riverains.

**Pour la Syrie, l'Euphrate est un fleuve international** et il doit y avoir respect des "droits acquis" et interdiction de tout aménagement qui modifierait le débit sans l'accord de l'ensemble des États riverains. Sur ce point la position syrienne est identique à celle de l'Irak. Par contre, elle s'en écarte sur un autre point : **elle soutient "l'unicité" du bassin versant du Tigre et de l'Euphrate**. En clair elle propose que le partage des eaux de l'Euphrate ne s'opère qu'entre la Syrie et la Turquie et que l'Irak se satisfasse d'une exploitation quasi exclusive des eaux du Tigre qui n'est qu'un fleuve frontalier pour elle.

**La Turquie soutient que les deux fleuves constituent un seul bassin et qu'ils sont transfrontaliers et non internationaux**. Un tel statut permettrait à la Turquie de gérer à sa guise les ressources disponibles des deux fleuves sans prendre en considération les demandes et les besoins de la Syrie et de l'Irak. La Turquie accepte pourtant de ne pas porter atteinte aux droits acquis antérieurs aux nouveaux projets hydrauliques. Elle a ainsi accepté de signer l'accord de 1987. Pour l'avenir sa position est nette : elle accepte de coopérer pour la gestion des eaux du Tigre et de l'Euphrate, à condition de se limiter à des projets précis. Mais elle n'est pas prête d'accéder à la demande de ses co-riverains de conclure un accord multilatéral sur des quotas de répartition. Cette attitude contribue à entretenir la tension dans la région. La Turquie soutient que les déficits en eau en aval sont liés à une mauvaise gestion et ne relèvent pas du domaine juridique. Les pays en aval doivent mettre en œuvre des techniques plus économes en eau. Elle soutient que l'accord de 1987 sur les quantités allouées à la Syrie est définitif et rejette les demandes conjointes de la Syrie et de l'Irak pour une augmentation des quotas à 700 m3/s.

Cette position risque d'être lourde de conséquences. Si on se fie à des estimations turques figurant dans le tableau 13, sans doute un peu gonflées, dans les décennies à venir les demandes d'eau ne feront que croître et l'on voit mal comment les demandes des pays en aval pourraient être satisfaites.

| Milliards de m3 | Euphrate | Tigre |
|---|---|---|

| | Potentiel | Utilisation souhaitée | Potentiel | Utilisation souhaitée |
|---|---|---|---|---|
| Turquie | 31,58 | 18,42 | 25,24 | 6,87 |
| Syrie | 4 | 11,3 | 0 | 2,6 |
| Irak | 0 | 23 | 23,43 | 45 |
| Total | 35,58 | 52,92 | 48,67 | 54,47 |

*Tableau 13 : Potentiel et demande d'eau des trois pays riverains du Tigre et de l'Euphrate*

Dans un cadre plus large, la politique turque est de plus en plus ambiguë. A la fin des années 80, elle se présentait comme le château d'eau de la région et on lui prêtait l'intention de céder une partie de ses eaux aux pays arabes (cf l'aqueduc de la Paix). Maintenant elle veut toute son eau pour elle. Cette réticence manifeste va de pair avec un désintérêt croissant vis à vis du Moyen Orient arabe. La Turquie commence à se considérer comme le «centre géopolitique d'une région en train d'émerger». Il y a plus d'avenir pour elle dans le développement de liens économiques et politiques avec les États de l'ex URSS notamment les turcophones.

**Plus que jamais la Turquie reste maître des eaux**.

# IV. Inégal partage dans le bassin du Jourdain

Les conflits pour l'eau sont particulièrement vifs au Proche Orient. Pour l'essentiel les rivalités se nouent à propos de l'utilisation des nappes souterraines et des eaux du Jourdain et de ses affluents (essentiellement le Yarmouk) dont le bassin versant se partage entre 4 États : le Liban, la Jordanie, la Syrie, Israël et les Territoires occupés. A cela s'ajoute en filigrane l'utilisation des eaux du Litani.

# 1. Quelques données de base

## 1.1 Des conditions naturelles médiocres

Toute une série de **reliefs calcaires sont disposés parallèlement au littoral :**

- au Nord en Syrie : le Jbel Ansarié (1 583 m) et le Jbel Zawiyé enserrent le fossé du Ghab dans lequel coule l'Oronte.
- Ce double alignement montagneux se prolonge au Liban avec le mont Liban en bordure du littoral (3 083 mètres) et, plus à l'est, l'Anti Liban qui se prolonge avec le mont Hermon (2 814 mètres). Entre Mont Liban et Anti Liban se loge la plaine de la Bekaa comprise entre 900 et 1 000 mètres d'altitude.
- Ces alignements montagneux s'abaissent en Israël et en Palestine et constituent les collines de Cisjordanie (1 000 m près d'Hebron). Le fossé dans lequel coule le Jourdain prolonge en quelque sorte la Bekaa. L'escarpement jordanien domine, à l'est, la vallée du Jourdain de ses 1 200 m.

A l'est de cette zone de collines ou de montagnes proches du littoral, s'étire une zone tabulaire en Syrie ou en Jordanie.

En raison de l'alignement du relief perpendiculairement aux **dépressions d'ouest** la région est bien arrosée. Il tombe bien souvent plus de 1000 mm sur les reliefs et même jusqu'à 1 500 sur le Mont Liban. En Palestine, les collines de Cisjordanie reçoivent de 500 à 700 mm. Mais en arrière des reliefs, plus à l'est, en Syrie et en Jordanie les pluies diminuent très rapidement (300 mm et moins). Les précipitations hivernales de type méditerranéen tombent sur les plus hauts reliefs sous la forme de neige ce qui autorise un certain stockage des eaux favorisé en outre par la nature calcaire des massifs. Cette frange littorale bien arrosée disparaît plus au sud : en Israël le Neguev est une zone semi aride : les précipitations sont inférieures à 200 mm. Il en est de même pour la vallée du Jourdain (Jéricho 166 mm)

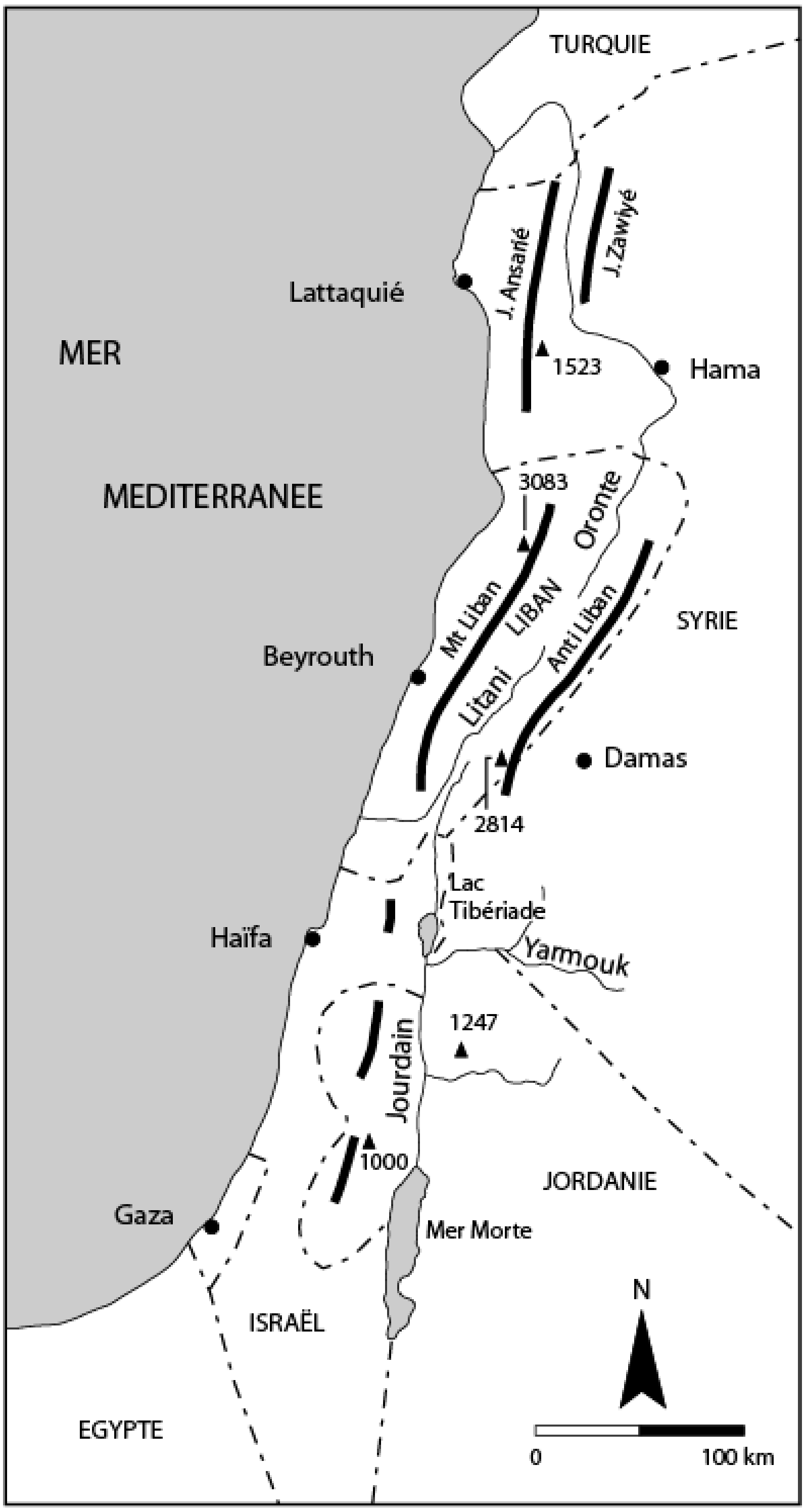

*Figure15 – Fleuves et montagnes au Proche-Orient*

Les ressources en eau sont relativement abondantes. De nombreux cours d'eau parcourent la région. En raison du cloisonnement du relief beaucoup sont très courts et, nettement guidés par l'orographie, s'écoulent dans un sens méridien. En dehors des nombreux torrents qui dévalent le flanc occidental du Mont Liban, trois organismes fluviaux jouent un grand rôle. Dans la plaine de la Bekaa coulent en sens inverse **l'Oronte et le Litani**. L'Oronte (570 km) coule vers le nord, pénètre en Syrie, franchit en gorge le Jbel Zawiyé et se jette en Méditerranée dans le golfe d'Iskenderun en Turquie. Le Litani est un fleuve entièrement libanais, il s'écoule vers le sud et se jette en Méditerranée au sud de Saida. Le **Jourdain** (360 km) s'écoule du nord au sud. Il prend naissance sur le mont Hermon et se dirige vers la mer Morte.

Les **nappes souterraines** jouent également un grand rôle et parfois leur exploitation est source d'intenses rivalités. La plus importante est la nappe que l'on rencontre en Cisjordanie (680 millions de m3). Une nappe littorale se localise dans le territoire autonome de Gaza et se prolonge plus au nord en Israël. Une troisième nappe se trouve dans le sous-sol du plateau jordanien. Enfin certains aquifères fossiles de la péninsule Arabique se prolongent dans le sous-sol de la Jordanie.

## 1.2 L'inégale répartition des ressources en eau entre les entités territoriales (tableau 5)

Le **Liban** est, de loin, le pays le mieux doté avec la Syrie. Ses ressources en eau douce sont estimées à 4,8 milliards de m3 soit 1 230 m3/an/personne, ce qui est une norme haute. La consommation totale annuelle du pays est de l'ordre de 900 millions de m3 : 550 provenant des eaux de surface et 350 des eaux souterraines. 80 % des eaux consommées sont destinées à l'irrigation. Le Liban est donc doté d'importantes réserves, il peut apparaître un peu comme un château d'eau dans ce Levant où beaucoup d'États souffrent de pénurie. Beaucoup de stratégies s'échafaudent à partir de ce constat, notamment autour du Litani. Toutefois, le Liban a des projets d'extension d'irrigation qui conduiraient à une augmentation substantielle de ses consommations d'eau.

La **Syrie** est, pour l'instant, assez bien pourvue en eau (1317 m3/an/personne). Une partie des sources du Jourdain se trouvent sur son territoire (le Golan) mais en fait la consommation du pays n'en dépend pas. Les ressources syriennes en eau proviennent de la partie méditerranéenne du pays avec l'Oronte et des nappes phréatiques et surtout de l'Euphrate.

Par contre, les **ressources jordaniennes** sont très précaires : le pays souffre actuellement de pénurie d'eau et la consommation par habitant est très faible (150 l/jour). Les ressources en eau douce sont de 158 m3/an/habitant : une norme très basse. Le pays consomme environ 880 millions de m3/an ce qui excède son potentiel renouvelable. L'eau consommée provient des origines suivantes :

- nappes fossiles non renouvelables : 210 millions de m3· La nappe Wadi-Sir utilisée s'étend sur le territoire jordanien jusqu'à 700 mètres de profondeur, elle se rattache aux aquifères fossiles de la Péninsule.
- les eaux de surface du Jourdain et du Yarmouk 360 millions de m3 (175 millions de m3 sont prélevés sur le Yarmouk, 105 sur les wadi du Jourdain)
- nappes renouvelables : 310 millions de m3.

**Israël** dispose d'un réseau national de distribution de l'eau depuis 1964. Il réalise l'interconnexion de l'eau à travers tout le pays à partir du lac de Tibériade. Le pays connaît de graves tensions : sa consommation est de l'ordre de 1 900 à 2 000 millions de m3 ce qui dépasse très largement les ressources dont le pays dispose à l’intérieur de ses frontières internationalement reconnues (1 400 millions de m3). Les ressources en eau douce d'Israël sont faibles : 190 m3/an habitant. L'origine de l'eau consommée provient des sources suivantes :

C'est toutefois dans les **Territoires Occupés** que la situation est la plus inquiétante. Les Palestiniens ne disposent que, très partiellement, de l'eau des nappes de Cisjordanie et de Gaza. C'est le pays où les quotités disponibles d'eau renouvelable sont les plus faibles : 200 m3/an/habitant si on ne prend pas en compte les prélèvements israéliens mais seulement 50 dans le cas contraire!

Ce rapide tableau met l'accent sur les inégalités de répartition dans un contexte de rareté. Les concurrences et affrontements régionaux se nouent autour des deux sources essentielles : les eaux du Jourdain et l'accès à la nappe souterraine de Cisjordanie.

# 2. L'inégale appropriation des eaux du Jourdain

## 2.1 Le fleuve

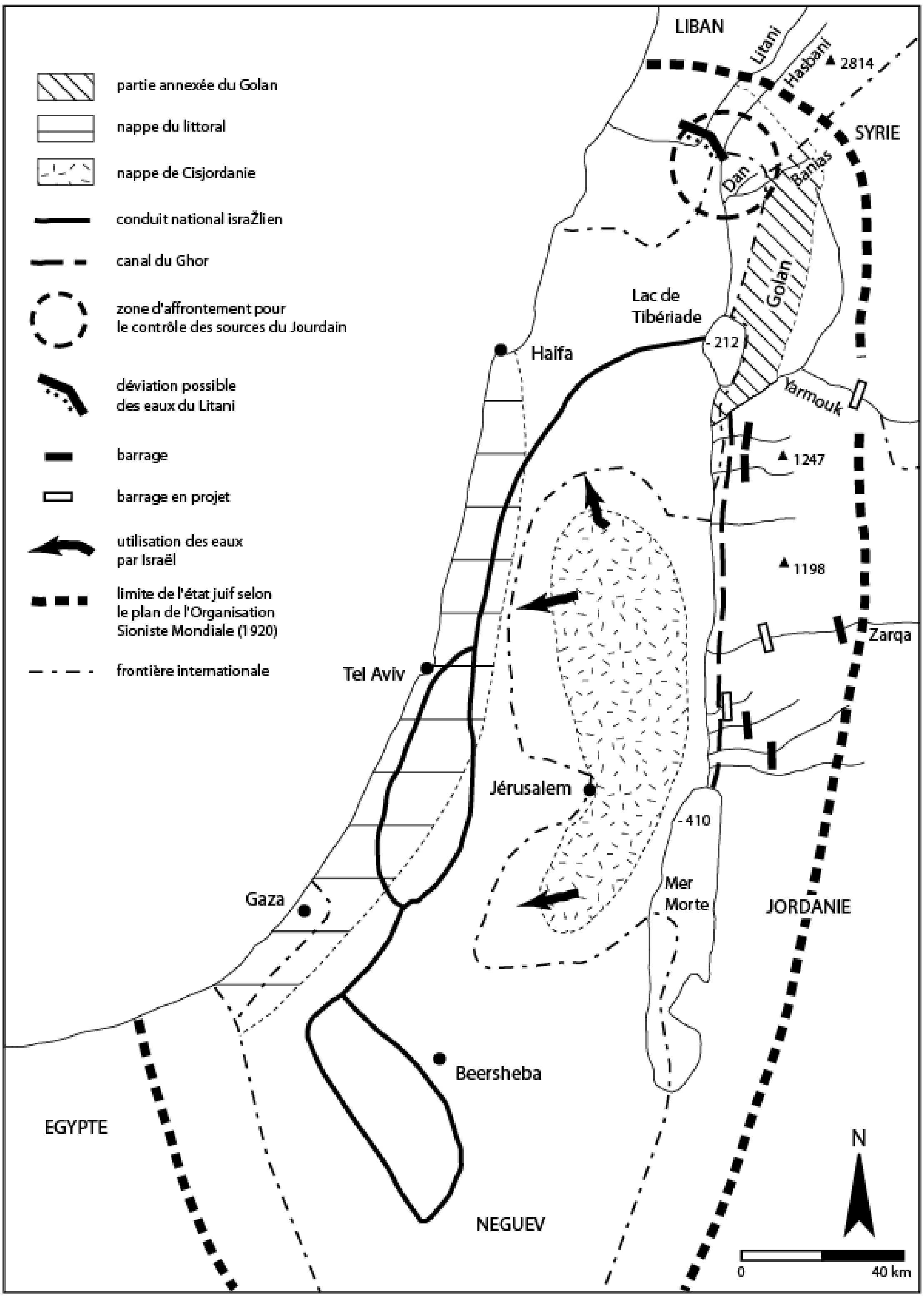

*Figure 16 – Inégal partage des eaux dans la vallée du Jourdain*

Le Jourdain est un petit fleuve (360 km) dont le bassin est partagé entre le Liban, la Syrie, la Jordanie et Israël. Il résulte de la réunion dans la dépression de Houle de trois rivières alimentées à la fois par les résurgences karstiques et la couverture neigeuse du Mont Hermon (2 814 m.) : le Hasbani, le Dan et le Banias. Le Hasbani s'écoule sur 21 km en territoire libanais et a un débit moyen annuel de 150 millions de m3; le Dan (12 km et 260 millions de m3) prend naissance à l'intérieur du territoire israélien, tandis que le Banias (30 km et 160 millions de m3) prend sa source dans le Golan syrien. Le débit moyen annuel après leur confluence est de l'ordre de 560 millions de m3. On peut constater que le débit de ces trois rivières, "les sources du Jourdain" est élevé. Il est bien plus important que ne le laisserait supposer l'étendue de leurs bassins respectifs. Ils bénéficient vraisemblablement, dans cette région calcaire, d'une importante circulation souterraine d'origine plus ou moins lointaine alimentant d'importantes résurgences. Le lac Houle, barré par une coulée basaltique, est en fait un marécage de très faible profondeur 3 à 5 m.) qui a été asséché par Israël en 1953.

Le haut Jourdain rejoint en un court trajet (17 km) à travers des gorges volcaniques, le lac de Tibériade, situé à 210 mètres au dessous du niveau de la mer. Il y déverse un volume annuel de 560 millions de m3. Long de 21 km, large de 12 et profond de 45 m, le lac de Tibériade occupe une superficie de 160 km2. Des sources souterraines partiellement salées y déversent 230 millions de m3/an. Mais l'évaporation est intense, les eaux se chargent de sel et au sortir du lac, le débit ne dépasse pas 500 millions de m3. Puis le Jourdain se dirige vers la Mer morte (à moins 410 mètres) à travers une vallée encaissée. Il décrit des méandres dans une plaine alluviale humide (le *zor*), dominée par tout un ensemble de terrasses sèches, ravinées en bad lands (le *ghor*). Au cours de ce trajet (320 kilomètres de méandres pour un tracé de 109 km à vol d'oiseau), le Jourdain reçoit des affluents de rive droite venant de Palestine mais surtout de rive gauche. Parmi eux, à 7 km au sud du lac de Tibériade, le Yarmouk (57 km), originaire de Syrie, a un débit abondant (500 millions de m3 annuels) et précieux car constitué d'eaux très peu salées. Les apports des sources et des oueds de rive gauche descendus de Jordanie sont de l'ordre de 600 millions de m3 : parmi eux l'oued Zarka (100 millions de m3) est le plus important. Au total, compte non tenu des prélèvements, le débit "naturel" à son arrivée dans la mer Morte serait de 1 450 millions de m3 (certains auteurs avancent 1 850!) En réalité en raison de l'importance des prélèvements opérés en amont, le Jourdain, en étiage permanent, n'est qu'un mince filet d'eau aisément franchissable. Ainsi, au sortir du lac de Tibériade, le débit “naturel” du fleuve est estimé à 500 millions de m3/an; il n’est en réalité que de 100! Les faibles apports du Jourdain ne compensent plus l’évaporation et le niveau de la mer Morte baisse (15 mètres entre 1955 et 2000 et un mètre par an actuellement). Sa superficie a diminué d’un tiers entre 1960 (950 km2) et 2006 (637 km2). Les industries israélienne et jordanienne d’extraction de sel et de mineraux aggravent la situation en prélevant dans la mer près de 200 millions de m3/an : deux fois plus qu’elle n’en reçoit actuellement. Elle est sérieusement menacée de disparition dans le demi-siècle.

Les eaux du Jourdain conditionnent la vie en Israël et en Jordanie. Arabes et Israéliens se disputent âprement leur utilisation. L'enjeu du Jourdain tient aux deux éléments suivants :

- 23 % de ce débit seulement est originaire d'Israël dans ses frontières de 1967.
- Le débit du Jourdain représente 2 fois le total des autres disponibilités en eau d'Israël et 3 fois celles de la Jordanie.

## 2.2 Le contrôle des eaux du Jourdain : une revendication sioniste permanente

Dès le projet de création du Foyer National juif, les militants sionistes ont souhaité contrôler l'ensemble des eaux du Jourdain et même celles du Litani. Au lendemain de la déclaration Balfour (1917), les représentants du mouvement sioniste ont demandé sans succès à la Grande Bretagne d'intégrer l'ensemble des sources du Jourdain dans la Palestine et de fixer la frontière nord sur le cours aval du Litani! En 1920, dans le même état d'esprit, le Président de l'organisation sioniste mondiale, Chaim Weizman, adresse au premier ministre anglais Lloyd George, une lettre dans laquelle il affirme que :

> *"les frontières (du Foyer national juif) ne sauraient être tracées exclusivement sur la base des limites historiques (= bibliques)...nos prétentions vers le Nord sont impérativement décidées par les nécessités de la vie économique moderne"*

Il ajoute ensuite :

> *"tout l'avenir économique de la Palestine dépend de son approvisionnement en eau pour l'irrigation et pour la production d'électricité, l'alimentation en eau doit provenir des pentes du Mont Hermon, des sources du Jourdain et du fleuve Litani. Nous considérons qu'il est essentiel que la frontière nord de la Palestine englobe la vallée du Litani sur une distance de 25 miles ainsi que sur les flancs ouest et sud du Mont Hermon."*

Il suggère donc que les frontières de la Palestine soient déterminées à partir de considérations hydrauliques. Au même moment, Chaim Weizman précise au ministre anglais des affaires étrangères l'importance considérable du Litani pour la Palestine :

> *"Même si la totalité du Jourdain et du Yarmouk se trouvaient inclus dans la Palestine, il n'y aurait pas assez d'eau pour satisfaire nos besoins. L'irrigation de la Haute Galilée et l'énergie nécessaire, fût-ce à une activité industrielle restreinte doit provenir du Litani. Si la Palestine se trouvait coupée du Litani, du Haut Jourdain et du Yarmouk, elle ne pourrait être indépendante au plan économique "*

Si ces demandes ne sont pas acceptées par la Conférence de Paris, la revendication n'en demeure pas moins. En 1944 Lowdermilk propose un plan, d'ensemble pour le développement des ressources en eau de la Palestine. Sa proposition vise à créer une "Jordanian Valley Authority" sur le modèle de la "Tennessee Valley Authority" aux États-Unis qui favoriserait le développement de fermes et d'industries pour permettre l'intégration de "*4 millions de réfugiés juifs venant d'Europe en plus des 1 800 000 Arabes et Juifs vivant déjà en Palestine et en Transjordanie*". A cet effet, le plan prévoit l'irrigation des terres de la vallée du Jourdain, la dérivation des eaux du Jourdain et du Yarmouk vers le désert du Neguev et l'utilisation des eaux du Litani.

Le plan de partage proposé par l'ONU avantage, de ce point de vue, les colons juifs. Les Juifs à cette époque possèdent 7% des terres et représentent 32 % de la population : ils se voient accorder 55% du territoire. L'État juif est aussi favorisé en termes d'accès aux ressources hydrauliques : le cours supérieur du Jourdain est compris dans son territoire. Tout au long des années cinquante, le désir des Israéliens de mettre la main ou de contrôler plus ou moins directement les terres et les eaux du Sud-Liban est resté vif. David Ben Gourion estimait ainsi qu'il était sage "*de pousser le Liban, c'est à dire les maronites de ce pays à proclamer un État chrétien!*"

## 2.3 Le Plan Johnston (1955) et son rejet

De nombreux plans de répartition des eaux entre Israéliens et Arabes sont mis à l'étude. La tension monte dans la région d'autant plus que des deux côtés débutent des chantiers pour mobiliser l'eau. Le plan Main de 1953, plus ou moins d'origine onusienne, pose le principe que les eaux du Jourdain ne doivent servir qu'à mettre en valeur sa vallée. Il ne concerne donc ni le désert du Neguev, ni le cours du Litani. Il est rejeté par les deux parties en raison de la gestion intégrée de la ressource qu'il supposait. Les Israéliens proposent alors (1954) le plan Cotton et les Arabes le plan du comité technique arabe. Les divergences sont énormes : le plan arabe préconise l'utilisation des seules eaux du Jourdain et uniquement dans la vallée du fleuve; le plan israélien Cotton intègre les eaux du Litani et prévoit l'irrigation de régions- notamment le Neguev- hors de la vallée du Jourdain. Les États Unis envoient en 1954 un émissaire spécial, Éric Johnston, pour tenter de dénouer le conflit. Après de longues tractations, ce denier reprend pour l'essentiel les propositions du plan Main de 1953.

**Le Plan Johnston (1955**) prévoit un partage des eaux du Jourdain et de ses affluents entre les pays riverains selon les quotas suivants : 56 % des eaux reviendraient à la Jordanie, 31 % à Israël, 10 % à la Syrie et 3% au Liban. Le plan, comme toutes les autres propositions antérieures, ne prend pas en compte les ressources souterraines.

| En hm3 | Liban | Syrie | Jordanie | Israël | Total |
|---|---|---|---|---|---|
| Hasbani | 35 | | | | 35 |
| Banias | | 20 | | | 20 |
| Jourdain | | 22 | 100 | 375 | 497 |
| Yarmouk | | 90 | 377 | 25 | 492 |
| Ghor | | | 243 | | 243 |
| Total | 35 | 132 | 720 | 400 | 1287 |
| % | 2,70 | 10,25 | 56 | 31,05 | 100 |

*Tableau 14 : Répartition des eaux du Jourdain et de ses affluents selon le plan Johnston*

Pour des raisons essentiellement politiques, le Conseil de la Ligue arabe tout comme Israël rejettent cette proposition qui, toutefois, restera une base de référence pendant des années pour le partage de l'eau dans la région. Le plan avait été, en effet, accepté par les comités d'experts.

Dès lors, Israël et les États arabes mettent en œuvre leurs propres projets. Israël entreprend la construction d'un grand conduit national dont la première tranche est achevée en 1964. Il est alimenté pour une large part par des pompages (plus de 400 millions de m3/an actuellement) effectués dans le lac de Tibériade. Israël utilise 16 à 17% de son énergie pour effectuer ces pompages. Le «**National Water Carrier» réalise l'interconnexion de l'eau à travers tout le pays** jusqu'au désert du Neguev (capacité 1 150 millions de m3). Ce conduit de 130 km (en partie enterré) permet le transfert des eau en dehors de leur bassin, ce que n'acceptent pas les riverains arabes. Cette eau est à la base du développement agricole d'Israël (178 000 ha irrigués soit 43% des terres). On a pu faire baisser la salinité des eaux du lac de 400 mg/l en 1950 à 220 en 1995 par le captage des sources salées de la rive occidentale qui se déversent en profondeur dans le lac et leur rejet dans le Jourdain.

De leur côté les **États arabes mettent à exécution des projets qui visent à détourner les eaux des sources du Jourdain** soit vers le Litani, soit vers le Yarmouk. Israël riposte en lançant des attaques contre les chantiers ouverts sur le Banias ou le Yarmouk. De son côté, dès 1957, la Jordanie entreprend la construction sur la rive gauche du Jourdain du **canal du Ghor** qui est long actuellement de 120 km qui serait alimenté grâce à la retenue (200 millions de m3) du barrage de Mukhaïba sur le Yarmouk. Israël s'oppose à la réalisation de ces barrages par des bombardements et la destruction des chantiers. Seule est tolérée la poursuite des travaux du canal d'irrigation du Ghor sur la rive gauche du Jourdain. Le canal Abdallah (120 km) est alimenté à partir de 1961 par la prise d'eau d'Adissiya (175 millions de m3/an) sur le Yarmouk. Le dispositif est complété par la mobilisation des eaux des oueds adjacents, et notamment celles du Zarqa sur lequel le barrage du roi Tahal, récemment surélevé autorise un réservoir de 90 millions de m3. Le barrage sur le wadi Arab apporte 20 millions de m3 et les autres barrages totalisent 105 millions de m3. L'opération d'aménagement est d'envergure. Plus de 140 000 personnes se sont installées dans cette région pionnière; 30 000 ha ont été transformés en terres irriguées, distribuées par lots de 3 à 4 ha à des colons ou mises en valeur par de grands exploitants privés. Sur 1% du territoire national, la vallée du Jourdain contribue pour plus du tiers à la production agricole (notamment pour les fruits et légumes, les plantations d'agrumes et de bananiers), grâce à l'introduction de techniques modernes (cultures sous serre, arrosage au goutte à goutte).

## 2.4 Extensions territoriales d'Israël et contrôle des eaux

La compétition pour l'eau fut encore ravivée avec la guerre de 1967. En occupant les hauteurs du Golan les Israéliens rendent impossible le projet arabe de dérivation des eaux du Jourdain supérieur vers le Yarmouk et ils contrôlent deux des sources du Jourdain (aux sources du Dan localisées dans le territoire de 1948, s'ajoutent les sources du Banias).

En occupant le triangle du Yarmouk, ils peuvent contrôler toute la partie aval du fleuve qui marque la limite entre le Jordanie et le Golan (ils en tirent environ 100 millions de m3/an) ainsi que la prise d'eau jordanienne qui alimente le canal du Ghor.

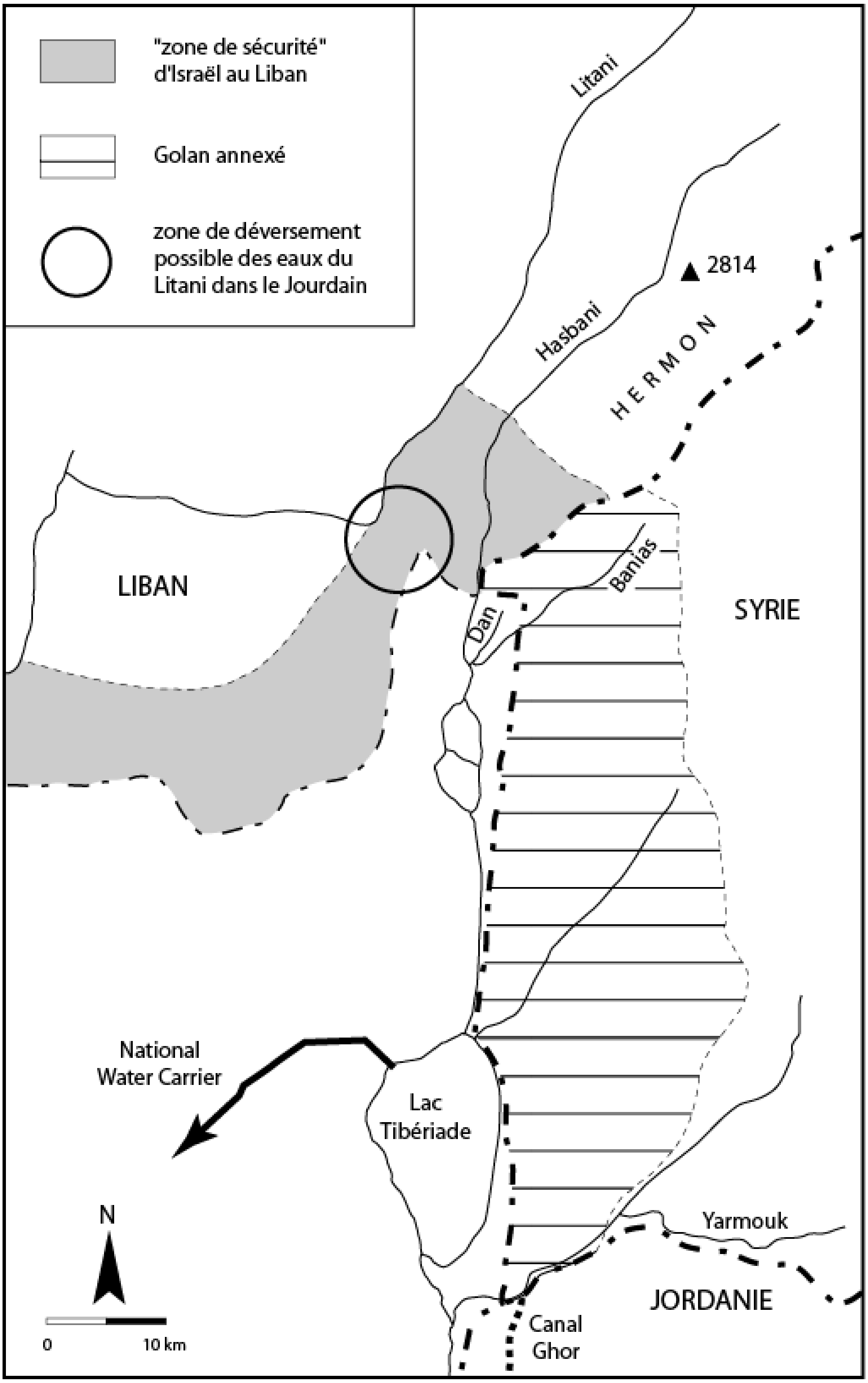

*Figure 17 – Les sources du Jourdain*

Alors que, jusqu'à une date récente, l'utilisation des eaux du Jourdain par les riverains était conforme aux quotas du plan Johnston, un dépassement par Israël est très sensible du fait de l'augmentation des pompages.

**Les eaux convoitées du Litani**

Le Litani (170 km) prend sa source dans la plaine de la Bekaa et draine, sur 2160 km2, le du territoire libanais. Son tracé est nord-sud sur une grande partie de son parcours, puis il s'infléchit vers l'ouest et le Litani va se jeter en Méditerranée un peu au nord de Tyr *(carte 16).* Ce petit fleuve côtier est relativement bien alimenté. Son bassin s'inscrit dans une zone où les précipitations dépassent 500 mm/an. Il bénéficie notamment des pluies importantes qui s'abattent sur la Montagne libanaise. Son module annuel moyen est de l'ordre de 900 millions de m3, plus de la moitié du débit naturel du Jourdain!

Les eaux du Litani ont longtemps été sous-exploitées dans un Liban relativement bien pourvu en ressources hydrauliques. A la veille de la guerre civile libanaise (1975), on irriguait 20 000 hectares dans le bassin du Litani essentiellement par des pompages dans les nappes phréatiques. Des plans ont été élaborés à l'époque pour une meilleure mobilisation des eaux du fleuve. Une extension de l'irrigation nécessita prélèvement de 230 millions de m3 sur le fleuve. On souhaitait aussi dériver, à des fins de production d'électricité, une partie des eaux (460 millions de m3) par deux tunnels sous la Montagne libanaise vers un petit fleuve côtier, le nahr al-Awali qui dévale le versant occidental du Mont Liban. Bref, la quasi totalité des eaux pouvait être utilisée.

La guerre civile libanaise n'a pas permis la mise en œuvre de ces projets et actuellement la plus grande partie des eaux du Litani se déverse directement dans la mer .... de quoi susciter bien des convoitises dans cette région où la pénurie se fait de p plus menaçante!

# 3. Les eaux souterraines disputées entre Israël et les Palestiniens

A l'issue de sa victoire de 1967, Israël est en mesure de contrôler la totalité des ressources souterraines de la région

## 3.1 Un enjeu majeur : les aquifères de Cisjordanie et de Gaza

**3.1.1** La **Cisjordanie occupée** (Judée au sud, Samarie au nord) est une région de collines (l'altitude peut atteindre 1 000 m). C'est une région arrosée : de 500 mm jusqu'à 700 mm. La structure géologique est constituée d'une couche de calcaire épaisse de 600 m. surmontant des roches imperméables. **L'aquifère** se situe dans les couches calcaires. Son aire de recharge est, pour l'essentiel, en territoire palestinien le long des pentes supérieures et des crêtes de la chaîne à plus de 500 m. d'altitude. La nappe phréatique est à une profondeur de 200 à 400 m. En raison de sa structure anticlinale, la nappe est drainée dans plusieurs directions. On distingue trois zones dans cet aquifère : *(carte 18)*

**L'aquifère occidental** s'écoule en Israël, sa capacité est de 350 millions de m3/an. On estime qu'il est alimenté à 70% par des pluies qui tombent en Cisjordanie occupée. Ce sont les premiers colons juifs qui ont commencé à l'exploiter de façon intensive à partir de 1930 pour irriguer les orangeraies en creusant des puits ou en captant des sources dans la partie aval. Aujourd'hui, l'aquifère (Yarkon-Tanimin) est drainé par des centaines de puits situés en deçà de la «ligne verte» c'est à dire à l'intérieur des frontières d'Israël qui exploite ainsi la quasi totalité de la ressource : 330 millions de m3, 20 seulement revenant aux Palestiniens.

**L'aquifère nord-oriental** qui part des environs de Naplouse s'écoule vers la vallée de Jezreel. Son débit normal est estimé à 130 millions de m3. Il est alimenté en totalité par les précipitations tombant en territoire occupé. Certains puits et sources sont depuis toujours utilisés par les Palestiniens habitant les villages de la région. Les premiers colons juifs se sont aussi servis de cette eau avant la création d'Israël en 1948. Actuellement, les Israéliens exploitent une grande partie de l'aquifère (110 millions de m3/an).

**L'aquifère oriental** a un débit potentiel de 200 millions de m3/an dont près de la moitié en eau saumâtre. Il est en totalité alimenté par les pluies de Cisjordanie et s'écoule vers le Jourdain. Son eau est utilisée depuis toujours par les agriculteurs palestiniens. Plus récemment, dans les Territoires occupés, les Israéliens ont creusé des puits de grande profondeur ce qui, parfois, a eu pour conséquence de réduire le débit des puits ou des sources traditionnellement utilisés par les Palestiniens. Les Israéliens pompent 40 millions de m3/an dans cet aquifère, les Palestiniens 80, le reste est inutilisé car bien souvent les eaux sont trop salées.

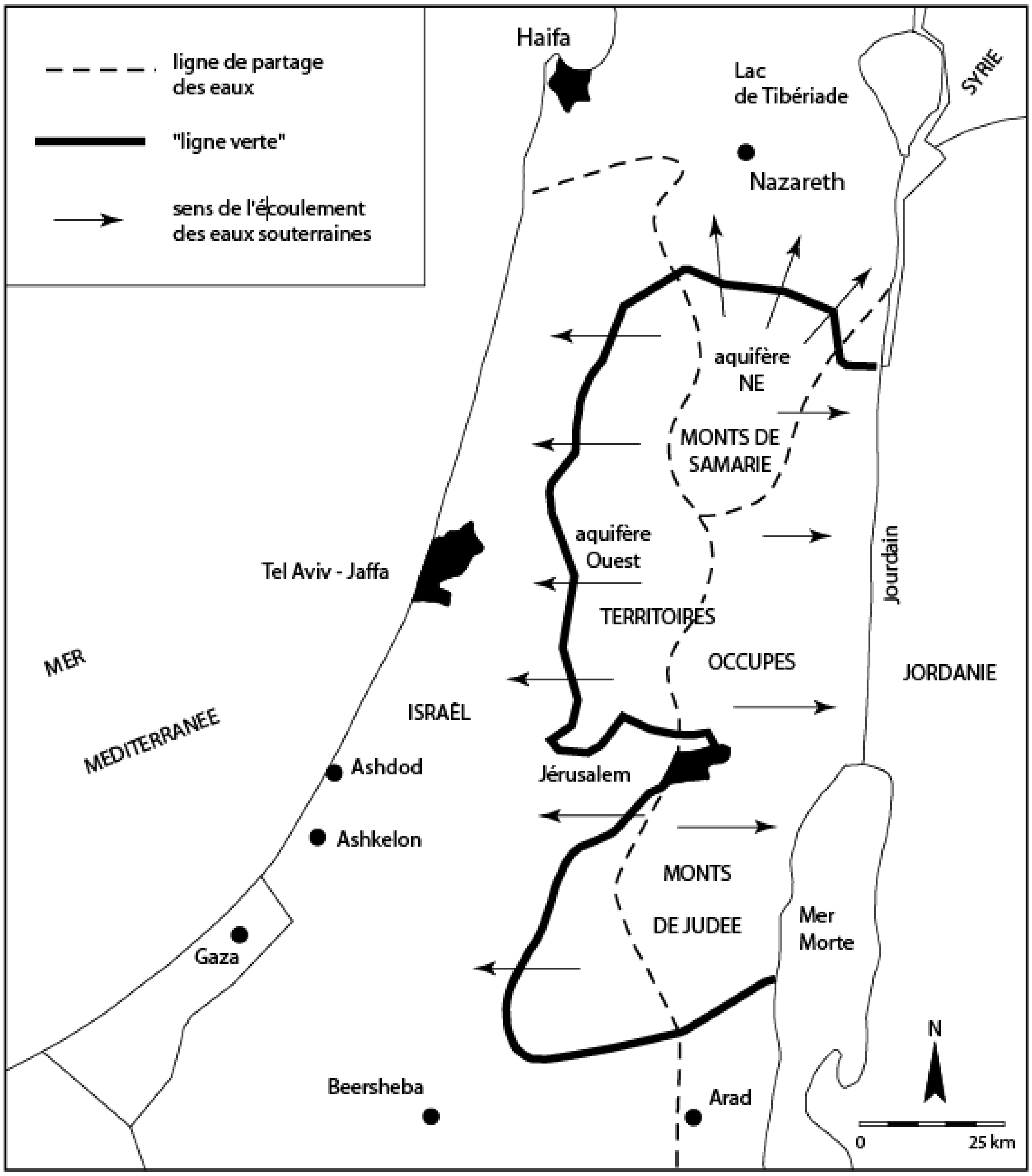

*Figure 18 - L'aquifère cisjordanien*

**3.1.2.L'aquifère côtier n'a pas l'ampleur de celui de la Cisjordanie**. La plaine côtière est formée de sables et de grès qui surmontent des marnes et des craies imperméables. Du Mont Carmel à Gaza gît une nappe phréatique (300 millions de m3), exploitée par 1600 puits, à une profondeur de 20 à 50 mètres. La partie de la nappe qui s'étend sur le territoire de Gaza a une capacité de 60 millions de m3 mais elle est surexploitée depuis des années à hauteur de 120 à 130 millions de m3 dont la moitié était destinée aux colonies agricoles avant l'évacuation de Gaza par Israël (2005).

## 3.2 L'exploitation des aquifères par Israël : le droit du plus fort

Cette exploitation des eaux souterraines n’est possible qu’avec la mise en place d’une législation discriminatoire. Sous le Mandat, l’eau était considérée comme une propriété privée. La législation israélienne, étendue aux Territoires occupés, la considère comme une propriété d'État : il faut une autorisation pour l’utiliser. Les eaux des Territoires occupés sont déclarées ressources stratégiques sous contrôle militaire par la puissance occupante. Le droit jordanien applicable avant 1967 à la Cisjordanie interdit le transport de l'eau d'un bassin de drainage à un autre sans autorisation. Actuellement, sous le régime israélien, toutes les eaux font partie d'un réseau national et peuvent par conséquent être transportées partout où le besoin s'en fait sentir que ce soit d'un bassin à un autre de la Cisjordanie ou le plus souvent de la Cisjordanie vers d'autres bassins à l'intérieur des frontières d'Israël comme la région du Néguev. **L'ensemble des ressources en eau de la Cisjordanie et de la bande de Gaza sont placées sous le contrôle de la *Mekorot*, la compagnie israélienne des eaux** qui gère 1,3 millard de m3 /an et distribue 90 % de l’eau potable. Les Palestiniens soulignent avec inquiétude le fait que dans tous les nouveaux projets mis en œuvre par Israël, les équipements essentiels de contrôle (réservoirs régionaux, vannes, stations d'observation) sont localisés à l'intérieur des implantations israéliennes.

**La culture irriguée palestinienne est sévèrement limitée** et le coût de l’eau distribuée par le réseau public, grâce à un jeu subtil de subventions, est 5 fois moins cher pour un agriculteur israélien que pour un agriculteur palestinien. En 2004, l’eau- quel que soit son usage- est cédée aux Palestiniens à 1 $ le m3, elle est proposée aux colons à 0,2 $ pour l’irrigation et à 0,5 pour l’usage domestique. Les Palestiniens n’ont pratiquement pas le droit de forer de nouveaux puits; par contre, les colons israéliens peuvent disposer de puits très profonds qui assèchent les puits arabes voisins, dont la profondeur est limitée. La consommation des Palestiniens est en outre très sévèrement contrôlée. La quantité d'eau disponible pour l'agriculture palestinienne a été gelée par les autorités à 90 millions de m3/an pour les 400 villages palestiniens. On estime actuellement que plus de 100 000 hectares de bonnes terres en Cisjordanie et 40 000 à Gaza pourraient être irriguées : elles ne le sont pas faute d'eau. Sous l’administration jordanienne (avant 1967), 27% des terres de Cisjordanie étaient irriguées contre seulement 6% acuellement. A l'inverse, les quantités d'eau attribuées à l'agriculture israélienne dans les Territoires occupés ont doublé au cours de la décennie 1980. En 1990 60 millions de m3 sont alloués aux 30 colonies agricoles de Cisjordanie!

En occupant la Cisjordanie, les Israéliens assurent en outre la recharge de leur nappe phréatique littorale surexploitée. Sur le potentiel en eau non saumâtre commun de la Cisjordanie et d'Israël (aquifères occidental et nord oriental) de quelque 480 millions de m3, Israël et les colons juifs en utilisent 91% soit 440 millions de m3 et les Palestiniens 9% seulement soit 40 millions de m3! En outre, dans l'aquifère oriental qui s'écoule vers la vallée du Jourdain, Israël pompe 30 millions de m3 utilisés par les colonies agricoles de la vallée du Jourdain. Au total, **Israël pompe chaque année 470 millions de m3 à partir des ressources souterraines de Cisjordanie soit le 1/4 de la consommation du pays!**

Comme l'indique un récent rapport des Nations Unies, l'application des lois israéliennes dans les Territoires occupés "a produit des modifications substantielles des usages légitimes de l'eau selon la législation de Gaza, des hauteurs du Golan et de la Cisjordanie". La politique en vigueur est conçue pour assurer prioritairement un approvisionnement en eau suffisant aux colons juifs et au réseau hydraulique israélien. Ce n'est qu'une fois ces priorités assurées que les droits et les besoins des Palestiniens sont pris en considération.

| en millions de m3 | Potentiel | Prélèvements totaux | Israël et colo | Palestiniens |
|---|---|---|---|---|
| **Nappe de Cisjordanie** | | | | |
| • aquifère occidental | 350 | 350 | 330 | 20 |
| • aquifère nord oriental | 130 | 130 | 110 | 20 |
| • aquifère oriental | 200* | 110 | 30 | 80 |
| s/Total Cisjordanie | **680** | **590** | **470** | **120** |
| *%* | | *100* | *80* | *20* |
| **Nappe de Gaza** | **60** | **120**** | **60** | **60** |
| % | | *100* | *50* | *50* |
| **Total** | **740** | **710** | **530** | **180** |
| % | | 100 | 75% | 25% |

* : 90 millions de m3 sont des eaux saumâtres

** : il y a surexploitation de la nappe (environ 60 millions de m3)

*Tableau 15 : L'exploitation des nappes souterraines de Cisjordanie et de Gaza*

Le pompage massif des eaux de Cisjordanie et de la bande de Gaza par les colons et par le gouvernement israélien a conduit à une baisse de la qualité des eaux. Si le pompage se poursuit sur une longue période le processus peut être irréversible. En Cisjordanie le phénomène reste encore ponctuel. On signale une augmentation de la salinité des eaux dans la région de Jéricho. La situation est beaucoup plus préoccupante dans la bande de Gaza où depuis des années les autorités pratiquent un surpompage. On estime qu'on tire quelque 120 à 130 millions de m3/an alors que la reconstitution naturelle de la ressource n'est que de 60 millions de m3. Le niveau hydrostatique s'abaisse de 15 à 20 cm par an. La nappe en bordure du littoral est envahie par la mer. L'eau pompée n'est plus, dans certains cas, propre à la consommation humaine. La salinité progresse chaque année de 15 à 20 mg/l et 70% des eaux souterraines y dépassent le niveau de salinité de 500 mg/l alors que la norme maximale préconisée par l'O.M.S. est de 250! Les experts israéliens font remarquer que si le pompage excessif se poursuit, les dommages qu'il provoque seront bientôt irréparables. L’absence d'égouts dans toute la bande de Gaza ajoute encore à la pollution.

Dès à présent, les ressources en eau (courantes et souterraines) de la région sont insuffisantes pour satisfaire les besoins de deux populations israélienne (7,3 millions d'habitants) et palestinienne (4 millions). Israël et les colons juifs dans les Territoires occupés accaparent une part disproportionnée de cette ressource rare dans le territoire de l'ancienne Palestine. Israël absorbe 86% des ressources en eau, les Palestiniens des Territoires occupés 10% environ et les colons 4%.

# 4. Un avenir préoccupant

L'eau est rare. La pénurie touche la région. Cette situation de rareté est d'autant plus mal ressentie que la répartition de la ressource est extraordinairement inégalitaire entre les trois partenaires/adversaires : Israël, la Jordanie et les Palestiniens.

## 4.1 Israël ou l'eau d'irrigation en question

Israël, déjà dans une situation difficile, doit trouver dans un proche avenir des quantités supplémentaires pour faire face à la poussée démographique attisée par la reprise de l'immigration et à la montée de l'urbanisation.

Actuellement, la consommation annuelle d’Israël est de l’ordre de 2 milliards de m3

- Les <u>ressources renouvelables</u> fournissent 1 400 millions de m3 (75% du total) :

- un peu plus du 1/3 revenant aux eaux de surface dont évidemment le lac de Tibériade
- un peu moins des 2/3 revenant aux eaux souterraines soit de l'aquifère côtier dans les sédiments quaternaires (20%), soit dans l'aquifère cisjordanien des calcaires mésozoïques (40%), le reste provenant d'aquifères locaux.

• Les autres ressources dites non conventionnelles (25%)

- 220 millions de m3 d'eaux retraitées
- 145 provenant de l'utilisation des eaux saumâtres pour des usages industriels
- 150 provenant d'eau de mer dessalée

Pour satisfaire ses besoins le pays surexploite occasionnellement, sa nappe littorale depuis longtemps, bien avant la mise en place du Grand Conduit. Depuis 1983, il a recours de façon constante à des eaux non renouvelables.

En outre, **pour les 2/3, la satisfaction des besoins d'Israël est assurée par des ressources provenant de l'extérieur des frontières de 1948** : 1/3 en gros venant de Cisjordanie et de la nappe de la bande de Gaza et 1/3 provenant du lac de Tibériade et du Yarmouk. On voit bien que la sécurité de l'approvisionnement en eau est une donnée essentielle dans la politique d'extension territoriale d'Israël. Le pays ne peut plus se passer, de ce point de vue, des Territoires occupés et du Golan. Par ailleurs de sérieux problèmes qualitatifs se posent : nitrification des eaux et pollution.

Quelles sont les solutions envisageables pour répondre à une demande qui va sans doute augmenter dans les prochaines années d'au moins 500 millions de m3 annuels? Les hypothèses suivantes peuvent être envisagées pour améliorer l'offre.

• recourir au dessalement des eaux de mer et des eaux saumâtres mais c'est une solution coûteuse. Actuellement deux usines (Ashkelon et Eilat) produisent 135 millions de m3/an. Cinq autres sont à l'étude qui porteraient le volume traité à 350 millions de m3 annuels à l'horizon 2013 et 750 Millions en 2020.

• augmenter les quantités d'eau traitée et recyclée (220 millions de m3 actuellement).

• importer de l'eau du Litani ce qui suppose un règlement géopolitique de la situation régionale qui paraît totalement exclu dans l'immédiat. Importer par voie maritime de l'eau turque provenant de la Rivière Manavgat qui dévale les pentes du Taurus pour se jeter dans la Méditerranée dans la plaine de Pamphylie. Des aménagements ont bien été effectués pour cette exportation. Le projet a connu un arrêt brutal en 2006 pour de multiples raisons, notamment l'aggravation de la situation géopolitique régionale.

Mais le pays pourrait aussi mettre en œuvre une **autre politique d'affectation de la ressource**. Le secteur agricole israélien est, de très loin, le plus fort consommateur : en 2004, 56% de la consommation israélienne soit en moyenne 1  milliard de m3/an pour irriguer 45% des terres agricoles (190 000 hectares). La progression de la consommation a été impressionnante entre 1947 et 1974, actuellement Israël semble parvenu à un palier. La part de l'irrigation qui représentait 82% de la consommation totale d'eau en 1962 n'en représente plus que 62% trente ans plus tard et 56% actuellement. Les superficies ont en effet peu progressé et surtout Israël a recours à des techniques sophistiquées d'irrigation et très économes (goutte à goutte, gestion informatisée de l'irrigation etc...). La charge d'irrigation par hectare est en moyenne de 6 100 m3 alors que dans les pays arabes voisins on atteint facilement le double. Il n'en demeure pas moins que la consommation agricole au regard des disponibilités paraît disproportionnée[2]. Face à un avenir pour le moins sombre où la consommation urbaine ne fera que croître, on peut se demander si ce n'est pas l'utilisation de l'eau à des fins agricoles qui est à revoir. Dans une économie à très forte dominante tertiaire, dans un pays très urbanisé, où la part de l'agriculture dans le PIB est très faible (4%), est-il judicieux de maintenir un secteur agricole qui consomme une telle part d'une ressource rare à des fins exclusives d'exportation (vergers d'agrumes)? La définition d'une nouvelle politique est très difficile car c'est remettre en cause un des mythes fondateurs de la création d'Israël, celui du retour à la terre promise et de sa mise en valeur.

| **en %** | **1962** | **1992** | **2004** |
|---|---|---|---|
| secteur agricole | 83 | 62 | 56 |
| secteur urbain | 13 | 31 | 38 |
| secteur industriel | 4 | 7 | 6 |

*Tableau 16 : Évolution de la consommation d'eau en Israël par grands secteurs*

A l'évidence, il y a contradiction entre les espoirs économiques d'Israël et ses ressources en eau. Les règlements politiques futurs ne peuvent pas faire l'impasse sur ces problèmes.

[2] La différence de consommation moyenne d'un Israélien et d'un Palestinien est faussement illustrée par l'image du premier plongent dans sa piscine juxtaposée à celle du second recueillant un maigre filet d'eau d'un robinet hors d'âge. C'est le secteur agricole sophistiqué et subventionné par les prix de l'eau qui explique fondamentalement la consommation élevée d'Israël.

## 4.2 Pour les Palestiniens, une situation dramatique

La question apparaît essentielle et, malheureusement, sans issue satisfaisante. **L'eau manque cruellement**. **La consommation palestinienne en Cisjordanie (compte tenu de la récupération des eaux de pluie) s'élève à 135 millions de m3 et à 65 à Gaza. Ce sont des volumes très faibles qui correspondent à une norme individuelle moyenne d'un peu plus de 50 m3/an/habitant en 2009. Elle est près de six fois inférieure à celle des Israéliens (290 m3/an/habitant)** L'approvisionnement en eau en Cisjordanie est assuré par des puits artésiens (382) qui fournissent environ 50 millions de m3/an, le reste provient des 295 sources de la région, des eaux de ruissellement et des citernes de récupération des eaux de pluie. La situation dans le territoire autonome de Gaza est aussi très critique. La nappe depuis longtemps surexploitée connaît des intrusions de l'eau de mer : l'eau fournie est à peine propre à la consommation humaine. Depuis l'abandon du contrôle israélien, la situation a empiré car les Palestiniens multiplient ces dernières années la construction de puits. La question de l'eau est cruciale pour l'avenir des Territoires Occupés. Il est peu probable qu'Israël puisse accepter une véritable indépendance des Territoires occupés quand on sait que le potentiel en eau actuellement exploité, commun à Israël et à la Cisjordanie est considérable. Sur un prélèvement annuel d'environ 590 millions de m3 en Cisjordanie et 120 à Gaza, 530 millions de m3, soit 75%, sont utilisés en Israël ou dans les colonies de Cisjordanie et de Gaza. *(tableau 15)*

Les éventuelles discussions à venir avec les Palestiniens risquent d'être très difficiles. Ils réclament 80% des ressources en eau de la Cisjordanie alors qu'ils n'ont accès pour l'instant qu'à 20%. Accepter cette revendication priverait Israël de 360 millions de m3 ce qui signifie une diminution de 20% des ressources actuellement disponibles. La fin des prélèvements Israéliens sur les Territoires occupés conduirait à un déclin inévitable de l'agriculture et de l'industrie dans l'État hébreu.

## 4.3 En Jordanie, la pénurie est là

La situation est très critique d'autant plus que la croissance urbaine, notamment l'approvisionnement de la capitale Amman, et la mise en valeur des terres exigent des quantités croissantes d'eau. La pénurie n'est pas à venir, elle est là. Ce pays de 5,7 millions d'habitants est en zone aride : 72% du territoire reçoit moins de 100 mm/an et 6% seulement plus de 300 mm. La ressource en eau douce est de l'ordre de 158 m3/an/habitant en 2007, elle s'abaissera à 115 en 2025. La Jordanie qui consomme actuellement 880 millions de m3 doit trouver 500 millions de m3 supplémentaires dans le court terme.

Les solutions envisagées reposent toutes sur de nouvelles ressources d'eau de surface car les nappes souterraines sont désormais largement surexploitées (on estime que le bilan cumulé des surpompages est déjà équivalent à une année de consommation!). Il faut compléter l'équipement des wadi sur l'escarpement qui domine le Jourdain mais les volumes à récupérer sont insignifiants.

Actuellement une usine de retraitement des eaux usées est en construction d'une capacité de 190 millions de M3/an. Mais tout l'avenir jordanien repose désormais sur la mobilisation des eaux du Yarmouk qui, pour l'instant, est loin d'être réalisée.

Jusque-là Israël s'était opposé par la force (bombardements) à tout équipement de ce fleuve venu de Syrie. Toutefois depuis 1961, grâce à la médiation américaine, la Jordanie peut alimenter le canal d'Abdallah à partir des eaux du Yarmouk (prise d'Adassiya). Israël tire de son côté 100 millions de m3 de ce cours d'eau depuis l'occupation du triangle du Yarmouk en 1967.

L'accord de paix de 1994 laissait entrevoir certaines possibilités qui ne se sont pas encore toutes vérifiées. Le traité de paix comporte les clauses suivantes :

- Israéliens et Jordaniens pompent dans les eaux souterraines près de la confluence entre Jourdain et Yarmouk et Israël s'engage à garantir 50 millions de m3/an à la Jordanie et à participer à des travaux sur le Yarmouk, en aval de la prise d'Adassiya, qui permettront de procurer au Royaume 100 millions de m3 supplémentaires (cette dernière clause relative aux travaux sur le Yarmouk n'est toujours pas honorée). Par ailleurs Israël se voit garantir les 100 millions de m3 annuels qu'elle tire de cette même zone depuis 1967.
- L'accord de paix prévoit des échanges intersaisonniers d'eau entre les deux pays.
- Le traité de paix israélo-jordanien reconnaît les droits du royaume hachémite sur les eaux du Jourdain dont les eaux sont actuellement entièrement exploitées par l'État hébreu. La Jordanie a commencé à recevoir de l'eau d'Israël après la construction d'une conduite reliant le lac de Tibériade au canal du roi Abdallah.

Enfin, pour la Jordanie, la construction du barrage de l'unité (Al Wahda) pourrait apporter une solution. Ce barrage pourrait former une retenue de 225 millions de m3 dont 120 reviendraient à la Jordanie (50 pour les seuls besoins d'Amman). La réalisation de cet ouvrage reste malheureusement suspendue à l'éventuel règlement de paix israélo-syrien car il implique une coordination entre les trois pays en matière hydraulique. Par ailleurs, la Jordanie peut nourrir des craintes concernant l'ampleur de la retenue car la Syrie prévoit de construire de nombreux barrages sur la partie amont du Yarmouk et de ses affluents (en 1992 : 150 millions de m3 sont déjà stockés).

## Conclusion

Au plan politique : le dossier de l'eau est de toute première importance dans un éventuel règlement de paix. Le problème le plus ardu à régler est celui du partage des eaux de l'aquifère cisjordanien.

Au plan des ressources : la région manque d'eau et en manquera de plus en plus d'eau car les ressources sont limitées et la demande ne fera qu'augmenter. La population pour l'ensemble du bassin du Jourdain pourrait avoisiner 20 millions d'habitants en 2025! Il faut s'attendre à une extraordinaire augmentation des besoins : un rapport de la Banque Mondiale (1994) prévoit que la demande vers 2040 pour Israël, les Territoires actuellement occupés et la Jordanie pourrait s'élever à 7 milliards de m3! De toute évidence les ressources du bassin du Jourdain sont insuffisantes. Le règlement ne peut être que régional par des transferts d'eau (on comprend une fois de plus tout l'intérêt porté au Litani libanais) ou par le recours coûteux à des ressources non conventionnelles : dessalement de l'eau de mer, recyclage des eaux usées.

# V. L’eau, la ville et les champs au Maghreb et ailleurs

## 1. Un potentiel médiocre : éléments d’évaluation

1.1 Les données d’ensemble

1.2 Une forte diversité régionale

## 2. Priorité à la grande hydraulique

2.1 Le Maghreb face au défi alimentaire

2.2 Des réalisations récentes très contrastées

2.3 L’essor de la moyenne et de la petite hydraulique

## 3. L'eau et la ville : une demande exponentielle?

3.1 La diversité de la consommation urbaine

3.2 La course aux captages

3.3 De très sévères concurrences entre secteurs utilisateurs

3.4 Pénurie et déstabilisation régionale : l'alimentation en eau de la région algéroise

## 4. Un avenir préoccupant

4.1 L’inexorable montée de la demande : un déficit à partir de 2010/2015

4.2 Une mobilisation accrue et une meilleure gestion du potentiel actuel

4.3 Peut-on compter sur une amélioration de l’offre?

## 5. Faire reverdir le désert?

5.1 Eau et irrigation dans le désert arabique

5.2 Les tentatives maghrébines

La question de l’eau dans les trois pays du Maghreb (Maroc, Algérie, Tunisie) se pose dans des termes spécifiques par rapport au reste du Monde Arabe. Ici le problème de l’eau, de sa répartition se place dans un cadre strictement national qui exclut les interférences d’ordre géopolitique. Aucun organisme fluvial ne traverse plusieurs pays. Par ailleurs, dans le contexte de rareté de la ressource, les États maghrébins ne disposent pas des ressources financières suffisantes qui leur permettraient de recourir largement, à l’instar des pays du Golfe, aux ressources non conventionnelles. Les problèmes de gestion sont donc essentiels et notamment l’allocation de l’eau aux différents utilisateurs. C’est sans doute dans les pays du Maghreb que les concurrences entre les usages urbains et agricoles sont les plus vives.

# 1. Un potentiel médiocre : éléments d’évaluation

## 1.1 Les données d’ensemble

Le bilan des ressources en eaux douces renouvelables et de leur utilisation dans chacun des trois pays figurent dans les tableaux ci dessous :

| **2009** | **Population 2** | **Ressources to** | | **Prélèvements annue** | | | **Utilisation** | | |
|---|---|---|---|---|---|---|---|---|---|
| | millions | Km3 | m3/hab | km3 | % res | m3/hab | Eau urba | Industrie | Irrigation |
| Algérie | 35,4 | 12 | 339 | 6 | 50 | 176 | 22 | 14 | 64 |
| Maroc | 32 | 29 | 906 | 12,5 | 42 | 394 | 6 | 3 | 91 |
| Tunisie | 10,4 | 4,6 | 442 | 3 | 65 | 294 | 10 | 6 | 84 |
| Total | 77,8 | 45,6 | 586 | 21,5 | 47 | 282 | 10 | 6 | 84 |

*Tableau 17 : Potentiel et utilisation des eaux douces au Maghreb*

| | **Potentiel** | **Eaux mobili** | **Eaux régularisables** | | | **Prélèvements** | |
|---|---|---|---|---|---|---|---|
| km3 | | | surface | souter. | total | km3 | % eaux régularis. |
| Algérie | 12 | 9,5 | 5 | 3,5 | 8,5 | 6 | 70 |
| Maroc | 29 | 21 | 12 | 4,5 | 16,5 | 12,5 | 75,8 |
| Tunisie | 4,6 | 3,8 | 1,7 | 1,7 | 3,4 | 3 | 88 |
| Total | 45,6 | 34,3 | 18,7 | 9,7 | 28,4 | 21,5 | 75 |
| | 100 | 75% | | | 62% | 47% | |

*Tableau 18 : Mobilisation des eaux douces au Maghreb*

Les données des tableaux peuvent s'analyser d'un double point de vue.

En considérant la **ressource hydrologique**, on distingue entre :

- les eaux de surface, le ruissellement qui avec 32 milliards de m3 représentent 70% du potentiel total estimé à 45,6 milliards de m3.
- les eaux souterraines pour lesquelles on ne retient que les nappes renouvelables. Ces nappes se localisent à des profondeurs très variables de quelques dizaines à quelques centaines de mètres. Le grand aquifère saharien de l'albien considéré comme fossile est exclu du décompte. Le potentiel des eaux souterraines est estimé à 14 milliards de m3 soit 30% du total.

Du point de vue de **l'économie et de la gestion de l'eau** on distingue entre trois notions :

- les ressources potentielles **:** constituées par les apports mesurés au niveau des stations hydrométriques ou calculées par des formules hydrologiques.
- les ressources mobilisables recouvrent la part des ressources potentielles maîtrisables par des ouvrages hydrauliques (barrages, lacs collinaires, stations de pompage).
- les ressources régularisables sont la partie des ressources mobilisables garanties à l'utilisation quelles que soient les conditions hydrologiques 9 années sur 10 pour l'AEP (alimentation en eau potable), 8 sur 10 pour les autres usages.

Avec une ressource en eau douce de 45,6 milliards de m3 soit 586 m3/an/habitant, le Maghreb n'apparaît pas comme particulièrement bien doté. Il se situe très en dessous de la norme de 1 000 m3/an/hab. qui détermine, on l'a vu, le seuil de pénurie. Avec l'accroissement attendu de la population (94 millions d'habitants en 2025), la norme tombera à 485 m3/an/hab. Ce potentiel n'est en réalité que très partiellement utilisable. Les volumes régularisables sur lesquels on peut compter véritablement tombent à 28 milliards de m3 seulement soit 62% des possibilités. C'est, pour **les eaux de surface**, la conséquence du régime des précipitations. La pluviométrie n'est pas négligeable *(carte 19)* mais les pluies sont très inégalement réparties : elles tombent essentiellement sur la frange littorale ou la chaîne de l'Atlas. Elles sont, en outre, très irrégulières d'une année sur l'autre et au cours de la même année se concentrent dans la proportion de 75% sur les trois mois d'hiver. Les ruissellements opposent des crues d'hiver spectaculaires et des étiages estivaux très creusés. Ce ne sont pas les meilleures conditions pour stocker et mobiliser les eaux courantes et une part importante de l'écoulement se déverse directement dans la mer sans qu'on ait pu l'utiliser! Des contraintes techniques (sites des barrages, recharge des nappes) ou économiques (coûts de mobilisation ou d'exhaure) rendent compte également du faible volume régularisé.

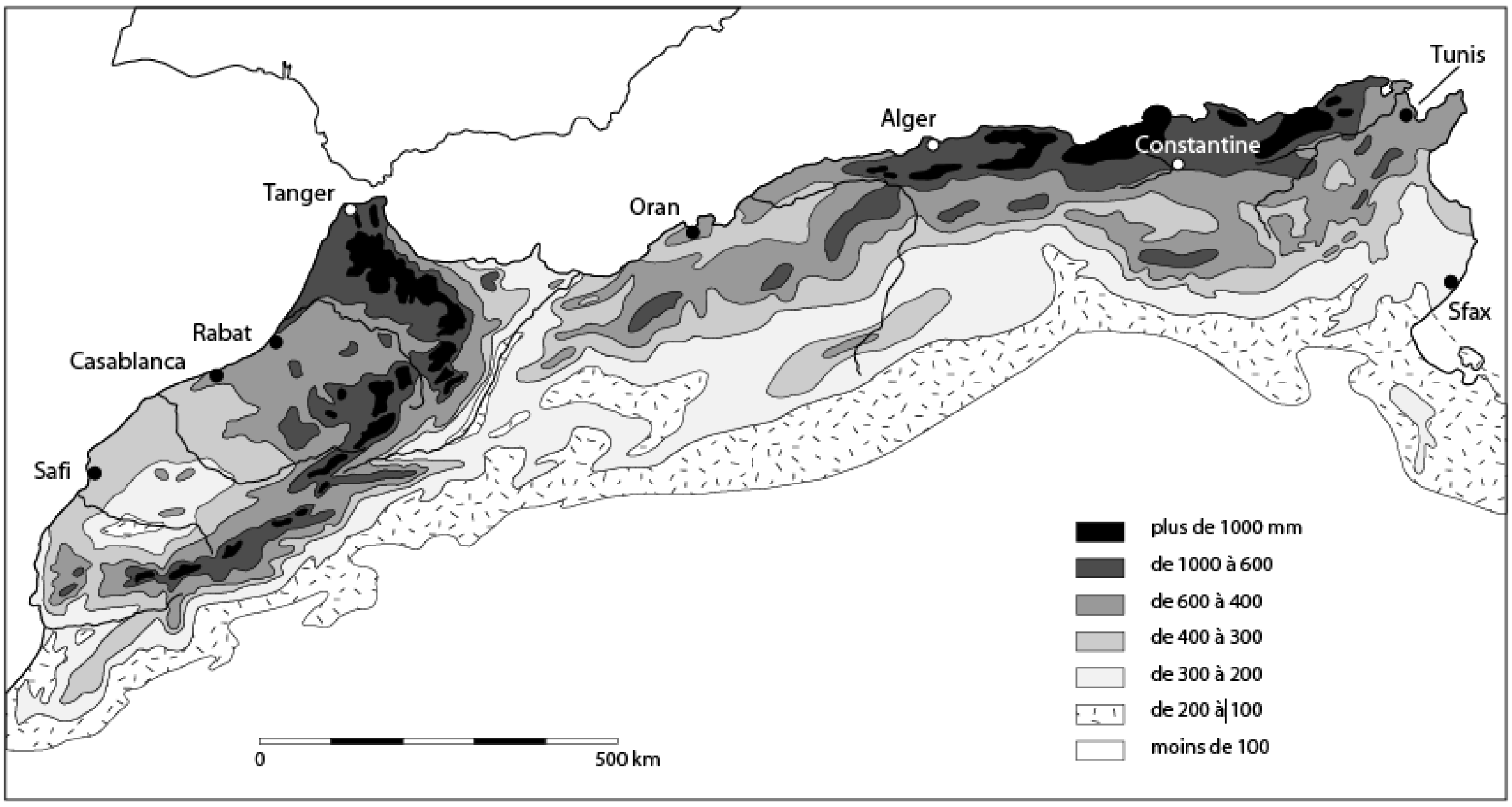

*Figure 19 – Les précipitations au Maghreb*

Les difficultés d'utilisation des eaux de surface ont amené à recourir depuis des siècles aux **ressources souterraines** : à elles seules elles représentent plus du tiers des volumes régularisables. Elles sont plus largement sollicitées en Algérie et en Tunisie qu'au Maroc. Dans le Tell cloisonné les nappes sont partout alimentées par les pluies qui tombent sur les versants. Au Sahara, où les écoulements sont absents, le recours aux nappes est général et celle du Continental terminal est connue depuis longtemps. Actuellement, ces nappes sont souvent surexploitées, les rabattements sont énormes et la ressource n'est pas renouvelée que ce soit en Tunisie où le biseau salé apparaît dans les nappes littorales du Cap Bon et du Sahel de Sousse, en Algérie où le même phénomène s'observe avec la nappe de Mitidja ou même au Maroc où la nappe de la plaine du Sous enregistre de très fort rabattements en raison de surpompages.

## 1.2 Une forte diversité régionale

Il faut tenir compte des spécificités géographiques des trois pays.

**1.2.1** On voit d'après ces chiffres le **net avantage du Maroc**. C'est le pays qui, avec 29 milliards de m3, dispose de la ressource la plus abondante (906 m3/an/hab). Il le doit à l'importance des précipitations qu'il reçoit de l'Atlantique. L'organisation en amphithéâtre du relief permet l'organisation d'un dispositif à la californienne. Les hautes montagnes de l'Atlas capitalisent les précipitations parfois sous la forme neigeuse et, en avant, se déploient plaines et plateaux parcourus par des oueds disposés régulièrement : ce sont là des conditions propices à des aménagement hydro-agricoles dont le pays a su tirer parti. Plus de la moitié est régularisable (16,5 milliards de m3) mais les perspectives d'avenir sont plus incertaines car le volume des prélèvements actuels (12,5 milliards de m3) ne laisse pas une grande marge de manœuvre : le pays prélève déjà les trois quarts des eaux régularisables.

**1.2.2** En **Algérie** nettement moins bien pourvue (potentiel de 12 milliards de m3 soit 339 m3/an/hab), on peut relever l'importance prise par les ressources souterraines. En fait l'essentiel des eaux de surface provient de petits bassins fluviaux, au débit peu abondant, tributaires de la Méditerranée. Les eaux souterraines proviennent de nappes phréatiques des bassins et vallées du Tell mais surtout des nappes plus profondes notamment celle du Continental terminal au Sahara. Deux traits caractérisent le potentiel algérien. Tout d'abord la très forte disparité entre l'Ouest, une région riche en plaines et bassins mais faiblement arrosée et l'Est montagneux où s'écoulent les principaux oueds du pays : le Rhummel (910 millions de m3), la Soummam (700), les petits fleuves côtiers constantinois (3 250) et l'Isser (520). Seul le Chelif (1 540 millions de m3) présente un débit notable dans l'Ouest. Ainsi, en Algérie autant l'eau manque cruellement dans l'Ouest autant, elle est souvent perdue pour toute utilisation dans l'Est. Autre trait marquant de l'hydraulique algérienne : la relative faiblesse des prélèvements par rapport aux volumes régularisables : c'est la traduction du retard pris par le pays dans la mobilisation de ses ressources.

**1.2.3** La **Tunisie** n'est pas mieux lotie. Son potentiel, est faible : 4,6 milliards de m3 soit 442 m3/an/habitant est déjà très sollicité. Le pays a fait un très gros effort pour régulariser toutes les ressources qui pouvaient l'être et les prélèvements représentent actuellement 88% du volume régularisable! Autant dire que pour l'avenir la marge dont dispose le pays est très étroite. En Tunisie, les ressources provenant des eaux souterraines sont équivalentes à celles fournies par les eaux de surface. 80% des eaux de surface se situent dans la frange montagnarde du Nord-Ouest du pays alors que 91% des besoins se concentrent dans la bande littorale du pays où se localisent villes, zones industrielles, activités touristiques. Des transferts importants sont effectués pour pallier cette situation à partir des ressources fournies par l'oued Medjerda (1 000 millions de m3), des oueds côtiers du Nord (550) ou de l'Ichkeul (265). Les nappes phréatiques, nombreuses dans le Nord et le Centre du pays sont déjà surexploitées : on extrait désormais chaque année 700 millions de m3 alors que les réserves renouvelables sont de 670 millions de m3. Les réserves en eaux profondes, essentiellement celles du Continental intercalaire, sont localisées dans le Sud. Elles sont moins sollicitées que les nappes du Nord et du Centre (59% du potentiel est mobilisé) mais l'exhaure requiert des techniques de forage complexes et coûteuses.

Ainsi les trois pays disposent de ressources en eau d'inégale importance et surtout à des niveaux d'accessibilité très variables : faiblesse du potentiel, irrégularité intersaisonnière ou interannuelle, mauvaise répartition régionale, coûts élevés de l'exhaure des nappes profondes, tels sont les principaux handicaps que doit surmonter le Maghreb. Or, la situation, déjà tendue, ira en s'aggravant en raison de la progression de la demande. Les pays maghrébins soucieux d'assurer leur sécurité alimentaire sont conduits à multiplier les superficies irriguées tandis que la consommation urbaine augmente à des rythmes impressionnants.

# 2. Priorité à la grande hydraulique

## 2.1 Le Maghreb face au défi alimentaire

Ces trente dernières années la demande alimentaire a augmenté dans des proportions considérables. C'est, en premier lieu, la conséquence de l'accroissement démographique lui-même qui, depuis 1960, est, en valeur moyenne, de l'ordre de 2,8 à 3% l'an soit un doublement des effectifs en 25 ans. L'amélioration du niveau de vie a considérablement amplifié les effets de la croissance démographique. A titre indicatif le PNB/an/habitant des pays maghrébins était de 3 à 400 $ au lendemain des Indépendances, il oscille aujourd'hui selon les pays entre 1 000 et 1 800. Dans ces sociétés, au faible niveau de vie initial, les dépenses alimentaires représentent de 40 à 50% des dépenses budgétaires des ménages voire davantage. Toute augmentation des revenus se traduit par une amélioration de la ration alimentaire. C'est ainsi, qu'en moyenne, pour l'ensemble de la population concernée la ration journalière était de 2 200 calories, elle est passée à 3 000 de nos jours. On comprend l'effet cumulatif de tous ces facteurs. **La demande de biens alimentaires croît ainsi à un rythme encore beaucoup plus soutenu que celui, déjà notable, de la démographie.**

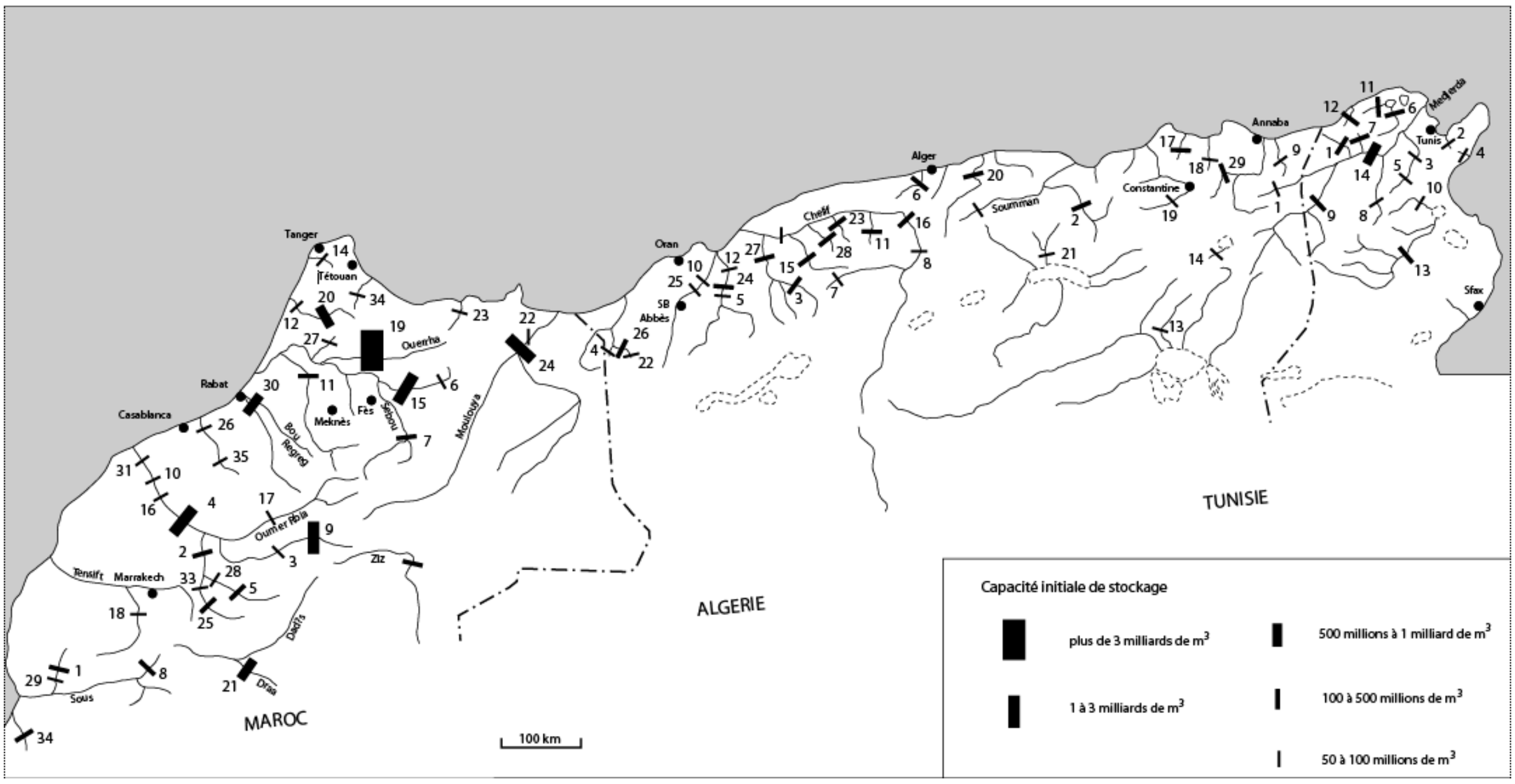

*Figure 20 – Les grands barrages au Maghreb*

| Maroc | Algérie | Tunisie |
|---|---|---|
| 1.Abdelmoumen | 1 Aïn Delia | 1.Ben Metir |
| 2 Agadir bou Achiba | 2 Aïn Zada | 2.Bezirk |
| 3 Aït Ouarda | 3 Bakhada | 3.Bir |
| 4 Al Massira | 4 Beni Bahdel | 4.Chiba |
| 5 Aït Chouarit | 5 Bou Hanifia | 5.El Kebir |
| 6 Aït Youb | 6 Bou Roumi | 6.Jouinine |
| 7 Allal el Fassi | 7 Bougara | 7.Kasseb |
| 8 Aoukouz | 8 Boughzoul | 8.Lakhmes |
| 9 Bine el Ouidane | 9 Cheffia | 9.Mellegue |
| 10 Daourat | 10 Cheurfas | 10 Nebhana |
| 11 El Kansera | 11 Deurdeur | 11 Sedjenane |
| 12 Garde du Loukkos | 12 Fergoug | 12 Sidi Barak |
| 13 Hassan Addakhil | 13 Foum el Gherza | 13 Sidi Saad |
| 14 Ibn Battouta | 14 Foum el Gueiss | 14 Sidi Salem |
| 15 Idriss 1er | 15 Gargar | |
| 16 Imfout | 16 Ghrib | |
| 17 Kasba Tadla | 17 Guenitra | |
| 18 Lalla Takerkoust | 18 Hammam Debdagh | |
| 19 M'jara | 19 Hammam Grouz | |
| 20 Makhazine | 20 Keddara | |
| 21 Mansour Eddahbi | 21 Ksob | |
| 22 Mechra Homade | 22 Melffrouch | |
| 23 Mohamed Ben Abdelkrim | 23 Oued Fodda | |
| 24 Mohamed V | 24 Ouizert | |
| 25 Moulay Youssef | 25 Sarno | |
| 26 Oued Meilah | 26 Sidi Abdelli | |
| 27 Ouezzane | 27 Sidi Mohamed Ben Aouda | |
| 28 Si Driss | 28 Sly | |
| 29 Sidi Driss | 29 Zardenas | |
| 30 Sidi Mohamed Ben Abdallah | | |
| 31 Sidi Said Maachou | | |
| 32 Thelat | | |
| 33 Timinoutine | | |
| 34 Youssef Ben Tachline | | |
| 35 Zemrane | | |

A cela s'ajoutent les effets de la **croissance urbaine accélérée.** La demande de biens alimentaires commercialisés augmente considérablement. Les nouveaux citadins modifient leur ration alimentaire et consomment des produits nouveaux. Ils délaissent peu à peu les modèles de consommation traditionnels pour adopter les modèles occidentaux : le pain fabriqué au blé tendre supplante les produits à base de blé dur comme le couscous ou la galette; la consommation de fruits, de légumes, d'œufs, de viande, progresse. On mesure donc les défis que doivent affronter les systèmes agricoles. Non seulement la demande augmente de façon impressionnante mais, en même temps, elle se diversifie et s'enrichit.

**La production agricole est loin de pouvoir satisfaire à cette croissance de la demande**. Ainsi, la production des céréales, aliment de base par excellence des Maghrébins est passée de 5,8 millions de tonnes (moyenne 61-65) à 9 millions pour la période 91-95 soit une progression de 1,56 alors que la population durant la même période 1961-1995 a triplé! Tous les États sont obligés, depuis de longues années, de recourir à des importations massives de produits vivriers de base qu'ils ne peuvent pas contrôler. Ils sont entrés en dépendance alimentaire car ce sont des importations dont ils ne peuvent se passer. Les tableaux suivants donnent une idée de la gravité de cette situation :

| Maghreb | tonnages en .000 | | | valeur en millions de | | |
|---|---|---|---|---|---|---|
| | 61/65 | 90/95 | 95/61 | 61/95 | 90/95 | 95/61 |
| céréales | 1088 | 10096 | 9,27 | 76 | 1547 | 20,3 |
| sucre | 668 | 1358 | 2,2 | 93 | 440 | 4,7 |
| huiles veget. | 93 | 573 | 6,1 | 26 | 272 | 10,4 |
| lait en poudre | 10 | 193 | 19,3 | 8 | 353 | 44 |
| Total | 1859 | 12220 | 6,57 | 203 | 2612 | 12,8 |

*Tableau 19 : Evolution 1961/95 des importations maghrébines pour 4 produits de base*

Les tonnages de produits alimentaires importés ont été multipliés par 6 en 35 ans et leur coût par 12! La situation est particulièrement sensible pour les céréales : les tonnages importés ont été multipliés par 9 et leur coût a augmenté de plus de 20 fois! Désormais le Maghreb importe plus de céréales qu'il n'en produit. On devine les conséquences de ces situations : les balances commerciales sont déséquilibrées et les produits alimentaires représentent selon les pays de 10 à 30 % des importations totales. Des trois pays c'est l'Algérie qui connaît les formes les plus sévères de dépendance puisque la pays importe actuellement 70% de son alimentation de base!

| millions de $ | 1970 | 1991/95 | 95/70 | importations 1995 | % imp alimentair |
|---|---|---|---|---|---|
| Algérie | 159 | 2758 | 17,3 | 9048 | 30,4 |
| Maroc | 141 | 1244 | 8,8 | 7337 | 16,9 |
| Tunisie | 85 | 679 | 8 | 6453 | 10,5 |
| Total | 385 | 4681 | 12,1 | 22838 | 20,4 |

*Tableau 20 : Évolution des importations totales et alimentaires au Maghreb 1970/95*

On comprend aisément la gravité de cette situation d'autant plus que toutes les prévisions concluent à une augmentation considérable de la demande de ces produits de base. Les perspectives admises donnent le vertige. Comment faire face à une demande de céréales qui passerait, en Algérie, de 4,4 millions de tonnes en 1985 à 7,1 en l'an 2 000 et 11,5 en 2025, à une demande de lait qui, au Maroc, passerait de 769 000 à 1 184 000 puis 1 951 000 tonnes, à une demande de sucre, en Tunisie, qui passerait de 186 000 tonnes à 251 000 et 351 000 tonnes ? C'est un véritable défi.

Faute de parvenir à une indépendance alimentaire irréalisable, les pays du Maghreb cherchent à **assurer une certaine sécurité alimentaire** et ont pour objectif d'améliorer leur production. Toutefois les moyens pour y parvenir sont fort limités. La conquête de nouvelles terres est impossible et trop de terres marginales sont déjà mises en culture. Bien plus, les terres cultivées ont tendance à diminuer devant la progression de l'érosion, de la désertification, d'une urbanisation et d'une industrialisation mal maîtrisées. Avec la croissance démographique, la terre arable disponible par habitant décroît telle une peau de chagrin.

| Hectare | Maroc | Algérie | Tunisie |
|---|---|---|---|
| 1967 | 0,51 | 0,52 | 0,88 |
| 1978 | 0,40 | 0,42 | 0,81 |
| 1990 | 0,28 | 0,27 | 0,56 |
| 2000 | 0,22 | 0,21 | 0,46 |

*Tableau 21 : Évolution de la terre cultivable par habitant dans les pays maghrébins 1967/2000*

Dans ce contexte de rareté grandissante de la terre et devant l'impossibilité de conquérir de nouveaux terroirs il n'existe qu'une solution : **l'intensification**. Pour tous les responsables, l'irrigation en est la voie royale. Par rapport à la culture en sec, l'irrigation améliore le produit brut à l'hectare dans des proportions considérables, permet l'augmentation des rendements, autorise l'introduction de nouvelles variétés, offre plus d'emplois à l'hectare ce qui peut être appréciable dans des campagnes où les paysans n'ont jamais été aussi nombreux en dépit de l'important exode rural qui s'est partout manifesté. Les avantages paraissent incontestables mais, en fait, le bilan, on le verra, mérite d'être nuancé d'autant plus que le seule voie retenue a été celle de la grande hydraulique dans laquelle se sont engagés tous les États des pays concernés. Ce choix technique repose sur l'équipement des cours d'eau par des barrages-réservoirs et la création de périmètres irrigués en aval. Il est pratiqué par l'État alors que la mobilisation de l'eau à partir des nappes phréatiques de faible profondeur est laissée le plus souvent à l'initiative privée.

En ce domaine les États indépendants ont d’une certaine façon reconduit la politique coloniale. A partir des années 20 la colonisation en Algérie d’abord, au Maroc ensuite et en Tunisie beaucoup plus tardivement (au cours des années 50) opte pour la construction de grands barrages-réservoirs. A la fin de la période les réalisations restent relativement limitées. L’Algérie devance les autres pays : 14 barrages ont été construits pouvant stocker 487 millions de m3 mais, en raison de l’irrégularité des débits, les volumes régularisables sont bien inférieurs : 250 millions de m3. Les superficies effectivement régularisées par la grande hydraulique n’excèdent pas 50 000 ha et le total des terres irriguées tous modes confondus 165 000! Au Maroc, 14 barrages sont construits mais seuls quatre d’entre eux comptent réellement en permettant le stockage de 1 900 millions de m3. La progression des superficies réellement irriguées est plus faible et ne porte que sur 38 000 ha en 1956. En Tunisie les réalisations sont encore plus tardives : 4 ouvrages construits après 1950 permettent de régulariser 175 millions de m3 et, en 1956, les périmètres sont à peine équipés.

## 2.2 Des réalisations récentes très contrastées

Actuellement, au Maghreb, **plus de 140 barrages offrent une capacité de stockage de l’ordre de 18 milliards de m3 et une régularisation de 8 milliards.**

**2.2.1** Incontestablement en la matière **le Maroc joue un rôle pionnier**. En 1974, à défaut de réaliser la réforme agraire, le Roi du Maroc lance le slogan du million de nouveaux hectares irrigués pour la fin du siècle. Programme ambitieux car l'objectif, compte tenu des superficies déjà irriguées en 1974, devait conduire à 1 150 000 hectares irrigués de façon pérenne (850 000 en grands périmètres et 300 000 en petite et moyenne hydraulique) et 300 000 ha en irrigation temporaire. L'État intervient massivement, priorité est accordée à l'agriculture et les crédits consacrés à l'irrigation représentent plus de la moitié de la somme allouée à l'ensemble du secteur agricole.

Le pays compte 85 barrages qui ont une capacité de stockage totale de l’ordre de 13 milliards de m3. En réalité il y a beaucoup d’ouvrages dont les capacités de stockage sont très faibles. Seuls I7 peuvent retenir plus de 100 millions de m3 et parmi ces derniers 5 sont véritablement de grands ouvrages dont les retenues dépassant le milliard de m3 totalisent les 3/4 des capacités du pays : le barrage de Bine el Ouidane construit dans les dernières années de la colonisation (1 400 millions de m3), le barrage Idriss 1er (1 207 millions de m3), le barrage Oued el Makhazine (1 000 millions de m3), le barrage Al Massira (2 774 millions de m3). Enfin le Maroc vient d’achever la construction d’Al Wahda (ou M'jara) sur un affluent du Sebou, capable de mobiliser 3 800 millions de m3 dont 1 700 régularisables. Les eaux régularisées par cet impressionnant dispositif marocain doivent être de l’ordre de 8 à 9 milliards de m3. Tout n'est pas destiné à l'irrigation, il faut aussi faire face à la demande des villes et des usines mais les champs accaparent pour l’instant la plus grande partie de la dotation. Ces grands barrages permettent d’irriguer environ 550 000 hectares soit une multiplication par 14 des superficies irriguées en grande hydraulique depuis l'Indépendance. Le Maroc compte 9 grands périmètres, certains atteignant près de 100 000 ha, ils sont les plus vastes du Maghreb. Cet impressionnant équipement pourrait faire du Maroc la «Californie du Maghreb» d'autant plus que l'effort va se poursuivre. "Plus une goutte d'eau à la mer", tel est le nouveau slogan. Dans quelques années le harnachement de la totalité des cours d'eau sera achevé.

**2.2.2** La **Tunisie** a opté aussi pour la politique des **grands barrages réservoirs**. Les premières décisions ponctuelles sont prises dans les années 60 et le dispositif d’ensemble est adopté au début des années 1970. Cette politique hydraulique est présentée comme une des grandes réalisations de la Tunisie indépendante d’autant plus, on l’a vu, que la construction des barrages s’accompagne de la construction de canaux et conduites pour transférer l’eau du Nord-Ouest fournisseur au littoral consommateur. Aux 4 barrages hérités de la colonisation, la Tunisie indépendante en ajoute 18 entre 1957 et 1995. La décennie 1980 est déterminante : au cours de cette période sont achevés des barrages plus importants (en 1982, Sidi Saad avec 210 millions de m3, et Sidi Sallem, avec 550 millions de m3; en 1983, Joumine avec 107 millions de m3) ainsi que le grand canal Medjerda-Cap Bon, long de 125 km terminé en 1984. Le total des capacités de stockage est de 2 100 millions de m3 dont 1 424 régularisables soit 83% des eaux de surfaces régularisables dans le pays. On mesure ainsi la faiblesse du potentiel mobilisable sans aucune comparaison possible avec le Maroc. Cet effort soutenu du pays a permis de faire progresser de façon significative les superficies irriguées en grande hydraulique : on est passé des quelque 15 000 ha effectivement irrigués au moment de l’Indépendance à 220 000 au milieu de la décennie 1990 : une multiplication par 15, comparable à la progression marocaine.

**2.2.3 L’Algérie apparaît en très net décalage par rapport à ses deux voisins.** Le pays a accumulé des retards. L’héritage colonial en matière de grande hydraulique était plus notable qu’au Maroc et en Tunisie. L’objectif du développement de la grande hydraulique a bien été affirmé comme dans les pays voisins. De nombreuses études ont été lancées mais les réalisations ont été peu nombreuses. De 1962 à 1980 seuls trois barrages nouveaux ont été construits (la Cheffia en 1965, Djorf Torba en 1969, Sidi Mohamed ben Aouda en 1970). Quelque 20 000 hectares nouveaux ont été mis sous irrigation mais les pertes de terres agricoles ont été plus importantes et en 1980 les superficies irriguées (49 000 hectares) étaient inférieures à celles de 1962! L'Algérie depuis a entrepris un vigoureux effort pour tenter de rattraper le retard. 15 barrages ont été construits depuis 1980; désormais, chaque année 1 ou 2 barrages sont lancés. En 2010, le pays compte 57 barrages en fonctionnement : ce sont des barrages de petite taille 18 seulement de plus de 100 millions de m3 de capacité; aucun pour l'instant ne dépasse un volume de 300 millions de m3 et parmi les 12 barrages en construction le plus grand atteindra 700 millions de m3 seulement. La retenue totale est de 5 200 millions de m3, mais les volumes régularisés sont beaucoup plus faibles (autour de 2 500). On est très loin des résultats marocains. Les retombées de cet effort de construction se font attendre avec un certain décalage sur les superficies irriguées. Actuellement les périmètres ne comptent pas plus de 75 000 hectares sous irrigation. Les superficies irriguées en grande hydraulique n'ont augmenté que de 47% depuis l'Indépendance! L'Algérie paie cher 20 ans d'attentisme et un redressement sensible n'est pas pour demain même si, en 2009, 13 nouveaux barrages sont en contruction.

Ces grands barrages, ces “cathédrales-hydrauliques" frappent l'imagination mais ne sont pas la panacée idéale. Fort coûteux pour les budgets nationaux, ils sont d'une rentabilité incertaine. A l'heure des premiers bilans, au Maroc, les résultats obtenus suscitent des appréciations contrastées. Les planificateurs et les ingénieurs qui ont conduit cette ambitieuse politique soulignent le capital d'expérience technique accumulé. En revanche, les économistes et les sociologues sont plus nuancés : pour les uns les gros investissements consacrés par l'État à la mise en valeur des périmètres ont surtout amplifié la stratification sociale dans les campagnes irriguées et accentué la prolétarisation des régions moins favorisées. D'autres soulignent la forte dépendance technologique et financière à laquelle le pays a dû consentir pour obtenir des résultats aussi spectaculaires. Enfin la contribution réelle de l'agriculture irriguée à l'indépendance alimentaire est elle-même contestée :«Les céréales et les légumineuses mis à part, il n'est pas exagéré de dire qu'on produit ce qu'on ne consomme pas et qu'on consomme ce qu'on ne produit pas» écrit un auteur marocain pour qui le résultat majeur de cette politique a été l'intégration accrue du Maroc au marché mondial.

La recherche d'un modèle d'intervention de moindre gabarit avec les retenues collinaires n'a pas pour autant donné les résultats escomptés parce que trop souvent initiée dans le cadre d'intervention étatique : 800 retenues collinaires ont été réalisées en Algérie, plusieurs centaines au Maroc. C'est dans la complémentarité entre une grande hydraulique, essentiellement étatique, d'une part et la petite et moyenne hydraulique d'autre part, où domine l'initiative privée qu'il faut trouver les solutions à venir.

## 2.3 L’essor récent de la moyenne et de la petite hydraulique

**La grande hydraulique** est constituée par les grands ensembles irrigables officiellement structurés en périmètres et dominés par de grands barrages-réservoirs.

**La moyenne hydraulique** concerne les ensembles de moins de 50 ha situés hors périmètres. Ils regroupent soit des zones d’irrigation collective (par exemple les aires d’irrigation en aval de retenues collinaires) soit des groupes d’exploitants privés regroupés en syndicats d’irrigation.

**La petite hydraulique** recouvre les irrigations individuelles effectuées à partir de puits, de pompages dans les oueds ou d’épandages de crues.

=

La grande hydraulique n’est pas la seule forme de mobilisation de l’eau agricole au Maghreb. La petite irrigation traditionnelle a des racines anciennes. Elle doit beaucoup aux techniques andalouses. Animée par des particuliers ou de petites collectivités paysannes, elle permet l’irrigation de parcelles de petites tailles à partir de sources de pompages dans les oueds et des îlots de cultures irriguées jalonnent tous les piémonts des montagnes maghrébines. La Tunisie était peut être le pays où l’éventail de ces techniques traditionnelles était le plus large. **Actuellement petite et moyenne hydraulique modernes apparaissent souvent comme une alternative -parfois encouragée par les autorités -face aux déboires rencontrés dans la grande hydraulique**. Elles ont trouvé des moyens qui expliquent leur expansion : le forage et la motopompe. Elles correspondent à l’appel à d’autres ressources que les cours d’eau. Si le pompage individuel au fil de l’eau est bien représenté, le plus souvent il s’agit d’exploiter des nappes phréatiques et des nappes plus profondes très nombreuses dans les vallées et plaines maghrébines.

Ces **nouvelles techniques d'irrigation ont connu un grand dynamisme avec l'émergence de nouveaux acteurs** : des paysans privés et ou de petits collectifs d'exploitants. Cet individualisme hydraulique que l'on peut opposer à l'encadrement étatique qui caractérise la grande hydraulique correspond bien aux aspirations de nouvelles générations d'agriculteurs. Le phénomène est apparu dans les années 60 mais il est monté en puissance au cours de la décennie 70. L'expansion de ces nouvelles formes d'irrigation correspond à une ouverture commerciale, à l'apparition de nouveaux marchés constitués par les villes en pleine croissance. Les cultures légumières, fruitières et fourragères sont les spéculations sur lesquelles reposent ces nouvelles formes d'agriculture irriguée.

L'importance de la petite et moyenne hydraulique peut se résumer à quelques chiffres. Elle s'étend sur 650 000 hectares et 54% des superficies irriguées au Maroc; 315 000 hectares soit 80% des superficies irriguée en Algérie, 240 000 ha et 52% en Tunisie.

**Ce mouvement pionnier donne un second souffle à l'irrigation : cela est particulièrement vrai en Tunisie et surtout en Algérie**. Il présente d'incontestables avantages par rapport à la grande irrigation : souplesse, rapidité de la mise en place, adaptation au marché, coût de l'hectare irrigué deux fois moins élevé. Cependant, après quelques années, on met nettement en évidence de sérieuses contreparties. Il n'y a aucune gestion globale de la ressource, très peu d'encadrement et partout il y a surexploitation des nappes. Cela est déjà vérifié au niveau national pour la Tunisie et les exemples se multiplient en Algérie et au Maroc : le cas le plus spectaculaire est fourni par le Souss marocain où la généralisation excessive des forages entre 1960 et 1980 a provoqué un rabattement de 20 à 40 mètres de la nappe.

| | 1 | 2 | | 3 | | 4 | | |
|---|---|---|---|---|---|---|---|---|
| en.000 ha | SAU | Superficie | 2/1 % | Grande hyd | 3/2% | Petite et moy. h | 4/2 % | épandages de crues |
| Algérie | 7850 | 390 | 5 | 75 | 19 | 315 | 81 | 45 à 165 |
| Maroc | 9920 | 1200 | 12 | 550 | 46 | 650 | 54 | 165 |
| Tunisie | 4952 | 460 | 9 | 220 | 48 | 240 | 52 | 100 |
| Maghreb | 22722 | 2050 | 9 | 845 | 41 | 1205 | 59 | |

*Tableau 22 : Les superficies irriguées au Maghreb en 1995*

On mesure bien l'importance de l'effort entrepris par les pays maghrébins au cours des trois dernières décennies. Partout l'hydraulique et notamment la grande hydraulique ont accaparé l'essentiel des investissements consacrées à l'agriculture. La progression des superficies irriguées est notable : un doublement en 30 ans : on passe de 1 000 000 d'hectares en 1961 à 2 050 000 en 1995. Les volumes consacrés à l’irrigation sont impressionnants : 84% des prélèvements soit environ 11,3 milliards de m3 pour le Maroc, 2,4 pour la Tunisie et 2,1 pour l’Algérie. Et pourtant la sécurité alimentaire est loin d’être assurée, il faut poursuivre l’effort alors que d’autres utilisateurs exigent désormais des volumes sans cesse croissants : la ville et l'industrie.

# 3. L'eau et la ville : une demande exponentielle?

Jusqu'à ces dernières années, l'irrigation des terres agricoles était un des usages quasi exclusifs de l'eau. La consommation urbaine ne paraissait pas poser de graves problèmes. Les données de la question ont au cours des trois dernières décennies radicalement changé. Pourquoi ?

## 3.1 La diversité de la consommation urbaine

Sous ce terme générique on englobe, en fait, de multiples usages de l'eau. Il faut compter avec la demande des citadins eux mêmes_soit pour la consommation humaine soit pour les besoins collectifs et publics. Cette demande augmente très fortement sous l'effet additionnel de plusieurs facteurs :

•**3.1.1** La **croissance des effectifs citadins** : ces trois dernières décennies on assiste à l'émergence du fait urbain. Les villes ont grandi à des taux moyens annuels de l'ordre de 5%. Désormais plus de la moitié de la population est citadine. Globalement, on peut estimer qu'il y a eu un doublement des effectifs tous les 15 ou 20 ans selon les pays. Il est vraisemblable que la progression continuera à se faire sentir. Le Maghreb des années 1965 comptait 9 millions de citadins, aujourd'hui ils sont 35 millions et, sans doute, ils seront entre 65 et 70 millions en 2025!

•**3.1.2** Il faut également compter avec **l'amélioration du niveau de vie, l’amélioration des conditions sanitaires et la diffusion de modes de vie occidentaux.** Autrefois la consommation par habitant et par jour était faible de l'ordre de 80 ou 100 litres. Elle a augmenté. Au Maroc, la dotation brute par habitant est passée de 80 litres/jour en 1960 à 163 en 1987. La consommation effective par tête serait de 100 litres à laquelle s’ajoute les énormes pertes en réseau : 30, 40% et dans bien des cas davantage. Évidemment ces estimations moyennes cachent de profondes inégalités à l’accès à l’eau : les quartiers aisés disposent de 200 à 300 litres, les quartiers en autoconstruction environ 100 et les bidonvilles doivent se contenter de 10 à 20 litres aux bornes fontaines. Les estimations prospectives qui sont faites sont bien souvent établies sur des consommations théoriques de 250 litres pour les années à venir. Pourtant on est encore loin des normes occidentales : près de 500 litres en France et parfois bien davantage aux États Unis.

•**3.1.3** Enfin **des citadins sont de plus en plus nombreux à être branchés sur réseaux**. Actuellement environ les 3/4 des ménages sont raccordés, les autres s'alimentent auprès de bornes-fontaines.

S'ajoutent aussi à la demande urbaine, les besoins d'eau liés à **l'industrialisation** qui a beaucoup progressé. En Algérie, les industries mises en place depuis un quart de siècle, orientées vers des activités de base, sont de très fortes consommatrices : la sidérurgie d'El Hadjar ou la zone pétrochimique d'Arzew consomment chacune autant qu'une ville moyenne. Au Maroc et surtout en Tunisie, il faut compter avec les consommations liées au développement de **l'activité touristique**. La consommation d'un touriste (par lit occupé) est estimé entre 600 et 900 litres/jour (tous usages confondus)

**La grande diversité de la consommation d'eau**

Selon des estimations de la FAO, une consommation de 15 000 m3 d'eau correspo
besoins de :

100 nomades et 450 têtes de bétail pendant 3 ans
100 familles rurales raccordées au réseau de distribution pendant 4 ans
100 familles urbaines dans un quartier populaire pendant 2 ans
100 clients d'un hôtel de luxe pendant 150 jours

On comprend dès lors l'extraordinaire progression des volumes consommés notamment dans les grandes métropoles. Les besoins annuels de l'agglomération algéroise étaient de 70 millions de m3 en 1970, 25 ans plus tard la consommation a triplé (212 millions de m3) et la demande (qui n'a pas pu être entièrement satisfaite) portait sur 312 millions. Même progression à Tunis : la consommation de la ville s'élevait à 45 millions de m3/an en 1970, 55 en 1981, 150 en 1990, on s'attend à une demande de 235 millions de m3 à la fin de la présente décennie. Au Maroc, le volume de l'eau distribuée dans les villes est passé de 260 millions de m3 en 1972 à 780 en 1992!

On rencontre toujours de très grandes difficultés à répondre à une telle demande. Dans un premier temps la ville cherche à se ravitailler en eau dans son environnement immédiat. Or, dans le Monde Arabe, la juxtaposition de la ville avec un environnement rural de qualité, de riches terroirs irrigués est générale. Cette solution se révèle bien souvent insuffisante et la ville va quérir son eau de plus en plus loin. Ce sont de massifs transferts qui sont mis en place : ils ont tendance à devenir la règle générale. Chaque organisme urbain domine un espace hydraulique dans lequel il puise l'eau qui lui est nécessaire. C'est sur ce mode de fonctionnement que, par exemple, s'est effectuée l'industrialisation des Hautes Plaines algériennes. Les grandes métropoles maghrébines s'alimentent à plus de 200 km : c'est le cas notamment de Casablanca et Tunis.

Mais de plus en plus les concurrences s’aiguisent entre les différents utilisateurs de l’eau. Ce sont des concurrences sauvages qui se déroulent entre secteur agricole, l’industrie et l’activité touristique. L'eau à destination agricole sert à des fins urbaines. Les arbitrages entre ville et campagne sont toujours prononcés en faveur de la ville. Cela se traduit par un recul de l'activité agricole parfois même son déclin dans des terroirs qui sont bien souvent exceptionnels.

## **3.2** La course aux captages

Pour assurer leur ravitaillement en eau, les villes exercent une emprise croissante sur leur environnement. Autour de chaque ville d’une certaine importance existe un “rayon hydraulique" qui détermine la zone d’alimentation de la cité.

)

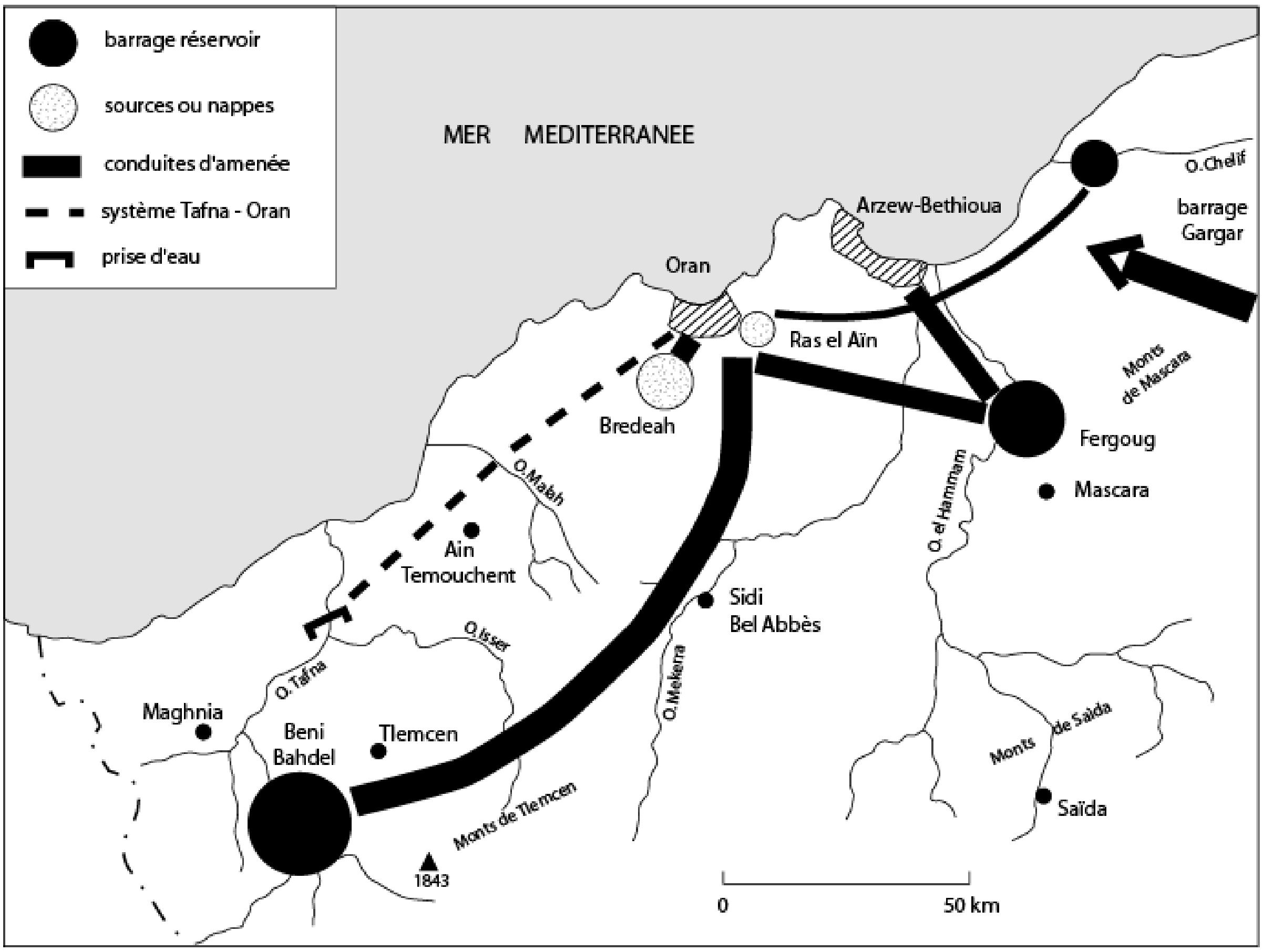

*Figure 21 – L'alimentation en eau d'Oran (d'après Maganiosc, 1991*

**3.2.1** L'**agglomération oranaise** (Maganiosc-Toubache 1991) présente un exemple tout à fait démonstratif de cette quête lointaine de l'eau. La ville est localisée dans la partie la plus sèche du littoral oranais : il ne tombe que 350 mm/an dans la plaine d'Oran où l'effet d'abri exercé par la péninsule Ibérique joue à plein. Plus au sud, les grands bassins telliens sont aussi faiblement arrosés. Dans les monts de Tlemcen, à près d'une centaine de kilomètres et les monts de Saida, l'altitude (1 800m) vient compenser les effets d'abri et les pluies peuvent atteindre 600 mm annuels. Ces montagnes plus humides fournissent l'eau nécessaire à la ville en pleine croissance. L'agglomération de 125 000 habitants en 1914 passe à 330 000 en 1966, 750 000 en 1987 et sans doute 850 000 actuellement. En outre, la périphérie oranaise s'est industrialisée au cours de la décennie 70 avec la création de petites zones industrielles en périphérie, l'aménagement de la grande base pétrochimique d'Arzew-Béthioua (1 500 ha et plus de 20 000 emplois) et, toujours dans l'orbite oranaise, la construction de deux complexes industriels autour de la baie voisine de Mostaganem : une sucrerie et une usine de pâte à papier. Au total, toute la zone oranaise industrialisée doit compter environ 1 000 000 d'habitants (estimations du recensement de 1997). Au cours de la période récente, besoins domestiques et besoins industriels se conjuguent et sont satisfaits de plus en plus difficilement. Les étapes pour assurer le ravitaillement ont été les suivantes :

- une alimentation proche avec la source de Ras el Aïn et des pompages dans la nappe à Bredeah ont satisfait le demande urbaine- de plus en plus mal- jusqu'en 1952. Aujourd'hui ces deux sources fournissent 16 millions de m3/an
- le recours à une alimentation plus lointaine date de 1952 à partir du barrage de Béni Bahdel à 175 km de la ville dans les monts de Tlemcen sur le cours supérieur de la Tafna Le stockage de Béni Bahdel doit à la fois permettre l'irrigation du périmètre irrigué voisin de Maghnia (8 millions de m3) et fournir la ville d'Oran (29 millions de m3). En fait cette alimentation, très sensible aux variations climatiques est assez aléatoire.
- en 1975, il est nécessaire de trouver des compléments : ils sont fournis par l'adduction du Fergoug depuis le barrage du même nom sur l'oued el Hammam qui dessert à la fois la nouvelle zone industrielle d'Arzew Bethioua pour 13 millions m3/an et 9 millions vers la ville d'Oran.
- enfin à l'embouchure du Chelif, le plus long fleuve algérien est mise en place au même moment une prise d'eau susceptible de fournir 15 millions de m3 à l'usine papetière qui vient d'être installée mais, en fait, le fonctionnement en sous capacité de cette unité industrielle permet de libérer 8 millions de m3 à destination de la ville d'Oran.

Cette course incessante aux captages est pourtant loin de satisfaire à la demande. A la fin de la décennie 80, la ville d'Oran alors peuplée de 750 000 habitants disposait de 61 millions de m3 annuels. Pendant toute cette période la dotation par habitant n'a en fait cesser de diminuer. En tenant compte des pertes en réseau la dotation par habitant est de 120 litres par habitant, (elle était de 187 litres en 1966) et la consommation domestique n'est que 80 litres/jour /habitant. Autant dire que la course aux captages se poursuit : on capte les eaux du bassin inférieur de la Tafna à partir d'une prise d'eau depuis 1992 et on fait appel au barrage de Gargar récemment construit sur l'oued Rhiou.

Le schéma oranais se retrouve partout en Algérie dès qu'il s'agit d'alimenter des villes d'une certaine importance. Certes, le rayon hydraulique d'Oran en raison de l'importance de la ville et de sa situation dans une zone faiblement arrosée est double de celui des autres villes. Constantine, la troisième ville du pays va chercher son eau à des dizaines de kilomètres. Il en est de même pour toutes les villes moyennes des Hautes Plaines dont la croissance urbaine doit beaucoup à l'industrialisation de la décennie 70.

**3.2.2** Au **Maroc** l'approvisionnement en eau de la **région casablancaise** (Berrada Sounni 1991) prend l'allure d'une véritable course de vitesse. Actuellement s'étend sur 180 km de Kénitra au nord en passant par Rabat et Casablanca un véritable axe urbain littoral qui se termine à Jorf Lasfar. Dans cette conurbation se concentrent environ 20% de la population marocaine soit 6 millions d’habitants, 40% des citadins (3 millions pour le Grand Casablanca, 700 000 à Rabat, 500 000 à Salé et 300 000 à Kénitra) et 70% des emplois industriels. En 1967, la consommation annuelle d'eau pouvait être évaluée à quelque 120 millions de m3, elle dépasse actuellement 600 millions de m3. Satisfaire une telle croissance de la demande exige une mobilisation constante et de plus en plus en plus lointaine de la ressource. Au cours du dernier demi-siècle, on est passé par les étapes suivantes :

- En 1934, le ravitaillement de la métropole casablancaise en cours de constitution s'effectue grâce aux captages souterrains de la Mamora près de Kénitra acheminés par la conduite de Fouarat qui longe le littoral sur plus de 100 km et fournit 32 millions de m3 annuels. Très rapidement, dès 1946, il faut faire appel à des compléments fournis par l'adduction de l'oued Mellah proche de Casablanca qui fournit 12 millions de m3 supplémentaires. En 1952, c’est le grand fleuve marocain l'Oum er Rbia qui est sollicité : depuis le barrage de Sidi Maachou une adduction fournit dès 1967 64 millions de m3 supplémentaires. Ainsi à cette date, l'agglomération de Casablanca consomme environ 120 millions de m3/an
- Les besoins en eau ne font que grandir et en 1970, un plan Directeur d'alimentation en eau est établi en collaboration avec l’O.M.S. : il concerne toute la zone Kénitra-Casablanca. Il vise essentiellement à utiliser les ressources du fleuve côtier, le Bou Regreg. Le barrage est construit et, en 1976, une nouvelle adduction d'eau fournit 130 millions de m3. En 1983, une deuxième adduction de 160 millions de m3 est mise en service.
- La construction, plus au sud, d'un important complexe de valorisation des phosphates à Jorf Lasfar remet en cause le plan directeur. L'Oum er Rbia est à nouveau mis à contribution et fournit 80 millions de m3. Ainsi en 1990, la consommation totale de la zone urbanisée de Kénitra à Safi peut elle être estimée à 490 millions de m3.

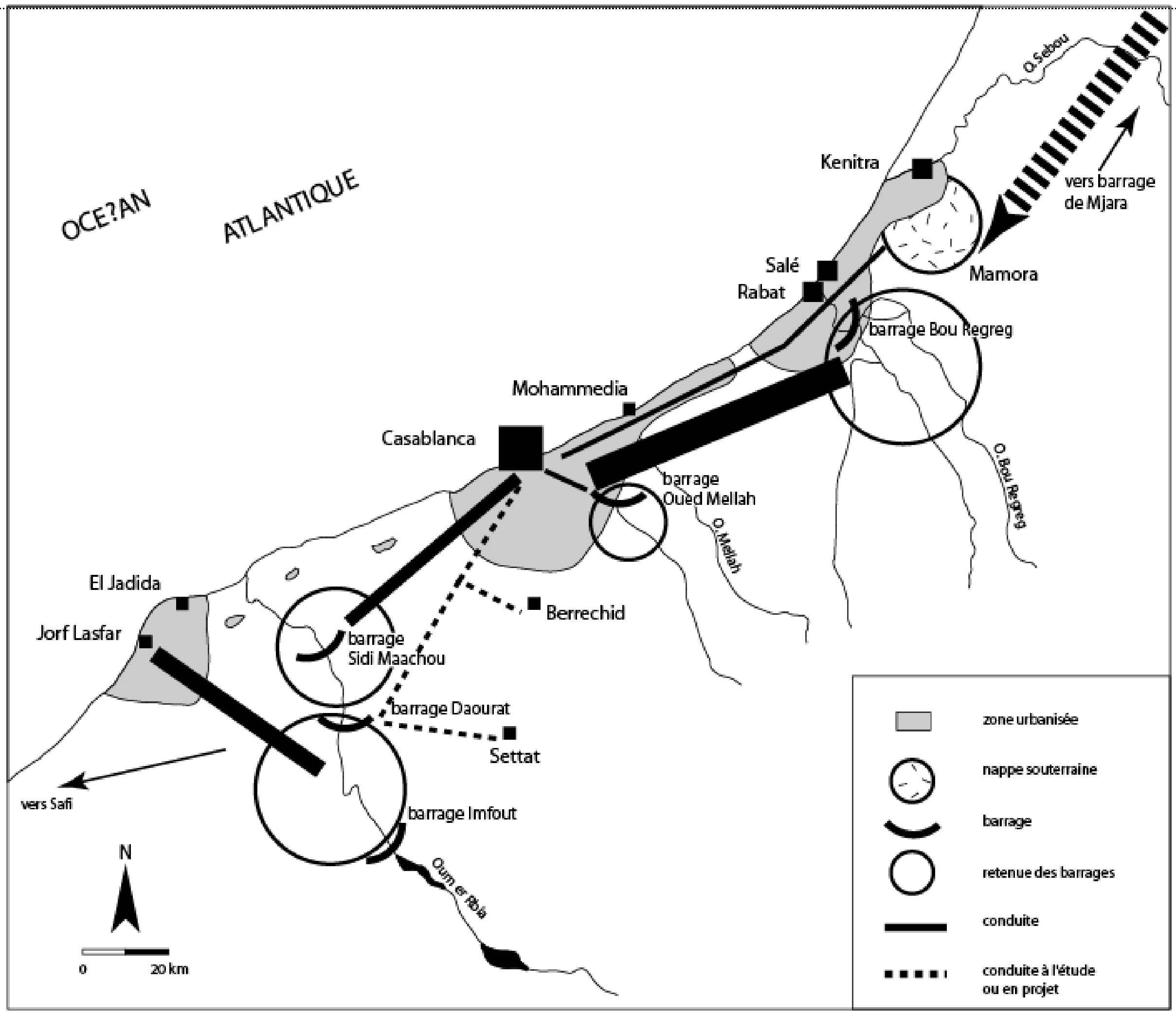

*Figure 22 – L'alimentation en eau de l'axe littoral marocain (d'après Bernada Sounni, 1993)*

- Les besoins sans cesse croissants conduisent à l'élaboration, avec l'aide de la Banque Mondiale, Figured'un nouveau schéma directeur couvrant la fin du siècle. Il repose sur la mobilisation de ressources supplémentaires à partir de l'Oum er Rbia et du Bou Regreg. Au cours de la décennie 90, le barrage de Daourat sur l'Oum er Rbia permettra, en 3 étapes, d'assurer un volume de 190 millions de m3 pour alimenter Safi, Jorf Lasfar, l'agglomération de Casablanca et aussi les villes moyennes de la périphérie (Settat, Berrechid). La première tranche est achevée. Le barrage du Bou Regreg surélevé permettra, par ailleurs, la réalisation d'une nouvelle adduction de 32 millions de m3/an vers Casablanca.

La croissance des besoins est donc impressionnante. Dans les prochaines années, ils vont s'élever à quelque 700 ou 800 millions de m3 et il faut songer à de nouvelles ressources. Une partie de la retenue gigantesque du dernier barrage réalisé celui de M'jara sur l'oued Ouergha dans le pays rifain permettra d'assurer les besoins à venir de l'immense conurbation marocaine. Cet exemple souligne bien l'engrenage dans lequel les autorités sont engagées. Les équipements nouveaux sont saturés en 5 ans. Même dans une région où des ressources disponibles existent des travaux très coûteux sont engagés. La satisfaction de la demande pose des problèmes techniques et financiers redoutables et l'avenir paraît menaçant car la croissance démographique de cette immense conurbation est loin d'être achevée. On ne peut par ailleurs s'empêcher de constater qu'en cette fin de siècle la satisfaction de la seule demande de la conurbation exige environ 5% des eaux régularisables et 7 % des prélèvements effectués dans le pays.

## 3.3 De très sévères concurrences entre secteurs utilisateurs

Les transformations récentes : industrialisation et développement touristique ont conduit à une demande supplémentaire d'eau. Partout au cours des trois dernières décennies des concurrences très vives entre secteurs utilisateurs ont vu le jour. Alors que le Maghreb a fait de gros efforts pour développer les superficies irriguées bien souvent la croissance urbaine, la construction des usines ou de zones touristiques se sont traduits par des consommations accrues satisfaites au détriment du secteur agricole ou de l'environnement. Deux exemples nous permettront de saisir la vigueur de cette concurrence.

**3.3.1** L'oasis de **Gabès** (Hayder 1991) est fort ancienne : on peut encore y visiter un petit barrage de l'époque romaine. Jusqu'à une date récente ses 1 080 ha étaient, non sans mal, sous irrigation. Dans ce Sud tunisien, l'oasis était irriguée par des pompages dans la nappe de la Jeffara (nappe artésienne captive dans les calcaires du Sénonien, qui donne naissance à de nombreuses sources). En 1970, on commence à manquer d'eau, notamment dans le nord de l'oasis. On pompe trop dans la nappe, le débit diminue, la salinité augmente. C'est dans cet environnement fragile que s'installe un **complexe industriel**. Le projet est conçu dans un programme global de développement du Sud notamment pour lutter contre l'émigration. Il aboutit à la réalisation autour d'un port de 100 ha d'un pôle industriel avec deux grandes filières :

- Une filière chimique : relié par une nouvelle voie ferrée au gisement de Gafsa, Gabès valorise la production locale de phosphates et développe des industries d'engrais phosphatés mais aussi d'acide phosphorique à partir du soufre importé.
- Une filière de matériaux de construction (briqueterie, carrelage) et surtout une cimenterie (600 000 tonnes/an de ciment et 100 000 de chaux).

- D'autres industries complémentaires sont nées dans la foulée : une sacherie pour le ciment, des ateliers de wagons etc...et, dans le même temps, l'urbanisation progresse considérablement.

La demande eu eau augmente. Ces industries sont fortes consommatrices, la concurrence s'aiguise de façon fort inégale. On utilise en la surexploitant la nappe locale. En 1974, le débit exhauré est de 51 millions de m3 annuels qui dépasse de très loin les potentialités. Il s'en est suivi une baisse du débit moyen par résurgence, un rabattement de la nappe de 2 à 3 mètres et localement une augmentation de la salinité de l'eau. Face à cette pénurie, on multiplia les sondages dont l'apport devait servir de complément aux sources et forages existants. La situation empira, le débit exhauré baissa considérablement passant de 51 millions de m3/an en 1975 à 42 en 1979. C'est évidemment l'agriculture qui fait les frais de cette politique : elle ne peut couvrir que 70% de ses besoins. Dans l'oasis le tour d'eau, qui était normalement de 15 à 20 jours, passe très vite à 30 et même 50 jours ! La gestion de l'eau se pose en termes nouveaux : d'un droit historique et d'une ressource gratuite, l'eau est devenue ainsi une ressource rare et coûteuse qui ne va pas tarder à devenir payante. C'est donc l'impasse : les ressources locales sont incapables de faire face aux nouveaux besoins.

On songe alors à faire appel à des ressources régionales. L'étude des Ressources en Eau du Sahara septentrional conduite sous l'égide de l'UNESCO a révélé les grandes potentialités des nappes du Sud. Il s'agit de la nappe du Continental intercalaire (albien). Sur ces bases est établi un plan de développement des ressources (Plan Directeur des Eaux du Sud) en eau qui se révèle assez optimiste. Il prévoit à l'horizon 2 000, la répartition suivante des 160 millions de m3/an nécessaires pour la région de Gabès : 107 pour l'agriculture, 60 pour l'industrie et la ville (dont 25 fournie par une usine de dessalement de l'eau de mer). Le projet est bien entré en phase de réalisation mais il ne sera pas poussé très loin car rapidement on s'est aperçu que la nappe dite locale supérieure est partiellement alimentée par le Continental intercalaire et il paraît très imprudent de pousser plus loin son exploitation. Le plan directeur trouve à ce niveau ses véritables limites et l'industrie continue à pomper dans la nappe locale les ressources en eau dont elle a besoin au détriment d'une agriculture vouée à la disparition.

Cet exemple met en évidence la position inégale des acteurs sociaux par rapport à cette rareté et par rapport à la loi. Par le gigantisme des installations et son coût (évalué à un milliard de dinars tunisiens en 1987), par le poids des sociétés en présence (des sociétés étatiques associées à des banques arabes du Golfe qui interviennent dans une région sensible du pays pour valoriser un minerai difficilement exportable) et le poids dans la ville et la région (5 000 emplois directs soit 20% des emplois de la ville auxquels s'ajoutent les emplois induits), ce complexe s'impose à la société locale comme un géant devant lequel il n'est pas facile de faire valoir ses droits. En somme l'équation du problème de l'eau se pose en ces termes : d'un côté une puissance industrielle, financière et politique et de l'autre un milieu rural traditionnel peu soudé, convoitant tous les deux une ressource rare : l'eau de la nappe locale.

**3.3.2** La Tunisie s'est lancée dans un vaste programme d'équipement touristique et attend de cette activité d'importantes recettes en devises. En 1970 on comptait 35 000 lits et 410 000 entrées touristiques et actuellement plus de 150 000 lits et près de 4 millions d'entrées annuelles. Parmi les équipements indispensables pour accompagner cette politique, il faut assurer les besoins en eau qui sont très importants. Un touriste balnéaire est aussi un très grand consommateur d'eau douce : on prévoit environ 900 litres/jour par lit occupé, cette norme incluant toutes les utilisations : piscines, douches, cuisines, arrosage des espaces verts. Cet essor touristique très rapide a entraîné de graves dommages pour les zones agricoles comme l'illustre bien le cas du **Cap Bon.** Dans cette région a été aménagé le plus grand complexe touristique de Tunisie : celui de Nabeul-Hammamet : avec 26 000 lits et 7 millions de nuitées par an, la première station touristique du Monde Arabe, de l'Afrique et du Sud de la Méditerranée.

Avant le boom touristique, autour des années 1970, une étude de la région montrait que les ressources en eau étaient déjà réduites et à peine suffisantes pour faire face aux besoins exigeants d'une riche agrumiculture et des cultures légumières. Le Cap Bon est au Maghreb une région de tradition maraîchère et agrumicole : les Andalous s'y implantèrent en grand nombre il y a des siècles et la colonisation française y développa de grandes propriétés. Le bassin versant est réduit et le pompage dans les nappes constituait le principal mode d'approvisionnement (*delou* à traction animale puis moto-pompe). Avec le haut niveau d'intensification de l'agriculture, la surexploitation des nappes était largement entamée. La nappe locale de Grombalia-Soliman est «un réservoir qui se vide». La proximité de Tunis a favorisé l'essor du maraîchage (16 000 ha) et des agrumes (12 000 ha) et on exploite déjà les ressources d'eau à l'excès au point de faire apparaître la salinisation par appel d'eau de mer dans les nappes. Les besoins en eau sont importants : ils sont chiffrés en 1970 à 130 millions de m3/an sur la base de 4 000 m3/ha pour les agrumes et 5 000 pour le maraîchage. Le transfert d'eau était la seule solution envisageable depuis les oueds du nord de la Tunisie. Effectivement le canal Mejerda-Cap Bon fut dimensionné pour un apport annuel de 163 millions de m3 depuis Sidi Salem dans un premier temps, puis 209 millions de m3 avec l'entrée en service des retenues de Joumine, Sedjenane et Sidi Barrak. Apparemment les besoins en eau d'irrigation du Cap Bon pouvaient être satisfaits.

En fait, l'expansion touristique va aller plus vite et récupérer à son avantage une part importante de l'eau prévue pour l'irrigation. Dans la région de Nabeul-Hammamet, l'eau du Nord est bien arrivée mais elle sert presqu'exclusivement à la consommation urbaine et touristique sans cesse croissante. L'agriculture irriguée est virtuellement morte, on ne parle plus de son sauvetage. Elle est devenue un secteur marginal. Le tourisme et l'urbanisation accélérée ont mis la main sur l'eau des nappes locales et accaparé une grande partie de l'eau transférée. Devant les difficultés de l'agriculture irriguée, la population se détourne progressivement du travail de la terre et s'oriente vers le secteur touristique. Beaucoup de vergers ou de jardins maraîchers sont occupés progressivement par les installations touristiques, le commerce, les routes. Très souvent, ils ne sont plus cultivés et donc transformés en friches sociales en attendant de servir de terrains de construction.

## 3.4 Pénurie et déstabilisation régionale : l'alimentation en eau de la région algéroise

L'étude de la question de l'eau dans la région algéroise présente un double intérêt :

- Tout se cumule pour faire de la région algéroise une très forte consommatrice d'eau : l'expansion urbaine avec la présence de la capitale algérienne, l'existence du plus riche terroir du pays avec la Mitidja dont de nombreux hectares sont sous irrigation, la multiplication récente des implantations industrielles.
- Le problème peut être analysé et étudié dans un cadre régional précis constitué par une chaîne de montagnes arrosée : l'Atlas Blidéen, le synclinal mitidjien et son importante nappe aquifère, la ride anticlinale du Sahel qui limite l'ensemble vers le nord en bordure de la Méditerranée. C'est pour les spécialistes une région hydrologique parfaitement circonscrite dont on connaît bien les potentialités.

Dès 1970, des études précises prévoyaient une très importante progression de la demande. Il faut compter avec une progression de la demande urbaine liée à la progression des effectifs urbains, au plus grand nombre de familles reliées au réseau de distribution et à l’augmentation de la consommation d’eau par habitant. En se fondant sur des estimations (surestimées on le verra) de la zone urbaine de 4 500 000 habitants, une consommation passant de 78 litres quotidiens à 250, les planificateurs avaient fixé un volume de 432 millions de m3 pour l’horizon 2 000 soit une demande urbaine multipliée par 6 en 30 ans (70 millions de m3 ont été consommés en 1970).

Le secteur agricole est un autre gros utilisateur. La Mitidja, le Sahel et le littoral comptent une SAU évaluée à 150 000 hectares. Dans cet ensemble, les cultures irriguées sont fort bien représentées notamment l'arboriculture (agrumes), les cultures légumières, les fourrages verts : 38 000 hectares environ sont sous irrigation dont 20 000 en Mitidja. L'évolution future de l'agriculture ne peut pas s'envisager sans une extension notable de l'irrigation pour répondre aux besoins du grand marché algérois tout proche. Les contraintes pédologiques sont très légères : pratiquement tous les sols peuvent porter des cultures irriguées. De très nombreuses études ont été conduites pour estimer les besoins en eau au cours des prochaines années et leurs conclusions sont très voisines. Elles peuvent se résumer de la façon suivante :

- en 1970, la consommation d'eau agricole est évaluée à 200/210 millions de m3 annuels.
- l'irrigation quasi totale de la plaine réclamerait quelque 800 millions de m3
- les experts pour des raisons évidentes de pénurie retiennent une hypothèse d'irrigation partielle qui nécessiterait une consommation annuelle de l'ordre de 400 millions de m3. C'est, sans aucun doute, une hypothèse minimale qui ne présente aucun caractère excessif si l'on veut bien prendre en compte l'inévitable augmentation de la demande de produits agricoles.

Enfin, il faut compter avec les besoins d'eau industrielle. En 1970, les usines réclamaient 17 millions de m3, à la fin du siècle la demande sera sans doute de 67 millions de m3.

Les prévisions sont impressionnantes. En 1970, les consommations cumulées- urbaine, agricole, industrielle- sont de l'ordre de 300 millions de m3 annuels. A la fin du siècle, la région algéroise aura besoin de volumes triples : quelque 900 millions de m3 : 432 pour la consommation urbaine, 400 pour l'agriculture et 67 pour l'industrie. La seule agglomération algéroise demande : 108 millions de m3 en 1980, 178 en 1985, 250 en 1990 et près de 400 en 2 000. Où trouver les m3 nécessaires ? Face à ces besoins grandissants quelles sont les potentialités?

L'eau n'est pas rare dans la région algéroise. Partout les totaux pluviométriques sont supérieurs à 650 mm, ils dépassent même le mètre sur les versants de l'Atlas Blidéen. Toutefois, dans leur répartition, les précipitations présentent les caractères du climat méditerranéen avec une double irrégularité. Une mauvaise répartition à l'intérieur de l'année agricole : les pluies sont de saison froide avec un maximum très net en décembre - janvier et 80 % tombent d'octobre à mars. L'irrégularité interannuelle est également très accusée : les variations peuvent atteindre le rapport de 1 à 4. Dans ces conditions, la mobilisation et la maîtrise de l'eau de l'eau passent nécessairement par un stockage important.

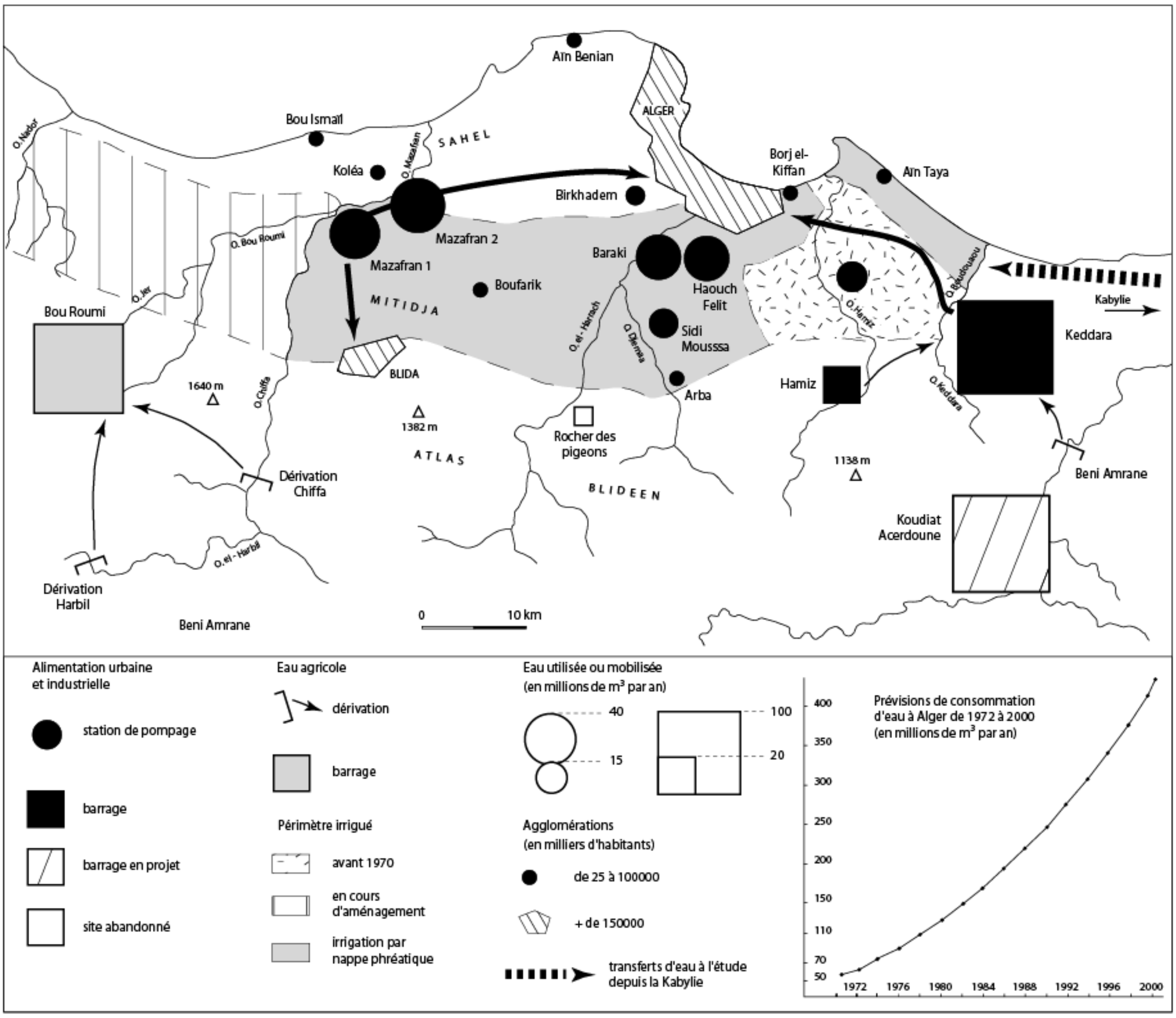

*Figure 23 – L'eau dans la région algéroise*

Le sous-sol de la plaine de la Mitidja constitue un important aquifère. Il est connu et exploité depuis longtemps. Le volume maximum mobilisable est estimé à 295 millions de m3. Vers les années 1970, son utilisation en année normale s'élève à 250 millions de m3 mais, en année sèche, il est nécessaire de puiser dans les réserves en raison du déficit de l'alimentation et de l'augmentation des prélèvements. Autrement dit la nappe est déjà à la limite de la surexploitation

Par contre, les possibilités offertes par les eaux de surface sont loin d'être pleinement utilisées. Quatre fleuves côtiers descendus des pentes de l'Atlas Blidéen traversent la plaine et le Sahel avant de se jeter dans la Méditerranée. Le bilan des écoulements annuels est bien connu. Il s'élève à 837 millions de m3 (Nador : 28, Mazafran : 389, Harrach : 370, Hamiz : 50). En 1970, une part infime de cette masse d'eau est utilisée : les pompages au fil de l'eau et les retenues par petits barrages de dérivation sont dérisoires et ne portent que sur 30 millions de m3. Il est donc possible de mobiliser d'importants volumes d'eau courante

En 1970, la consommation totale annuelle - tous usages confondus- annuelle s'élève à quelque 300 millions de m3. La nappe mitidjienne en fournit l'essentiel avec 254 millions de m3 (85%), le débit pérenne 30 millions (10%), les retenues 16 millions (5%). La plaine est parsemée de puits et de sondages pour l'irrigation. L'eau urbaine provient avant tout de deux forages situés en Mitidja orientale : Baraki et Haouch Felit. Ce sont eux qui assurent la quasi totalité des 200 000 m3 nécessaires à l'alimentation journalière de la capitale (70 millions de m3 annuels). Ainsi, à l'aube des années 70, face à la montée prévisible de la demande, l'incertitude est grande. La nappe est à la limite de son exploitation maximale : on ne peut guère y puiser que 45 millions de m3 supplémentaires sans risque d'une très sérieuse surexploitation. Comment passer en 30 ans d'une mobilisation de 300 millions de m3 à 900?

Des possibilités existent pour parvenir à une mobilisation des eaux satisfaisante. Les opportunités offertes reposent sur la construction rapide de barrages sur les fleuves côtiers.

- Le projet le plus important est constitué par le barrage de Sidi Brahim sur l'oued Bou Roumi à environ 13 km au sud d'El Affroun en Mitidja occidentale. Ce barrage aura une capacité effective de réservoir de 141 millions de m3 et pourra fournir environ 120 millions de m3 (volume régularisable). Le seul débit de l'oued Bou Roumi ne peut suffire à alimenter une telle retenue. Des dérivations de l'oued Jer, de l'oued Chiffa et de l'oued Harbil aménagées en galeries souterraines à travers la montagne seront nécessaires pour atteindre le volume d'eau suffisant.
- Sur l'oued Harrach, près de Hammam Melouane, le barrage du Rocher des Pigeons pourrait retenir 140 millions de m3 et fournir un volume régularisable de 110 millions de m3 annuels.
- Le dévasement du barrage du Hamiz pourrait faire passer son volume régularisable de 13 à 19 millions de m3 annuels.
- Un aménagement complexe dit du Hamiz-Keddara est projeté dans les confins orientaux des piedmonts de la Mitidja. Il comporte deux éléments. La construction du barrage de Keddara qui permettrait de régulariser le débit de l'oued Keddara lui-même. Aux eaux du Keddara (25 millions de m3) s'ajouterait une dérivation provenant de l'oued Hamiz (en amont du barrage existant). L'ensemble permettrait de régulariser 62 millions de m3. En outre, un très important complément serait fourni par l'oued Isser qui coule dans les montagnes de Kabylie. Le barrage de Beni-Amrane permettrait d'effectuer des prises au fil de l'eau qui pourraient atteindre 80 millions de m3 annuels. Au total, l'ensemble Hamiz-Keddara pourrait offrir 142 millions de m3 annuels.
- Enfin, à une échéance sans doute plus lointaine, on prévoit la construction dans la partie amont de l'oued Isser du barrage de Koudiat Acerdoun (140 millions de m3 annuels).

Ainsi, les possibilités sont réelles et ces interventions permettraient d'offrir dans les années à venir les millions de m3 supplémentaires. En 1970, la situation est tendue, voire critique. Tous les rapports d'experts insistent sur la nécessité d'agir très rapidement. Il faut se lancer sans retard dans la construction des barrages. Tout délai en la matière ne pourrait se traduire que par une surexploitation dangereuse de la nappe. C'est précisément la situation dans laquelle se trouve actuellement la région algéroise !

**Les retards de réalisation ont été énormes et, dans le même temps, il a fallu faire face aux besoins grandissants de l'agglomération algéroise. Le ravitaillement en eau de la capitale est devenu une priorité, voire un enjeu**. Et pourtant, pendant toute cette période, aucun barrage n'a été mis en eau, des retards aux lourdes conséquences ont été enregistrés. C'est donc la nappe mitidjienne seule qui a pu assurer les m3 nécessaires à la métropole voisine au cours de la décennie 1970. Les capacités de production des anciens champs de captage de Haouch Felit et de Baraki ont été améliorées mais surtout de nouveaux forages ont été mis en service dans le Bas Mazafran en 1973 et 1979. En 10 ans, les pompages nouveaux portent sur 210 000 m3/jour (77 millions annuels). A ce bilan, il faut ajouter, pour mémoire, un prélèvement de quelque 45 millions de m3 annuels pour permettre le ravitaillement de la ville de Blida. Cela ne suffit pas. La Mitidja orientale est aussi mise à contribution. Deux nouveaux champs sont ouverts : celui du Hamiz fournit 40 000 m3/jour, celui de Sidi Moussa 30 000. Ces 70 000 m3 quotidiens (26 millions/an) permettent tout juste de faire face à l'augmentation de la consommation des premières années de la décennie 80.

C'est donc une véritable course de vitesse qui est engagée. Au cours de ces années la consommation urbaine n'a pu être satisfaite qu'au détriment de l'eau agricole : telle est la conséquence des retards enregistrés dans la construction des barrages. En 1985, on prélève 150 millions de m3/an de plus qu'en 1970 dans la nappe mitidjienne. La surexploitation est désormais largement entamée. Les résultats ne se font pas attendre dans les exploitations. L'eau manque. L'abaissement du niveau de la nappe est général. Il s'est fait sentir dès 1974 et s'est accentué depuis. Le rabattement est particulièrement net en Mitidja centrale où il est toujours supérieur à 5 mètres et atteint parfois 10 mètres. En 1981, 20% des puits sont devenus inutilisables car ils ne sont plus assez profonds. Beaucoup ont une tranche d'eau inférieure à 5 mètres. En été, la situation est particulièrement critique et les vergers manquent cruellement d'eau. Le seul palliatif, pour l'instant, réside dans l'approfondissement des puits existants : il connaîtra très rapidement ses limites.

Et pourtant Alger est bien mal ravitaillé! La ville, dans les années 80, ne dispose que de 150 millions de m3/an. Le paradoxe veut que, dans la capitale, se côtoient pénurie et gaspillage. L'alimentation s'effectue dans de très mauvaises conditions. Les coupures d'eau sont quotidiennes et affectent désormais tous les quartiers. La moindre défaillance technique dans un réseau inadapté entraîne de très longues interruptions. En 1975, 12 000 heures de coupure ont été enregistrées, 14 000 en 1976, 18 000 en 1978. La pénurie s’installe. En 1980, une étude de l’O.M.S. indique que le taux de satisfaction des besoins n’est que de 50%. Le volume moyen consommé réellement par habitant est de 48 litres/habitant/jour alors que la dotation livrée en réseau est de 96 litres/jour. On mesure bien à la fois la faiblesse de la consommation et l’ampleur des pertes en réseau. Elles sont estimées à 50% dont 40 en amont des compteurs des usagers. Un effort vient d'être entrepris pour rénover les équipements, il sera de longue haleine : 20 ans devraient être nécessaires pour renouveler l'ensemble du réseau.

La construction des barrages trop longtemps différée n'est lancée qu'au cours de la décennie 1980. Le système Hamiz - Keddara terminé depuis 1987 devrait fournir en principe 140 millions de m3 supplémentaires à la capitale mais les quantités effectivement livrées sont bien inférieures : 70 millions de m3. Le barrage du Rocher des Pigeons est pour l’instant abandonné : les assises géologiques du site retenu paraissent trop fragiles. Autre déception, la mise en chantier du barrage de Koudiat Acerdoun, sur l'oued Isser n'est pas programmée avant la fin de la décennie 1990.

Sur le site de Bou Roumi, les choses paraissent mieux engagées. La construction du barrage a débuté à la fin des années 1970 et les travaux sont terminés. Dans l'esprit des planificateurs, l'eau du barrage devrait être exclusivement à destination agricole. A cette fin, on doit aménager un périmètre irrigué en Mitidja occidentale : El Moustakbel. Il paraît particulièrement judicieux de prévoir l'extension de l'irrigation dans cette partie de la plaine. C'est ici que le secteur agricole a le moins souffert des concurrences exercées par les villes et l'industrie. C'est dans cette zone que les investissements pourraient être valorisés le plus rationnellement. Les superficies actuellement irriguées sont restreintes (7 000 hectares) ; elles devraient être portées avec l'aménagement du périmètre à 25 000 hectares. Plusieurs phases d'aménagement sont prévues. Dans la première où seules seront disponibles les eaux du Bou Roumi, on équipera 8 000 hectares. Ce n'est que plus tard, quand seront creusées les galeries d'amenée des oueds Jer, Chiffa, Harbil que l'on portera les superficies irriguées à leur maximum prévu : 25 000 hectares.

La situation est de plus en plus tendue dans la capitale. Le manque d'eau ne fait que s’aggraver au cours des années 90. La dotation par habitant diminue : ainsi on passe de 1493 litres par abonné en 1987 à 996 en 1992. La pénurie est particulièrement mal ressentie au cours des années de mauvaise pluviométrie. En 1995, la demande algéroise est évaluée à 320 millions de

m3/an et l'offre n'est que de 210 millions de m3! Dans le court terme on envisage deux solutions pour tenter de pallier la crise :

- On prévoit de massifs transferts depuis la Kabylie. Les ouvrages de Souk el Tleta, Taksebt et Jemaa Tizra devront être réalisés pour livrer 294 millions de m3 supplémentaires.
- Enfin, il est évident que l'affectation agricole de l'eau du Bou Roumi ne sera pas maintenue. Elle ira dans un avenir proche aider au ravitaillement d'Alger.

De l'exemple qui vient d'être analysé peuvent se dégager un certain nombre de conclusions.

- On mesure toute l'acuité du problème de la fourniture d'eau dans une région à croissance démographique et urbaine forte et où progresse l'industrialisation. Faute d'une mobilisation à temps voulu des concurrences sévères se développent. Comme partout ailleurs l'usine et la ville l'emportent sur la campagne. De toute évidence, l'agriculture mitidjienne ne disposera pas de l'eau nécessaire à la reconversion de son système agricole et à l'indispensable intensification. Sera-t-elle seulement assurée, dans quelques années, de conserver les dotations qui étaient les siennes en 1970? On peut très légitimement en douter.
- La surexploitation de la nappe est maintenant bien engagée et, semble-t-il, de façon irréversible. Elle est d'abord la conséquence immédiate des retards énormes pris dans la construction des barrages. Elle se trouve accentuée par l'abandon de certains projets. En ce domaine, les conclusions des hydrogéologues sont formelles, une certaine surexploitation temporaire est possible au prix de très forts rabattements. Mais, en ce cas, le répit offert n'est que de 30 ans. Le modèle de simulation auquel se réfèrent les experts autorise une exploitation temporaire maximale de 315 millions de m3/an. Actuellement, avec la mise en exploitation de tous les nouveaux champs captants, le niveau de 350/360 millions de m3 annuels est atteint : il est bien supérieur au volume acceptable. Le biseau salé pénètre dans la nappe en bordure du littoral à l'est d'Alger.
- Une dernière remarque s'impose. En dépit d'une bonne pluviométrie, la région algéroise est incapable de faire face à ses besoins. Des transferts sont indispensables. De ce point de vue, les solutions algéroises ne sont guère différentes de celles adoptées pour les autres métropoles notamment au Maghreb. Enfin, à terme, il paraît indispensable de dégager d'autres ressources : notamment le recyclage des eaux urbaines et industrielles et le dessalement de l'eau de mer. Le recyclage n'a pas été envisagé concrètement jusque là. Par contre, il existe des projets pour le dessalement de l'eau de mer. Le coût très élevé de cette technologie retarde, pour l'instant sa mise en œuvre.

## A Alger la pénurie, c'est pour les pauvres!

- La politique de rationnement appliquée par les autorités fait apparaître d'extraordinaires inégalités dans la distribution de l'eau et dans la dotation des ménages. Il existe des quartiers, peu nombreux, où l'alimentation est continue : tel est le cas du quartier chic de Hydra où se concentrent belles villas et représentations étrangères, d'autres où l'eau ne coule dans les robinets que la nuit, d'autres où elle n'apparaît qu'un jour sur deux. Le plus souvent l'eau est distribuée à certains moments de la journée. Si certains quartiers disposent de l'eau 18 heures par jour (Garidi, Kouba, Birmandreis), d'autres sont moins favorisés : 6 heures ou moins à Bab el Oued, Casbah, Baraki, Belfort. Cette distribution inégalitaire est aussi intermittente : aucune règle précise n'est appliquée. Il faut prendr quand elle arrive quel que soit le moment de la journée ou de la nuit!

- Les indispensables stratégies de stockage de la part des ménages accentuent s'il en est encore besoin les inégalités. Des villas peuvent être équipées de plusieurs points d'eau alors que dans la Casbah il n'existe qu'un seul robinet collectif dans beaucoup d'immeubles de ce quartier populaire et surpeuplé. Les appartements sont le plus souvent encombrés d'instruments hétéroclites où l'eau est en réserve à moins de disposer de salle d'eau avec baignoire. Les citadins les plus riches se sont équipés de réservoirs placés sur le toit des immeubles ou des villas. Les capacités de stockage peuvent ainsi varier de façon considérable de quelques dizaines de litres par ménage à plus de 200! L'eau à faible pression ne parvient qu'avec difficulté dans les derniers étages des immeubles en hauteur si nombreux à Alger : pour être mieux ravitaillé il faut s'équiper de surpresseurs, si on a les moyens de le faire! La revente de l'eau dans les zones sans réseau que l'on rencontre encore dans certains quartiers de villes comme le Caire est inexistante à Alger.

- L'inégalité de la consommation d'eau est criante. Dans la Casbah, elle est en moyenne d'une vingtaine de litres/habitant/jour. Elle peut dépasser 470 litres/jour/habitant dans les quartiers favorisés... un niveau comparable à celle des citadins des pays industrialisés!

- C'est la femme qui au sein des ménages doit gérer cette pénurie. Les arrivées d'eau sont si fantaisistes que parfois elles peuvent gêner une activité professionnelle. Toute la vie de la ménagère se déroule au rythme de l'eau et si elle arrive la nuit il faut se réveiller. Beaucoup de robinets restent ouverts pour entendre l'annonce, du précieux liquide... ce qui est parfois la cause d'inondations entre voisins!. A Alger où le raccordement au réseau d'eau potable est généralisé, les coupures sont la forme moderne de la corvée d'eau.

- Cette situation est très mal vécue, elle dure depuis des années dans une métropole où la densité par logement est en moyenne de 8 ce qui signifie que les situations où 12 personnes sinon plus s'entassent dans un logement exigu ne sont pas rares et parfois un logement peut être partagé entre deux ou trois familles. On mesure ce que peuvent être des journées sans eau dans une telle promiscuité.

*d'après Chikhr Sai*

La distribution de l’eau dans l’agglomération algéroise est catastrophique et suscite les plus vifs mécontentements de la population. Le gouvernement qui dispose des ressources nécessaires fait appel à des sociétés étrangères. Depuis février 2009 une station de dessalement (200 000 m3/jour) assure une distribution plus satisfaisant de l’eau potable dans l’agglomération.

# 4. Un avenir préoccupant

Les très fortes tensions et concurrences qui viennent d’être analysées vont sans aucun doute prendre une tout autre dimension dans les décennies à venir. Les ressources en eau renouvelables du Maghreb vont nettement placer les pays maghrébins dans une situation difficile à compter de 2025/2030. Avec une disponibilité de 538 m3/an/habitant le Maghreb se trouvera placé pratiquement au seuil dit de pénurie selon les normes internationales admises (500 m3/an)... et dans le même temps la demande aura considérablement augmenté. Dans de telles conditions, les inévitables périodes de sécheresse qui sont le lot des terres maghrébines risquent de tourner à la catastrophe.

| | 1997 | | 2025 | |
|---|---|---|---|---|
| | population : millions | m3/an/habi | population : millions | m3/an/habi |
| Algérie | 29,8 | 402 | 43 | 278 |
| Maroc | 28,2 | 1028 | 39 | 746 |
| Tunisie | 9,3 | 494 | 12 | 380 |
| Maghreb | 67,3 | 677 | 94 | 485 |

*Tableau 23 : Les disponibilités en eau renouvelable dans les pays du Maghreb en 1997 et 2025*

## 4.1L’inéxorable montée de la demande :déficit à partir de 2010/2015

Les facteurs qui rendent compte de la pression actuelle seront toujours à l’œuvre. Au cours de la décennie 1990, les trois pays ont tenté des prévisions sinon de consommation d’eau, du moins de demande d'eau à l'horizon proche des années 2010/2015. Les données de base à l'établissement de ces prévisions sont partout les mêmes : on se place dans l'hypothèse d'une poursuite de la politique d'irrigation avec donc une augmentation des superficies irriguées et surtout on tient compte de l'évolution probable et même certaine de la population urbaine. Partout différentes hypothèses sont avancées avec un maintien du niveau de consommation urbaine proche de la situation présente (150 litres/jour/habitant) c'est l'hypothèse basse alors qu'avec l'hypothèse haute on se fonde sur une dotation journalière de 250 litres. Pour ne pas trop alourdir la présentation nous nous bornerons à évoquer les hypothèses basses.

En Algérie tellienne (Sahara exclu), l'objectif de 500 000 hectares irrigués pour 2010 nécessitera un volume d'eau agricole de quelque 3,2 milliards de m3. Les besoins urbains et industriels sont évalués pour cette période à 2,2 milliards de m3, donc au total 5,4 milliards de m3 soit une augmentation de 60% de la consommation par rapport au début de la décennie 1990 (3,4 milliards de m3). Satisfaire ces besoins ne paraît pas insurmontable car le volume algérien régularisable est de 8,5 milliards de m3 mais atteindre cet objectif suppose une politique soutenue de mobilisation des eaux et aucun relâchement dans l'effort entrepris depuis les premières années de la décennie 1980.

Au Maroc la poursuite du programme d'irrigation notamment en grande hydraulique et les besoins urbains exigeront la mobilisation de la quasi totalité des ressources actuellement régularisantes : entre 16 et 17 milliards de m3. La marge de manœuvre marocaine en dépit de l'abondance de la ressource est bien plus restreinte qu'en Algérie.

C'est en Tunisie que la situation sera la plus tendue, le niveau de la demande excédera sans doute le volume actuellement mobilisable de 3,4 milliards de m3. Certains auteurs ont même tenté des prévisions pour 2040 d'après lesquelles, la seule satisfaction des besoins urbains absorberait la totalité des ressources du pays!

## 4.2 Une mobilisation accrue et une meilleure gestion du potentiel actuel

Face à cette situation l'effort de mobilisation des eaux doit être poursuivi. Tous les programmes prévoient la construction de nouveaux barrages ou de nouveaux forages. Les inégalités régionales dans la répartition des eaux vont inexorablement se creuser et on devra procéder à de massifs transferts d’eau interrégionaux. Ces transferts sont déjà en place en Tunisie sur une grande échelle : ils reposent sur trois aménagements

- le canal Medjerda-Cap Bon
- le canal Djoumine-Ellil
- le canal El Barak-Sedjane

Le système doit être amélioré, le harnachement du Nord, mieux alimenté, sera complété. L’interconnexion est en cours d’achèvement entre le Nord et le littoral Est du pays jusqu’ à Sfax. Au delà de cette dernière ville, il n’est pas prévu d’interconnecter les régions les plus méridionales du pays. Le Sud restera un système autonome dont les très modestes ressources de surface seront complétées par des prélèvements dans les nappes déjà surexploitées et dans les grands aquifères d’eau fossile.

Le Maroc connaît déjà des transferts d’eau qui restent cependant, dans la majorité des cas, circonscrits dans un cadre régional et portent sur de petites distances. Actuellement 2 milliards de m3 sont ainsi transférés. Dans les années à venir, ce sont des transferts interrégionaux beaucoup plus longs qui devront être mis en place. La balance ressources /consommations entre les différentes régions marocaines à l’horizon 2020 est présentée dans le tableau suivant :

| en millions de m3 | Ressources | Demande | +/- |
|---|---|---|---|
| Région rifaine | 1545 | 1052 | +493 |
| Moulouya | 1670 | 1816 | -146 |
| Sebou | 4768 | 3916 | +852 |
| Bou Regreg | 852 | 902 | -50 |
| Oum Errabia | 4067 | 4869 | -802 |
| Tensift | 1221 | 1630 | -409 |
| Souss | 777 | 1175 | -398 |
| Massaa | 144 | 185 | -41 |
| TOTAL | 15044 | 15545 | |

*Tableau 24 : Balance régionale de l’eau au Maroc à l’horizon 2020*

Dès 2010 d’importants transferts, certains sur plusieurs centaines de kilomètres donc fort coûteux, devront être réalisés selon deux axes : nord-sud et est-ouest. Au total, le volume des eaux transférées sera alors de plus de 4 milliards de m3 soit le 1/4 des ressources totales régularisables du pays.

Des déficits régionaux aussi marqués apparaîtront également en Algérie. Pour alimenter essentiellement les villes des transferts sont déjà organisés vers Constantine, Oran, ils vont s'amplifier et prendre une grande ampleur pour desservir la région algéroise.

**Dans ce contexte, économiser l'eau, mieux utiliser les ressources existantes est un impératif absolu. Il convient de réduire autant que possible les pertes en toute nature que l'on peut observer**. Dans les réseaux urbains ce sont les mêmes constatations qui peuvent être faites : les pertes en réseau sont aussi importantes que les quantités consommées en Algérie, les pertes sont de l'ordre de 40% au Maroc. Dans certains cas, les pertes peuvent s'élever jusqu'à 60% de l'eau distribuée alors que dans un système moderne les pertes ne doivent pas excéder 10%. Les économies faites pourraient être substantielles : ainsi en Tunisie en 1993 on estime que les pertes en réseau se sont élevées à 85 millions de m3!. Des constatations analogues peuvent être faites dans les zones irriguées où d'importantes fuites sont constatées. En Tunisie les bénéfices cumulés que l'on peut attendre de la disparition des pertes en réseau aussi bien dans les villes que les campagnes irriguées sont évalués à 700 millions de m3/an soit le quart des consommations totales du pays en 1990! Ce type de constat est fait depuis longtemps mais les politiques à mettre en œuvre tardent en raison des délais de réalisation et de l'investissement que représentent ces travaux. (Tableau 25)

**Améliorer les techniques d'irrigation peut aussi conduire à de spectaculaires économies**. Il n'y a pas comme en Orient dans la vallée du Nil ou en Mésopotamie, un gaspillage des eaux par excès des charges d'irrigation. Au contraire, la plupart des bilans nationaux ou internationaux s'accordent pour estimer que trop souvent les charges d'irrigation sont insuffisantes. Il n'empêche que l'introduction de techniques plus économes par exemple l'aspersion au lieu de la submersion des parcelles ou le goutte à goutte pourraient conduire à de substantielles économies et à plus d'efficience. Le Maroc a initié un programme en ce domaine : 120 000 hectares en grande hydraulique sont équipés en aspersion. Le goutte à goutte est installé sur 12 000 hectares dans le Souss. On a calculé que ces innovations permettent d'améliorer de 6% l'efficience tout en permettant d'économiser 270 millions de m3 d'eau. De même une meilleure maîtrise du drainage et la réutilisation d'eaux de drainage faiblement salées pourraient permettre des gains importants. Des études tunisiennes montrent que ces technologies pourraient autoriser une réutilisation de 200 millions de m3/an

**Économiser l'eau, c'est aussi mieux contrôler, mieux maîtriser la demande**. Faire payer l'eau à son coût réel pourrait selon tous les experts conduire à une forte réduction de la demande. Les pays maghrébins sortent difficilement d'une conception minière de l'utilisation de l’eau considérée comme un bien naturel, gratuit et inépuisable. Pendant des années la sous tarification voire la gratuité a été la règle. Le prix de l’eau est partout très bas et, surtout, présente de très fortes inégalités entre régions et secteurs d’utilisation. De timides mesures ont été prises récemment afin de rapprocher le prix facturé et le coût réel de l'eau. L'eau potable est devenue plus chère depuis les années 1980. Le prix moyen de l'eau potable en 1992 est de 0,56 $ US le m3 en Tunisie, 0,36 au Maroc et 0,14 en Algérie. La politique tarifaire vise à répartir au mieux les coûts tout en décourageant les gaspillage. Le système de tarification adopté par les Marocains en 1977 est significatif de ce point de vue : un tarif social pour les petits consommateurs (moins de 24 m3/trimestre), un tarif au coût réel pour la tranche 24-60 m3, un tarif pénalisant les gros consommateurs (plus de 60 m3/trimestre), enfin une tarification préférentielle pour favoriser l'industrie. Son application à la ville de Casablanca en 1992 donne les résultats suivants pour les ménages : 0,14 $US, 0,40, 0,59 et 0,36 pour l'industrie. On reste, toutefois, encore très loin du coût réel de l'eau. A Kénitra, par exemple, le mètre cube d'eau revient à 1,70 DH mais est revendu à 0,75 DH pour la première tranche, 1,32 pour la seconde et 1,83 pour la troisième. Les tarifs pratiqués pour l'eau d'irrigation sont encore plus éloignés des coûts réels. Même si l'on exclut l'amortissement des barrages, on est loin de couvrir le coût complet de l'eau qui comprend les dépenses de production, de transport, de maintenance des réseaux etc...Le tarif moyen appliqué au Maroc est de 0,02 $US et 0,05 en Tunisie!

En Algérie après plus d'un quart de siècle de gel du prix de l'eau, les premiers changements apparaissent en 1983. Le prix du m3 d'eau urbaine est fixé alors à un dinar alors que le prix de revient est de 6,35 Da. La dynamique de réajustement des prix n'est réellement enclenchée qu'à partir de 1991 et surtout de 1993 avec la mise en place d'une nouvelle structure tarifaire qui comporte un relèvement du prix de l'eau et s'inspire de la démarche marocaine.

Cette ressource rare qu’est l’eau ne peut donner lieu à gaspillage. A l’ère de la planification par l’offre doit succéder celle de la planification de la demande, de la consommation. Il est possible de beaucoup économiser aussi bien dans les quartiers urbains que dans les campagnes irriguées.

## 4.3 Peut-on compter sur une amélioration de l’offre?

Le traitement des eaux usées pourrait offrir selon de nombreux experts d’importantes possibilités notamment pour l’eau d’irrigation. Tout en préservant l’environnement, le traitement des eaux usées peut constituer une source supplémentaire permanente. En ce domaine les pays maghrébins sont très en retard, or, les opportunités sont indéniables avec la croissance des consommations urbaines que nous avons relevées. On estime que le volume des eaux usées susceptibles d'être traitées sont dans les années 1990 de l'ordre de 600 millions de m3/an en Algérie, 700 millions au Maroc. Pour l'instant seule la Tunisie s'est lancée dans cette voie et retraite environ 180 millions de m3/an.

Une meilleure maîtrise de l'environnement aurait des effets positifs pour assurer la mobilisation des eaux. Au Maghreb, tout se conjugue pour donner aux phénomènes érosifs une ampleur particulière. Les sols fragiles, trop souvent développés sur de fortes pentes ne résistent pas aux agressions climatiques, à l'irrégularité et à la brutalité des pluies d'autant plus que de trop fortes densités ont conduit à une intense déforestation. Les ruissellements sont ravageurs. L'érosion est si intense, les transports solides dans les cours d'eau si massifs qu'ils en affectent même les données hydrologiques. Cela se traduit par des pertes continues de sols arables et par un très important envasement des barrages.

- En Algérie les 37 barrages en fonctionnement en 1990 dont la capacité de stockage initiale était de 3 900 millions de m3 ont déjà perdu 11% de leur capacité initiale (430 millions de m3) et en 2010 l'envasement atteindra 24% (930 millions de m3)! Les barrages les plus anciens portent les marques les plus fortes : le barrage du Fergoug est envasé à 100%, celui du Ksob à 40%, celui du Zardenas à 40%.
- La situation est identique au Maroc. La capacité de retenue de l'ensemble des barrages atteint 13 milliards de m3 mais chaque année s'accumulent 50 millions de m3 de sédiments. En 1990, on évaluait l'envasement total à 900 millions de m3.

On peut procéder à des surélévations pour maintenir le volume initial d'eau stockée mais cette opération rencontre des limites. Préserver efficacement l'environnement, prévenir l'érosion suppose un traitement, un reboisement des bassins versants. Les techniques de Défense et Restauration des Sols (DRS) sont au point depuis plus d'un demi-siècle. De très nombreux programmes ont été lancés, de multiples opérations engagées mais les résultats sont restés très modestes sinon dérisoires. L'adhésion des populations paysannes, toujours plus nombreuses, n'a jamais pu être obtenue, il est pourtant indispensable à la réussite de tels projets. Ainsi, à l'heure où les pays maghrébins parviennent à domestiquer leurs cours d'eau, à multiplier les barrages, ils découvrent que leurs bassins versants, leurs montagnes n'ont jamais été autant dégradés. Il paraît tout à fait illusoire d'attendre des résultats significatifs en ce domaine. Il est probable que la dégradation du patrimoine écologique ira en s'accentuant dans les années à venir.

On a tenté, pour la Tunisie, de faire le bilan des économies d'eau qui pourraient être réalisées en tenant compte des différentes options qui viennent d'être développées.

Tableau 25 : Gains potentiels de récupération des eaux en Tunisie en millions.

| | |
|---|---|
| Traitement des eaux usées | 200 |
| Suppression des pertes en réseau (urbain ou agricole) | 700 |
| Amélioration du drainage | 200 |
| Mesure de protection de l’environnement | 430 |
| Total | 1530 |

Au premier abord les effets semblent spectaculaires. Les gains potentiels en s'élevant à 1 500 millions de m3 représentent la moitié de la consommation actuelle du pays. Mais on voit combien ces chiffres sont illusoires. Ils ne prennent en compte que l'aspect technique de la question, que des possibilités technologiques. En fait il faut surtout évaluer les possibilités d'acceptation des populations, des sociétés. Toutes ces innovations sont en grande partie inapplicables : elles supposent des capacités d'investissement hors de portée des économies ou des paysans maghrébins. Elles supposent aussi de nouvelles pratiques de la part des sociétés paysannes qui ne peuvent pas être introduites du jour au lendemain. Le plus probable est que la demande en eau continuera à augmenter dans de très fortes proportions au cours des décennies à venir et que cette demande ne pourra être satisfaite que très partiellement. Plus que jamais l'heure est à la concurrence entre les différents acteurs sociaux. On devine aisément que la solution adoptée consistera à satisfaire en priorité -toujours aussi mal et peut être de plus en plus mal- la demande citadine au détriment de l'agriculture pour l'instant le plus grand consommateur. Il faudrait adopter de toute urgence des technologies plus économes permettant à la fois de réduire les volumes utilisés et d'améliorer l'efficience des systèmes de culture, faute de quoi la sécurité alimentaire sera de plus en plus aléatoire. Le Maghreb est bien entré dans une ère de pénurie! Plus encore que dans les autres régions du Monde Arabe l'eau, dans les terres maghrébines, est bien un enjeu du développement!

# 5. Faire reverdir le désert?

Forte de ses richesses énergétiques et de ses eaux profondes, l’Algérie confrontée à la dramatique insuffisance de la production alimentaire a lancé au cours des années 80 un programme de mise en valeur des terres désertiques. Le projet avait été conçu d’ailleurs bien auparavant. Dès les années 60, dans des périmètres expérimentaux ou des fermes pilotes on avait tenté d’étendre des formes d’agriculture désertique moderne à côté de l’utilisation traditionnelle des eaux. Tout était fondé sur une ressource en eau d’irrigation fournie par des forages profonds, initiés par l’État, dans les deux nappes du Continental terminal et du Continental intercalaire. A vrai dire les choses n’avaient pas été poussées très loin et pour l’essentiel les cultures sahariennes s’effectuaient dans le cadre habituel (cultures traditionnelles d’oasis et petits périmètres hérités de la colonisation). La généralisation de la motopompe avait pourtant introduit un changement notable dans les techniques d’irrigation.

En fait, l’exploitation de l’eau des grands aquifères à grande échelle a été d’abord pratiquée en Arabie saoudite et les expérimentations maghrébines libyenne et algérienne sont postérieures.

## 5.1 Eau et irrigation dans le désert arabique

Il y a peu de temps, l’Arabie était un pays en état de totale dépendance alimentaire. L’afflux des travailleurs étrangers, le mouvement d’exode vers les villes, la très forte augmentation du revenu par habitant ont considérablement gonflé la demande alimentaire. Le secteur agricole traditionnel n’était pas en mesure de répondre à de telles sollicitations. À partir de 1980, une vigoureuse politique a été engagée pour tenter de parvenir à l'autonomie alimentaire. Le développement agricole est devenu un objectif prioritaire. La *Saudi Arabian Agricultural Bank* en a été le principal instrument financier. Cet effort massif a porté sur un certain nombre de productions : blé, riz en toute priorité mais aussi les dattes, poulets et œufs, légumes et produits laitiers. Les résultats sont impressionnants. En 10 ans (1980-90), les superficies agricoles sont passées de 150 000 à près de 2 000 000 ha. La croissance de la production de blé, avec des rendements de pointe de 60 à 80 qx/ha, a été tout aussi spectaculaire : elle est passée en quelques années de 40 000 à 3 300 000 t, soit trois fois la consommation nationale. L’Arabie saoudite est devenue exportatrice de blé depuis 1985 à destination des pays du Golfe mais reste forte importatrice d'autres céréales pour l'alimentation du bétail (6 millions de tonnes en 1991). Elle assure une très grande partie de sa consommation en légumes, fruits, poulets et produits laitiers.

De tels résultats n’ont été possibles que par une très grande concentration de moyens. Quatre zones ont été équipées grâce à l'utilisation de l'eau de nappes profondes : près du Golfe, non loin des champs d’hydrocarbures, dans la plaine du Hasa; à Harad à 80 km à l'est de Riyad; à Hail dans le pourtour du Grand Nefoud; enfin, à Wadi el Dawasir aux confins du Rub el-Khali.

Des espaces agricoles supplémentaires ont été aménagés avec la construction de 84 barrages réservoirs dans les montagnes de l'Asir (en Tihama le long de l'oued Jizan et à l'intérieur près de Najran). Cet essor agricole en plein désert repose sur un apport d'eau continu : 16 heures par jour. L'irrigation se pratique par d'immenses pivots d'aspersion de 400 à 500 mètres de long qui effectuent une rotation complète en 16 ou 18 heures. Le survol en avion donne à voir le spectacle splendide de quelque 3 000 pastilles d'irrigation verdoyant sous le soleil du désert. Cette agriculture a pour cadre de grandes exploitations de plusieurs centaines d'hectares. La mise en place d'une infrastructure impressionnante a été nécessaire : routes, aéroports, systèmes d'irrigation, stations de pompage. Enfin, les Saoudiens ont recours à un matériel considérable et font grand usage d'engrais et pesticides.

Cette expérimentation à très grande échelle pose toutefois des problèmes. Elle est fondée sur une technologie raffinée et entièrement importée. La main-d'oeuvre agricole est elle-même étrangère pour plus de 80%. Les investissements consentis par l'État sont considérables et le prix de revient est très élevé : six fois le cours mondial pour le blé. Dans ces conditions, les aides consenties par l'État sont considérables. En 1992, les subventions de toute nature pour soutenir la production (subventions directes, prix réduit du carburant, achats à des prix garantis) se sont élevées à 2,5 milliards de $. Mais, signe des temps, en 1996, les subventions ont baissé et les autorités ne devraient acheter que 2 millions de tonnes de blé, plafond de l'autosuffisance permettant la satisfaction de la consommation nationale et la constitution de stocks stratégiques. Cette agriculture est le type même d'une agriculture «minière». Elle repose sur le recours massif à l'eau d'irrigation fossile, dont les nappes ne se renouvellent plus. Ce sont des milliards de mètres cubes d'eau qui sont consommés annuellement (2,4 milliards de m3 en 1980, 8 en 1986, 15 en 1992). En 1992, l'Arabie Saoudite a extrait de son sous-sol, 35 fois plus d'eau que de pétrole. À ce rythme, des risques d'épuisement sont réels et les dégâts provoqués à l'environnement deviennent d'année en année plus préoccupants. Le surpompage entraîne déjà un rabattement et une salinisation des eaux. La pollution, liée à la forte utilisation des engrais et pesticides devient notable.

A l'évidence, la spectaculaire expérience saoudienne n'est possible qu'en raison de l'abondance des pétrodollars. Le modèle est difficilement exportable et pourtant d'autres pays moins bien dotés ont voulu l'imiter.

Le modèle fondé sur l'exploitation d'eaux fossiles ne peut sans doute pas perdurer et , en tout cas, le recours à ces eaux fossiles ne peut pas prendre plus d'ampleur. Pour autoriser une sécurité alimentaire à une population en pleine croissance (35 millions d'habitants en 2025, 50 en 2050), le royaume engage une nouvelle politique : le contrôle de terres agricoles soit par la location soit par l'achat en dehors de la Péninsule. L'Arabie vient d'acquérir ainsi 1 600 000 hectares en Indonésie !

## 5.2 Les tentatives maghrébines

**5.2.1** La **Libye** (carte 6) s'est engagée à peu près au même moment dans un projet de même inspiration : 50 000 hectares sont irrigués en plein désert à Sarir à 600 km au sud de Benghazi, 100 000 ont été initiés à Koufra pour la culture du bersim et l'élevage du mouton. Les premiers résultats ont été obtenus en 1980. Les moyens mobilisés sont de même nature qu'en Arabie saoudite : gestion minière de l'eau, recours à la technologie étrangère, investissements considérables et le prix de revient du blé produit est comme en Arabie bien plus élevé que le cours mondial. Il semble toutefois que les performances soient inférieures aux résultats obtenus en Arabie. On enregistre une certaine déception : les possibilités de cette culture en plein désert sont limitées. La position de ces zones à 1 800 km du littoral où se concentre la population, un environnement hostile (chaleur, sable) rendent difficiles la maîtrise de la production et surtout la fixation d'une population agricole : entreprises étatiques ou fermes individuelles n'attirent guère les Libyens. A ce projet actuellement plus ou moins abandonné s'est substitué le grand chantier de la Rivière artificielle qui consiste à transporter l'eau où se trouvent, sur les plains littorale, les bonnes terres agricoles et les hommes. *(carte 6).*

**5.2.2 L'Algérie s'est aussi lancée dans l'agriculture par pivot en zone désertique.** Une étude avait déjà été conduite dans les années 70 à la ferme expérimentale de Gassi Touil près Hassi Messaoud. A partir de 1986, le projet confié aux Américains redémarre : on utilise aussi bien l'eau de la nappe miocène du Continental terminal que celle du Continental intercalaire (albien). Les résultats économiques n'apparaissent pas très concluants mais la mise en valeur est un nouveau slogan lancé par les autorités pour pallier le déficit des terres du Nord. Les terres du Sahara vont devenir la "Californie de l'Algérie". Rien de moins. Des dispositions foncières sont prises permettant de créer une propriété privée sur les terres désertiques (loi de l'accession foncière). C'est le déclic, le point de départ d'un véritable engouement. Deux formes de mises en valeur se juxtaposent désormais. Il existe une phase pionnière paysanne utilisant les ressources en eau peu profonde mais en abandonnant les techniques d'exhaure traditionnelles au profit de la motopompe. Ces nouveaux paysans ne visent plus à l'autosubsistance mais au marché et les paysans ont rompu avec les structures d'encadrement communautaires. Ces nouvelles exploitations se concentrent dans un petit nombre de régions : la région de Biskra, le Souf, l'oued Rhir, le Touat, certaines périphéries urbaines.

La mise en valeur pionnière capitaliste est encore plus récente (décennie 90). Des promoteurs agricoles venus bien souvent des villes investissent, créent des exploitations de plusieurs dizaines voire plusieurs centaines d'hectares, vont chercher l'eau grâce à de profonds forages qui atteignent l'albien et lancent une agriculture à la saoudienne spécialisée dans les céréales irriguées. Il est trop tôt pour arrêter un bilan économique mais il semble, qu'après l'engouement initial, la réalité soit plus amère.

Par contre, déjà un bilan peut être dressé pour l'environnement : ces nouvelles activités ont fait apparaître de sérieuses contraintes. Il existe désormais un important problème de drainage : paradoxalement le désert souffre dans certains cas d'un excès d'eau! Les eaux du Sahara, faute d'un débouché sur la mer, fonctionnent en circuit fermé. Traditionnellement les populations pour leurs besoins domestiques ou agricoles consommaient peu d'eau et évacuaient les eaux usées par l'intermédiaire de drains ou de puits perdus vers le sous-sol alluvial ou sableux. Il y avait auto-épuration naturelle et les eaux rejoignaient la nappe phréatique. Il y avait un système fermé en équilibre avec lui-même. En forant dans les nappes plus profondes d'une part, en utilisant massivement la motopompe hors de la nappe phréatique d'autre part, il y a un apport substantiel d'eau supplémentaire. Ces volumes supplémentaires regagnent après usage la nappe phréatique qui enfle, engorge les terrains, asphyxie les plantes et contribue à la salinisation des terrains inondés. L'urbanisation accélérée qu'a connu le Sahara, qui compte des villes de plus de 100 000 habitants, a encore accru les consommations d'eau et accentué le phénomène. Faute de drainage organisé à grande échelle, des palmeraies ont été transformées en marécages tel a été longtemps le cas de l'oued Rhir dont le drain principal n'a été achevé qu'en 1985. Le Souf connaît une situation inquiétante aussi. El Oued la ville principale est (bien) alimentée par des pompages dans le Continental intercalaire. La nappe phréatique est détruite par les eaux usées urbaines et dans un rayon d'une vingtaine de km les palmeraies sont moribondes.

Avec les expériences maghrébines il faut évidemment évoquer le risque d'épuisement de la nappe sur le long terme faute d'une gestion patrimoniale de la ressource. L'argument est aussi valable pour le transfert libyen, organisé de façon massive pour alimenter le nord du pays. Au Sahara algérien, la situation mérite par contre d'être sérieusement nuancée. Elle l'est d'ailleurs par des études récentes entreprises dans le cadre du “Projet des oasis de l'an 2 000" (Dubost 1992).

En 1990, la situation des deux grands aquifères (continental intercalaire et continental terminal) est présenté de la façon suivante par les hydrogéologues :

| m3/an | Alimentation | Sorties | Destockage |
|---|---|---|---|
| Continental intercalaire | 252 | 724 | 472 |
| Continental terminal | 582 | 897 | 315 |
| Total | 834 | 1621 | 787 |

Pour les scientifiques, les deux aquifères qui communiquent entre eux peuvent être pratiquement considérés comme une monocouche. L'alimentation des aquifères est extrêmement faible : elle est fournie par les infiltrations en provenance des montagnes de l'Atlas saharien. Les sorties sont constituées par les exutoires naturels vers le golfe de Gabès et les prélèvements effectués pour alimenter les villes et l'irrigation. Il y donc un destockage de l'ordre de 800 millions de m3/an qui apparaît minime par rapport à l'ampleur du stock.

Les hydrogéologues estiment, que l'on peut pomper dans les aquifères sans provoquer de graves rabattements : ils avancent le chiffre de 630 millions de m3/an pour le Continental terminal et 790 pour le Continental intercalaire. A ces ressources s'ajoutent 630 millions de m3 apportés par tous les oueds périphériques venus essentiellement de l'Atlas saharien et pour des valeurs très minimes du Hoggar ou du Tassili. Bref, les ressources sur lesquelles on peut raisonnablement compter s'élèvent à un peu plus de 2 milliards de m3.

Sur ce bilan, il faut tenir compte de la consommation urbaine saharienne prévisible (160 millions de m3/an); de l'eau actuellement utilisée pour l'irrigation (1 260 millions de m3 pour 40 000 hectares) si bien que la marge de manœuvre est très restreinte : elle ne porte que sur quelque 600 millions de m3 permettant une mise en valeur de 20 000 hectares supplémentaires.

A ces données s'ajoute la contrainte pédologique : très peu de sols peuvent être irrigués au Sahara et très souvent la localisation de ces sols ne correspond pas aux régions où se trouve l'eau. Ainsi perspectives économiques et médiocrité des potentialités naturelles se conjuguent. Il est très peu probable que l'agriculture saharienne suivent les traces de l'agriculture saoudienne et il est encore plus illusoire de transformer le Sahara en une nouvelle Californie!

# Conclusion

**L'eau dans le Monde Arabe** est bien, en fin de compte, **enjeu du développement futur**. La situation actuelle est lourde de menaces, source d'inquiétude.

La rareté **est réelle même si elle côtoie trop souvent le gaspillage et l'utilisation inconsidérée de la ressource.** Paradoxalement on rencontre dans le Monde arabe, le Bédouin dont la consommation, sans doute la plus faible du Monde, se limite à quelques litres quotidiens et le citadin de la péninsule Arabique qui utilise à profusion une eau fort chère pour l'entretien de ses jardins d'agrément et dont le niveau de consommation dépasse celui du citadin occidental. Il n'empêche la pénurie s'observe partout et, dans un contexte où la mobilisation de ressources nouvelles est incertaine, elle ne fera que s'aggraver avec l'accroissement démographique attendu au cours des prochaines décennies. En 2025, à l'exception de l'Irak, du Soudan, et  du Liban, les peuples arabes se trouveront tous en dessous du seuil de pénurie (1000 m3/an/habitant) et un tiers d'entre eux se situeront en dessous du seuil critique de 500 m3!(tableau 4) Il serait sans doute imprudent de prolonger plus loin l'extrapolation, mais notons toutefois, qu'en 2025 la transition démographique sera loin d'être achevée. Les démographes situent la population stationnaire à un niveau qui devrait être voisin de 700 millions d'habitants au milieu du siècle à venir.

Dans cette conjoncture, le **maintien de la politique actuelle d'irrigation incite à un nouvel examen.** La région consacre plus de 88% des volumes prélevés à l'irrigation des champs, la proportion la plus élevée du Monde. L'objectif d'autosuffisance alimentaire a sous-tendu l'important effort de mobilisation des eaux accompli au cours des dernières décennies. Mais n'est-on pas arrivé au terme d'une évolution? Le constat est là : il reste relativement peu d'eau renouvelable à mobiliser. Les eaux souterraines sont déjà trop souvent sollicitées à l'excès. Augmenter le volume régularisable des eaux courantes ne peut donner que des résultats limités. La ressource ne peut être entièrement mobilisée en raison de l'irrégularité climatique. Il existe donc une limite technique et, financièrement, le coût risque d'être excessif du fait des rendements décroissants. Le mot d'ordre de sécurité alimentaire - jusque là bien mal précisé- se substitue désormais à celui de l'autosuffisance alimentaire jugée partout irréaliste. Par ailleurs, les révisions de priorité dans les allocations d'eau au profit du secteur urbain et au détriment de l'agriculture, déjà constatées s'imposent chaque année avec davantage de force. Dans les pays arabes les contraintes d'accès à l'eau d'irrigation sont telles qu'il est illusoire d'escompter une augmentation des superficies irriguées comparable à celle qui a été constatée ces trois dernières décennies. La production agricole évoluera peu et une part croissante du revenu devra être consacré au financement des importations alimentaires.

La **dépendance des pays arabes ne cessera de s'aggraver**. Elle tient à des données géographiques : la position d'aval de grands pays : l'Égypte, l'Irak, la Syrie regroupant 100 millions d'habitants et 40% de la population de la zone sera, dans les années à venir, de plus en plus inconfortable et incertaine. Les quantités d'eau dont ils disposent n'ont aucune chance d'augmenter, bien au contraire. Par ailleurs, la dépendance économique à l'égard du reste du Monde risque de s'amplifier avec l'augmentation des importations alimentaires difficiles à supporter pour des économies fragiles et faiblement diversifiées. On a calculé que les **importations actuelles de biens alimentaires** correspondaient à une consommation **annuelle d'eau de 50 milliards de m3** que l'on peut donc considérer comme une importation déguisée d'eau. Ces fâcheuses perspectives supposent une reconversion des économies bien difficile à réaliser.

**L'urgence est évidemment d'économiser la ressource** dont les pays arabes dispose et de recourir éventuellement à d'autres sources d'approvisionnement. Les recommandations des experts convergent toutes mais on voit bien mal quel pourrait être leur champ d'application dans le contexte actuel de la société arabe. Le coût du dessalement de l'eau de mer a, certes, baissé mais il reste encore trop élevé pour envisager une utilisation autre qu'urbaine et industrielle. Pour l'essentiel -l'eau affectée à l'irrigation- la lutte contre les énormes gaspillages et pertes diverses ne peut être que de longue haleine mais elle pourrait conduire à de substantielles économies. Le recours aux eaux retraitées reste pour l'instant limité et l'accès à cette ressource est relativement coûteux. Toutes ces propositions, techniquement parfaitement justifiées, supposent une certaine évolution des mentalités toujours longue à s'imposer et surtout d'énormes investissements. Pour la période 1996/2005, la Banque Mondiale a proposé un plan global d'amélioration pour la région arabe concernant aussi bien la lutte contre les gaspillages, le retraitement des eaux, l'amélioration de la desserte et la protection de l'environnement. Il exige un investissement de 60 milliards de $, ce qui est à peu près, l'équivalent du produit national annuel d'un pays comme l'Egypte!

Les milieux économistes, les organismes internationaux souhaitent pour résoudre la crise au niveau mondial réguler l'accès à la ressource par le biais du marché. Au premier Forum mondial de l'eau qui s'est tenu -symboliquement- à Marrakech en mars 1997, spécialistes et représentants des grands organismes internationaux estiment que **l'eau deviendra une marchandise de type nouveau, non plus la ressource illimitée mais une matière première stratégique au même titre que le pétrole**. L'idée d'un prix mondial fait son chemin. Il s'agirait davantage d'organiser l'échange pour favoriser une répartition à peu près équitable que de laisser jouer un mécanisme d'essence libérale. Mais on mesure bien les difficultés de mise en place d'un tel marché. L'objectif, s'il est réalisable, est de long terme et on voit mal comment le Monde Arabe dans ses structures hydrauliques actuelles pourrait y adhérer dans l'immédiat. Or, l'urgence est là, les menaces sont pour demain... et les solutions techniques, économiques ou politiques risquent de rester longtemps en suspens.

Dans la configuration géopolitique actuelle du Monde Arabe et, plus particulièrement, du Moyen-Orient, l'eau est devenue un élément déterminant dans les stratégies des régimes en place. Enjeu économique, l'eau est désormais un enjeu de politique nationale ou internationale. Il existe bien une hydropolitique. Les grands ouvrages réalisés pour la mobilisation de la ressource sont autant de justification pour conforter les équipes aux commandes : le barrage d'Assouan et le lac Nasser en Égypte, Tabqa en Syrie, le troisième fleuve pour l'Irak de Saddam Hussein, le grand fleuve artificiel pour la Libye de Kadhafi. Dans les rapports inter-étatiques, la tentation est forte pour les pays d'amont d'exercer des pressions, voire des chantages sur les pays d'aval. En Palestine, l'eau est un élément fondamental du maintien de la domination israélienne sur la Palestine occupée. Journalistes et hommes politiques rivalisent pour forcer le trait et on évoque, de plus en plus fréquemment, une "guerre de l'eau". Déjà, il y a 20 ans en 1978, dans cette Égypte qui se sent menacée, le président Anouar al-Sadate déclarait "Si quelqu'un fait quelque chose qui puisse nuire à notre approvisionnement, nous n'hésiterons pas à entrer en guerre, car c'est une question de vie ou de mort". Quelques années plus tard le ministre des Affaires Étrangères égyptien, Boutros-Ghali n'hésite pas à déclarer"La prochaine guerre dans notre région concernera l'eau. Pas la politique". Pour les Nations-Unies, “la situation peut entraîner une série de catastrophes locales et régionales et des confrontations pouvant conduire à une crise mondiale. Comme le pétrole, l'eau pourrait devenir motif de guerre ou de paix”. Le Monde Arabe en situation de “stress ou de contrainte hydrique” peut-il affronter efficacement ces menaces. Ce n'est pas sûr. Il est à craindre que le pire soit à venir!

# Bibliographie

## Ouvrages

# Articles

# Parmi les revues